\documentclass[fogscopy,onehalfspacing,11pt]{ubcdiss}

\usepackage{checkend}	
\usepackage{graphicx}	
\usepackage{epsfig}	
\usepackage{amsmath}
\usepackage{MnSymbol}
\usepackage{slashed}
\usepackage{fancyvrb}
\usepackage{fancyhdr}
\usepackage{verbatim}
\usepackage{color}

\usepackage{booktabs}

\usepackage{listings}
\definecolor{dkgreen}{rgb}{0,0.6,0}
\definecolor{gray}{rgb}{0.5,0.5,0.5}
\definecolor{mauve}{rgb}{0.58,0,0.82}

\usepackage[printonlyused]{acronym}

\usepackage{color}
\definecolor{greytext}{gray}{0.5}

\usepackage{comment}

\usepackage[numbers,sort&compress]{natbib}

\usepackage[compact]{titlesec}
\titleformat*{\section}{\singlespacing\raggedright\bfseries\Large}
\titleformat*{\subsection}{\singlespacing\raggedright\bfseries\large}
\titleformat*{\subsubsection}{\singlespacing\raggedright\bfseries}
\titleformat*{\paragraph}{\singlespacing\raggedright\itshape}

\usepackage[format=hang,indention=-1cm,labelfont={bf},margin=1em]{caption}

\usepackage[hyphens]{url}
\usepackage[bookmarks,bookmarksnumbered,%
    citebordercolor={0.8 0.8 0.8},filebordercolor={0.8 0.8 0.8},%
    linkbordercolor={0.8 0.8 0.8},%
    urlbordercolor={0.8 0.8 0.8},%
    pagebackref,linktocpage%
    ]{hyperref}
\usepackage{breakurl}
\newcommand{\ie}{i.e.,\ }	

\DeclareUrlCommand\DOI{}
\newcommand{\doi}[1]{\href{http://dx.doi.org/#1}{\DOI{doi:#1}}}



\def\lesssim{\stackrel{<}{\sim}}

\newcommand{\be}{\begin{equation}}
\newcommand{\ee}{\end{equation}}
\newcommand{\ba}{\begin{eqnarray}}
\newcommand{\ea}{\end{eqnarray}}

\newcommand{\pderiv}[2]{\frac{\partial #1}{\partial #2}}
\newcommand{\create}[1]{a_{\vec{ #1}}^\dagger}
\newcommand{\uncreate}[1]{a_{\vec{ #1}}}
\newcommand{\npm}{n^{\pm}}

\newcommand{\mpm}{m^{\pm}}
\newcommand{\bra}[1]{\langle #1|}
\newcommand{\ket}[1]{|#1\rangle}
\newcommand{\braket}[2]{\langle #1|#2\rangle}
\newcommand{\suminf}[1]{\sum_{#1 = 0}^\infty}

\newcommand{\PlMass}{M_{\rm Pl}}
\newcommand{\trpsq}{1-\rm{Tr}(\rho^2)}
\newcommand{\mode}[2]{g^{#2}(#1,\tau)}
\newcommand{\modedot}[2]{\pderiv{\mode{#1}{#2}}{\tau}}
\renewcommand{\vec}[1]{ {\bf #1}}
\def\rmmat#1{{\hbox{\rm #1}}}
\def\rmscr#1{\rmmat{\scriptsize #1}}
\def\ie{{\em i.e.}\ }
\def\p{\partial}
\def\d{{\rm d}}

\def\pp#1#2{\frac{\p #1}{\p #2}}
\newcommand{\deriv}[2]{\frac{d #1}{d #2}}
\newcommand{\laplace}{{ \nabla}^2}
\newcommand{\ddt}[1]{\frac{\partial ^2 #1}{\partial t^2}}
\newcommand{\ddS}[1]{\frac{d ^2 #1}{d S^2}}
\newcommand{\Jp}{{\bf J}_p}
\newcommand{\del}{\nabla}

\newcommand{\curl}[1]{{ \nabla}\times #1}
\newcommand{\dds}[1]{\frac{{\rm d} #1}{{\rm d}S}}
\def\rmmat#1{{\hbox{\rm #1}}}
\def\rmscr#1{\rmmat{\scriptsize #1}}
\def\ie{{\em i.e.}\ }
\def\pp#1#2{\frac{\p #1}{\p #2}}
\newcommand{\lag}{{\cal L}}
\def\d{{\rm d}}
\newcommand{\bfi}{{\bf B}}
\newcommand{\efi}{{\bf E}}

\newcommand{\Aphi}{A_\phi(\rho)}
\newcommand{\flr}{f_\lambda(\rho)}
\newcommand{\bhat}[1]{\hat{\bf #1}}
\newcommand{\bhattext}[1]{${\bf \hat{#1}}$}
\newcommand{\subscripts}{\omega, k_z,m_l,\sigma^3}
\newcommand{\csubscripts}{\chi,m_l,\sigma^3}
\newcommand{\cylGreen}{G_{\subscripts}(\rho,\rho')}
\newcommand{\BESSO}[1]{#1_{m_l}(\sqrt{\chi^2-m^2}\rho)}
\newcommand{\mean}[1]{\langle#1\rangle}
\newcommand{\sun}{\ensuremath{\odot}}
\newcommand{\Paulispin}{\frac{1}{4} {\rm tr}e^{\frac{1}{2}\int_0^T d\tau \sigma_{\mu \nu}F^{\mu \nu}(x_{\rm CM}+x(\tau))}}

\DefineVerbatimEnvironment{code}{lstlisting}{fontsize=\small}
\DefineVerbatimEnvironment{example}{Verbatim}{fontsize=\small}
\def\bc{\begin{code}}
\def\ec{\end{code}}

\newenvironment{summary}
{
	\begin{center}
    \begin{tabular}[h!]{p{1.0\textwidth}}
    \\\hline
}
{        \\\hline
    \end{tabular}
	\end{center}
}

\title{Nonperturbative Quantum Field Theory in Astrophysics}

\author{Daniel Paul Mazur}
\previousdegree{B. Sc (high hons), University of Regina, 2006}

\degreetitle{Doctor of Philosophy}

\institution{The University Of British Columbia}
\campus{Vancouver}

\faculty{The Faculty of Graduate Studies}
\department{Physics}
\submissionmonth{September}
\submissionyear{2012}

\hypersetup{
  pdftitle={NPQFT in Astrophysics},
  pdfauthor={Daniel Paul Mazur},
  pdfkeywords={Mazur, Dissertation, Quantum field theory, Astrophysics}
}

\begin{document}


\maketitle




\chapter{Abstract}

The extreme electromagnetic or gravitational fields associated with some astrophysical objects 
can give rise to macroscopic effects arising from the physics of the quantum vacuum. Therefore, 
these objects are incredible laboratories for exploring the physics of quantum field 
theories. In this dissertation, we explore this idea in three astrophysical scenarios.

In the early universe, quantum fluctuations of a scalar field result in the generation 
of particles, and of the density fluctuations which seed the large-scale structure 
of the universe. These fluctuations are generated through quantum processes, but 
are ultimately treated classically. We explore how a quantum-to-classical transition 
may occur due to non-linear self-interactions of the scalar field. This mechanism is 
found to be too inefficient to explain classicality, meaning fields which do not 
become classical because of other mechanisms may maintain some evidence of their 
quantum origins.

Magnetars are characterized by intense magnetic fields. In these fields, the quantum 
vacuum becomes a non-linear optical medium because of interactions between 
light and quantum fluctuations of electron-positron pairs. In addition, there is a 
plasma surrounding the magnetar which is a dissipative medium. We construct a 
numerical simulation of electromagnetic waves in this environment which is non-perturbative in 
the wave amplitudes and background field. This simulation reveals a new class of waves 
with highly non-linear structure that are stable against shock formation.

The dense nuclear material in a neutron star is expected to be in a type-II superconducting
state. In that case, the star's intense magnetic fields will penetrate the core and crust
through a dense lattice of flux tubes. However, depending on the details of the 
free energy associated with these flux tubes, the nuclear material may be in a type-I 
state which completely expels the field. We compute the quantum corrections to the 
classical energies of these flux tubes by creating a new, massively parallel Monte-Carlo 
simulation. The quantum contribution tends to make a small contribution which adds 
to the classical free energy. We also find a non-local interaction energy with a 
sign that depends on the field profile and spacing between flux tubes.


\acresetall
\cleardoublepage


\chapter{Preface}
\label{ch:preface}
\acresetall

This dissertation includes reprints of the following previously published material:

\begin{itemize}
	\item{Chapter \ref{ch:inflaton}: Mazur, Dan and Heyl, J. S. \emph{Phys. Rev. D 80, 023523} (2009).  
	Copyright 2009 by the American Physical Society.}
	\item{Chapter \ref{ch:travellingwaves}: Mazur, Dan and Heyl, J.S. \emph{MNRAS 412, 2} (2011)}
\end{itemize}

Some additions and changes have been made to the original manuscripts while preparing 
this dissertation to improve the writing clarity
and to add more details to the descriptions of the methods and results.

The majority of the research described in this dissertation, as well 
as the majority of the work 
involved in preparing the manuscript was performed by the first author listed 
above (DM).  Both the research and the writing were supervised and directed by the co-author
listed above (JSH).  The scope of the research program and the methods 
employed were developed collaboratively by both DM and JSH.
Section \ref{sec:estim-sizes-lambda} was derived from a draft written by 
JSH.  All of the analytical and numerical calculations described in 
this dissertation were performed by DM, 
except where citations are made in the text.  

\cleardoublepage

\tableofcontents
\cleardoublepage	


\listoffigures
\cleardoublepage	



\chapter{Glossary}

\acrodef{UI}{user interface}
\acrodef{UBC}{University of British Columbia}

\begin{acronym}[$\mathrm{ScQED}$]
\acro{AXP}[$\mathrm{AXP}$]{anomalous X-ray pulsar}

\acro{CMB}[$\mathrm{CMB}$]{cosmic microwave background}
\acro{CPU}[$\mathrm{CPU}$]{central processing unit}
\acro{CUDA}[$\mathrm{CUDA}$]{compute unified device architecture}

\acro{GNU}[$\mathrm{GNU}$]{\acs{GNU}'s not unix}
\acro{GPU}[$\mathrm{GPU}$]{graphics processing unit}
\acro{GSL}[$\mathrm{GSL}$]{\acs{GNU} scientific library}

\acro{LCF}[$\mathrm{LCF}$]{locally constant field}

\acro{MPI}[$\mathrm{MPI}$]{message passing interface}
\acro{MRI}[$\mathrm{MRI}$]{magnetic resonance imaging}

\acro{npQFT}[$\mathrm{npQFT}$]{non-perturbative quantum field theory}

\acro{ODE}[$\mathrm{ODE}$]{ordinary differential equation}

\acro{QCD}[$\mathrm{QCD}$]{quantum chromodynamics}
\acro{QED}[$\mathrm{QED}$]{quantum electrodynamics}
\acroplural{QED}[$\mathrm{QED}$]{Quantum electrodynamics} 
\acro{QFT}[$\mathrm{QFT}$]{quantum field theory}
\acroplural{QFT}[$\mathrm{QFTs}$]{quantum field theories}

\acro{ScQED}[$\mathrm{ScQED}$]{scalar quantum electrodynamics}
\acro{SGR}[$\mathrm{SGR}$]{soft gamma repeater}

\acro{WLN}[$\mathrm{WLN}$]{worldline numerics}

\end{acronym}

%

\textspacing		


\chapter{Acknowledgments}

Firstly, I would like to thank my research supervisor, Jeremy S. Heyl, for his mentorship 
and support throughout my graduate education.

Thanks to the members of my supervisory committee for their helpful guidance and for 
providing input on this thesis.

I would like to thank my parents for their financial support and encouragement throughout 
all stages of my education.

Thanks to my research group, office mates, and close friends for countless research-related 
and welcome distracting conversations over the years, especially:  Anand Thirumalai, Ramandeep Gill, Kelsey 
Hoffman, Alain Prat, Sanaz Vafaei, Lara Thompson, Mya Warren, Conan Weeks, Gili Rosenberg, Francis-Yan 
Cyr-Racine, Stephanie Flynn, and Laura Kasian. 




Finally, many thanks to Margery Pazdor for her patience, encouragement, and countless 
kindnesses over the past few years.

This work was supported in part by the Natural Science and Engineering Research Council 
of Canada (NSERC).


\mainmatter

\acresetall	

%

\chapter{Introduction}
\label{ch:Introduction}
\acresetall

Some astrophysical environments, such as the early Universe, or the intense fields and 
hot plasmas near a neutron star, are so extreme that they can dramatically alter the properties
of the quantum vacuum. As a result, the behaviour of familiar particles and their interactions 
can change qualitatively in these environments. In some cases, these vacuum effects can 
have significant impacts, even at the scales of classical physics, or the 
largest distance scales in astrophysics.

Such effects are understood quantitatively using \acp{QFT} in external 
classical fields. 
\acp{QFT} are frameworks for modelling quantum mechanical fields that arose out of the 
need to unite quantum mechanics with special relativity~\cite{peskin1995introduction,
itzykson2006quantum}.  \acp{QFT} have become 
one of the corner stones of modern physics research for providing the quantitative 
framework used to understand particle and condensed matter physics. The elementary 
particles of the Standard Model 
are identified with quantized fluctuations of relativistic quantum fields.  
Interactions between these fundamental particles are also described by the theory and 
result in quantum descriptions of the fundamental forces. The stable or 
meta-stable bound states of a \ac{QFT} may be identified with composite particles 
such as hadrons or mesons. \acp{QED}, the \ac{QFT} within the standard model 
which describes electromagnetic interactions is arguably the most 
precisely tested theory in the history of science. 
Furthermore, \ac{QFT} lends itself to studying the 
interactions and statistics of large numbers of fluctuations that provide the 
quantitative framework for studying condensed matter physics.

One of the great conceptual developments arising from the development of \acp{QFT} was 
a better understanding of the vacuum, \ie what exists after all matter has been removed 
from a region of space. This development builds on a rich history 
of scientific discovery influencing the concept of `vacuum'. 
Ancient Greek philosophers such as Plato had difficulty conceiving 
that `nothing' could exist somewhere~\cite{genz2001nothingness}. 
In medieval Europe, some Catholic leaders may have even considered the 
concept heretical~\cite{grant1981much}. 
However, in the 17th century, many new ideas in thermodynamics developed,  
providing Evangelista Torricelli with the conceptual framework needed to build a laboratory 
vacuum at the top of a mercury barometer and explain it in terms of gas pressure in 1643. 
Later, during the development of 
electromagnetic theory in the 19th century, scientists reasoned that mercury barometers 
were evacuated of gases but must still contain luminiferous ether because light 
could propagate through the evacuated region. This concept was eventually abandoned 
largely because of the 1887 Michelson-Morley experiment null result and the development 
of special relativity. When Einstein developed general relativity, 
the concept of space itself was abstracted 
into a bundle of worldlines with no material properties.

The modern conception of the vacuum began to form with the introduction of the 
Dirac equation in 1930~\cite{Dirac01021928}
and the subsequent development of \ac{QFT}. Dirac's equation 
predicts solutions which come in pairs with both negative and positive frequencies. 
In order to prevent electrons in the theory from descending without limit
into lower and lower energy states
by emitting photons, he postulated that the vacuum state of the system 
was one in which the infinite negative frequency states were all filled, and none of the 
positive frequency states were filled. The infinite 
sea of negative frequency particles was called the Dirac sea. Based on this conception 
of the vacuum, Dirac correctly predicted the existence of the positron, the anti-matter 
partner of the electron. A particle could be liberated from the Dirac sea, leaving a positive 
frequency particle, and a hole in the negative energy sea which was identified with the 
positron. Thus, Dirac's vacuum could decay into particle-antiparticle pairs. As 
\ac{QFT} was further developed, this conception of the vacuum was further clarified.
In \ac{QFT}, the vacuum state is obtained by removing all physical particles 
from a region and it is identified with ground state of the quantum field theory.
Remarkably, it can be lively and have many of the properties we attribute to materials.

In an interacting \ac{QFT}, the Hamiltonian typically does not commute with the particle number operator.
So, we expect there to be an uncertainty relationship between energy and particle content of any state.  
This means that any interacting quantum system prepared in its ground state will later be in a superposition 
of states with arbitrary numbers of particles in each mode.  Physicists often describe this situation 
heuristically in terms of pairs of ``virtual" particles spontaneously coming into existence and annihilating 
each other a short time later. However, these quantum fluctuations are, by definition, not directly observable.  
They are important, though, for their direct role in computations in perturbative \ac{QFT} and for 
their indirect role in explaining several important observations: spontaneous atomic emission,
the Casimir effect~\cite{PhysRev.73.360}, and pair emission in heavy ion collisions
~\cite{PhysRevD.39.1330}\footnote{In general, 
these observations can also be described without any reference to vacuum fluctuations. 
For example, see ~\cite{2005PhRvD..72b1301J}}.

Some of the most dramatic effects arising from interactions between quantum and external fields 
are inaccessible to perturbative \ac{QFT}.  An early example of this is the Schwinger mechanism,  
where electron-positron pairs are emitted from an 
intense electric field~\cite{Schwinger:1951}. This phenomenon
cannot be understood as a perturbative sequence of discrete interactions with the 
external field to any order.  Other effects such as solitons and instantons are also non-perturbative aspects of 
a \ac{QFT} which may be important in the presence of external fields~\cite{rajaraman1982solitons}.  
Even when studying 
perturbative phenomena, the weak-field expansion is an
asymptotic series that generally fails to converge for some large 
value of the external fields. For this reason, it is particularly interesting to study 
the quantum field theories in external backgrounds nonperturbatively. 

In this thesis, we explore the physics of the \ac{QFT} vacuum states in the presence of 
astrophysically relevant external fields. To do this, we make use of 
numerics and special methods which are described and developed 
in the relevant chapters. We focus specifically on scalar 
fields in the rapidly expanding gravitational field of inflation and 
Dirac fields in the intense magnetic fields of magnetars. In the 
case of inflation, the gravitational interaction generates physical particles from the 
vacuum which then perform the remarkable trick of seeding the galaxies 
and the largest structures in astrophysics.  Near a magnetar, the magnetic 
fields are so intense that they have a significant interaction with 
the \ac{QED} vacuum, and hence we may think of the vacuum as a dense optical medium with 
its own unique physical properties. So, the central theme of this dissertation is 
to investigate the link between the microscopic physics of the quantum vacuum
and the macroscopic behaviour of astrophysical systems.

\section{Outline of Thesis}
I have organized this thesis into three main parts. Part \ref{pt:cosmology} deals with the 
emergence of classical behaviour in the large-scale structures of cosmology. Part \ref{pt:EMwaves}
explores electromagnetic waves near the surfaces of highly magnetic stars. Finally, part \ref{pt:fluxtubes}
discusses magnetic flux tubes such as those in neutron star crusts and interiors. Each part begins with a chapter 
of introductory material relevant to the other chapters in that part, so that the introductory 
material for this thesis is distributed between chapters \ref{ch:classicality}, \ref{ch:magnetizedvacuum}, 
and \ref{ch:fluxtubes}.

In part \ref{pt:cosmology} (chapters \ref{ch:classicality} and  \ref{ch:inflaton}), 
I explore the role played by quantum fluctuations in the 
inflationary epoch of our Universe's early history.  In the standard picture of inflationary
cosmology, the natural evolution of vacuum fluctuations in the early Universe leads to the 
creation of large-scale structure.  But there is an open question of how the quantum 
fluctuations developed the very classical characteristics displayed by the galaxies.  
Since it is absurd to believe that observing a new galaxy collapses the wavefunction 
describing the density distribution of the early quantum field, 
we want to explore how the observed classicality likely developed in this system.  The research 
described in chapter \ref{ch:inflaton} explores the possibility that nonlinear interactions 
inherent within the quantum field itself could lead to a quantum system which to us appears 
classical through the process of decoherence.  I use a toy model of the scalar field 
which allows us to directly track particle production and interactions so that we can 
exactly compute the entanglement entropy and compare this with other measures of classicality 
discussed in the literature.

Magnetars are a class of neutron star characterized by unusually intense magnetic fields.  These magnetic 
fields can be so large that the Larmor radius of the electron becomes smaller than its Compton 
wavelength, so that \ac{QED} effects are expected to be important.  We may incorporate the 
effects of these magnetic fields into our description of \ac{QED} using an effective action 
approach.  In this approach, we integrate over the effects of the electron-positron quantum 
fluctuations so that we may describe the average properties of the quantum vacuum 
in terms of the magnetic field alone, without any reference to electronic degrees of freedom.

In part \ref{pt:EMwaves} (chapters \ref{ch:magnetizedvacuum} and \ref{ch:travellingwaves})
we consider the non-linear electrodynamics 
of the \ac{QED} vacuum in intense magnetic fields near the 
magnetosphere of a magnetar, incorporating 
the dispersive effects of a plasma. Media which are both 
dispersive and non-linear often display interesting travelling 
wave phenomena, and we explore this possibility through a numerical 
model in chapter \ref{ch:travellingwaves}.
 
In the dense nuclear matter in the cores of neutron stars, we expect neutrons and positrons to 
form superconducting materials.  In part \ref{pt:fluxtubes}, 
I turn my attention to 
studying the intense magnetic vortices which may be present in neutron star crusts and interiors.  In this 
case, the path integral over fermionic degrees of freedom in the effective action takes the form of 
a functional determinant in very intense magnetic fields which vary significantly on the 
Compton wavelength scale.  The determinants in this situation are very difficult to compute analytically, 
so we discuss several numerical approaches that could be applied to the problem. In chapter 
\ref{ch:greensfunc}, I develop a new technique for computing the functional determinant in cylindrically 
symmetric magnetic fields based on Green's function methods. In chapter \ref{ch:WLNumerics}, I explore 
a more established Monte Carlo technique called \ac{WLN} that approximates a functional integral over fermion loops 
as an average over a cloud of discrete loops. 
A discussion of the uncertainties in this technique is given in chapter \ref{ch:WLError} 
and involves several interesting subtleties that have not been given a proper treatment in the literature. 
Finally, in chapter \ref{ch:periodic}, I apply the \ac{WLN} 
technique to computing the effective actions in a cylindrically symmetric toy model of a dense lattice 
of flux tubes.

The final part of this dissertation is the conclusion, chapter \ref{ch:conclusions}. This chapter provides a
part-by-part summary of the main contributions from this thesis and suggests areas for future 
work.

\part{Classicality in Cosmology}
\label{pt:cosmology}
\chapter{Classicality of Large-Scale Structure}
\label{ch:classicality}
\acresetall

\begin{summary}
In this chapter, I will introduce cosmological inflation and the evolution of galaxies 
and galactic clusters in the Universe from initial density perturbations. 
One of the conceptual problems with the inflation paradigm is that the density 
perturbations are believed to begin from quantum fluctuations, but are later treated 
classically. Some progress has been made in understanding this quantum-to-classical 
transition, but some open questions remain. One powerful tool for understanding 
the boundary between classical and quantum physics is a framework called decoherence, 
which is introduced in this chapter along with a discussion of the implications 
of decoherence on the evolution of the Universe.
\end{summary}

Despite the massive success of quantum mechanics in describing physics at small distance scales, 
it can be safely neglected when describing physics at a very wide range of larger 
distance scales.  The physics of an atom is simply different from the physics 
of our day-to-day experience as macroscopic beings. 
However, the boundary separating quantum physics from classical 
physics is not well understood.  When a quantum system is measured 
with a classical detector, the quantum wavefunction appears to collapse instantaneously, 
destroying quantum information about the system. However, this scenario 
seems to violate unitarity, a postulate of quantum mechanics that forbids the destruction of information in a 
closed system. This creates an open problem of fundamental physics 
which has been dubbed `the measurement problem'.

Large-scale structure is the term used in cosmology to describe the organization 
of matter in the Universe on galactic scales up to the scales of superclusters and 
filaments.  In the modern standard model of cosmology, large-scale structure is 
initially seeded by fluctuations of the quantum vacuum in a period of the early 
Universe known as inflation.  However, it is difficult to believe any suggestion that 
the distribution of matter in the Universe remains in its original
quantum superposition on the largest scales of astrophysics until 
human astronomers collapse the wavefunctions by observing the positions of the galaxies.  
So, large-scale structure provides a very dramatic 
example of a quantum system that makes a transition to being a classical system 
 without a well-defined measurement event to 
explain the collapse of the wavefunction.

This chapter elaborates on the above ideas and provides a technical introduction 
to chapter \ref{ch:inflaton}.

\section{Cosmological Inflation}

Cosmological inflation, a postulated period of rapid-expansion in the Universe's early
history, is currently a well-accepted component of the standard model of cosmology. While 
accelerated cosmological expansion had already been discussed~\cite{1980PhLB...91...99S}, the
idea generated great interest
 in 1981 when Alan Guth \cite{1981PhRvD..23..347G} showed that it 
addressed three major 
problems with the standard big bang cosmological models of that time~\cite{1982PhRvL..48.1220A,
1982PhLB..108..389L}:
\begin{itemize}
	\item The flatness problem: As the Universe expands, the energy density due to curvature 
	decreases more slowly than the energy densities of matter and radiation. This means that 
	a small curvature in the early Universe would be expected to become dominant in the later 
	Universe. Instead, we observe that the energy density of the present Universe is consistent 
	with it being flat, implying that the early Universe was somehow fine-tuned for precise flatness
	~\cite{1994PhRvD..49.3830H}.
	\item The horizon problem: If we observe two points near the cosmological horizon on opposite 
	sides of the sky, the light from each of those points is reaching us for the first time, and 
	for the first time bringing us into causal contact with those points. However, we would naively 
	expect that those points have never been in causal contact with each other, since they are each 
	further from each other than they are from us. Nevertheless, we observe homogeneity and isotropy 
	throughout the sky. For example, those two distant points share the same \ac{CMB} temperature, 
	implying that they must have been in thermal (and therefore causal) contact at some point in the past
	~\cite{1995PhRvD..51.5347L, 1994PhRvD..49.3830H}.
	\item The magnetic-monopole problem: Most grand unified theories predict the copious production 
	of magnetic monopoles at high temperatures, such as those of the early Universe~\cite{1978PhLB...79..239Z}.
	These theories 
	predict that the present Universe should contain a significant density of magnetic monopoles, 
	however our experimental searches have not discovered any, indicating that the density is much lower 
	than predicted. In fact, inflation tends to dilute many types of predicted but unobserved relics
	~\cite{1994csot.book.....V, 1995RPPh...58..477H}.
	
\end{itemize}
Inflation solves these problems by postulating a period of rapid expansion in the Universe's early 
history. An inflationary period would smooth out inhomogeneities, and anisotropies, 
drive the curvature toward precise flatness, and dilute away exotic relic particles, like magnetic monopoles
\cite{kolb1990early}. 
The resulting Universe takes on a very simple effective initial
state which is dominated by the fields that drive the 
inflationary phase.

The simplest models of inflation postulate a scalar field
called the inflaton, $\phi(x)$,
which varies slowly with 
time so that the energy is dominated by the potential term(s), $V(\phi)$.
In units where $\hbar = c = 1$, we have

\be
	\left({\dot{\phi}}\right)^2 \ll V(\phi).
\ee

The inflaton is usually imagined to be in a metastable false vacuum 
state, slowly rolling down a shallow potential throughout the 
inflationary epoch~\cite{1982PhLB..108..389L,1982PhRvL..48.1220A}.
In this case, the inflaton field contributes to cosmological evolution 
like a cosmological constant with

\be
	\Lambda \approx 8\pi G V(\phi).
\ee
This leads to a nearly exponential expansion of the scale factor, $a(t)$ with 
(coordinate) time.

\be
	a(t) \sim \exp{(\sqrt{\Lambda/3}t)}
\ee

Inflation was initially proposed as a solution to the problems 
mentioned at the beginning of this section by smoothing out the 
initial conditions of the early Universe. Unfortunately, inflation 
as a theory of initial conditions is difficult to test as there 
are very few observable signals which could falsify its initial 
condition predictions~\cite{1997cdc..conf..399L, 1997cdc..conf..265A}. 
However, inflation also provides a framework 
for understanding the origin of structure in the Universe. In this
role it is a strongly predictive theory, where different models of 
inflation lead to different predictions of observable cosmological 
structures. For this reason, 
recent research interest on inflation focuses primarily on the 
relationship between inflation and the formation of structure
in the Universe.

\subsection{Origin of Large-Scale Structure}

The large-scale structure that we observe today is a natural consequence of the
gravitational collapse of initial density perturbations in the early Universe.
From cosmological perturbation theory~\cite{Mukhanov:1990me}, we can predict the nature of the 
structure that we observe today from some initial perturbations which 
collapse classically into structure under the gravitational instability.
To high precision, our Universe is consistent with Gaussian density perturbations which 
become frozen-in at horizon exit~\cite{2011ApJS..192...18K}.  
Gaussian perturbations are exactly the prediction of the density 
perturbations that arise from quantum fluctuations of the inflaton 
field.

To see the prediction that inflation makes for the density perturbations,
we begin by separating the inflaton field into an unperturbed part and 
a perturbation:

\be
	\phi(\vec{x},t)=\phi(t)+\delta\phi(\vec{x},t).
\ee
In the conformal-time formalism, we define the conformal time, $\tau$ so that 
$d\tau = dt/a$.  During inflation, this means that we have 
\be
	\tau = -(aH)^{-1}.
\ee

We define a new field to describe the perturbations:
\be
	u(\vec{x},\tau) = a(\tau)\delta \phi(\vec{x},\tau).
\ee
The Fourier expansion of this field is 
\be
	u(\vec{x},\tau) = \left(\frac{1}{2\pi}\right)^{3/2}
	\int d^3k u(\vec{k},\tau)e^{i\vec{k}\cdot \vec{r}}.
\ee
The fluctuations well before the mode exits the horizon can be analyzed 
in terms of a set of creation and annihilation operators appropriate 
for flat spacetime~\cite{Liddle:2000}
\be
	\label{eqn:flatmode}
	u(\vec{k},\tau) = \frac{1}{\sqrt{2k}}\left(e^{-ik\tau}a(\vec{k}) 
	+ e^{ik\tau}a^\dagger(-\vec{k})\right).
\ee

In discussing the evolution of large-scale structure, it is useful to know 
the spectrum of density inhomogeneities in the inflaton field well after 
the modes exit the horizon. This information is encoded in the
power spectrum of the field, $\mathcal{P}_\phi$. The power spectrum
is defined so that the mean-square field is

\be
	\mean{\phi^2(\vec{r})} =\int_0^\infty \mathcal{P}_\phi(p) \frac{dp}{p}.
\ee
So, we have
\be
	\mathcal{P}_\phi(p) = \frac{p^3}{2 \pi^2} \bra{0}~|\phi(\vec{p})|^2~\ket{0}.
\ee
where $\phi(\vec{p})$ is the Fourier transformed scalar field. In terms of our 
new field, $u(\vec{x},\tau)$, and its Fourier transform, the power spectrum is
\be 
	\mathcal{P}_\phi(p) = \frac{\mathcal{P}_u(k)}{a^2} 
	= \frac{k^3}{2 \pi^2 a^2}\bra{0}~|u(\vec{k},\tau)|^2~\ket{0}.
\ee

To evaluate the power spectrum, 
we would like to evaluate this quantity well after horizon exit when 
the mode experiences a curved spacetime and 
the equation (\ref{eqn:flatmode}) is therefore no longer appropriate. In general, 
we describe the field in terms of creation operators and a mode function:

\be
	u(\vec{k},\tau) = w(k,\tau)a(\vec{k}) + w^*(k,\tau)a^\dagger(-\vec{k}).
\ee
Well after horizon exit, the form of the mode function which gives us appropriate 
creation and annihilation operators is ~\cite{Liddle:2000}
\be
	w(k, \tau) = -\frac{1}{\sqrt{2k^3}}(i-k\tau)\frac{e^{-ik\tau}}{\tau}.
\ee
This description of the field in terms of creation and annihilation 
operators allows us to evaluate the vacuum expectation value in 
the definition of the power spectrum if we recall the basic properties of the 
creation and annihilation operators:
\be
	[a(\vec{k}), a^\dagger(\vec{k}')] = \delta^3(\vec{k}-\vec{k}');
\ee
\be
	[a(\vec{k}), a(\vec{k}')] = 0;
\ee
\be
	a(\vec{k}) \ket{0} = 0.
\ee
We can now find the vacuum expectation value,
\be
	\bra{0}~|u(\vec{k},\tau)|^2~\ket{0} = |w^2(k,\tau)| = \frac{1}{2k^3\tau^2} = \frac{a^2H^2}{2k^3}.
\ee

We finally find that the fluctuations of the inflaton result in a scale 
invariant power spectrum:
\be
	\mathcal{P}_\phi = \left(\frac{H}{2\pi}\right)^2,
\ee
where $H$ is evaluated at horizon exit $k=aH$.  The scale invariance 
insures that the Fourier components of the perturbations are uncorrelated.
Thus, inflation with linear perturbations makes a clear 
prediction of Gaussian inhomogeneities.  This prediction is confirmed to 
high precision in \ac{CMB} measurements \cite{2011ApJS..192...18K} and 
constitutes one of the most important observational
tests of the inflationary paradigm~\cite{2009astro2010S.158K}. 
Of course, non-linearities 
may still occur in the \ac{CMB} from relaxing the assumption that the perturbations 
are linear and allowing them to self-interact and to interact with other fields.

\section{Classicality of the Vacuum Fluctuations}

The Fourier coefficients of the inflaton 
field, $\delta \phi_\vec{k}$, are not eigenstates of the Hamiltonian, 
so in the quantum vacuum state these coefficients are in a 
superposition; we cannot know what value their measurement will yield 
until we perform a measurement which collapses the superposition 
onto a particular eigenvalue of $\delta \phi_\vec{k}$.
Well before the mode exits the horizon, this is exactly what we 
expect since the vacuum state is a quantum state in every sense.
Well after horizon exit, however, we have an interpretation problem.

By observing large-scale structure, we make measurements of 
the density perturbation, and therefore of
$\delta \phi_\vec{k}$ well after horizon exit.  However, 
we treat these measurements as classical and assume that 
each galaxy arrived at its current location along a classical 
trajectory.  But this leads us into a cosmological Schr\"{o}dinger's
cat paradox.  Quantum mechanics seems to suggest that cats and 
Universes are in a superposition until we measure them and 
collapse their wavefunctions, even though that conclusion seems absurd
for macroscopic objects.

The evolution of classicality in cosmological perturbations has 
been studied from a wide variety of perspectives. The research 
can be placed into two main categories. The first category deals with the 
evolution of field modes as closed-systems. From this perspective, 
inflation causes general initial quantum states to become very 
peculiar quantum states at the end of inflation. The mode is driven 
toward a state whose phase space occupies the minimum uncertainty 
allowed by the Heisenberg principle, and which is squeezed into a 
highly elongated ellipse with negligible width~\cite{Kiefer:2008ku}. This state is a
very extreme example of a squeezed state and closely reproduces 
classical stochastic behaviour. More heuristically, the field 
produces a large particle content during inflation and these 
particles form a condensate that behaves classically because the 
canonical momentum and coordinate operators approximately commute 
when the particle occupation number, $n$, becomes large
(since $\langle [a,a^\dagger] \rangle =  \mathcal{O}(1/n)$)~\cite{PhysRevD.57.2138}.

While the above explanation may be sufficient to explain the 
emergence of classicality in the modes, it is not the complete story. 
Even in a highly squeezed state, the quantum coherence within the system is 
preserved and there is still an interpretation problem associated with measuring the 
density perturbations. The concept of decoherence is generally invoked 
in discussions of these remaining issues~\cite{1998AnP...510..137K}.
The field of interest will be interacting with other fields, and experiencing 
non-linear self-interactions. Even if we neglect any interactions, different 
spatial regions of the field will become entangled inside and outside of the 
horizon. These interactions can have a decohering effect on 
the field modes of interest and produce entropy
\cite{1988PThPh..79..442S, 1989PhRvD..39.2912H, 1989PhRvD..39.2924P}. 
The second main category of research investigates the role that 
these decohering interactions have on the classicality of the field modes.

\section{Decoherence and the Quantum to Classical Transition}

Both classical and quantum systems may be described as stochastic 
distributions of states.  However, interference terms 
arising from quantum superpositions and entanglement
are unique to quantum 
systems and do not occur in classical mechanics.  For example, 
consider a quantum mechanical transition probability:

\ba
	\label{eqn:classquantterms}
	|\braket{\psi}{\phi}|^2 &=& \sum_i |\psi_i^* \phi_i|^2 + \sum_{i,j; i\ne j} \psi_i^*\psi_j \phi_j^* \phi_i \nonumber \\
	&=& \mbox{Classical terms} + \mbox{Quantum interference terms}.
\ea
The expanded probability includes both classical terms that would 
also be present in a stochastic distribution, as well as 
additional interference terms that characterize the 
quantum nature of the system.

The key result of decoherence is that these quantum interference 
terms, and the quantum information they represent, are destroyed 
by interactions with complicated degrees of freedom if we do not 
observe those interactions.  

Consider a quantum state interacting 
with the complicated degrees of freedom in a classical environment, 
such as a detector.  In that case, we must sum over all possible 
states of the environment:

\ba
	|\braket{\psi}{\phi}|^2 \rightarrow \sum_j | \braket{\psi}{\phi,\epsilon_j}|^2 
	= \sum_j | \sum_i 	\psi_i^* \braket{i}{\phi}\braket{\epsilon_i}{\epsilon_j}|^2.
\ea
The nature of the interaction with the environment leads to 
a phenomenon called environment-induced superselection 
(or Einselection for short)
\cite{2003RvMP...75..715Z}. Only a few 
states of the quantum system will be robust against frequent monitoring 
by the environmental degrees of freedom, and these states are called 
pointer states.  Einselection occurs when the states of the 
environment, which correspond to the pointer states, become orthogonal:

\be
	\braket{\epsilon_i}{\epsilon_j} \approx \delta_{ij}.
\ee
Then, we have
\be
	|\braket{\psi}{\phi}|^2 \rightarrow \sum_i |\psi_i^* \phi_i|^2.
\ee

So, we see that the environment has destroyed the quantum phase information.  
More precisely, the phase information has been hidden in the unobserved 
degrees of freedom of the environment through a unitary evolution.  
The resulting system is indistinguishable from a classical 
stochastic distribution.  This decoherence occurs on a time scale called the 
decoherence time which for most macroscopic environments is typically many 
orders of magnitude faster than any other dynamic time scale~\cite{2003quant.ph..2044Z}.
So, decoherence 
results in a nearly-instantaneous destruction of superposition and entanglement and 
produces a distribution which is consistent with classical physics.  In this framework, 
wave-function collapse is a natural, unitarity-preserving
consequence of the interaction between a quantum 
system and a complicated environment. 

Because quantum superposition and entanglement are incompatible with 
classical physics, decoherence is a necessary condition for the 
emergence of classicality.

\subsection{Measures of Decoherence}

An operator, $\rho$, called the density matrix is useful for characterizing quantum 
mixed-states and the decoherence of quantum 
systems~\cite{landau1977quantum, schlosshauer2008decoherence}:

\be
	\rho = \ket{\psi}\bra{\psi}.
\ee
In this picture of the quantum state, the quantum phase information 
(recall equation (\ref{eqn:classquantterms}))
is stored in the off-diagonal elements in the pointer basis.  In general, 
we consider the density matrix of the states belonging to the system while 
the environment's states are unobserved.   We account for this by 
defining the reduced density matrix, $\rho_{\mbox{red}}$: the density matrix of the combined 
environment and system with the environment's degrees of freedom traced 
out,

\be
	\rho_{\mbox{red}} = \sum_i \braket{\epsilon_i}{\psi}\braket{\psi}{\epsilon_i}.
\ee

If the resulting reduced density matrix is in a pure quantum state, the 
eigenvalues will be $\{1,0,0,...\}$.  However, if there is mixing 
between the system of interest and the environment, the reduced 
density matrix will have several positive eigenvalues which sum to unity.
Therefore, a useful measure of the entanglement between the system and 
the environment is the entanglement (or von Neumann) entropy~\cite{schlosshauer2008decoherence},

\be
	S = -{\rm Tr}(\rho_{\mbox{red}} \ln \rho_{\mbox{red}}).
\ee
For a pure state, $S(\rho_{\mbox{red}})$ is zero, and it increases 
with the amount of entanglement between the reduced system and the 
environment.

Computing the entropy for complicated quantum systems using this expression 
can be difficult analytically, and intensive computationally
since it requires diagonalizing large matrices. So, it is often practical to find 
alternative measures of decoherence which may be easier to compute. 
Several examples of these which have been used in the context of 
classicality in cosmology will be discussed in section \ref{sec:entanglementmeasures}. 
One of our goals in chapter \ref{ch:inflaton} will be to compare these alternative 
measures of entanglement with the entanglement entropy for a scalar field 
during inflation.

\section{Discussion}

In this chapter, I have briefly reviewed the important ideas behind the puzzle 
of the classicality of large-scale structure. The modern standard model of 
cosmology includes an inflationary period which helps to explain the isotropy, and 
flatness of the Universe that we observe today. During the inflationary period, 
fluctuations of a quantum scalar field become the initial density perturbations 
around which matter gravitationally collapses to eventually 
form the large-scale structure that we observe today. Initially, the fluctuations 
are quantum and finally, they are classical. In between, we are confronted with an 
open problem of quantum mechanics: how exactly does the quantum-to-classical
transition take place?

In the next chapter, I will introduce a new toy model
that allows us to exactly evolve certain 
modes of a scalar field during inflation. This method has two interesting 
features. First, the mechanism leading to the decoherence of the field can be 
easily understood in terms of interactions between particles exchanging quantum 
information between modes. The second feature of this technique is that one 
can compute several different measures of decoherence including the exact 
entanglement entropy. Thus, we use the model to compare different measures 
of decoherence and entanglement in the context of inflation.

\chapter{Creation of Entanglement Entropy During Inflation}
\label{ch:inflaton}
\acresetall

\vspace{-2cm}{\renewcommand{\thefootnote}{} \footnote{This chapter contains only minor changes 
from the published manuscript: Mazur, Dan and Heyl, 
J. S. \emph{Phys. Rev. D 80, 023523} (2009).  
	Copyright 2009 by the American Physical Society.}}
 \addtocounter{footnote}{-1}

\begin{summary}

The density fluctuations that we observe in the universe today are 
thought to originate from quantum fluctuations produced during a 
phase of the early universe called inflation. By evolving a wave 
function describing two coupled Fourier modes of a scalar field 
forward through an inflationary epoch, we demonstrate that nonlinear 
effects can result in a generation of entanglement entropy between 
modes with different momenta in a scalar field during the inflationary 
period when just one of the modes is observed. Through this mechanism, 
the field would experience decoherence and appear more like a classical 
distribution today; however the mechanism is not sufficiently efficient 
to explain classicality. We find that the amount of entanglement entropy 
generated scales roughly as a power law $S \propto \alpha^{1.75}$, 
where $\alpha$ is the coupling 
coefficient of the nonlinear potential term. We also investigate how the 
entanglement entropy scales with the duration of inflation and compare 
various entanglement measures from the literature with the von Neumann entropy. 
This demonstration explicitly follows particle creation and interactions 
between modes; consequently, the mechanism contributing to the generation 
of the von Neumann entropy can be easily seen.
\end{summary}

\section{Introduction}

Most modern cosmological models include a period in the universe's
history called inflation during which the scale parameter increased
exponentially with the proper time of a comoving observer.  This
period was originally introduced to address the horizon and flatness
problems of cosmology ~\cite{1981PhRvD..23..347G}.  More recently,
however, research on inflation has been toward understanding structure
formation
\cite{1982PhRvL..49.1110G,1982PhLB..117..175S,1985PhRvD..31.1792L}.
The distribution of galaxies and clusters that we observe in the
universe today are thought to have originated from fluctuations of a
quantized field created during inflation~\cite{1981JETPL..33..532M,Hawking:1982}.
A thorough review of structure formation and inflationary cosmology
can be found in Liddle and Lyth~\cite{Liddle:2000}.

Despite their quantum mechanical origins, the late-time evolution of
these fluctuations is treated in a classical framework.  It is
therefore important to understand the quantum-to-classical transition
made by these fluctuations (for a recent review, see
~\cite{Kiefer:2008ku}).  The classicality of a quantum system is often
discussed in the context of decoherence.  That is, as a quantum system
interacts with unobserved environmental influences, that system loses
quantum coherence and begins to behave as a classical statistical
distribution.

The quantized field may of course be the inflaton itself, which drives
the inflation of the universe, or it could be another quantized field
that produces density fluctuations as in curvaton models or the
gravitational field.  It is possible, in principle, that
non-classical correlations from an inflationary period in our
universe's history may one day be observed.  But this depends on the
decoherence that the scalar or tensor field has experienced since the
beginning of inflation. Several authors have investigated decoherence
of the density fluctuations by calculating the entropy of cosmological
perturbations created during inflation
~\cite{2000PhRvD..62d3518K,2005PhRvD..72d5015C,2006PhRvD..74b5001C,
  2007JCAP...11..029P,campo:065044,2008PhRvD..78f5045C}.

It has been suggested~\cite{Burgess:2006} that decoherence is unlikely
to occur during inflation because the Bunch-Davies state occupied by
the scalar field during inflation is similar to the Minkowski vacuum.
Because the ordinary Minkowski vacuum does not decohere, we would not
expect to see any decoherence from a scalar field during inflation.
In the particle-based picture adopted for the present analysis, it
becomes clear that the scalar field does undergo decoherence when
the potential is non-linear.  

Since decoherence is a necessary condition for the emergence of
classicality in a quantum system ~\cite{Zurek:1993}, non-linearities in
the scalar field help to explain the classical matter distribution
that we observe today.  This simple model demonstrates that this
entropy generation can occur during inflation itself and does not
depend on the reheating process at the end of 
inflation~\cite{1994PhRvL..73.3195K, PhysRevD.56.3258}; therefore,
the results are perhaps most interesting for cosmological scalar
fields that do not participate in reheating.  For such fields, the
non-linear interactions do not generate a sufficient amount of
decoherence to result in classicality for the fields.

Here, we examine the case where certain modes of a field play the role
of the environmental influence and cause decoherence when a
non-linearity in the potential allows the modes to
interact~\cite{Lombardo:1995,Burgess:2006,Martineau:2006}.  We discuss
a simulation that was performed to compute the entanglement entropy
between such modes in a very transparent model that follows particle
creation and the interaction between modes during the inflationary
period.  The entropy is computed as inflation progresses to
demonstrate the decoherence of a scalar field.  

Computing the entanglement entropy of a large quantum system is a
computationally difficult task since it involves diagonalizing the
density matrix.  To evaluate several possible expediencies, we have
compared our results to other measures of entanglement and
correlations between modes.  We have found that the other measures
considered share a similar qualitative behaviour with the
entanglement entropy and can be much easier to compute.  Therefore,
for some applications, these measures may be useful as stand-in
quantities in simulations where the entanglement entropy is too costly
to compute.  We verify several efficient methods to characterize the
entropy.

\section{Cosmological Scalar-Field Evolution}

We would like to investigate the evolution of a scalar field in an
isotropic, homogeneous, flat spacetime.  The analysis for this
situation is covered extensively in part I, chapter 6 of Mukhanov et
al. ~\cite{Mukhanov:1990me}.  The relevant metric for this evolution is
\begin{equation}
	\label{eq:inflFRW}
	ds^2 = a^2(\tau)(d\tau ^2 - d{\bf x}^2).
\end{equation} 
where $\tau$ is the conformal time, which is related to the comoving time 
by $dt=a(\tau) d\tau$, and ${\bf x}$ is a comoving displacement.  For
simplicity we will take $a(\tau)=-(H\tau)^{-1}$ (pure de Sitter
expansion) during inflation.

The evolution of a scalar field $\phi$ is governed by its Lagrangian
$\mathcal{L}$.  The lowest-order Lorentz-invariant expression
containing up to first derivatives is
\begin{equation}
	\label{eq:inflsfl}
	\mathcal{L} = \frac{1}{2} g^{\mu \nu} \partial_\mu \phi \partial_\nu\phi - V(\phi).
\end{equation}
For simplicity we will neglect the mass of the scalar field
during inflation ($m \ll H$).  We include a non-linearity in the
potential that couples the Fourier modes of the field.  Even if the
field itself is free, its self-gravity will introduce an interaction
potential of the form ~\cite{Burgess:2006, Martineau:2006}
\begin{equation}
	\label{eq:inflpotential}
	V=\lambda \PlMass \phi^3.
\end{equation}
Although the $\phi^3$ potential is generally unstable, one should
interpret this as an effective potential to account for the
gravitational self-interaction, so the instability is not surprising
because the gravitational self-interaction is generally unstable.

\subsection{Mode Coupling During Inflation}

For this analysis, we choose to use a simple model in which the universe
contains only particles with four possible momenta: $\pm \vec{k}$ and
$\pm 2\vec{k}$.  Given this requirement, we construct a Hamiltonian which
incorporates a coupling term between these two Fourier modes so that
we can observe the effect this non-linearity has on the entanglement 
between modes during inflation.

The creation and annihilation operators satisfy the following
commutation relations
\begin{equation}
	\label{eq:inflcommute}
	[\uncreate{k},\create{k'}]=\delta^{(3)}(\vec{k}-\vec{k'})
\end{equation}
\begin{equation}
	\label{eq:inflcommute2}
	[\create{k},\create{k'}] = [\uncreate{k},\uncreate{k'}]=0.
\end{equation}

Including our potential term (\ref{eq:inflpotential}), the action for
the field is
\begin{equation}
   S = \frac{1}{2}\int d^4x \sqrt{-g}\biggr [\partial_{\mu} \phi
   \partial^{\mu}\phi 
        + \lambda \PlMass \phi^3\biggr].
\end{equation}

Following the steps outlined by ref.~\cite{Heyl:2006fb}, we arrive at
the following expression for the action.
\begin{equation}
	S = \frac{1}{2}\int d^4x a^2\left\{ \left[\pderiv{\phi}{\tau}\right]^2
	- [\nabla \phi^2] 
	+ a^2\lambda \PlMass  \phi^3\right\}
\end{equation}

If we make the substitution $u=a\phi=\frac{1}{(2 \pi)^{3/2}}
\int d^3k u_{k}(\tau)e^{i \vec{k}\cdot \vec{r}}$, the action becomes
\ba
	S &=& \frac{1}{2}\int d\tau d^3k \biggr[ \left|\pderiv{u_k}{\tau}\right|^2
	- (k^2+m_{\rm eff}^2)|u_k|^2 \nonumber \\
	& &- \frac{ \lambda \PlMass}{(2 \pi)^{1/2}a} \int{d^3k' d^3k''
	u_{\vec{k}} u_{\vec{k}'} u_{\vec{k}''}\delta^{(3)}(\vec{k} + \vec{k}' + \vec{k}'')}
	\biggr]
\ea
where the effective mass is $m_{\rm eff}^2 = -2\frac{Q}{\tau^2}$, 
\begin{equation}
	Q \equiv \frac{1}{(1+3w)^2}\left[(1-3w)\right]
\end{equation}
and $w=p/\rho$ is the equation of state parameter.

The Hamiltonian is, then,
\begin{equation}
	\label{eq:Ham1}
	H = \frac{1}{2}\int d^3k\left[\left|\pderiv{u_{\vec{k}}}{\tau}\right|^2
	+ (k^2+m_{\rm eff}^2)|u_{\vec{k}}|^2 + \frac{ \lambda \PlMass}{\sqrt{2 \pi}a}
	\int d^3k' u_{\vec{k}}u_{\vec{k}'}u_{-(\vec{k}+\vec{k}')}\right].
\end{equation}
In general, we have $u_{\vec{k}} = g(k,\tau)\uncreate{k} + g^* (k,\tau)
\create{-k}$.  Putting this into to the Hamiltonian, (\ref{eq:Ham1}), and 
neglecting terms that do not conserve energy in flat spacetime gives
\begin{eqnarray}
	\label{eq:Ham2}
	H &=& \frac{1}{2}\int d^3k\biggr[\left(\left|\modedot{k}{}\right|^2 +|\mode{k}{}|^2(k^2+m_{\rm eff}^2)\right)
	(\create{k}\uncreate{k} + \uncreate{k}\create{k}) \nonumber \\
	& &+ 
	\left(\modedot{k}{*2} +\mode{k}{*2}(k^2+m_{\rm eff}^2)\right)\create{-k}\create{k} \nonumber \\
	& &+
	\left(\modedot{k}{2} +\mode{k}{2}(k^2+m_{\rm eff}^2)\right)\uncreate{k}\uncreate{-k}  \nonumber \\ 
	& &+  \frac{ \lambda \PlMass}{\sqrt{2 \pi}a}
	\big[\mode{k}{}\mode{k}{}\mode{2k}{*}\create{2k}\uncreate{k}\uncreate{k} \nonumber \\
	& &+
		\mode{k}{*}\mode{k}{*}\mode{2k}{}\uncreate{2k}\create{k}\create{k}\big].
	\biggr]
\end{eqnarray}

The mode function is normally chosen to be 
\begin{equation}
	\label{eq:usualmode}
	\mode{k}{} = -\frac{1}{\sqrt{2 k^3}}(i-k\tau) \frac{e^{-ik\tau}}{\tau},
\end{equation}
as this choice satisfies the equation of motion for the free field
during a de Sitter phase and because it simplifies the Hamiltonian to
one that commutes with the number operator since, when $Q=1$,
\begin{equation}
\modedot{k}{2}+\mode{k}{2}(k^2+m_{\rm eff}^2)=0.
\end{equation}
However, this choice is not practical for our calculation because the
scalar field is not free; therefore, this choice does not satisfy the
field equation of motion, and in fact it complicates the Hamiltonian
because, for example
\begin{equation}
	\mode{k}{}\mode{k}{}\mode{2k}{*} = \frac{1}{4 k^{9/2}\tau} (2(k\tau)^3-3i(k\tau)^2-i)
\end{equation}
does not have a simple dependence on $\tau$ and the simplifications provided by 
(\ref{eq:usualmode}) are lost.

We would like to know the amount of entropy at the end of inflation during radiation domination.  
The usual way to proceed is to select the mode function ($\ref{eq:usualmode}$) 
and use this to determine the equation of motion for the scalar field during 
inflation.  We would then determine the Bogoliubov 
coefficients at the transition from inflation to radiation domination.  
After performing the transformation, we would compute 
the amount of entropy from the transformed density matrix.

However, we can simplify the problem by instead choosing a mode function 
that describes the system during radiation domination and use this 
mode function to compute the entire evolution.
The choice of function $g(k,\tau)$ is flexible due to the vacuum
ambiguity and is related to choosing the set of states that the
creation and annihilation operators act upon.  Any choice will provide
us with a complete basis with which we can describe any state of the
field.  The arbitrariness of the mode function is also discussed in~\cite{2007JCAP...02..031A}.
  
For us, it is most prudent to choose the simple function
\begin{equation}
\label{eq:modefunction}
g(k,\tau)=\frac{1}{\sqrt{2k}}e^{-ik\tau},
\end{equation}
which defines the vacuum
both during radiation domination and for scales much smaller than the
horizon even during the de Sitter phase.  Thus, we can make a very natural 
connection between our initial state and our final state.  
The choice is as arbitrary as choosing
to perform a calculation in classical mechanics in a rotating frame
rather than an inertial frame.

Correctly interpreting the wavefunction where (\ref{eq:modefunction})
is inappropriate (\ie~ after horizon exit during a de Sitter phase)
would require a Bogoliubov transformation, but for our purposes we do
not require this.  We are only interested in calculating the entropy
after the transition to radiation domination where our choice of mode
function corresponds to the usual creation and annihilation operators
for this background.  Therefore, we avoid transformations entirely
since we already have the required description of our wavefunction.

 With the choice (\ref{eq:modefunction}), the Hamiltonian is not constant in
time even without the non-linear couplings.  In particular the mass
depends on time; this choice is similar in spirit to the calculations
of Guth and Pi~(\cite{1982PhRvL..49.1110G}).  Heyl~\cite{Heyl:2006fb}
has shown for a free scalar field that this choice gives the same
results as the standard function $g(k,\tau)$ and we refer the reader
to that article for a more thorough discussion of the technique.
 
Choosing to use ($\ref{eq:modefunction}$), we have
\begin{equation}
u_{\vec{k}} = \frac{1}{\sqrt{2 k}}(e^{-ik \tau}\uncreate{k} 
	+ e^{ik \tau}\create{-k}) .
\end{equation}
The nonlinear terms in the Hamiltonian provide a coupling mechanism
between the modes of interest.  To perform the integral over $d^3k'$
in (\ref{eq:Ham1}), we neglect the effect of the coupling on the modes
that are not considered in our simulation and treat the functions
$u_\vec{k}$ as constant on a spherical shell surrounding the momenta,
$\vec{k}'$, that we are interested in.  For
$u_{\vec{k}'}=\rm{const.}$ on spherical shells of constant volume
around $\vec{k}$ and $2\vec{k}$, the integral becomes
\begin{equation}
\int d^3k' u_{\vec{k}}u_{\vec{k}'}u_{-(\vec{k}+\vec{k}')} \rightarrow Vk^3u_{\vec{k}}u_{\vec{k}}u_{-2\vec{k}},
\end{equation}
where $V=\frac{4}{3}\pi\left(\frac{4}{1+\sqrt[3]{2}}\right)^3
\approx23$ is a (somewhat arbitrary) geometrical constant.

Making this substitution, we arrive at the final form of the Hamiltonian.
\ba
	\label{eq:inflHam}
	H &=& \int d^3\vec{k}\biggr[\left(k-\frac{Q}{\tau ^2 k} \right)(\create{k}\uncreate{k}+
	\uncreate{k}\create{k}) -\frac{Q}{\tau ^2 k}
	( \uncreate{-k}\uncreate{k} e^{-2ik\tau} + \create{-k}\create{k} e^{2ik\tau})\nonumber \\
	& & + \frac{\lambda V k^{3/2} \PlMass}{4\sqrt{2\pi}a}(\create{2k}\uncreate{k}\uncreate{k} 
        + \uncreate{2k}\create{k}\create{k})\biggr].
\ea
This Hamiltonian is similar to that used by ref.~\cite{Heyl:2006fb},
generalized to allow for the interactions between Fourier modes.

Here, the two terms multiplied by the factor $\lambda$ are responsible
for the annihilation of two particles from the $\vec{k}$ mode into a
single particle from the $2\vec{k}$ mode and the decay of an
$2\vec{k}$ mode particle into two $\vec{k}$ mode particles,
respectively. 
As the two modes of the field exchange particles with each
other, we expect that entanglement entropy will be generated in either
of the modes observed individually.

We wish to use this Hamiltonian to evolve Fock space wavefunctions
representing the number of particles in each of four modes: Those with
$m^+$ particles with momentum $2\vec{k}'$, $m^-$ particles with
momentum $-2 \vec{k}'$, $n^+$ particles with momentum $\vec{k}'$, and
$n^-$ particles with momentum $-\vec{k}'$.
\begin{eqnarray}
	\label{eq:inflwavefunctionlong}
	\ket{\psi} &=& \suminf{m^+} \suminf{m^-} \suminf{n^+} \suminf{n^-} B_{m^+, m^-, n^+, n^-}(\tau)
	\left(\frac{(\create{2k'})^{m^+}}{\sqrt{m^+}[\delta^{(3)}(2\vec{k}'-2\vec{k}')]^{\frac{m^+}{2}}}\right)
	\nonumber \\
	& & \times
	\left(\frac{(\create{-2k'})^{m^-}}{\sqrt{m^-}[\delta^{(3)}(2\vec{k}'-2\vec{k}')]^{\frac{m^-}{2}}}\right)
	\left(\frac{(\create{k'})^{n^+}}{\sqrt{n^+}[\delta^{(3)}(\vec{k}'-\vec{k}')]^{\frac{n^+}{2}}}\right)
	\nonumber \\
	& & \times
	\left(\frac{(\create{-k'})^{n^-}}{\sqrt{n^-}[\delta^{(3)}(\vec{k}'-\vec{k}')]^{\frac{n^-}{2}}}\right)
	\ket{0} \\
	&=& \suminf{m^+, m^-, n^+, n^-}\!\!\!\!\!\!\!\!\!\!\!\!
	B_{m^+, m^-, n^+, n^-}(\tau)
	\ket{m^+,2\vec{k}'}\ket{m^-,-2\vec{k}'}\ket{n^+,\vec{k}'}\ket{n^-,\vec{-k}'}.~~~
\end{eqnarray}
Whenever possible, we will use simplified notation such as
\begin{equation}
	\label{eq:inflWavefunc}
	\ket{\psi} = \sum_{\npm,\mpm=0}^\infty B_{\npm,\mpm}(\tau) \ket{\mpm}\ket{\npm}.
\end{equation}

In order to evolve the wavefunction forward in time, we replace $\tau$
with a new variable, $x=-1/(k\tau)$.  The equation of motion is then
found from $i\frac{d}{d\tau}\ket{\psi}= H \ket{\psi}$, left multiplied
by $\bra{\npm, \pm \vec{k}}\bra{\mpm , \pm 2\vec{k}}$. The following
identities are needed to evaluate $H\ket{\psi}$:
\begin{eqnarray}
	\label{eq:inflcreateidentity1}
	\create{k}\uncreate{k}\ket{\mpm}\ket{\npm} &=& [m^+\delta^{(3)}(\vec{k}-2\vec{k}') 
	+ n^+ \delta^{(3)}(\vec{k} -\vec{k}')  \nonumber \\
	& & + m^- \delta^{(3)}(\vec{k}+2\vec{k}') +n^-\delta^{(3)}(\vec{k}+\vec{k}')]
	\ket{\mpm}\ket{\npm} ~~~\\
	\label{eq:inflcreateidentity2}
	(\create{k}\uncreate{k}+\uncreate{k}\create{k})\ket{\psi} &=& 
	(2\create{k}\uncreate{k}+Z)\ket{\psi} \\
	\label{eq:inflcreateidentity3}
	\uncreate{-k}\uncreate{k} \ket{\mpm}\ket{\npm}&=&\sqrt{m^+m^-}(\delta^{(3)}(\vec{k}-2\vec{k}') \nonumber \\
	& &+ \delta^{(3)}(\vec{k}+2\vec{k}'))\ket{\mpm-1}\ket{\npm}  \nonumber\\
	& &  + \sqrt{n^+n^-}(\delta^{(3)}(\vec{k}-\vec{k}') \nonumber \\
	& &+ \delta^{(3)}(\vec{k}+\vec{k}'))\ket{\mpm}\ket{\npm-1} \\
	\label{eq:inflcreateidentity4}
	\create{-k}\create{k} \ket{\mpm}\ket{\npm}&=&\sqrt{(n^++1)(n^-+1)}(\delta^{(3)} (\vec{k}-\vec{k}')\nonumber \\
	& & +\delta^{(3)}(\vec{k}+\vec{k}'))\ket{\mpm}\ket{\npm+1}\nonumber \\
	& &  + \sqrt{(m^++1)(m^-+1)}(\delta^{(3)}(\vec{k}-2\vec{k}') \nonumber \\
	& & +\delta^{(3)}(\vec{k}+2\vec{k}'))\ket{\mpm+1}\ket{\npm} \\
	\label{eq:inflcreateidentity5}
	\uncreate{2k}\create{k}\create{k} \ket{\mpm}\ket{\npm} &=&
	\sqrt{m^+(n^++1)(n^++2)}\delta^{(3)}(2\vec{k}-2\vec{k}')\nonumber \\
	& & \times \ket{m^+-1}\ket{m^-}\ket{n^++2}\ket{n^-} \nonumber \\
	& &
	+\sqrt{m^-(n^-+1)(n^-+2)}\delta^{(3)}(2\vec{k}+2\vec{k}')\nonumber \\
	& & \times \ket{m^+}\ket{m^--1}\ket{n^+}\ket{n^-+2} \\
	\label{eq:inflcreateidentity6}
	\create{2k}\uncreate{k}\uncreate{k} \ket{\mpm}\ket{\npm} &=&
	\sqrt{n^+(n^+-1)(m^++1)}\delta^{(3)}(\vec{k}-\vec{k}')\nonumber \\
	& & \times \ket{m^++1}\ket{m^-}\ket{n^+-2}\ket{n^-} \nonumber \\
	& & +\sqrt{n^-(n^--1)(m^-+1)}\delta^{(3)}(\vec{k}+\vec{k}')\nonumber \\
	& & \times \ket{m^+}\ket{m^-+1}\ket{n^+}\ket{n^--2},
\end{eqnarray}
where $Z=[\uncreate{k},\create{k}]=\delta^{(3)}(\vec{k}-\vec{k})$ is
an infinite constant (related to the renormalization of the vacuum energy).

After some algebra, we find the time evolution of the states is given by 
\begin{eqnarray}
	\label{eq:inflEOM}
	i \frac{d}{dx}A_{\mpm,\npm}(x) &=&
	-\frac{Q}{2}\biggr[\biggr(\sqrt{m^+ m^-}A_{\mpm-1,\npm} 
	+\sqrt{n^+n^-}A_{\mpm,\npm -1}\biggr)e^{-2i\gamma/x}\nonumber \\ 
	& &   +\biggr(\sqrt{(n^+ +1)(n^-+1)}A_{\mpm,\npm+1} \nonumber \\
	& &+\sqrt{(m^++1)(m^-+1)}A_{\mpm+1,\npm}\biggr)e^{2i\gamma/x}\biggr]
	\nonumber \\
	& &     +\frac{\alpha}{x^3}\biggr[
	\biggr(\sqrt{(n^+-1)(n^+)(m^++1)}A_{m^++1,m^-,n^+-2,n^-} \nonumber\\
	& & +   \sqrt{(n^--1)(n^-)(m^-+1)}A_{m^+,m^-+1,n^+,n^--2}\biggr) \nonumber\\
	& &     +
	\biggr(\sqrt{(m^+)(n^++1)(n^++2)}A_{m^+-1,m^-,n^++2,n^-}\nonumber\\  
	& & +\sqrt{(m^-)(n^-+1)(n^-+2)}A_{m^+,m^--1,n^+,n^-+2}\biggr)\biggr],
\end{eqnarray}
where the matrices $A$ and $B$ are related by a phase transformation
\begin{equation}
A_{\mpm,\npm}(x)= e^{-i (m^++m^-+n^++n^-+Z) (\gamma(x)-1)/x}
B_{\mpm,\npm}(x),
\end{equation}
with $\gamma(x)=2+Qx^2$.  The dimensionless constant
$\alpha$ has the value $\frac{\lambda V H}{8 \sqrt{2\pi k \phi}}$.  To
arrive at equation (\ref{eq:inflEOM}), we have ignored terms that
involve modes $\pm \frac{1}{2}\vec{k}$ and $\pm 4\vec{k}$ since we are
not concerned with how these modes evolve for our present purposes.

We begin the simulation for small values of $x$, well before the modes
cross outside the Hubble length.  At such a time, there has been a
negligible amount particle production, so our initial wavefunction is
simply the Fock vacuum, $\ket{\psi}_i = \ket{\mpm=0}\ket{\npm=0}.$ In
the limit of $\frac{k}{a} \ll H$ or $x \ll 1$, this initial condition
corresponds to the Bunch-Davies vacuum.  During
vacuum-energy-domination, the equation of state parameter, $w$, equals $-1$.
Therefore, neglecting the mass of the scalar field, the value of $Q$
is unity.


\subsection{Entanglement Measures}
\label{sec:entanglementmeasures}

Discussions of decoherence rely on the notion of an environment: a
collection of degrees of freedom that interacts and becomes entangled
with the system of interest.  Our model is naturally separated into
modes with different magnitudes of momentum.  Noting that the
entanglement entropy does not depend on our choice of which set of
modes is the environment and which is the system, we identify the
modes with momentum $\pm 2\vec{k}$ with the environmental degrees of
freedom and the modes with momentum $\pm \vec{k}$ to be the system.

This choice represents an entanglement due to coarse graining the
internal degrees of freedom of the scalar field based on scale.  One
can think of the coarse graining as either being due to practical
limitations in the observations that can be made or as physical
limitations such as a mode being entangled with a mode with a
wavelength greater than the horizon size.  The latter case is
discussed in ~\cite{Martineau:2006}.

\begin{figure}[ht]
	\centering
		\includegraphics[width=8.5cm]{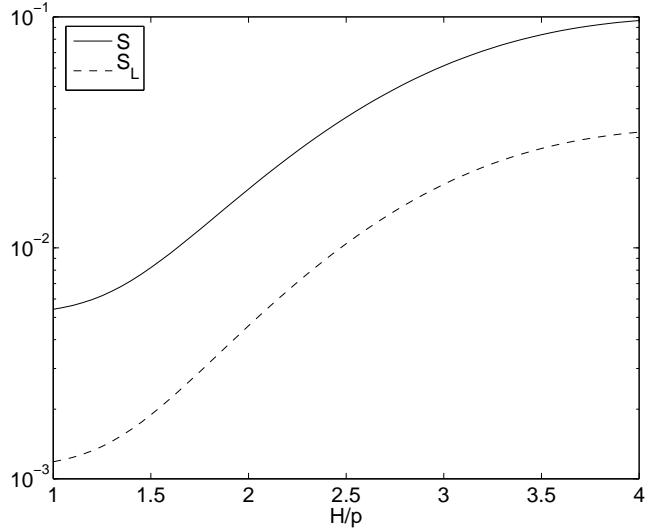}
                \caption[Evolution of entropy and linear entropy]
				{The evolution of entanglement entropy, $S$,
                  and linear entropy, $S_L=|1-{\rm Tr}(\rho^2)|$, as
                  the mode stretches past the horizon for
                  $\alpha=0.2$.  This demonstrates that the
                  non-linearities in the inflaton potential are
                  capable of producing entanglement entropy between
                  the coupled modes.  Also note that $S$ scales
                  monotonically with $S_L$.}
	\label{fig:Sandtrrho2}
\end{figure}

We measure the entanglement between modes using two different
entanglement measures.  The first of these is the entanglement or von
Neumann entropy.  The other is the linear entropy.  While the former
is more common, the latter is easier to compute and scales
monotonically with the entanglement entropy.  Figure
\ref{fig:Sandtrrho2} shows a comparison between these two measures for
$\alpha=0.2$.

The density matrix of the above described system is 
\begin{eqnarray}
	\label{eq:infldensity}
	\rho &=& \ket{\psi}\bra{\psi}  \\
	&=&\!\!\!\!\!\!\!\!
	\sum_{\mpm,\npm,m'^\pm, n'^\pm=0}^\infty\!\!\!\!\!\!\!\!\!\!\!\!\!\!\!
	B_{\mpm,\npm} B_{m'^\pm,n'^\pm}^\dagger
	\ket{\mpm}\ket{\npm}\bra{n'^\pm}\bra{m'^\pm},~~~~~~~
\end{eqnarray}
and we assume that the modes with momentum $2k$ are inaccessible to
measurement.  This gives rise to a reduced density matrix obtained
from tracing over the unobserved degrees of freedom:
\begin{eqnarray}
	\label{eq:inflreddensity}
	\rho_N &=& {\rm Tr}_M \rho = \sum_{m''^\pm =0}^\infty
	\braket{m''^\pm}{\psi}\braket{\psi}{m''^\pm} \\
	\label{eq:inflreddensity2}
	&=& \!\!\!\! \sum_{n'^\pm =0}^\infty \sum_{n^\pm=0}^\infty \left( \sum_{\mpm=0}^\infty
	B_{\mpm,\npm}B_{\mpm,n'^\pm}^\dagger\!\!\right) \ket{\npm}\bra{n'^\pm}.~~~~~
\end{eqnarray}

The von Neumann entropy is then a measure of the entanglement between
the $N$ system and the unobserved $M$ system.
\begin{equation}
	\label{eq:vNentropy}
	S=-{\rm Tr}(\rho_N \ln\rho_N)=-\sum_{i=1}^{N}\rho_i \ln \rho_i,
\end{equation}
where the $\rho_i$'s are the eigenvalues of the reduced density
matrix, $\rho_N$.  A system with a finite Hilbert space spanned by $N$
basis states will have a maximum entropy $S_{\rm max} = \ln{N}$.

The linear entropy, $S_L=\trpsq$, is often used as a stand-in for the
entanglement entropy since it can be computed more easily and in our
case contains the same qualitative information,
\begin{equation}
{\rm Tr}(\rho^2) = \sum_{i=1}^\infty \sum_{j=1}^i 
\begin{cases}
 2|\rho_{i j}|^2 & \text{if $j\ne i$,}
\\
|\rho_{i j}|^2 &\text{if $j=i$.}
\end{cases}
\end{equation} 
A system with a finite Hilbert space spanned by $N$ basis states will 
have a maximum linear entropy $S_{L,{\rm max}} = (N-1)/N$.  

From figure \ref{fig:Sandtrrho2}, we can see that this quantity is
nearly proportional to the entropy.  We will present the results both
in terms of entanglement entropy and $S_L$.

\subsection{Thermal Entropy and Classicality}

The amount of entropy generated can be compared to the entropy of a
thermal system that contains the same average number of particles.
For a thermal system, the entropy is
\begin{equation}
S_{\rm th} = -\sum_{n=1}^\infty \rho_{n,{\rm th}} \ln \rho_{n,{\rm th}}
\end{equation}
where the thermal density matrix is given by

\begin{equation}
\rho_{n,{\rm th}} = \frac{e^{-\beta E_n}}{\sum_{n'=1}^\infty e^{-\beta E_{n'}}}
\end{equation}
and $n'$ labels the Fock states.  Since the energy is $m=n^++n^-$,
each $n'$ state is $m+1$ times degenerate, the partition function can
be written

\begin{equation}
\sum_{n'=1}^\infty e^{-\beta E_{n'}} = \sum_{m=0}^\infty (m+1)e^{-\beta m} = \frac{1}{(e^{-\beta}-1)^2}.
\end{equation}  

\begin{figure}[ht]
	\centering
		\includegraphics[width=8.5cm]{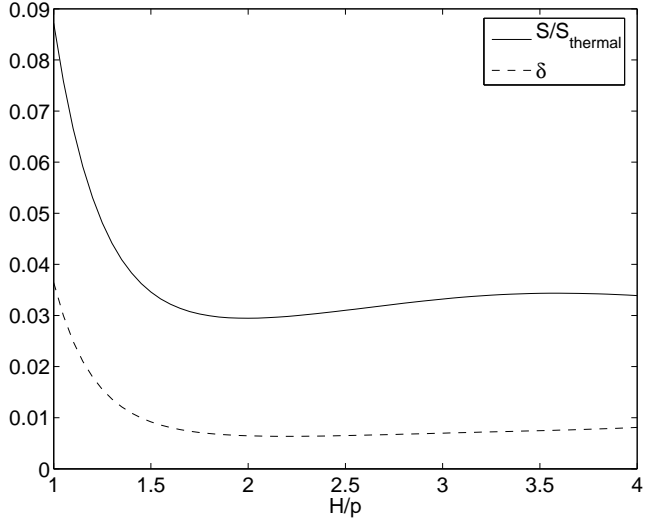}
                \caption[Information lost and the separability parameter]
				{The fraction of information lost due to
                  tracing out the unobserved degrees of freedom,
                  defined by equation (\ref{eqn:Ilost}), and the
                  separability parameter $\delta$ (equation~\ref{eqn:delta2})
                  as the observed mode stretches past the horizon for
                  $\alpha = 0.2$.  By the end of the simulation,
                  $I_{\rm lost}$ appears to have leveled off to a
                  constant few percent.  For a separable (classical)
                  system, we would expect $I_{lost}$ to grow to at
                  least 0.5 to the right of the graph.  If the system
                  were to become classical, we would expect $\delta$
                  to grow to $1$.  However, it too levels off to less
                  than a percent.}
	\label{fig:Ilost}
\end{figure}

Using the relation
\begin{equation}
\label{eq:delta}
\langle n \rangle = \sum_{n'=0}^\infty n \rho_{n',\rm{th}} = \sum_{n=0}^{\infty} n(n+1)e^{-\beta n}(1-e^{-\beta})^2
\end{equation}
we can eliminate $\beta$ for $\langle n\rangle$ using
\begin{equation}
e^{-\beta} = \frac{\langle n\rangle}{2+\langle n\rangle}
\end{equation}
where $\langle n\rangle$ is the average number of particles in the
reduced system.  Finally, we can write the thermal entropy as
\begin{equation}
S_{\rm th}(\langle n \rangle) = -\sum_{m=0}^\infty (m+1)\frac{4\langle n\rangle^m}{(2+\langle n\rangle)^{m+2}}
\ln \left(\frac{4\langle n\rangle^m}{(2+\langle n\rangle)^{m+2}}\right).
\end{equation}
This quantity allows us to compare the entropy generated due to the
coupling with the total energy of a thermal system at the same
temperature.  For example, if the information content of a system is
defined as $I=S_{\rm th}-S$ then the relative information lost from
the system due to the non-linear coupling term is
\begin{equation}
	\label{eqn:Ilost}
I_{\rm lost} = 1-\frac{I}{I_{\rm max}}=\frac{S}{S_{\rm th}}.
\end{equation}  
Figure \ref{fig:Ilost} shows that the rate of information loss due to
the coupling is roughly the same as the rate of particle production.

Another approach for determining the classicality of a quantum system 
is to determine the conditions under which the subsystems can be considered 
separable states. 
Campo and Parentani argue that for Gaussian states at the threshold of
separability and for $\langle n \rangle \gg 1$, the entanglement
entropy between modes will be one half the entropy of the thermal
state~\cite{campo:065044}.  Since our states are not Gaussian, there
is no known general separability condition.  However, from the lack of
growth in the information loss function shown in Figure
\ref{fig:Ilost}, we can see that the Gaussian separability condition
is unlikely to occur as $\langle n \rangle$ grows much larger than $1$
at times greater than can be shown on the figure.  Therefore, these
types of non-linear interactions alone are likely insufficient to
cause the system to appear classical.

Another measure of separability used by Campo and Parentani is the
parameter $\delta$ defined by the equation
\begin{equation}
	\label{eqn:delta2}
   |c|^2 = n(n+1-\delta),
\end{equation}
where $n={\rm Tr}(\rho \create{k} \uncreate{k})=\langle n^+ \rangle$
and $c={\rm Tr}(\rho \uncreate{k} \uncreate{-k})$.  The parameter,
$\delta$, is a measure of the correlations between the $\vec{k}$ and
$-\vec{k}$ modes.  For Gaussian density matrices, it can be shown that
separability occurs when $\delta=1$.  The value of $\delta$ measured
for our model is shown alongside the information loss function in
Figure \ref{fig:Ilost}.  In both cases, the measures flatten out after
the modes leave the horizon and fail to grow as one would need for
non-linearities to explain the classicality of the quantum state.  We
can generalize the definition of $c$ to measure the correlation
between modes of different magnitudes of momenta in our system
\begin{equation}
d \equiv {\rm Tr}(\rho \uncreate{-2k} \uncreate{k} \uncreate{k}).
\label{eqn:d-def}
\end{equation}
Although the interpretation of this quantity or $\delta$ is not as
clear cut as for Gaussian density matrices, we find that both are
useful and convenient tracers of the entanglement entropy.

\subsection{Estimating the Sizes of \texorpdfstring{$\lambda$}{lambda}
 and \texorpdfstring{$x_{\rm final}$}{x-final}}
\label{sec:estim-sizes-lambda}

In order to match our above analysis with reality, we would like to
make order of magnitude estimates for the parameters $\alpha$ in
equation (\ref{eq:inflHam}) and the final value of the $x$ at the end
of inflation, $x_{\rm final}$.

For fluctuations in a scalar field other than the inflaton, the value
of $\lambda$ is essentially arbitrary; however, the gravitational
self-interaction of the field provides a strict lower bound.  Burgess,
et al. \cite{Burgess:2006} give an estimate of this self-interaction,
\begin{eqnarray}
\lambda_g &\approx& \frac{48}{(2
  \epsilon)^{3/2}}\left(\frac{H}{\PlMass}\right)^2  
= \frac{128\pi}{(2
  \epsilon)^{3/2}}  \left(\frac{M}{\PlMass}\right)^4  
\\
&\approx& 6 \times 10^{-16} \left ( \frac{\epsilon}{0.01} \right
)^{-3/2} \left ( \frac{M}{10^{14} {\rm GeV}} \right )^4,
\end{eqnarray}
where $M^4$ is the vacuum energy associated with the scalar field, and 
$\epsilon = \frac{\PlMass^2}{2}\left(\frac{V'}{V}\right)^2$ is a slow-roll parameter 
which may be larger than $1$ if the scalar field is not the inflaton.  We 
have included a possible matter-dominated period following
the end of inflation from scale factor $a_{\rm EI}$ to $a_{\rm RH}$
before reheating and taken $a$ to be the value of scale factor at the
end of inflation.    

The parameter $\alpha$ was introduced in equation (\ref{eq:inflHam}) to 
replace 
\begin{equation}
   \alpha = \frac{\lambda V H}{8 \sqrt{2\pi k \PlMass}}.
\end{equation}
So, if we take, for example, a mode of size $\omega = c k \sim 0.1 {\rm Hz}=
5\times10^{-45}\PlMass$ today, we arrive at an estimate for $\alpha$ due to 
gravitational self-interactions. 
\begin{equation}
	\alpha_g \approx 2 \times 10^{-3} \left ( \frac{\epsilon}{0.01}
        \right )^{-3/2} \left ( \frac{M}{10^{14} {\rm GeV}} \right )^6
        \left ( \frac{\omega}{0.1~{\rm Hz}} \right )^{-1/2}.
\end{equation}  
If the scalar field in question is the inflaton field, the
gravitational self-interaction will dominate over self-coupling
interactions.

The analysis here has assumed that reheating is quick and efficient
\cite{1997PhRvD..56.3258K,2004PhDT.......414Z}, but in principle the
end of inflaton may be followed by a period of matter domination from
scale factor $a_{\rm EI}$ to $a_{\rm RH}$ before reheating.  With this
generalization, the comoving Hubble rate at the end of inflation is
\begin{eqnarray}
a_{\rm EI} H &=&  \left ( \frac{\pi^2}{30} g_r \frac{a_{\rm EI}}{a_{\rm
      RH}}  \right )^{1/4} \left (\frac{8\pi}{3}\right )^{1/2}
\frac{T_0 M}{\PlMass} \\
&=& 6.3 \left (  g_r \frac{a_{\rm EI}}{a_{\rm
      RH}}  \right )^{1/4} \frac{M}{10^{14} {\rm GeV}}   {\rm MHz},
\end{eqnarray}
where $M^4$ is the vacuum energy associated with the inflaton field,
($\approx \lambda \PlMass^4/4$) and $g_r$ is the number of
relativistic degrees of freedom at the end of reheating where the
photon counts as two.  The value of
$x_{\rm final}$ (at the end of inflation) for the comoving scale
$a_{\rm EI} H$ is simply unity and for other scales we have
\begin{equation}
x_\rmscr{final}=\frac{a_{\rm EI} H}{\omega} = 6.3\times 10^7 g_r^{1/4}  \frac{M}{10^{14} {\rm GeV}}
\frac{0.1~\rmmat{Hz}}{\omega}.
\label{eq:21}
\end{equation}
Consequently although the correlations are present on all scales, they
are most obvious on the comoving scale of the Hubble length at the end
of inflation (\ie really small scales).  On these small scales the
density fluctuations are well into the non-linear regime today, but
tensor fluctuations, gravitational waves (GW), would still be a loyal
tracer of these correlations.  Inflationary tensor perturbations were 
first calculated in ~\cite{Starobinsky:1979}.

The expression given in equation (\ref{eq:21}) is very uncertain.  Typically
today's Hubble scale is assumed to pass out through the Hubble length
during inflation after about 50$-$60 $e-$foldings \cite{Liddle:2000};
equation (\ref{eq:21}) gives 56 $e-$foldings before the end, so the
centihertz scale would pass through the Hubble length 12$-$22
$e-$foldings before the end of inflaton.  However, the former number
is highly uncertain.  For example, if inflation occurs at a lower
energy scale or if there is a epoch of late ``thermal inflaton''
\cite{1996PhRvD..53.1784L,1995PhRvL..75..398D,1996NuPhB.458..291D},
the number of $e-$foldings for today's Hubble scale could be as low as
25 \cite{Liddle:2000}, yielding $x_\rmscr{final} \ll\ 1$ for $ck \sim
0.1$~Hz.

Because the simulation increases in complexity as particles are
produced (see figure \ref{fig:Fockvsx}), we are confined to keeping
$x_{\rm final}\sim O(1)$.  So, even though $\alpha$ may be small in
reality, there may be sufficient time during inflation for even a
small non-linearity to produce a great deal of entanglement entropy
because of very large values of $x_{final}$.

\section{Results}

\begin{figure}[ht]
	\centering
		\includegraphics[width=8.5cm]{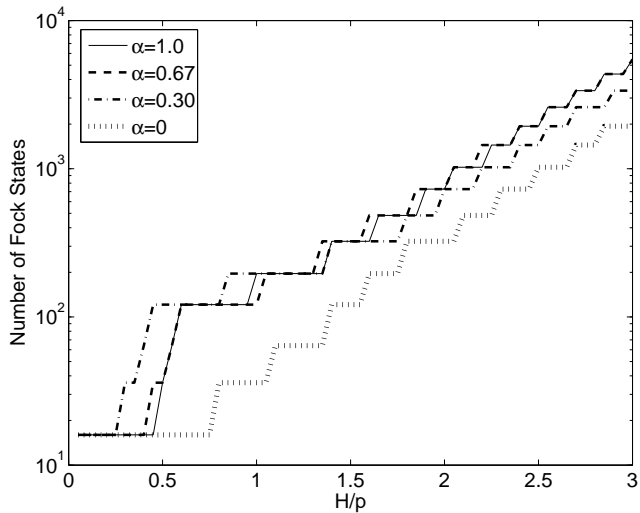}
	\caption{The number of Fock states associated with the reduced system.}
	\label{fig:Fockvsx}
\end{figure}

We would like to investigate how the amount of entropy generated in a
single mode scales with the coupling strength and the duration of
inflation (i.e.\ $\alpha$ and $x_{\rm final}$).  Figure
\ref{fig:Sandtrrho2} explicitly shows the creation of entanglement
entropy for $\alpha=1$ as the universe undergoes its inflationary
phase.  The horizontal axis, $x=H/p$, is the physical size of a mode
with respect to the horizon scale.  The entanglement entropy increases
less quickly than exponentially, which would be a straight line on the
figure.  Unfortunately, as was mentioned previously, the computational
size of the problem prevents us from simulating far past horizon
crossing because the number of particles becomes too large.  Figure
\ref{fig:Fockvsx} shows how many Fock states are in the reduced system
at each time step in the simulation.  The number of states being
integrated is this number to the $3/2$ power, and the number of
entries in the density matrix is the square of this number.

\begin{figure}[ht]
	\centering
        \includegraphics[width=8.5cm]{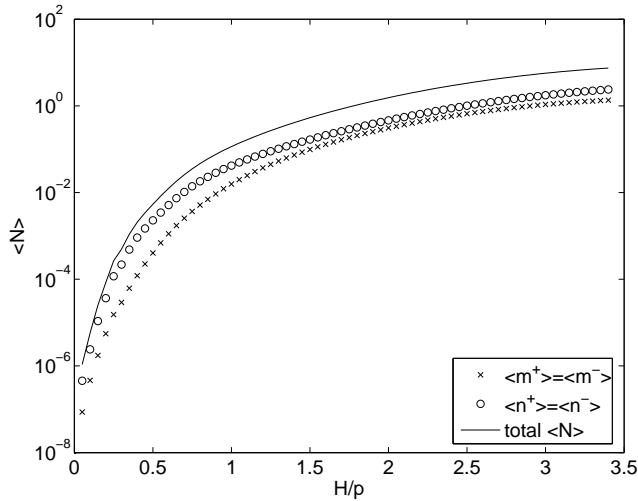}
	\caption[Evolution of particle numbers]
	{Evolution of the average particle numbers for each
          mode for $\alpha=1.0$}
	\label{fig:Nvsx}
\end{figure}

\begin{figure}[ht]
	\centering
		\includegraphics[width=8.5cm]{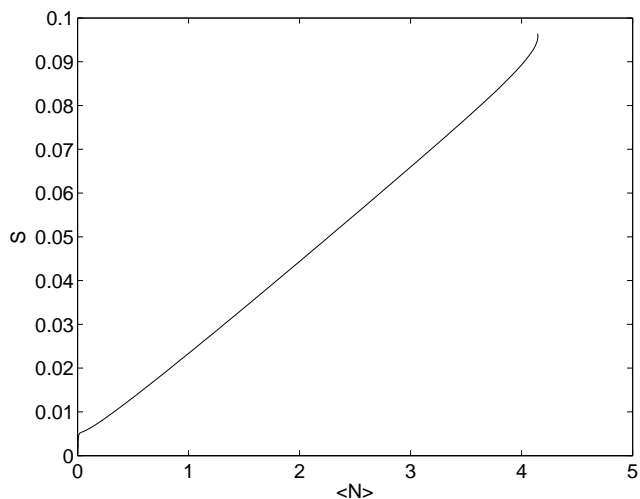}
                \caption[Entropy versus particle number]
				{Entanglement entropy vs. average particle
                  number for $\alpha=0.2$}
	\label{fig:SvsN}
\end{figure}

The evolution of particles in the system is shown in figure
\ref{fig:Nvsx}.  Our results are consistent with those found in
Heyl~\cite{Heyl:2006fb} and show a nearly exponential evolution of the
average particle number.  Moreover, we can look at the evolution of
each mode separately.  For $\lambda=0$, each mode evolves
according to the same equations of motion, and in this case, there is
no difference between the rate that each of the modes evolves.
However, the nature of the interaction between the modes is not
symmetric because the decay of a single $M$ mode particle results in 2
$N$ mode particles and therefore the interaction results in an
increased rate of production of $N$ mode particles, relative to the
$M$ mode.  Figure \ref{fig:SvsN} shows how the entanglement entropy 
scales with average particle number when $\alpha=0.2$.  

\begin{figure}[ht]
	\centering
		\includegraphics[width=8.5cm]{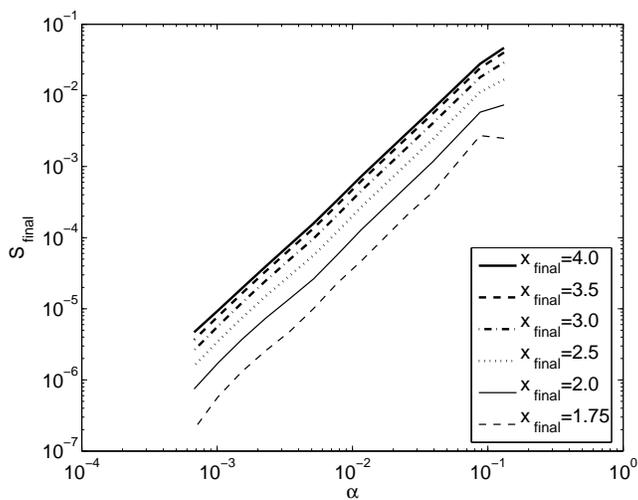}
	\caption[Entropy versus nonlinear coupling]
	{von Neumann entropy vs. $\alpha$ for various values of $x_{\rm final}$}
	\label{fig:bigalphaplot}
\end{figure}

We performed the simulation for a variety of values for the coupling,
$\alpha$, spanning several orders of magnitude.  Figure
\ref{fig:bigalphaplot} shows entropy generation as a function of
$\alpha$ for a variety of inflation durations $x_{\rm final}$.  From
this plot, we can see that $S_{\rm final}$ scales roughly as a power
law in $\alpha$.  Most of the $\alpha$ dependence can be removed by
dividing $S_{\rm final}$ by $\alpha^{1.75}$.  Doing this also helps to
illustrate how $S_{\rm final}$ scales with $x_{\rm final}$.  As
expected, there is no entropy generated without the coupling terms
(i.e.\ when $\alpha=0$).  In this case, there is no communication
between modes of the scalar field and they evolve independently.

\begin{figure}[ht]
	\centering
		\includegraphics[width=8.5cm]{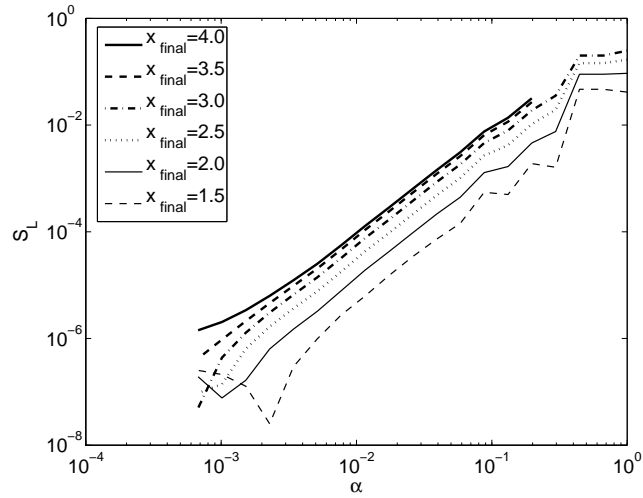}
	\caption[Linear entropy versus nonlinear coupling]
	{$S_L$ vs. $\alpha$ for various values of $x_{\rm final}$}
	\label{fig:bigalphaplotp2}
\end{figure}

As was mentioned earlier, $S_L$ is a useful stand-in for $S$ that can
be computed faster than $S$.  Figure \ref{fig:bigalphaplotp2} echoes
the previous results in terms of $S_L$ instead of $S$.  In this case,
$\trpsq$ scales more like $\alpha^2$ instead of $\alpha^{1.75}$.
However, both $S_L$ and $S$ demonstrate the same qualitative
behaviour.

\begin{figure}[ht]
	\centering
		\includegraphics[width=8.5cm]{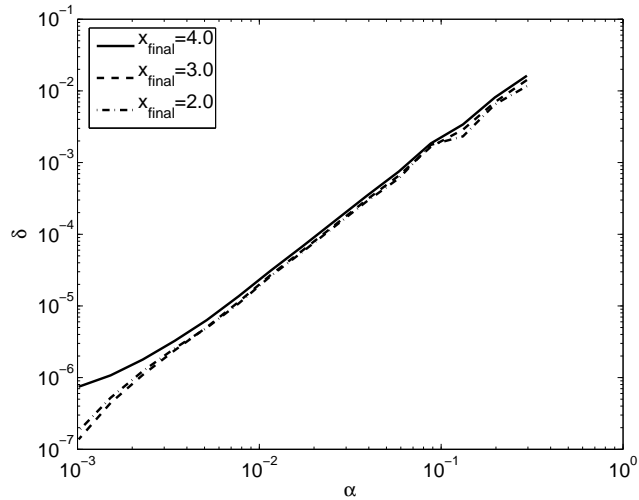}
                \caption[Entanglement versus nonlinear coupling]
				{$\delta$, defined in equation
                  (\ref{eqn:delta2}) scales with $\alpha$ in much the
                  same way as $S_L$, but is less costly to compute.}
	\label{fig:CPdeltavsl}
\end{figure}

\begin{figure}[ht]
	\centering
		\includegraphics[width=8.5cm]{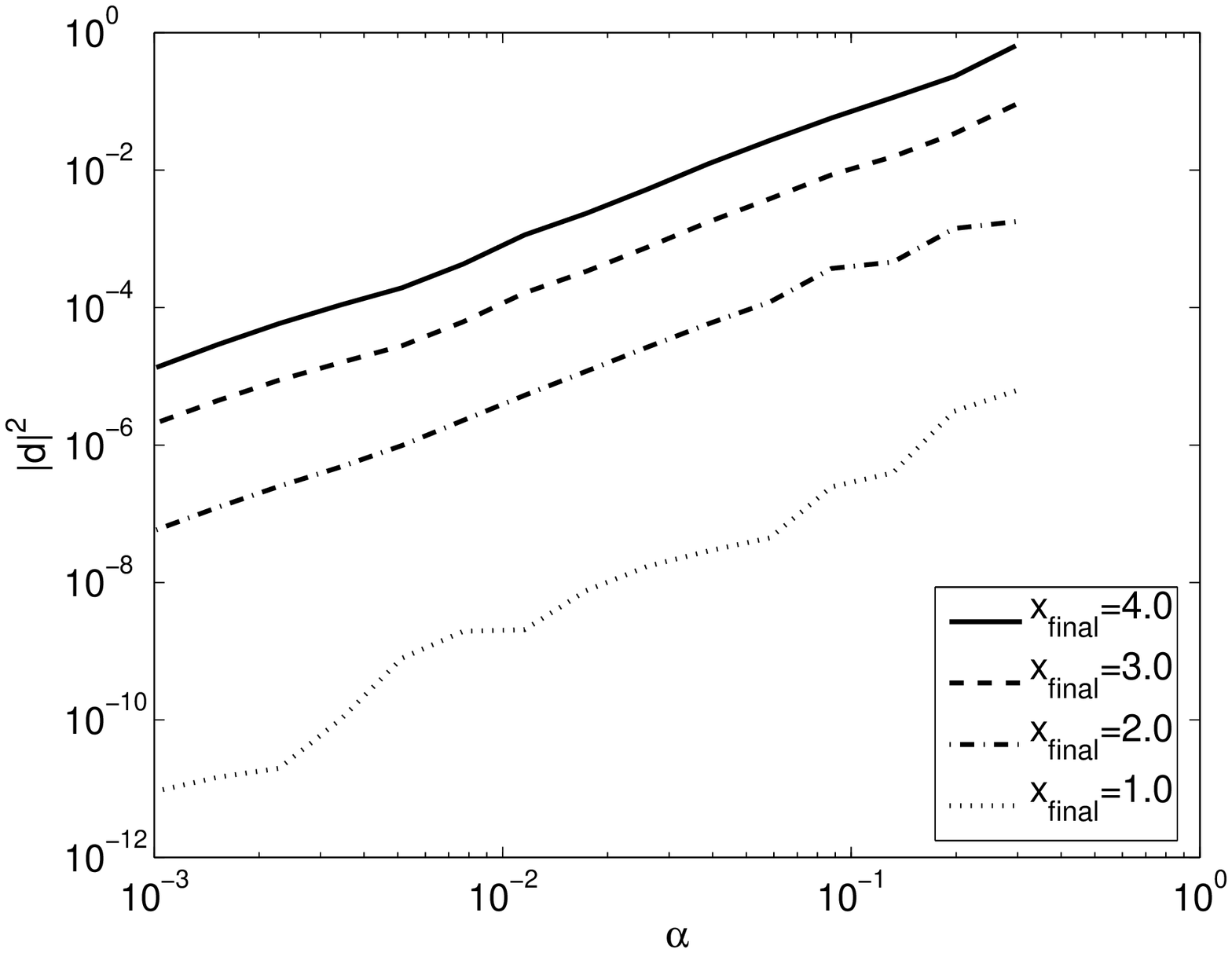}
                \caption[Entanglement versus nonlinear coupling]{$|d|^2 = |{\rm Tr}(\rho
                  \uncreate{2k}\uncreate{k}\uncreate{k})|^2$ scales
                  with $\alpha$ in much the same way as $S$, but is
                  less costly to compute.}
	\label{fig:CPdvsl}
\end{figure}

In addition, we have found other useful stand-ins for the entanglement
entropy that are easier to compute and scale similarly with $\alpha$.
The $\delta$ parameter, defined in (\ref{eqn:delta2}), scales roughly
like an $\alpha^2$ power law much like $S_L$.  Figure
\ref{fig:CPdeltavsl} shows the power law behaviour of this function.
Additionally, if we use a simple measure of correlation between
entangled modes, $|d|^2$ (equation~\ref{eqn:d-def}), we find that its
scales like $\alpha^{1.85}$ (see figure \ref{fig:CPdvsl}), and so can be
a useful stand-in for the von Neumann entropy, $S \propto
\alpha^{1.75}$.

In the real universe, we are dealing with small values of $\lambda$
(and therefore of $\alpha$)
and very large values of $x$.  However, the simulation outlined in
this chapter is limited because its computational complexity increases
dramatically as particles are produced, even for small values of the
coupling, $\alpha$.  Moreover, for small values of $\alpha$, the
production of entropy is too small to be meaningful.  While the
dependence of $S$ on $\alpha$ nearly follows a power law, there is no
simple relation describing the dependence of $S$ on $x_{\rm final}$.
The value $S_L$ is approximately proportional to $\alpha^2 x_{\rm final}^3$
over a wide range of $\alpha$ and the modest range $x$ probed by the
simulations;
therefore, very roughly, we can write the scaling law as
$S_L \propto \alpha^{2} x_{\rm final}^3$ where
\begin{equation}
S_L \approx 10^{16} g_r^{3/4} \left ( \frac{M}{10^{14} {\rm
      GeV}} \right )^{15} \left ( \frac{\omega}{0.1~{\rm Hz}} \right
  )^{-4} .
\end{equation}
Of course, only values of $S_L$ less than unity make sense, so a
larger value from the fitting formula indicates that $S_L$ is very
close to one.  However, a value of $S_L < 1$ is obtained
by lowering the mass scale of inflation below
\begin{equation}
M < 8\times 10^{12} \left ( \frac{\omega}{0.1~{\rm Hz}} \right
)^{-4/15} 
{\rm GeV};
\end{equation}
therefore, if the energy scale of inflation is low, the quantum states
of fluctuations at $\omega \sim 0.1$~Hz will remain coherent despite
the non-linear coupling.


The simulation was checked for consistency in several ways.  First, we
traced the probability throughout the simulation measured both by the
sum of squares of the matrix elements $\sum_{\mpm,\npm = 0}^\infty
A_{\mpm,\npm}$ and the trace of the density operator.  
Both of these quantities were conserved
to a few parts in $10^{-7}$.  Moreover, we estimated the level of
numerical error by rerunning the simulation with a variety of phase
rotations multiplying the initial wavefunction.  The standard
deviation of the results from these numerical changes in the initial
conditions give us an idea of the level of numerical error in the
simulation, which were typically at the level of one part per
thousand.


\section{Conclusions}

In this chapter we have developed a model in which two modes of a scalar
field evolve during inflation and we have computed the entanglement
entropy between them.  The entanglement entropy generated between
observed and unobserved modes in the inflaton field give the
appearance that entropy is being produced, even though the scalar
field remains in an overall pure state.  The preceding results clearly
show that non-linearities in the inflaton potential give rise to a
generation of entanglement entropy between observed modes and
unobserved modes in a scalar field during inflation.  This entropy is
an additional source to that caused by coupling to external degrees of
freedom \cite{Kiefer:2006je}, entanglement between the inside and
outside of the horizon \cite{Sharman2007} and that which is created
during reheating after inflation has ended.

We have attempted to extrapolate the results of our simulation to the
real universe.  The relevant parameters determining the amount of
entropy generated via non-linearities are the strength of the coupling
$\alpha_g \sim 10^{-3}$ and the scale of the fluctuation at the end of
inflation given by the dimensionless parameter $x_{\rm final} \sim
10^{7}$.  The entanglement entropy was found to scale like
$\alpha^{1.75}$ for a fixed $x_{\rm final}$.  The dependence of $S_L$
on $x_{\rm final}$ for a given value of $\alpha$ is not as
straightforward, but $S_L \propto x_{\rm final}^3$ over a short range
of $x_{\rm final}$ values.  Based on these rough scaling patterns, we
estimate that non-linearities due to gravity and inflaton
self-coupling are insufficient to decohere modes that spend only a
few Hubble times at super-horizon scales.  In particular, if the
energy scale of inflaton is less than $10^{13}$~GeV, fluctuations 
at about 0.1~Hz may remain coherent.

We found two measures of the decoherence related to the correlations
between modes of different momenta provide a faithful estimate of the
entanglement entropy in our model --- one of these measures is new to
this work ($d$) and specifically probes the non-linear coupling
between modes.  In particular these estimates are very inexpensive to
calculate as compared to the von Neumann entropy and should prove
useful for more detailed models of entropy generation.

It is usually assumed that the main contribution to the entropy
observed in the density perturbations is generated during reheating,
when the inflaton decays.  However, the analysis demonstrates that
entropy can be generated independently of reheating, provided there is
even a small non-linearity in the scalar potential; therefore, the
results are applicable to scalar fields that do not participate in
reheating.  For example, the gravitational wave background can be
treated as a pair of scalar fields, so even tensor fluctuations may
contribute to the entropy and the classicality of the distribution of
density perturbations in this way and observations of the
gravitational wave background at high frequency could reveal the
quantum mechanical origin of density fluctuations.

\part{Electromagnetic Waves Near Magnetars}
\label{pt:EMwaves}
\chapter{The Magnetized Vacuum}
\label{ch:magnetizedvacuum}
\acresetall

\begin{summary}
In magnetic fields near or above the critical field strength, $B_k = 4.4\times 10^{13}$ Gauss, 
the Larmor radius of the electron becomes comparable to its Compton wavelength. 
In this regime, the vacuum fluctuations of \ac{QED} interact significantly 
with the background magnetic field, altering the physical properties of the vacuum. 
We may take these effects into account by defining a new classical theory with a 
Lagrangian that effectively accounts for the quantum fluctuations by averaging over 
them. The resulting theory describes the quantum vacuum with a strong background 
field as a nonlinear optical medium. In this chapter, I review the history and 
theory behind this idea.
\end{summary}

The purpose of this chapter is to introduce the interesting physics behind 
magnetic fields near to and exceeding a critical value determined by the 
electron mass and electric charge, $B_k = \frac{m^2}{e} = 4.4 \times 10^{13}$ Gauss.
Magnetic fields have important physical effects even at the scale of Earth's weak 
magnetic field, which is typically less than one Gauss. Here on Earth, the magnetic 
field is strong enough to magnetize ferromagnetic materials. This effect is used 
for navigation in both human technologies and by some animals, 
for example~\cite{magnetoreception}. Large 
fields used in research labs (for \ac{MRI}, for example) 
are on the order of $10^4$ Gauss~\cite{wikiMRI}.
The quantum critical field is well beyond the strength of fields that we can achieve 
in labs. Fortunately, extreme magnetic fields in excess of the critical field strength 
are believed to exist around neutron stars. The emissions from these neutron stars 
are strongly influenced by their incredible magnetic fields.

Quantum effects can alter the structure 
of the vacuum when we include a background magnetic field near the quantum critical 
field strength.
For example, in two-dimensional \ac{QED}$_{2+1}$, 
a classically stable uniform magnetic field is unstable to the formation 
of inhomogeneities when QED effects are taken into account~\cite{PhysRevD.52.R3163}.
This result motivates the question of whether \ac{QED}$_{3+1}$ also prefers more interesting 
structures over a strong, uniform magnetic field.  If so, what might these structures look like?

Here, we wish to study these questions by looking at the effective classical theory 
that is obtained by averaging over quantum effects to the one-loop level. 
This picture results in an effective action of \ac{QED} where the quantum effects appear 
as correction terms to the familiar classical Maxwell action. These correction terms 
destroy the linearity of Maxwell's equations, and allow light waves to interact 
with one another. So, the quantum vacuum in the presence of 
a strong electromagnetic field behaves like a non-linear optical medium that may be 
capable of supporting novel electromagnetic structures like solitons.

Solitons are local wave excitations that can travel undisturbed for considerable distances. 
They commonly appear in a wide variety of wave equations displaying 
 nonlinear and dispersive behavior. Electromagnetic solitons are known to 
exist in certain nonlinear optical 
materials where they are called light bullets~\cite{2005opso.book.....T}.
 It is therefore possible that solitons can be found in the systems similar to the magnetar 
magnetosphere model described above, where there is a strong magnetic background field and 
a plasma.
Recently, perturbative methods revealed electromagnetic 
waves in a magnetized plasma with slowly varying 
soliton solutions~\cite{1979JETP...49...75K, cattaert:012319}.

\section{Strong-Field \texorpdfstring{\acs{QED}}{QED}}

Research into the interactions between the electronic vacuum and external 
electromagnetic fields predates modern \ac{QFT} and was among the earliest 
applications of \ac{QED}. In 1936, Heisenberg and Euler~\cite{heisenberg-1936-98} and 
Weisskopf~\cite{weisskopf1936kongelige} independently derived the effective action due to 
electronic fluctuations from electron-hole theory. In 1951, Schwinger re-derived 
the result from the new theory of \ac{QED}~\cite{Schwinger:1951}. The 1960s and 
early 1970s saw considerable progress in understanding the one-loop corrections 
to the classical action in \ac{QED}~\cite{PhysRev.136.B1540, RevModPhys.38.626, 
Adle71, doi:10.1139/p88-114, greiner2003quantum}.

\ac{QED} is the quantum field theory describing the interactions between 
electrons (and positrons) and light.  The electrons are described by a Dirac spinor 
field, $\psi(x)$, and the photon field is the vector field, $A_\mu(x)$. 
In each case, the quantized fluctuations of the field are identified with 
the corresponding particles.
The Lagrangian is~\cite{weinberg1966quantum, peskin1995introduction, greiner2003quantum}
\be
	\lag = \bar{\psi}(i\slashed{\partial} + e\slashed{A} -m)\psi 
	-\frac{1}{4}F_{\mu \nu}F^{\mu \nu}.
\ee
The Feynman slash indicates a vector-index contraction with the Dirac 
gamma matrices, $\slashed{A} \equiv \gamma^\mu A_\mu$ and $\bar{\psi}$ is
defined by $\bar{\psi} \equiv \psi^{\dagger}\gamma^0$. 
We may incorporate the effects of a classical background 
electromagnetic field, $\tilde{A}_\mu^0(\vec{q})$, through 
a new interaction with the Feynman rule

\be
	-ie\gamma^\mu \tilde{A}_\mu^0(\vec{q}).
\ee
This interaction dresses all of the fermion propagators, including internal 
lines.

The background field may be thought of as an average over all 
possible quantum fluctuations of a quantum field composed of a large 
number of photons. 
For weak background fields, the new interaction may be treated perturbatively.
However, when the fields exceed $\frac{1}{2}\frac{m^2}{e}\approx 2\times 10^{13}\mbox{G}$, 
the perturbation series fails to converge.

\section{Neutron Stars and Extreme Magnetism}

\begin{figure}[h]
	\centering
		\includegraphics[width=10.5cm]{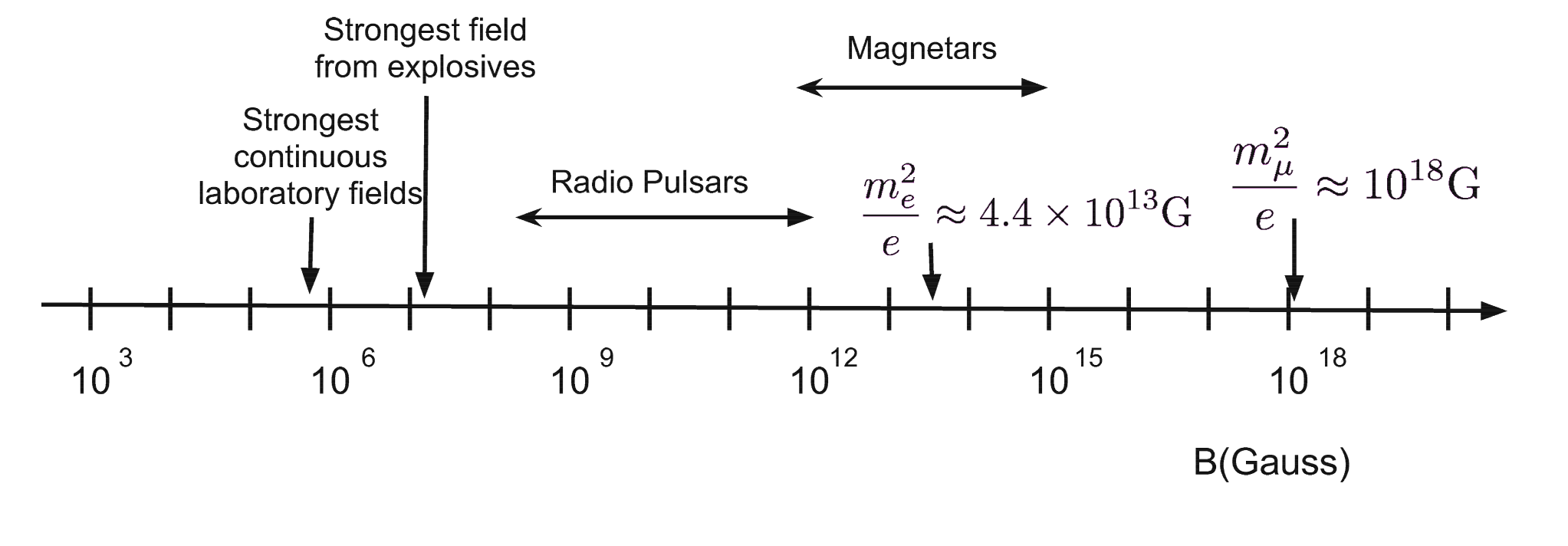}
                \caption[Magnetic field orders of magnitude]
				{Orders of magnitudes of magnetic fields 
				in terrestrial experiments and astrophysical objects.}
	\label{fig:MagneticFieldOOM}
\end{figure}

The extreme strengths of the quantum critical fields make effects from the 
quantum vacuum difficult to probe in terrestrial experiments.  Very large 
magnetic fields up to $10^{19}$G can be 
created in heavy ion collisions~\cite{PhysRevLett.36.517, 2009IJMPA..24.5925S}. 
Unfortunately, experimental access to these fields is limited 
since they only exist in microscopic volumes. In general, 
the largest macroscopic, continuous magnetic fields that can be created in the lab are 
on the order of tens of Teslas, or tens of thousands of Gauss
~\cite{Patten1999}.  Pulsed magnetic fields of several kiloteslas can 
be produced with the use of explosives~\cite{823621}.  However, these fields are
less than a millionth the strength of the \ac{QED} 
critical field (see figure \ref{fig:MagneticFieldOOM}).  Luckily, 
nature occasionally provides us with very strong astrophysical magnets.

A few times per century in our galaxy, a massive star will reach the end of its 
nuclear fuel supply and will no longer have sufficient thermal pressure to 
support itself against gravitational collapse.  The energy from the collapse 
may blow away the outer layers of the star in a supernova explosion, leaving 
behind an ultradense core.  
In most of these supernova explosions, the remaining
core becomes a neutron star, supported from further collapse by the degeneracy pressure 
of its nucleons, and often with an intense magnetic field, approaching or 
exceeding the quantum critical field.  Therefore, our galaxy provides 
us with real astrophysical laboratories for exploring the physics of the magnetized vacuum.

The idea of a neutron star was first proposed by Landau in 1932~\cite{landau32}, 
and in 1934, Baade and
Zwicky suggested that one could result from the supernovae of a star with a massive iron 
core exceeding the Chandrasekhar mass~\cite{baade34}. In this case, the gravitational pressure of 
the collapsing core would exceed the electron degeneracy pressure, causing a collapse 
to incredible densities until the neucleon degeneracy pressure eventually stabilized the star.

\begin{figure}[h]
	\centering
		\includegraphics[width=10.5cm]{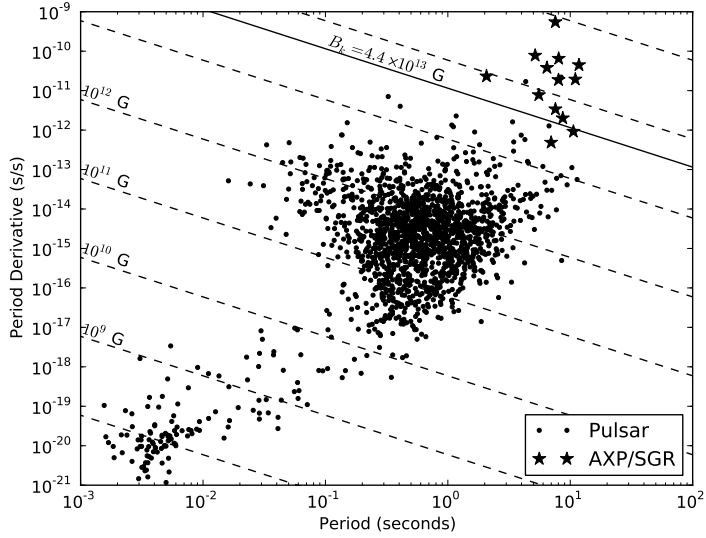}
                \caption[Pulsar period-period derivative diagram]
				{The $P$-$\dot{P}$ diagram for the 
				$\sim$1800 known pulsars. The magnetars (\acp{AXP} and \acp{SGR})
				occupy 
				the top right region of the diagram and are labelled with stars. 
				The magnetic fields indicated by equation (\ref{eqn:pulsarB})
				are shown with diagonal dotted lines, and the line representing the 
				critical field is shown as a solid line.
				This data was collected from the 
				ATNF Pulsar Catalogue\cite{ATNFDatabase, 2005AJ....129.1993M}}.
	\label{fig:ppdot}
\end{figure}

In 1968, Bell and Hewish observed an unusual steadily-pulsing astrophysical radio signal, 
marking the discovery of a new class of stars called `pulsars'~\cite{1968Natur.217..709H}. 
Pulsars were quickly identified as strongly magnetized neutron stars by Pacini~\cite{1967Natur.216..567P}
and Gold~\cite{1968Natur.218..731G}. Soon afterwards, this link was solidified with 
strong observational evidence when a pulsar was discovered near the Crab Nebula
\cite{Staelin27121968}.
The radio pulsations from the new class of stars
were explained by relativistic plasma velocities in the magnetosphere leading to 
charged particles emitting acceleration radiation beamed along the magnetic axis. 
When the magnetic axis and the rotation axis are misaligned, the result is a lighthouse beacon 
of radio emission consistent with the observations\footnote{A review of the emission processes can 
be found in Lyne and Graham-Smith~\cite{lyne2006pulsar}.}. 
To date, more than 2000 pulsars have been identified~\cite{2005AJ....129.1993M} (see figure \ref{fig:ppdot}).

The incredible surface magnetic field strengths of neutron stars arises partly because
the magnetic flux flowing through the surface of the progenitor star becomes 
frozen in the core as it collapses.  The mass-radius ($m-R$) relation for a neutron
star supported by non-relativistic degeneracy pressure is 

\be
	R \approx 4.5\frac{\hbar^2}{G m_p^{8/3}} M^{-1/3}.
\ee
Putting the canonical neutron star mass of $M\approx 1.4 M_\sun$ ~\cite{thorsett-1999-512}
into this expression 
yields an expected radius of $R \sim 1.2 \times 10^4\mbox{m}$, roughly the 
size of a city.  During the collapse to this astronomically tiny radius, 
the star is strongly ionized with free electrons and protons 
and is nearly a perfect conductor.  
The magnetic field lines are frozen in the star.  
The constant flux condition requires

\be
	\Phi = \int \vec{B} \cdot d \vec{A} \sim \vec{B} \cdot \vec{A} = {\rm constant.}
\ee
So, when the radius shrinks from a typical solar radius of $10^9$m to 
$10^4$m, the magnetic field strength at the surface is amplified by 
10 orders of magnitude.  Through this effect alone, neutron stars 
are expected to have magnetic fields of $10^{12}-10^{13}$ Gauss.

The magnetic fields of astrophysical neutron stars can be 
inferred from observations of pulsars.  The most common 
way of doing this is by equating the observed spin-down 
power with magnetic dipole radiation~\cite{melia2009high}.  For example, 
consider a crude model in which the pulsar is a rigidly rotating sphere with a 
dipole magnetic field.
For a pulsar with mass $M$, radius $R$, period $P$, 
and period derivative, $\dot{P}$, we have

\be
	\dot{E}_{{\rm spin-down}} = -\frac{8}{5} \pi^2 MR^2 \frac{\dot{P}}{P^3}.
\ee
This spin-down energy loss is a consequence of the energy radiated away as dipole radiation,
\be
	\dot{E}_{{\rm dipole}} = -\frac{8 \pi^4}{3c^3}\frac{B_0^2 R^6}{P^4}.
\ee
Equating these expressions gives an expression for the surface magnetic field in terms of
$P$ and $\dot{P}$
\be
	\label{eqn:pulsarB}
	B_0 = \left(1.3 \times 10^{19}~\frac{\rm G}{\sqrt{\rm s}}\right)
		\sqrt{P \dot{P}}.
\ee

The magnetic fields inferred from these measurements are largely 
consistent with the estimate we made above (see figure \ref{fig:ppdot}).
However, there 
are two related classes of pulsars which appear to have even 
larger magnetic fields.  These are \acp{AXP} and \acp{SGR}, 
the so-called magnetars.  
These pulsars generally have periods and period derivatives 
which imply magnetic fields in the range $10^{14}-10^{16}$ Gauss.

A favoured explanation for such intense fields is that they 
are generated by a convective dynamo mechanism in the first 
few seconds of the proto-neutron star's life~\cite{1993ApJ...408..194T}. 
The neutron star fluid is a very good conductor as it contains 
free protons and electrons.  If the star is born rotating very 
quickly, and is differentially rotating, 
the magnetic field lines can be dragged through the 
conductive neutron star fluid in convection currents.  The magnetic 
field is built up through this dragging process.  This is 
similar to the dynamo mechanisms which generate magnetic fields 
in the Earth and Sun.  However, if it works efficiently in 
newborn neutron-stars, it can generate magnetic fields of up to 
$10^{16}$ Gauss~\cite{1993ApJ...408..194T}.

The inferred magnetic field strengths can also be checked for 
consistency against other lines of observational evidence
(for a review, see \cite{2006csxs.book..547W, 2006RPPh...69.2631H}).  
In the cases of some pulsars, spectral features can be seen 
which imply large magnetic fields ($10^{13}-10^{15}~{\rm G}$) 
if they are interpreted as proton cyclotron resonances.
For a few X-Ray pulsars, electron cyclotron spectral 
features have been detected implying magnetic fields 
up to $10^{13}$ G.  There are strong theoretical arguments 
that the emissions from \acp{SGR} and \acp{AXP} are likely 
powered by fields exceeding the quantum critical field
\cite{1995MNRAS.275..255T, 1996ApJ...473..322T}.  Similarly, 
the giant flares and bursts from \acp{SGR} have been 
argued to be consistent with highly-magnetized stars
\cite{2001ApJ...561..980T, 2006csxs.book..547W}

The McGill \acs{AXP}/\acs{SGR} catalog
\cite{McGillCatalog}
lists 23 confirmed or candidate magnetars at the time of writing: 
11 \acp{SGR} (7 confirmed, 4 candidates), and 12 \acp{AXP} (9 confirmed, 3 candidates).
A thorough, recent review of the physics of highly magnetized 
neutron stars can be found in \cite{2006RPPh...69.2631H}.

\section{The \texorpdfstring{\acs{QED}}{QED} Effective Action}
\label{sec:qedea}
The effective action can be viewed as a quantum-corrected expression of 
the classical action which averages over all possible quantum fluctuations 
of the field.  Thus, it provides a means of interpolating between the 
classical and quantum regimes.  The quantum correction terms 
destroy the linearity of Maxwell's equations, so the quantum vacuum 
state in the presence of external fields resembles a non-linear optical 
medium which can be polarized and magnetized. In this section, we will 
derive the effective action of \acs{QED}\footnote{This derivation follows 
sections 11.3 and 11.4 of Peskin and Schroeder~\cite{peskin1995introduction} 
and chapter 16 of Weinberg~\cite{weinberg1966quantum}, but has been made specific 
to the case of \ac{QED}.}.

Consider a function, $E[J^\mu, \bar{\eta}, \eta]$, representing the energy of the vacuum 
as a function of external sources, $J^\mu$, $\bar{\eta}$, and $\eta$. The sources $\bar{\eta}$
and $\eta$ are anti-commuting (Grassman) variables. We may use this 
functional to express a partition functional:

\ba
Z[J^\mu, \bar{\eta}, \eta] &=& e^{-iE[J^\mu, \bar{\eta}, \eta]/\hbar} \\
  & =&\int \mathcal{D}A_\mu \mathcal{D}\psi \mathcal{D} \bar{\psi}
  \exp\left( \frac{i}{\hbar}\int d^4x(\mathcal{L} + J^\mu A_\mu + \bar{\eta}\psi
  +\bar{\psi}\eta)\right).~~~
  \label{eqn:qedpartition}
\ea
There is a strong analogy with statistical mechanics.  Here, the 
energy functional is the analog of the Helmholtz free energy.  We 
would like to define the effective action, $\Gamma[A_\mu^0]$,
so that stable quantum 
states of the theory are solutions to

\be
	\frac{\delta \Gamma[A_\mu^0]}{\delta A_\mu^0(x)}\biggr|_{J^\mu=0} =0
\ee
where $A_\mu^0(x)=\bra{0}A_\mu(x)\ket{0}$ 
is a weighted average of the field configuration over 
all possible quantum fluctuations.  We refer to this field as a classical 
field.

This problem is analogous to finding the most probable thermodynamic 
state in a thermally fluctuating background.  This state is a minimum 
of the Gibbs free energy, $G$, which is related to the Helmholtz free 
energy, $F$, by a Legendre transformation

\be
	G=F-V\pderiv{F}{V}\biggr|_T = F+PV.
\ee
Analogously, the effective action is defined in terms of a Legendre 
transformation of the energy functional

\be
	\label{eqn:LegendreTrans}
	\Gamma[A_\mu^0, \bar{\psi}^0, \psi^0] = -E[J^\mu, \bar{\eta},\eta]
	-\int d^4 y (J^\mu(y)A_\mu^0(y) + \bar{\eta}(y)\psi^0(y) +\bar{\psi}^0(y)\eta(y)).
\ee

We compute the energy functional by expanding the fields about their classical values (\ie
$A_\mu(x) = A_\mu^0(x) + \Delta A_\mu(x)$):

\ba
E[J_\mu, \bar{\eta},\eta] & = &\int d^4x\left(\lag[A_\mu,\bar{\psi},\psi] +J^\mu A_\mu + \bar{\eta}\psi + \bar{\psi}\eta \right) \\
   & = & \int d^4x\left(\lag[A_\mu^0,\bar{\psi}^0,\psi^0] +J^\mu A^0_\mu + \bar{\eta}\psi^0 + \bar{\psi}^0\eta \right) \nonumber \\
   &  & +\int d^4x \biggr[ \Delta A_\mu(x) \left(\frac{\delta \lag}{\delta A_\mu(x)} + J^\mu \right) \nonumber \\
   & & + \left(\frac{\delta \lag}{\delta \psi(x)} + \bar{\eta}\right)\Delta \psi(x)
   + \Delta \bar{\psi}(x)\left(\frac{\delta \lag}{\delta \bar{\psi}(x)} + \eta\right)\biggr] \nonumber \\ 
   & & +\frac{1}{2}\int d^4x d^4y \biggr[ \Delta A_\mu(x)\Delta A_\nu(y) 
   \frac{\delta^2 \lag}{\delta A_\mu(x) \delta A_\nu(y)} \nonumber \\
   & & + \Delta \bar{\psi}(x) \frac{\delta^2 \lag}{\delta \bar{\psi}(x) \delta \psi(y)}
   \Delta \psi(y) + \Delta A_\mu(x) \frac{\delta^2 \lag}{\delta A_\mu(x) \delta \psi(x)}
   \Delta \psi(x) \nonumber \\
   & & +\Delta \bar{\psi}(x)\frac{\delta^2 \lag}{ \delta \bar{\psi}(x)\delta A_\mu(x)}
   \Delta A_\mu(x) \biggr]  + ...
\ea

In order to have an effective action which is independent of the external currents, we 
must find a relationship between the currents and the classical fields.  Here we will 
promote the result from lowest order perturbation theory to a requirement connecting 
the currents $J^\mu$, $\eta$, and $\bar{\eta}$ to the classical fields 
$A_\mu$, $\bar{\psi}$ and $\psi$.  That is, the fields and currents must obey 
the classical field equations.  For example,

\be
	\label{eqn:fieldeqns}
	\frac{\delta \lag}{\delta A_\mu}\biggr|_{A_\mu = A_\mu^0} + J_\mu =
	\bar{\psi}e \gamma^\mu \psi + J^\mu - \partial_\nu F^{\nu \mu} = 0.
\ee
We may imagine this step as replacing the currents in the energy functional with 
whatever currents are required to satisfy (\ref{eqn:fieldeqns}) exactly, with the 
relationship between the two currents being determined order-by-order in perturbation theory.
In this case, the first-order derivative terms vanish.  If we terminate the series 
at the second-order derivative terms (the one-loop approximation), the energy functional 
is a Gaussian integral, 
which we can evaluate by treating the functional derivatives as infinite dimensional matrices 
in a field-configuration space:

\ba
E[J_\mu, \bar{\eta},\eta] & = & \int d^4x\left(\lag^0 +J^\mu A^0_\mu + \bar{\eta}\psi^0 + \bar{\psi}^0\eta \right) \nonumber \\
   & & +i \hbar {\rm ln} \int \mathcal{D}A_\mu\mathcal{D}\psi \mathcal{D} \bar{\psi}
   \exp\frac{i}{2\hbar} \int d^4x d^4y\biggr[\nonumber \\
   & &   \Delta A_\mu(x)\Delta A_\nu(y) 
   \frac{\delta^2 \lag}{\delta A_\mu(x) \delta A_\nu(y)} \nonumber \\
   & & + \Delta \bar{\psi}(x) \frac{\delta^2 \lag}{\delta \bar{\psi}(x) \delta \psi(y)}
   \Delta \psi(y) \biggr]  \\
   &=& -\int d^4x\left(-\frac{1}{4}F^0_{\mu \nu} F^{0,\mu \nu} + J^\mu A_\mu^0\right)
   \nonumber \\
   & &-\frac{i\hbar}{2}{\rm ln~ Det}\left[-\frac{\delta^2 \lag}
	{\delta A_\mu(x) \delta A_\nu(y)}\right] \nonumber \\
	& &+\frac{i\hbar}{2}{\rm ln~ Det}\left[-\frac{\delta^2 \lag}
	{\delta \bar{\psi}(x) \delta \psi(y)}\right]
	+{\rm Constant~Terms}.
\ea
The different signs in front of the 
functional determinants in the 
above equation arise due to the difference between Gaussian 
integration involving Grassman versus standard complex variables:
\be
	{\rm Grassman:} ~\int d\vec{\eta} d\vec{\psi} 
	\exp(-\eta_\mu^* \psi_\nu b^{\mu \nu}) = {\rm det} (b);
\ee
\be
	{\rm Standard:} ~\frac{1}{\pi}\int d\vec{\theta} d\vec{\phi} \exp(-\theta_\mu \phi_\nu b^{\mu \nu}) 
	= \frac{1}{{\rm det}(b)}.
\ee

The Legendre transformation, (\ref{eqn:LegendreTrans}), 
eliminates the $J^\mu$ dependent terms and we are left with

\ba
	\Gamma[A_\mu^0] &= &\int d^4x\left(-\frac{1}{4}F^0_{\mu \nu} F^{0,\mu \nu}\right)
	+\frac{i \hbar}{2} {\rm ln~ Det}\left[-\frac{\delta^2 \lag}
	{\delta A_\mu(x) \delta A_\nu(y)}\right] \nonumber \\
	& &-\frac{i \hbar}{2} {\rm ln~ Det}\left[-\frac{\delta^2 \lag}
	{\delta \bar{\psi}(x) \delta \psi(y)}\right]
	+{\rm Constant~Terms}.
\ea

The functional determinants in the above expression are divergent, so we 
must renormalize the expression.  We therefore require that the effective 
action vanishes when the classical action vanishes.  So, we subtract off 
two terms corresponding to the functional determinants evaluated at $A_\mu^0=0$.
The term arising due to the photon field then vanishes at the one-loop order since the 
photon fluctuations do not interact with the background field except through the fermion 
loop. We are left with

\be
	\Gamma[A_\mu^0] = \int d^4x\left(-\frac{1}{4}F^0_{\mu \nu} F^{0, \mu \nu}\right)
	-i \hbar {\rm ln~Det}\left[\frac{\slashed{p}+e\slashed{A}^0
	-m}{\slashed{p}-m}\right].
\ee

We may put the one-loop effective action into a more manageable form by using the 
linear algebra result, ${\rm ln~Det} A = {\rm Tr~ln}A$.  Additionally, we may 
simplify the operator using invariance under the charge conjugation operator, 
$C(\slashed{p} + e\slashed{A}^0)C^{-1} = -(\slashed{p}+e\slashed{A}^0)^{T}$.  Then,

\be
	{\rm Tr~ln}\left[\frac{\slashed{p}+e\slashed{A}^0-m}{\slashed{p}-m}\right]
	={\rm Tr~ln}\left[\frac{\slashed{p}+e\slashed{A}^0+m}{\slashed{p}+m}\right]
	=\frac{1}{2}{\rm Tr~ln}\left[\frac{(\slashed{p}+e\slashed{A}^0)^2-m^2}{\slashed{p}^2-m^2}\right],
\ee
where the final expression is half the sum of the first two expressions.  
The effective action is now in its most useful, fully general form:

\be
	\label{eqn:generalEA}
	\Gamma[A_\mu^0] = \int d^4x\left(-\frac{1}{4}F^0_{\mu \nu} F^{0, \mu \nu}\right)
	-i \hbar {\rm Tr~ln}\left[\frac{(\slashed{p}+e\slashed{A}^0)^2
	-m^2}{\slashed{p}^2-m^2}\right].
\ee
\subsection{Proper-Time Formalism}
The proper-time formalism is a useful trick for regularizing the 
functional determinants in terms of a fictional proper-time parameter 
while preserving the gauge and Lorentz invariance of the expressions~\cite{Fock37, Schwinger:1951}. 
We may express the logarithm appearing in the effective action in terms of 
an integral using the following identity:

\be
	\ln \frac{a}{b} = \int_0^\infty \frac{dT}{T}\left[\exp{(iT(b+i\epsilon))}
		-\exp{(iT(a+i\epsilon))} \right].
\ee
Employing this identity, we may write
\ba
	\label{eqn:proptimelag}
	{\rm Tr~ln}\left[\frac{\slashed{p}+e\slashed{A}^0
	-m}{\slashed{p}-m}\right] &=& -\frac{1}{2}\int d^4x \int_0^\infty
	\frac{dT}{T}e^{-iTm^2} \nonumber \\
	& &\times {\rm tr}\biggr( \bra{x}e^{iT(\slashed{p}
	+e\slashed{A}^0)^2}\ket{x} 
	- \bra{x}e^{iT\slashed{p}^2}\ket{x}\biggr)
\ea
where previously ${\rm Tr}$ signified a trace over both 
spin and coordinate degrees of freedom, we use ${\rm tr}$ to 
signify that the trace is now only over spinorial components.

Beyond its function as a useful technical tool, the proper-time formalism 
also provides an intuitive picture of the functional determinants.
The object $\bra{x}e^{iT(\slashed{p}+e\slashed{A}^0)^2}\ket{x}$ 
can be thought of as a transition amplitude in standard quantum 
mechanics for a fictitious particle with space-time coordinates 
evolving in proper-time, $T$, under the 
influence of a ``Hamiltonian", $-(\slashed{p}+e\slashed{A}^0)^2$, 
that depends only on the $T$-independent electromagnetic field. 
Thus, we have expressed the 
functional determinant in terms of the motion of a
``particle" that obeys a $T$-dependent 
Schr\"{o}dinger-like equation in the given $T$-independent electromagnetic field.

\subsection{Effective Lagrangian in a Uniform Field}

The effective Lagrangian can only be evaluated analytically 
for a small number of field configurations. The homogeneous 
field case is the simplest of these. We begin by considering the 
Lorentz frame and coordinate system 
where the classical electric and magnetic fields 
are parallel and point in the $x^3$-direction. The magnitudes of the 
fields in this frame provide us with a set of Lorentz scalar
field invariants, $a\equiv |\vec{E}|$ and $b\equiv |\vec{B}|$.

The operator of interest can be separated into a scalar and a
fermion term,

\be
  -(\slashed{p}+e\slashed{A}^0)^2 =(p_\mu+eA^0_\mu)^2 
	- \frac{1}{2}e\sigma^{\mu \nu} F_{\mu \nu},
\ee
where $\sigma^{\mu \nu} = \frac{i}{2}[\gamma^\mu, \gamma^\nu]$.
This means that the trace factorizes.

First, consider the scalar factor,
\be
	{\rm tr}\left(\bra{x}e^{iT(p_\mu+e A^0_\mu)^2}\ket{x}\right).
\ee
In this coordinate system and frame, we may choose a gauge 
so that $A^3 = -ax^0$, and $A^1 = -bx^2$. The relevant operator 
then expands as
\ba
	(p_\mu+e A^0_\mu)^2 &=& p_0^2-p_2^2-(p_1+ebx_2)^2 -(p_3+eax_0)^2 \\
	& = & {\rm exp}\left(-i\frac{p_2 p_1}{eb}-i\frac{p_0 p_3}{ea}\right)
		\left[ p_0^2 - p_2^2 - (ebx_2)^2 -(eax_0)^2\right] \nonumber \\
	& &\times	{\rm exp}\left(i\frac{p_2 p_1}{eb}+i\frac{p_0 p_3}{ea}\right),
\ea
where we have used the shift operator, $e^{-ip_1c}f(x_1)e^{ip_1c} = f(x_1+c)$.
The scalar factor can then be evaluated by taking the trace in momentum 
space,
\ba
	{\rm tr}\left(\bra{x}e^{iT(p_\mu+e A^0_\mu)^2}\ket{x}\right) & = &
	\int\frac{dp_3 dp_1}{(2\pi)^4} dp_0 dp_0' dp_2 dp_2' \nonumber \\
	& &\times {\rm exp}\left[i(p_0'-p_0)\left(x_0+\frac{p_0}{ea}\right)\right] \nonumber \\
	& &\times {\rm exp}\left[i(p_2'-p_2)\left(x_2+\frac{p_2}{eb}\right)\right] \nonumber \\
	& &\times \bra{p_0}{\rm exp}[iT(p_0^2-e^2a^2x_0^2)]\ket{p_0'} \nonumber \\
	& & \times \bra{p_2}{\rm exp}[iT(p_2^2+e^2b^2x_2^2)]\ket{p_2'} \\
	&=& \frac{e^2ab}{(2\pi)^2}\int_{-\infty}^\infty dp_0
	\bra{p_0}{\rm exp}[iT(p_0^2-e^2a^2x_0^2)]\ket{p_0'} \nonumber \\
	& & \times \int_{-\infty}^\infty dp_2 \bra{p_2}{\rm exp}[iT(p_2^2+e^2b^2x_2^2)]\ket{p_2'}.
\ea

The remaining traces can be evaluated by using the known eigenvalues 
of the harmonic oscillator. For example, we evaluate the second one:
\ba
	\lefteqn{\int_{-\infty}^\infty dp_2\bra{p_2}{\rm exp}
		[-iT(p_2^2+e^2b^2x_2^2)]\ket{p_2}} \nonumber \\
	&=&{\rm Tr}e^{-iT(p^2+e^2b^2x^2)} \\
	&=&\sum_{n=0}^{\infty} {\rm exp}\left[-2iTeb\left(n+\frac{1}{2}\right)\right] \\
	&=&\frac{1}{2i\sin{(Teb)}}.
\ea
The other trace proceeds the same way, except with $b\rightarrow ia$, producing 
a sinh() function. 

The eigenvalues of $\frac{i}{2}eT\sigma^{\mu \nu} F_{\mu \nu}$
are $\pm eT(a\pm ib)$. So, the fermion factor is
\be
	{\rm tr}\left(\bra{x}{\rm exp}\left(\frac{i}{2}eT\sigma^{\mu \nu} F^0_{\mu \nu}\right)\ket{x}\right)
		=4 \cosh{(eTa)}\cos{(eTb)}.
\ee

Finally, we may write the complete expression for the trace:

\be
	{\rm tr}\left(\bra{x}e^{iT(\slashed{p}+e\slashed{A}^0_\mu)^2}\ket{x}\right) = 
		-i\frac{e^2ab}{(2\pi)^2}\coth{(eTa)}\cot{(eTb)}.
\ee
This expression is easily evaluated in the case where the fields vanish:
\ba
	\lefteqn{{\rm tr}\left(\bra{x}e^{iT(\slashed{p}+e\slashed{A}^0_\mu)^2}\ket{x} -
		\bra{x}e^{iT\slashed{p}^2}\ket{x}\right)} \nonumber \\
	& & = -\frac{i}{(2\pi)^2}
		\left(e^2ab\coth{(eTa)}\cot{(eTb)}-\frac{1}{T^2}\right).
\ea
Therefore, we arrive at an expression for the effective Lagrangian,
\ba
	\mathcal{L}_{\rm eff} &=& -\frac{1}{4}F^0_{\mu \nu}F^{0,\mu \nu} \nonumber \\
		& & + \frac{\hbar}{4(2\pi)^2}\int_0^\infty \frac{dT}{T}e^{iTm^2}
		\left[e^2ab\coth{(eTa)}\cot{(eTb)} -\frac{1}{T^2}\right].
\ea
The integral is divergent for small values of the proper time, $T$. 
This reflects the usual ultra-violet divergence in \ac{QED}. 
Correspondingly, the divergence is proportional to the classical 
term and can be absorbed with a field strength renormalization.
The renormalized Lagrangian is
\ba
	\mathcal{L}_{\rm eff} &=& -\frac{1}{4}F^0_{\mu \nu}F^{0,\mu \nu} \nonumber \\
		& & + \frac{\hbar}{4(2\pi)^2}\int_0^\infty \frac{dT}{T}e^{-iTm^2}
		\biggr[e^2ab\coth{(eTa)}\cot{(eTb)}\nonumber \\
		& & -\frac{1}{T^2}-\frac{1}{3}e^2(a^2-b^2)\biggr].
\ea
We may express this more simply by Wick rotating~\cite{peskin1995introduction}:
\ba
	\label{eqn:HEWSab}
	\mathcal{L}_{\rm eff} &=&\frac{a^2-b^2}{2} \nonumber \\
		& & +\frac{\alpha}{8\pi^2}B_k^2\int_0^\infty \frac{d\zeta}{\zeta}e^{-\zeta}
		\biggr[\frac{ab}{B_k^2}\cot{\left(\zeta \frac{a}{B_k}\right)}\coth{\left(\zeta \frac{b}{B_k}\right)} 
		\nonumber \\
		& &-\frac{1}{\zeta^2}-\frac{1}{3}\frac{(a^2-b^2)}{B_k}\biggr],
\ea
where we have substituted $\zeta = Tm^2$ after performing a Wick rotation.

This expression is the Heisenberg-Euler-Weisskopf-Schwinger 
one-loop effective Lagrangian~\citep{heisenberg-1936-98,
  weisskopf1936kongelige, Schwinger:1951}. It is quite 
useful in practice because it describes the local 
energy densities of fields that vary slowly relative 
to the electron Compton wavelength scale, $\lambda_e = 2\times10^{-12}$m.
As such, the constant field effective Lagrangian provides the 
leading order term in the popular derivative expansion technique, 
which expands the effective action in derivatives of the fields
\cite{PhysRevD.40.4202,ISI:000083342100008,Masso2002566}.
It is therefore an excellent approximation for sufficiently slowly varying fields.

\subsection{Wave Propagation in the \texorpdfstring{\acs{QED}}{QED} Vacuum}

In intense electromagnetic fields, photons and 
electromagnetic waves will interact significantly with 
fermion vacuum fluctuations.  These effects can be taken into account, 
on average, by treating the vacuum as a non-linear optical medium
whose properties are given by the effective Lagrangian.  In this 
section we will derive the dielectric and inverse magnetic 
permeability tensors for the \ac{QED} vacuum.

We start with the Heisenberg-Euler-Weisskopf-Schwinger~\citep{heisenberg-1936-98,
  weisskopf1936kongelige, Schwinger:1951} one-loop effective Lagrangian, 
equation (\ref{eqn:HEWSab}), rewritten in terms of the field invariants
\ba
	K&=&\left(\frac{1}{2}\varepsilon^{\lambda \rho \mu \nu} 
	F_{\lambda \rho}F_{\mu \nu}\right)^2=-(4\efi \cdot \bfi)^2\\
	I&=&F_{\mu \nu}F^{\mu \nu} = 2(|\bfi|^2-|\efi|^2).
\ea 

\ba
	\lag_0 & = & -{\frac{1}{4}} I, \label{eq:lag0def} \\
	\lag_1 & = & {\frac{e^2}{h c}} \int_0^\infty e^{-\zeta} 
	\frac{\d \zeta}{\zeta^3} \left \{ i \zeta^2 \frac{\sqrt{-K}}{4} \times
	\frac{ 
	\cos \left ( \frac{\zeta}{B_k} \sqrt{-\frac{I}{2} + i \sqrt{K}} \right ) 
	}{
	\cos \left ( \frac{\zeta}{B_k} \sqrt{-\frac{I}{2} + i \sqrt{K}} \right ) } 
	\right . \\*
	\nonumber
	& & ~~ \left . 
	\frac{ \cos \left ( \frac{\zeta}{B_k} \sqrt{-\frac{I}{2} + i \frac{\sqrt{-K}}{2}} \right ) +
	\cos \left ( \frac{\zeta}{B_k} \sqrt{-\frac{I}{2} - i \frac{\sqrt{-K}}{2}} \right ) }{
	\cos \left ( \frac{\zeta}{B_k} \sqrt{-\frac{I}{2} + i \frac{\sqrt{-K}}{2}} \right ) -
	\cos \left ( \frac{\zeta}{B_k} \sqrt{-\frac{I}{2} - i \frac{\sqrt{-K}}{2}} \right ) } 
	+ |B_k|^2 + \frac{\zeta^2}{6} I \right \}.
	\label{eq:lag1def}
\ea

Consider the expansion of the non-linear term in the Lagrangian about $K=0$.

\be
	\lag_1 = \lag_1(I,0) + K \left . \pp{\lag_1}{K} \right |_{K=0} +
	\frac{K^2}{2} \left . \frac{\partial^2 \lag_1}{\partial K^2} \right
	|_{K=0} + \cdots
	\label{eq:lag1exp}
\ee
We may write each of the derivatives in terms of an analytic function of 
$1/\xi$ where

\be
	\xi=\frac{1}{B_k}\sqrt{\frac{I}{2}}.
\ee

\ba
	\lag_1(I,0) & = & \frac{e^2}{h c} \frac{I}{2}
	X_0\left(\frac{1}{\xi}\right), \\
	\left . \pp{\lag_1}{K} \right |_{K=0} & = & \frac{e^2}{h c}
	\frac{1}{16 I} X_1\left(\frac{1}{\xi}\right) \\
	\left . \pp{^2\lag_1}{K^2} \right |_{K=0} & = &
	\frac{e^2}{h c} \frac{1}{384 I^2} X_2\left(\frac{1}{\xi}\right)
\ea
Explicit expressions for these functions are given in ref. 
\cite{1997PhRvD..55.2449H} and in section \ref{sec:tensors}. 
Since we want to describe the vacuum as an optical medium, we should 
evaluate the macroscopic fields arising from the non-linear terms 
of the effective Lagrangian. The macroscopic electromagnetic 
fields are

\be
\vec{D}=\pderiv{\lag}{\vec{E}}, ~~ \vec{H} = -\pderiv{\lag}{\vec{B}}.
\ee
The response from the vacuum can now be characterized in terms 
of vacuum dielectric and permeability tensors:

\be
	D_i = \epsilon_{ij}E_j, ~~ H_i = \mu'_{ij}B_j,
\ee
where 
\ba
	\label{eqn:epij}
	\epsilon_{ij} &=& \delta_{ij} - 4 \pderiv{\lag_1}{I} \delta_{ij} - 32
	\pderiv{\lag_1}{K} B_i B_j, \\
	\label{eqn:muij}
	\mu'_{ij} &=& \delta_{ij} - 4 \pderiv{\lag_1}{I} \delta_{ij} + 32
	\pderiv{\lag_1}{K} E_i E_j. 
\ea

We may take
\be
	\pderiv{\lag_1}{K}=\pderiv{\lag_1}{K}\biggr|_{K=0} =
	\frac{e^2}{hc}\frac{1}{16I}X_1\left(\frac{1}{\xi}\right)
	\label{eqn:dlagdK}
\ee
since the factor $B_iB_j$ is already of order $\sqrt{K}$, and 
we may neglect any larger orders in $K$.  The other derivative 
is given by 

\ba
	\pderiv{\lag_1}{I}&=&\frac{e^2}{2hc}\biggr(X_0\left(\frac{1}{\xi}\right)
	-\frac{1}{\xi}X'_0 \left(\frac{1}{\xi}\right) \\
	& &-\frac{K}{8I^2}X_1\left(\frac{1}{\xi}\right) 
	 -\frac{K}{8I^2\xi}X'_1\left(\frac{1}{\xi}\right) \biggr)
	 \label{eqn:dlagdI}
\ea
where the Lagrangian and its derivatives have been expressed in terms 
of a set of analytic functions given by equations (\ref{eqn:x0anal}-\ref{eqn:x2anal})
and 
\be
	X_n'(x) = \frac{\d X_n(x)}{\d x}.
\ee

Putting equations (\ref{eqn:dlagdK}) and (\ref{eqn:dlagdI}) into 
equations (\ref{eqn:epij}) and (\ref{eqn:muij}), we arrive at 
an expression for the vacuum dielectric and inverse magnetic permeability 
tensors:

\be
	\label{eqn:inroepsilon}
	\varepsilon^{i j} = \Delta^{i j} - \frac{\alpha}{2
          \pi}\left[\frac{2}{I} X_1\left(\frac{1}{\xi}\right)+
	\frac{12 K}{I^3}X_2\left(\frac{1}{\xi}\right)\right]B^i B^j,
\ee

\be
	\label{eqn:intromu}
	(\mu^{-1})^{ij} = \Delta^{i j} +  \frac{\alpha}{2
          \pi}\left[\frac{2}{I} X_1\left(\frac{1}{\xi}\right)+
	\frac{K}{12 I^3}X_2\left(\frac{1}{\xi}\right)\right]E^i E^j,
\ee
where

\ba
	\Delta^{i j} &=& \delta^{i j}\Bigg [1 + \frac{\alpha}{2 \pi} \Bigg(-2 X_0\left(\frac{1}{\xi}\right) + \frac{1}{\xi}X_0^{(1)}\left(\frac{1}{\xi}\right) \nonumber \\ 
		& & + \frac{K}{4I^2}X_1\left(\frac{1}{\xi}\right) + \frac{K}{8 I^2\xi}X_1^{(1)}\left(\frac{1}{\xi}\right)\Bigg) \Bigg]. 
\ea 
These expressions will be useful when they 
appear as equations (\ref{eqn:epsilon}) and (\ref{eqn:mu}).

\chapter{Travelling Electromagnetic Waves in the Magnetosphere of a Magnetar}
\label{ch:travellingwaves}
\acresetall

\vspace{-2cm}{\renewcommand{\thefootnote}{} \footnote{This chapter contains only minor changes 
from the published manuscript: Mazur, Dan and Heyl, J.S. \emph{MNRAS 412, 2} (2011)}}
 \addtocounter{footnote}{-1}
\begin{summary}
We compute electromagnetic wave propagation through the magnetosphere 
of a magnetar. The magnetosphere is modelled as the quantum electrodynamics 
vacuum and a cold, strongly magnetized plasma. The background field and 
electromagnetic waves are treated non-perturbatively and can be arbitrarily 
strong. This technique is particularly useful for examining non-linear 
effects in propagating waves. Waves travelling through such a medium typically 
form shocks; on the other hand we focus on the possible existence of waves 
that travel without evolving. Therefore, in order to examine the non-linear 
effects, we make a travelling wave ansatz and numerically explore the 
resulting wave equations. We discover a class of solutions in a homogeneous 
plasma which are stabilized against forming shocks by exciting non-orthogonal 
components which exhibit strong non-linear behaviour. These waves may be an 
important part of the energy transmission processes near pulsars and magnetars.
\end{summary}

\section{Introduction}
\label{sec:introduction}
The magnetosphere of a magnetar is a particularly interesting
medium for the propagation of electromagnetic waves.  Magnetars are
characterized by exceptionally large magnetic fields that can be
several times larger than the quantum critical field
strength~\citep{mereghetti-2008}.  Because the magnetic fields are so
large, the fluctuations of the vacuum of \ac{QED} influence the propagation of light.
Specifically, the vacuum effects add nonlinear terms to the wave
equations of light in the presence large magnetic fields.  In
addition, the magnetosphere of a magnetar contains a plasma which alters
the dispersion relationship for light.  Because of the unique optical
conditions in the magnetospheres of magnetars, they provide excellent
arenas to explore nonlinear vacuum effects arising due to quantum
electrodynamics.

The influence of \ac{QED} vacuum effects from strong magnetic fields in the
vicinities of magnetized stars has previously been studied by several
authors.  The combined \ac{QED} vacuum and plasma medium is discussed in
detail in the context of neutron stars in \citet{meszarosbook}.  Vacuum
effects have been found to dominate the polarization properties and
transport of X-rays in the strong magnetic fields near neutron stars
\citep{PhysRevD.19.3565, PhysRevLett.41.1544, 1981ApJ...251..695M,
  1980ApJ...238.1066M, 1983ZhETF..84.1217G}.  Detailed consideration
of magnetic vacuum effects is therefore critical to an understanding
of emissions from highly magnetized stars.

Most studies of waves in systems including plasmas or vacuum effects
approach the problem perturbatively, which limits the applicability of
their results.  The purpose of the present paper is to examine the
combined impact of the \ac{QED} vacuum and a magnetized plasma using
nonperturbative methods to fully preserve the nonlinear interaction
between the fields.  Neutron stars may be capable of producing very
intense electromagnetic waves, comparable to the ambient magnetic
field.  For example, a coupling between plasma waves and seismic
activity in the crust could produce an Alfv\'en wave with a very large
amplitude~\citep{1989ApJ...343..839B,1995MNRAS.275..255T}. Even
  if they may not be produced directly, electromagnetic waves
  naturally develop through the interactions between Alfv\'en waves.
  To lowest order in the size of the wave, Alfv\'en waves do not
  suffer from shock formation \citep{Thom98} whereas electromagnetic
  waves do \citep{heyl1998electromagnetic,Heyl98mhd}. Therefore, how
  to stabilise the propagation of the the latter is the focus of this
  paper.

If such a magnetospheric disturbance results in electromagnetic waves
of sufficiently large amplitude and low frequency, then
nonperturbative techniques are required to characterize the wave.  The
importance of studying such a system nonperturbatively is particularly
well illustrated by the fact that some nonlinear wave behaviour is
fundamentally nonperturbative, as is generally the case with
solitons~\citep{rajaraman1982solitons}.  In order to handle the
problem nonperturbatively, we choose to study waves whose spacetime
dependence is described by the parameter $S=x-vt$, where $v$ is a
constant speed of propagation through the medium in the
$\bhat{x}$-direction.  In the study of waves, one normally chooses the
ansatz $e^{i(\omega t - {\bf k}\cdot{\bf x})}$.  However, in this
picture, a numerical study would typically treat the self-interactions
of the electromagnetic field by summing the interactions of finitely
many Fourier modes.  So, this ansatz conflicts with our goal of
studying the nonlinear interactions to all orders.  In contrast, a
plane wave ansatz given by $S=x-vt$ allows us to study a simple wave
structure to all orders without any reference to individual Fourier
modes.


We model a magnetar atmosphere by including the effects of arbitrarily
strong electromagnetic fields using a \ac{QED} one-loop effective
Lagrangian approach.  These effects are discussed in
section~\ref{sec:tensors}.  Plasma effects are included by assuming
free electrons moving under the Lorentz force without any
self-interactions.  The model is that of a cold magnetohydrodynamic
plasma and is discussed in section~\ref{sec:plasma}.  We have
  also assumed that the medium is homogeneous in agreement with the
  travelling-wave ansatz.  Of course the actual situation is more
  complicated with a thermally excited plasma \citep[e.g.][]{Gill09BW}
  and inhomogeneities --- the latter can result in a whole slew of
  interesting interactions between the wave modes
  \citep{Heyl99polar,Heyl01qed,Heyl01polar,2003PhRvL..91g1101L} that
  are especially crucial to our understanding of the thermal radiation
  from their surfaces, but these are beyond the scope of this paper.


The formation of electromagnetic shocks is expected to be an important
phenomenon for electromagnetic waves in the magnetized vacuum since
electromagnetic waves can evolve discontinuities under the influence
of nonlinear interactions~\citep{PhysRev.113.1649,
  zheleznyakov1982shock, heyl1998electromagnetic}.  Such shocks can
form even in the presence of a plasma~\citep{PhysRevD.59.045005}.  In
this study, through our explicit focus on travelling waves, 
we examine an alternate class of solutions to the wave equations 
 which do not suffer this fate.  Instead,
they are stabilized against the formation of discontinuities by
dispersion.  These waves travel as periodic wave trains
without any change to their form, such as wave steepening or shock
formation.  Waves such as these may contribute to the formation of
pulsar microstructures~\citep{1987STIN...8816622C,2001ApJ...558..302J}.

\section{Wave Equations}
\label{sec:wave-equations}
\subsection{The Maxwell's Equations}
\label{sec:maxwells-equations}
The vacuum of \ac{QED} in the presence of large magnetic fields can be
described as a non-linear optical medium~\citep{1997JPhA...30.6485H}.
We also choose to treat the effect of the plasma on the waves through
source terms $\rho_p$ and $\Jp$. Therefore, we begin by considering
Maxwell's equations in the presence of a medium and plasma sources.
In Heaviside-Lorentz units with $c=1$, Maxwell's equations can be used
to derive the wave equations
\ba
\label{eqn:Maxwell1}
\laplace {\bf D} - \ddt{{\bf D}} &=& -\curl{(\curl{({\bf D}-{\bf E})})}+  \nonumber \\
	& &\pderiv{}{t} (\curl{({\bf B}-{\bf H})}) 
	+{\bf \nabla}\rho_p + \pderiv{{\bf J_p}}{t} \\
\label{eqn:Maxwell2}
\laplace {\bf H} - \ddt{{\bf B}} &=& -\del(\del \cdot({\bf B}-{\bf H}))- \nonumber \\
	& &\pderiv{}{t}(\curl{({\bf D}-{\bf E})}) - \curl{{\bf J_p}}.
      \ea 
For clarity, we will avoid making cancellations or dropping vanishing terms.
We define the vacuum dielectric and inverse magnetic permeability tensors as
follows \citep{Jack75}
\be
D_i = \varepsilon_{ij} E_j,~~H_i = \mu^{-1}_{ij} B_j.
\ee
In the next few sections we build a model describing travelling waves 
in a magnetar's atmosphere from these equations.

\subsection{Vacuum Dielectric and Inverse Magnetic Permeability Tensors}
\label{sec:tensors}
In this section, we describe our model of the \ac{QED} vacuum in strong
background fields in terms of vacuum dielectric and inverse magnetic
permeability tensors.  These are most conveniently described in terms
of two Lorentz invariant combinations of the fields:
\ba
K&=&\left(\frac{1}{2}\varepsilon^{\lambda \rho \mu \nu} 
F_{\lambda \rho}F_{\mu \nu}\right)^2=-(4\efi \cdot \bfi)^2\\
I&=&F_{\mu \nu}F^{\mu \nu} = 2(|\bfi|^2-|\efi|^2).
\ea 
In order to examine the nonlinear effects of the vacuum
nonperturbatively, we wish to use vacuum dielectric and inverse
magnetic permeability tensors which are valid to all orders in the
fields.  Analytic expressions for these tensors were derived by 
\citet{1997JPhA...30.6485H} for the case of wrenchless
fields ($K=-(4\efi \cdot \bfi)^2=0$) from the
Heisenberg-Euler-Weisskopf-Schwinger~\citep{heisenberg-1936-98,
  weisskopf1936kongelige, Schwinger:1951} one-loop effective Lagrangian in
\citet{1997PhRvD..55.2449H} and expressed in terms of a set of analytic
functions:
\ba
X_0(x) & = & 4 \int_0^{x/2-1} \ln(\Gamma(v+1)) \d v 
+ \frac{1}{3} \ln \left ( \frac{1}{x} \right )\nonumber \\
& &~~+ 2 \ln 4\pi - 4 \ln A-\frac{5}{3} \ln 2 \nonumber \\
& & ~~ - \left [ \ln 4\pi + 1 +  \ln \left ( \frac{1}{x} \right ) \right ] x \nonumber \\
& &~~+ \left [ \frac{3}{4} + \frac{1}{2} \ln \left ( \frac{2}{x} \right )
\right ]
x^2
\label{eqn:x0anal} \\
X_1(x) & = & - 2 X_0(x) + x X_0^{(1)}(x) + \frac{2}{3} X_0^{(2)} (x) -
\frac{2}{9} \frac{1}{x^2}
\label{eqn:x1anal} \\
X_2(x) & = & -24 X_0(x) + 9 x X_0^{(1)}(x)\nonumber \\
& &~~ + (8 + 3 x^2) X_0^{(2)}(x)
+ 4 x X_0^{(3)}(x) \nonumber \\
& & ~~ - \frac{8}{15} X_0^{(4)}(x) + \frac{8}{15}
\frac{1}{x^2} + \frac{16}{15} \frac{1}{x^4}.
\label{eqn:x2anal}
\ea 
where
\be
X_0^{(n)}(x) = \frac{\d^n X_0(x)}{\d x^n}
\ee
and
\be
{\rm ln}A = \frac{1}{12}- \zeta^{(1)}(-1) = 0.248754477.
\ee
The tensors we need are derived in \citet{1997JPhA...30.6485H}, except
that we have kept terms up to linear order in the expansion about
$K=0$ instead of dealing with the strictly wrenchless case.  Our
analysis therefore requires that $K \ll B_k^4$. We have
\be
	\label{eqn:epsilon}
	\varepsilon^{i j} = \Delta^{i j} - \frac{\alpha}{2
          \pi}\left[\frac{2}{I} X_1\left(\frac{1}{\xi}\right)+
	\frac{12 K}{I^3}X_2\left(\frac{1}{\xi}\right)\right]B^i B^j
\ee

\be
	\label{eqn:mu}
	(\mu^{-1})^{ij} = \Delta^{i j} +  \frac{\alpha}{2
          \pi}\left[\frac{2}{I} X_1\left(\frac{1}{\xi}\right)+
	\frac{K}{12 I^3}X_2\left(\frac{1}{\xi}\right)\right]E^i E^j
\ee
where

\ba
	\Delta^{i j} &=& \delta^{i j}\Bigg [1 + \frac{\alpha}{2 \pi} \Bigg(-2 X_0\left(\frac{1}{\xi}\right) + \frac{1}{\xi}X_0^{(1)}\left(\frac{1}{\xi}\right) \nonumber \\ 
		& & + \frac{K}{4I^2}X_1\left(\frac{1}{\xi}\right) + \frac{K}{8 I^2\xi}X_1^{(1)}\left(\frac{1}{\xi}\right)\Bigg) \Bigg], 
\ea 
the fine-structure constant is $\alpha=e^2/4 \pi$ in these
units where we have set $\hbar = c =1$,
and 
\be
	\xi=\frac{1}{B_k}\sqrt{\frac{I}{2}}.
\ee
Equations (\ref{eqn:epsilon}) and (\ref{eqn:mu}) define our model for
the \ac{QED} vacuum in a strong electromagnetic field.  Because we will
focus on photon energies much lower than the rest-mass energy of the
electron, we have treated the vacuum as strictly non-linear.  It is
not dispersive.  The treatment of the dispersive properties of the
vacuum would require an effective action
treatment~\citep[e.g.][]{1995PhRvD..51.2513C,1995PhRvD..52.3163C}
rather than the local effective Lagrangian treatment used here.

\subsection{Plasma}
\label{sec:plasma}
To investigate travelling waves, we choose our coordinate system so
that the $\bhat{x}$-direction is aligned with the direction of
propagation.  Then, the spacetime dependence of the fields and sources
is given by a single parameter $S\equiv x-vt$ where $v$ is the
constant phase velocity in the $\bhat{x}$-direction of the travelling
wave.  At this point, we are choosing to work in a specific Lorentz
frame.

We model the plasma as a free electron plasma which enters the wave
equation through the source terms $\rho_p$ and $\Jp$.  For
electromagnetic fields obeying the travelling wave ansatz, the sources
must also obey the ansatz.  Then, $\rho_p$ and $\Jp$ are functions
only of $S$.  We therefore treat them as an additional field which is
integrated along with the electromagnetic components of the field.

In order to perform the numerical \ac{ODE} integration for the currents, 
we wish to find expressions for $\dds{\rho_p}$
and $\dds{\Jp}$.  For travelling waves, the continuity 
equation is 
\be
	\label{eqn:continuity}
	-v \deriv{\rho_p}{S} + \delta^{i x} \deriv{J_p^i}{S} = 0.
\ee where we are using index notation to label our explicitly 
cartesian $\{x, y, z\}$ coordinate system.  Repeated indices are summed. However, 
whenever $x$ or $z$ appears as an index, it is 
fixed and does not run from 1 to 3.

We use equation (\ref{eqn:continuity}) to rewrite the source terms from equation (\ref{eqn:Maxwell1})
\be
	\label{eqn:plasma1}
\partial^i \rho_p + \pderiv{{ J}^i_p}{t} = \frac{1}{v}\frac{d J_p^i}{d S}(\delta^{i x}-v^2).
\ee Similarly, the source term from equation (\ref{eqn:Maxwell2}) is
\be
	\label{eqn:plasma2}
(\curl{J_p})^i = \varepsilon^{ i x k} \left( \frac{d J_p}{dS} \right)^k.
\ee

To find an expression for $\frac{d \Jp}{dS}$ we express the current as
an integral over the phase-space distribution of the electrons 
and linearize the plasma density.
\be
	\label{eqn:Jpdef}
\Jp = \int{f({\bf p}_p) e {\bf v}_p d^3{\bf p}_p} \approx \bar{\gamma}ne \bar{\bf v}_p
\ee where $\bar{\bf v}_p \equiv \frac{\int f({\bf p}_p) {\bf
    v}_p d^3 {\bf p}_p}{\bar{\gamma} n}$
and $n$ is the mean electron density in the plasma in the reference frame where $\bar{{\bf v}}_p=0$.
It is important to note the distinction between the mean plasma speed, $\bar{ v}_p$, and 
the propagation speed of the wave, $v$.
The Lorentz factor $\bar{\gamma}\equiv\frac{1}{\sqrt{1-\bar{v}_p^2}}$ accounts for a relativistic increase in the plasma density
 since $d^3 {\bf x} = \gamma^{-1} d^3 {\bf x}'$.

Next, we take a time derivative of the current and express this in terms of a
three-force acting on the plasma:

 \ba
 \label{eqn:dJdt2}
 \pderiv{{\bf J}_p}{t} &=& n e \pderiv{\bar{\gamma} \bar{{\bf v}}_p}{t} \nonumber \\
		& =&  \frac{n e}{m} {\bf F}.
\ea

Noting that $\Jp$ is a function only of $S$, 
we insert the Lorentz force ${\bf F} = e(\efi + \bar{{\bf v}}_p \times \bfi)$
 and arrive at an expression that can be substituted into the source terms,
equations (\ref{eqn:plasma1}) and (\ref{eqn:plasma2}), to find $\Jp(S)$:

\ba
\label{eqn:dJdS}
\frac{d {\bf J}_p}{dS} = -\frac{1}{v}\frac{e }{m} (e n\efi + \bar{\gamma}^{-1}{\bf J}_p \times \bfi).
\ea 
The equations we have given above describe a cold, relativistic,
magnetohydrodynamic plasma.  Forces on the plasma arising due to
pressure gradients and gravity are neglected.  Moreover, in our
simulations, we neglect the forces on the plasma due to the magnetic
field.  This approximation is suitable if the plasma in question is a
pair plasma, or for wave frequencies much less than the cyclotron
frequency.  

As we will show in section \ref{sec:results} (see figure
 \ref{fig:freqvsspeedEz}), the field configurations 
generated in our simulation vary over timescales similar to the
inverse of the plasma frequency; the latter is 
about 9 orders of magnitude smaller than the 
cyclotron frequency.  This observation allows us to justify 
some aggressive assumptions regarding the plasma response.
As mentioned above, we may neglect forces on the plasma from 
the magnetic field, and quantum effects are expected to be 
small far away from 
the cyclotron resonance \citep{meszarosbook}.

We also neglect thermal effects, since the influence of the 
electromagnetic fields will dominate over thermal motion.  
For strong background magnetic fields such that 
$\frac{eB}{m} \gg kT$, the electrons 
will be confined to the lowest Landau level, restricting 
thermal motion perpendicular to the background magnetic field.  
As we will elaborate on in section \ref{sec:results},
the greatest nonlinear effects occur for waves with an 
electric field component oriented along the background magnetic 
field.  Because we are interested in waves with amplitudes 
comparable to $B_k$, thermal motion is negligible 
relative to the dynamics induced by the wave.  
We are therefore justified in 
neglecting thermal effects in every direction
for the cases of greatest interest. 

If one combines equation~(\ref{eqn:dJdt2}) with
equation~(\ref{eqn:Maxwell1}), one sees that any nonlinearity in this
treatment must originate with the dielectric and permeability tensors
--- any non-linearity that may originate from the plasma itself has been
neglected \citep[{\it c.f.}][]{1979JETP...49...75K,cattaert:012319}. The
plasma is modelled as strictly dispersive. 
In section~\ref{sec:WFL},
we confirm that this method of describing the plasma is consistent
with standard accounts in the weak-field, small-wave limit.  

\subsection{Travelling Wave \texorpdfstring{\acsp{ODE}}{ODEs}}
 
The wave equations for travelling waves, are found by combining
Maxwell's equations (\ref{eqn:Maxwell1}) and (\ref{eqn:Maxwell2}) with
the continuity equation, (\ref{eqn:continuity}).  We also make the
plane-wave approximation, and assume that the fields and sources are
described by the parameter

\be
	\label{eqn:S}
	S=x-vt.
\ee
The equations governing travelling wave propagation are

\be
\label{eqn:ODE1}
\ddS{\psi^i(S)}=\frac{1}{v}\dds{J^i(S)}(\delta^{i x} - v^2)
\ee
and 
\be
\label{eqn:ODE2}
\ddS{\chi^i(S)}=-\varepsilon^{i x j} \dds{J^j(S)}.
\ee
  The 
auxiliary vectors $\psi^i$ and $\chi^i$ are related to the electric and magnetic fields:
\ba
\label{eqn:u}
\psi^i(S)&=&(1-v^2)\varepsilon^{i j} E^j + \delta^{ix}(\varepsilon^{x j} E^j-E^x)- \nonumber \\
& & ~~(\varepsilon^{i j}E^j-E^i) + \varepsilon^{i x
  k}v(B^k-(\mu^{-1})^{k j}B^j) \\
\label{eqn:v}
\chi^i(S)&=&((\mu^{-1})^{i j} B^j-v^2 B^i) + \delta^{ix}(B^x-(\mu^{-1})^{x j}B^j) - \nonumber \\
& & ~~\varepsilon^{i x k}v(\varepsilon^{k j}E^j-E^k).
\ea 
Equations (\ref{eqn:ODE1}) through (\ref{eqn:v}) define a set of
coupled ordinary differential equations that can be integrated to
solve for the travelling electric and magnetic fields.  Solving these
equations requires that we have at hand the vacuum dielectric and
inverse magnetic permeability tensors as well as an expression for
$\frac{d \Jp}{d S}$.  These were discussed in sections
\ref{sec:tensors} and \ref{sec:plasma} respectively.

\subsection{Weak Field, Small Wave Limit}
\label{sec:WFL}
In this section, we would like to demonstrate that our equations
reduce to standard expressions in the case of small background fields
and small electromagnetic waves.  To make this comparison, it is also
prudent to assume waves have a spacetime dependence like $e^{i(\omega
  t - {\bf k}\cdot{\bf x})}$ instead of $x-v t$.  By making this
change, we can compare our other assumptions with those made in
standard textbook accounts directly.

Under the assumption that the fields and currents have the standard
plane-wave spacetime dependence, we can make the replacements
$\pderiv{}{t} \rightarrow -i\omega$ and $\nabla \rightarrow i {\bf
  k}$.  We can then write a second expression for $\pderiv{{\bf
    J}}{t}$,

\be
\label{eqn:dJdt1}
\pderiv{\Jp}{t} = -i \omega \Jp.
\ee

Setting equations (\ref{eqn:dJdt1}) and (\ref{eqn:dJdt2}) equal, we get

\be
\label{eqn:ainJ}
\frac{e}{m}(\varepsilon^{i j k} J^j_p B^k) + i \omega J^i_p = \frac{e^2}{m} n E^i,
\ee 
where we have used the nonrelativistic approximation, which is appropriate
for small waves.  We will now specialize to a background magnetic field pointing in 
the $\bhat{z}$-direction.  We can then write this background field in terms of the 
cyclotron frequency $\omega_c = \frac{eB}{m}$.  The density $n$ determines the 
plasma frequency $\omega_p^2 =n \frac{e^2}{m}$.  Then, we can write 
equation (\ref{eqn:ainJ}) as
\be
\label{eqn:ainJ2}
\left[ -\omega_c \varepsilon^{i j z} -  i \omega \delta^{j i}\right] J_p^j = \omega_p^2 E^i.
\ee

We can solve this for $\Jp$ by writing a matrix equation

\be
\Jp =  \left( \begin{array}{ccc}
-i \omega &  -\omega_c & 0 \\
\omega_c & -i\omega & 0 \\
0 & 0 & -i\omega \end{array} \right)^{-1} 
\omega_p^2 {\bf E},
\ee 
inverting the matrix gives
\be
\label{eqn:Jmatrix}
\Jp =  \left( \begin{array}{ccc}
\frac{i \omega}{\omega^2-\omega_c^2} & \frac{-\omega_c}{\omega^2-\omega_c^2} & 0 \\
\frac{\omega_c}{\omega^2-\omega_c^2} & \frac{i \omega}{\omega^2-\omega_c^2} & 0 \\
0 & 0 & i/\omega \end{array} \right) 
\omega_p^2 {\bf E}.
\ee

Finally, we would like to use equation (\ref{eqn:Jmatrix}) to write the right hand side of 
equation (\ref{eqn:Maxwell1}) in the small wave limit in a manner we can interpret as 
a dielectric tensor.

Ignoring (for now) the contributions from the vacuum, 
and assuming an approximately homogeneous plasma density, 
equation (\ref{eqn:Maxwell1}) simplifies to
\ba
\label{eqn:RHSMax1}
\laplace {\bf E} - \ddt{{\bf E}} &=& \pderiv{\Jp}{t} \nonumber \\ 
	&=& - i\omega\Jp .
\ea 
If our macroscopic field is to obey $\laplace {\bf D} - \ddt{{\bf D}}=0$,
we can insert equation (\ref{eqn:Jmatrix}) into (\ref{eqn:RHSMax1}) to obtain the 
following expression for the dielectric tensor due to plasma effects:

\be
\varepsilon^{(p)}_{i j} = \left( \begin{array}{ccc}
1-\frac{\omega_p^2}{\omega^2-\omega_c^2} & -i\frac{\omega_c}{\omega}\frac{ \omega_p^2}{\omega^2-\omega_c^2} & 0 \\
i\frac{\omega_c}{\omega}\frac{ \omega_p^2}{\omega^2-\omega_c^2} & 1-\frac{\omega_p^2}{\omega^2-\omega_c^2} & 0 \\
0 & 0 & 1-\left(\frac{\omega_p^2}{\omega^2}\right) \end{array} \right).
\label{eq:2}
\ee 
This expression is in agreement with the cold plasma dielectric tensor given in \citet{meszarosbook}.
As noted above, our analysis neglects the off-diagonal (Hall) terms, as is appropriate for pair plasmas 
or waves with frequencies much less than the cyclotron frequency.

Because the vacuum effects are added explicitly in the form of dielectric and magnetic permeability tensors,
we only need to confirm that the weak field limits of our expressions agree with the standard results.
This confirmation is done explicitly in \citet{1997JPhA...30.6485H}.

In the weak field limit, the tensors given by equations (\ref{eqn:epsilon}) 
and (\ref{eqn:mu}) are
\ba
\varepsilon_{ij}^{(v)} &=& \delta_{ij} + \frac{1}{45 \pi} \frac{\alpha}{B_k^2}
\left [ 2(E^2-B^2) \delta_{ij} + 7 B_i B_j \right ], \\
\mu^{-1 (v)}_{ij} &=& \delta_{ij} + \frac{1}{45 \pi} \frac{\alpha}{B_k^2}
\left [ 2(E^2-B^2) \delta_{ij} - 7 E_i E_j \right ].
\ea 
  In the case of a weak background magnetic field pointing in the $\bhat{z}$-direction,
these become
\ba
\varepsilon_{ij}^{(v)} &=& \left( \begin{array}{ccc}
1-2\delta & 0 & 0 \\
0 & 1-2\delta & 0 \\
0 & 0 & 1+5\delta  \end{array} \right)
\label{eq:1}
\\
\mu_{ij}^{-1 (v)} &=& \left( \begin{array}{ccc}
1-2\delta & 0 & 0 \\
0 & 1-2\delta & 0 \\
0 & 0 & 1-6\delta \end{array} \right),
\ea 
with 

\be
\label{eqn:delta}
\delta = \frac{\alpha}{45 \pi} \left( \frac{B}{B_k} \right) ^2.
\ee 
Again, this result agrees with the vacuum tensors given in \citet{meszarosbook}.

In this limit, we may simply add together the contributions to the dielectric tensor 
from the plasma and the vacuum according to

\be
\varepsilon_{i j} = \delta_{i j} + (\varepsilon_{i j}^{(p)}-\delta_{i j}) + (\varepsilon_{i j}^{(v)}-\delta_{i j}),
\ee 
with $\mu_{i j}^{-1}$ given entirely by the vacuum contribution.

We have thus recovered the standard result for a medium consisting of a plasma and the QED
vacuum in the weak field, small wave limit.

\section{Solution Procedure}
In total, there are 15 coupled non-linear \acp{ODE} which must be
integrated to produce a solution.  Equations (\ref{eqn:ODE1}) and
(\ref{eqn:ODE2}) define the electric and magnetic fields as functions
of $S$.  In addition, we must simultaneously integrate equation
(\ref{eqn:dJdS}), which gives the plasma current as a function of $S$.
Each of these equations has three spatial components.
Initial conditions are given for each of the six field components, and
the six derivatives of the field with respect to $S$.  The equations
describing the electromagnetic fields do not depend on the initial
values of the current, but in principle these are the three remaining
initial conditions.

\begin{figure}
	\centering
		\includegraphics[width=8.5cm]{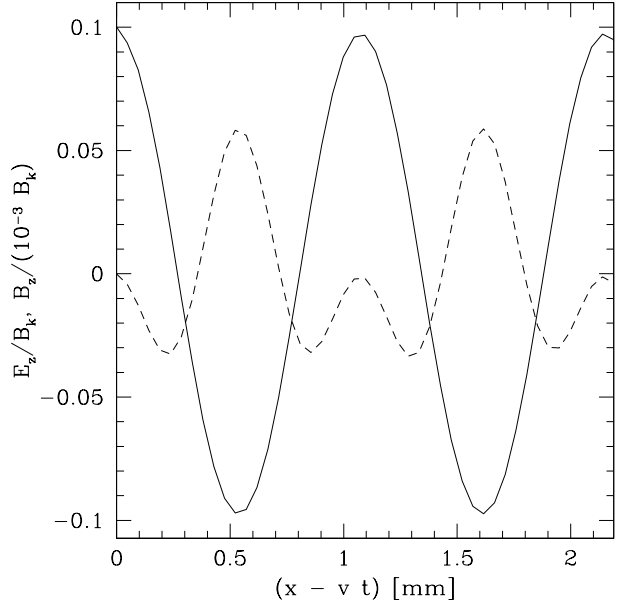}
                \caption[Nonlinear electromagnetic travelling wave]
				{A comparison between the $\bhat{z}$
                  components of the electric (solid) and magnetic
                  fields (dashed) showing the nonorthogonal
                  stabilizing wave. On larger scales, the $B_z$
                  component is seen to have a periodic envelope
                  structure as in figure \ref{fig:envelope}}.
	\label{fig:EzBzcompare}
\end{figure}

\begin{figure}
	\centering
		\includegraphics[width=8.5cm]{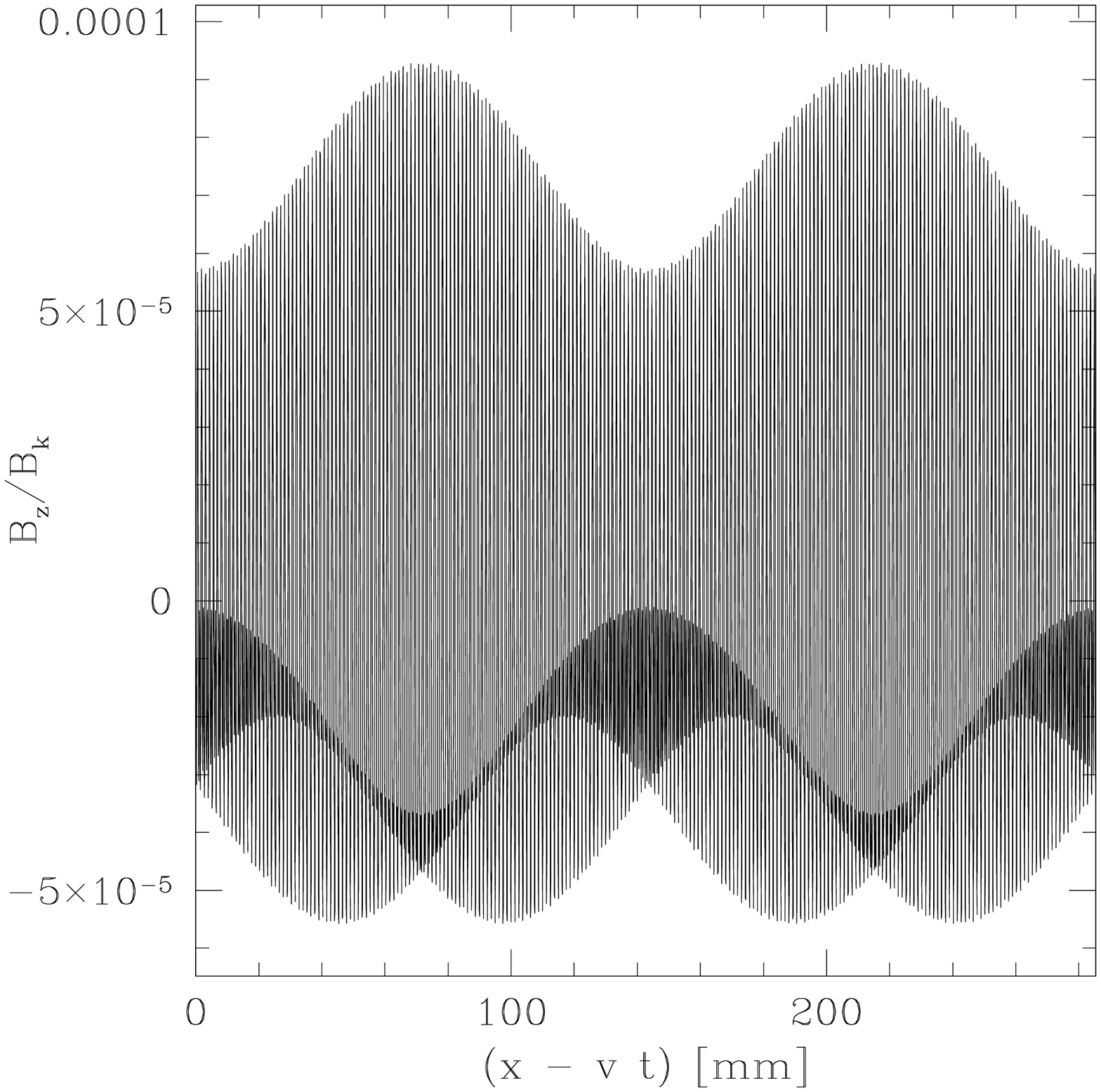}
                \caption[Travelling wave at larger scales]
				{The $B_z$ wave component as shown
                  in figure \ref{fig:EzBzcompare} has an envelope
                  structure when viewed at larger scales.  In this
                  case, the $E_z$ wave component has a nearly constant
                  amplitude of $0.1 B_k$.}.
	\label{fig:envelope}
\end{figure}
The \acp{ODE} are solved using a variable stepsize Runge-Kutta method.  In
order to translate between the $\efi$-and-$\bfi$ picture and the ${\bf
  \psi}$-and-${\bf \chi}$ picture, equations (\ref{eqn:u}) and
(\ref{eqn:v}) must be solved numerically at each time step, including
the first step when the initial conditions are given.

Tables of numerical values of the functions defined by equations
(\ref{eqn:x0anal}) to (\ref{eqn:x2anal}), as well as their
derivatives, were computed in advance and these were used to
interpolate the values needed in the simulation using a standard cubic
spline interpolation algorithm.  In producing these tables,
expressions for the weak and strong field limits were used in the
appropriate regimes as this reduced the numerical errors.

Aside from the initial conditions for the fields and derivatives,
there is one parameter in the model which must be selected.  The
density of the plasma, given by $n$ in equation(\ref{eqn:Jpdef}) is
chosen to be
\be
n = 10^{13}\mathrm{cm}^{-3}.
\ee
This value corresponds to the Goldreich-Julian 
density~\cite{1969ApJ...157..869G, lorimer2004handbook} for a star of
period $P\sim 1$~s, $\dot{P}\sim 10^{-10}$.  The uniform background
field is taken to equal the quantum critical field strength 
\be
	B_k = \frac{m^2}{e} = 4.413 \times 10^{13} \rmmat{G}.
\ee

\section{Results}
\label{sec:results}

This study focuses on the case of waves propagating transverse to a
large background magnetic field.  We have already chosen the direction
of propagation to be the $\bhat{x}$-direction through our definition
of $S$ in equation~(\ref{eqn:S}).  We now choose the background magnetic
field to point in the $\bhat{z}$-direction.  In this situation, the
largest non-linear effects occur when there is a large amplitude wave
in the $\bhat{z}$ component of the electric field.  This is quite
natural.  The values of both the dielectric tensor of the plasma
equation (\ref{eq:2}) and the weak-field limit of the vacuum dielectric
tensor equation (\ref{eq:1}) differ most from unity for this component; for
strong fields the index of refraction (as well as its derivative with
respect to the field strength) is largest for vacuum propagation in
this mode \citep{1997JPhA...30.6485H}.  Furthermore, the dominant
three-photon interaction (\ie photon splitting) couples photons with the electric field
pointed along the global magnetic field direction with photons whose
magnetic field points along this direction.  The three-point
interaction for photons whose electric field is perpendicular to the
magnetic field with parallel photons vanishes by the $CP-$invariance
of QED~\citep{Adle71}.  Therefore, we focus on initial
($S=0$) conditions in which the dominant component of the electric
field points along in the $\bhat{z}$-direction and that of the
magnetic field along the $\bhat{y}$-direction.

In the classical vacuum, these initial conditions correspond to
transverse, linearly polarized sine-wave solutions that travel at the
speed of light.  However, when the wave amplitudes are large, and there
is a strongly magnetized plasma, we find that there is a deviation from
normal transverse electromagnetic waves.  In particular, in order to
remain stable, a wave with large $E_z$ and $B_y$ field components must
also excite waves in the $E_y$ and $B_z$ fields.  These stabilizing
wave components exhibit strong non-linear characteristics (see figures
\ref{fig:EzBzcompare} and \ref{fig:envelope}).  The symmetries of the
wave equation require a close correspondence between the $E_y$ and
$B_z$ waveforms as well as between the $E_z$ and $B_y$ waveforms.  For
simplicity, only one of each is plotted in the figures.  The field
strengths are given in units of the quantum critical field strength,
$B_k$.

As is apparent from figure~\ref{fig:EzBzcompare} the dominant electric
field along the direction of the external magnetic field is
essentially sinusoidal.  Subsequent figures will show that there is a
small harmonic component.  The waveform for the dominant magnetic
field component is similar.  On the other hand, the magnetic field
along the direction of the electric field (the non-orthogonal
component) is smaller by nearly four orders of magnitude and obviously
exhibits higher harmonics.  In particular if one expands the scale of
interest (figure~\ref{fig:envelope}), the magnetic field exhibits
beating between two nearby frequencies with similar power.

In order to examine the harmonic content of the waveforms, we perform
fast Fourier transforms (FFTs) on the signals produced in the
simulations.  We present the results in terms of power spectra
normalized by the square amplitudes of the electric field of the
waves.  In these plots, the horizontal axis is normalized by the
frequency with the greatest power in the electric field, so that
harmonics can be easily identified.

\begin{figure}
	\centering
		\includegraphics[width=8.5cm]{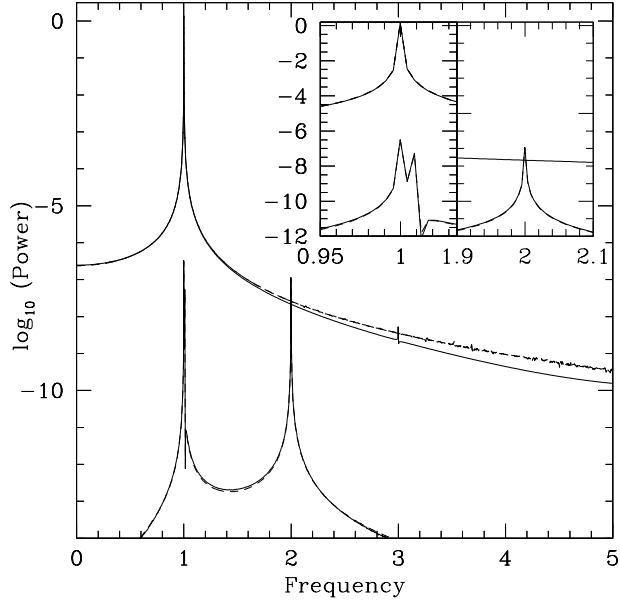}
                \caption[Travelling wave power spectrum]
				{The solid curve depicts power spectra of the
                  electric (upper) and magnetic fields (lower) along the
                  $\bhat{z}$-direction for the solution depicted in
                  Figs.~\ref{fig:EzBzcompare} and~\ref{fig:envelope}.
                  The inset focusses on the fundamental and the first
                  harmonic.  The dashed curve follows the power
                  spectrum of a single sinusoid for the electric field
                and three sinusoids for the magnetic field. Near the
                peaks the dashed curve is essentially
                indistinguishable from the solid one.}
	\label{fig:PScompare}
\end{figure}

Figure~\ref{fig:PScompare} depicts the power spectra of the electric and
magnetic fields along the $\hat{z}$-direction for the wave depicted in
Figs.~\ref{fig:EzBzcompare} and~\ref{fig:envelope}.  The conclusions
gathered from an examination of the waveforms are born out by the power
spectra.  In particular the electric field is a pure sinusoidal
variation to about one part in ten thousand -- the power spectrum of a
pure sinusoid is given by the dashed curved.  The duration of the
simulation is not an integral multiple of the period of the sinusoid,
resulting in a broad power spectrum even for a pure sinusoid.
The power spectrum of the magnetic field follows the expectations
gleaned from the waveforms.  In particular the fundamental and the
first harmonic are dominant, with the first harmonic having about
one-third the power of the fundamental.  If one focusses on the
fundamental, one sees that two frequencies are involved.  The envelope
structure is produced by a beating between the fundamental and a
slightly lower frequency with a similar amount of power as the first
harmonic.   Over the course of the simulation the envelope exhibits
two apparent oscillations; the lower frequency differs by two
frequency bins, so it is resolved separately from the fundamental, as
shown in the inset.

\begin{figure}
	\centering
		\includegraphics[width=8.5cm]{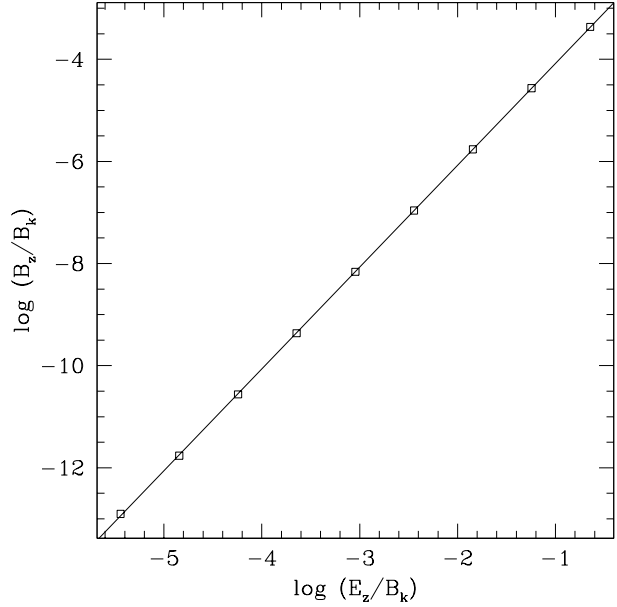}
                \caption[Amplitude of $B_z$ versus amplitude of $E_z$]
				{Amplitude of the $B_z$ component plotted against the amplitude of the 
		$E_z$ component for a $B_0 = B_k$ background field.
                The line is the best-fitting power-law relation. The slope is consistent with 
		a scaling exponent equal to two.
              }
	\label{fig:ampvsamp}
\end{figure}

As the amplitude of the electric field increases the non-linear and
non-orthogonal features of the travelling wave increase.
figure~\ref{fig:ampvsamp} shows that the strength of the non-orthogonal
magnetic field increases as the square of the electric field, a
hallmark of the non-linear interaction between the fields.  For the
strongest waves studied with $E_z \approx 0.2 B_k$ (the rightmost
point in the figure), the magnetic field, $B_z$, is about $10^{-4} B_k$, nearly
one-percent of the electric field.  The amplitude of the
non-orthogonal magnetic field is given by
\be
B_z = 0.008 B_k \left ( \frac{E_z}{B_k} \right )^2
\label{eq:3}
\ee
for $B_0=B_k$.  The coefficient is coincidentally very close to
three-quarters of the value of the fine-structure constant.  It
increases with the strength of the background field.

\begin{figure}
	\centering
		\includegraphics[width=8.5cm]{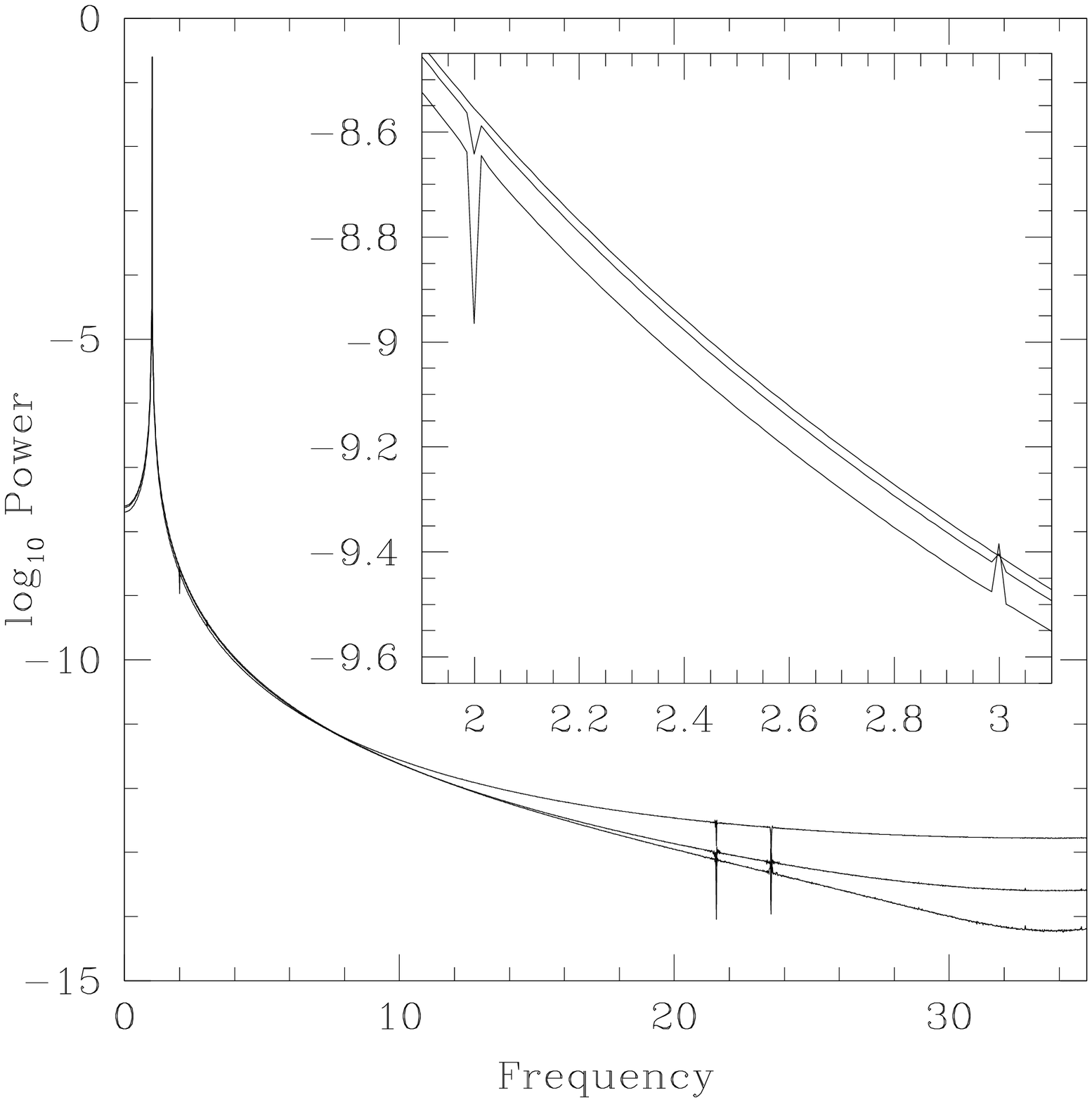}
                \caption[Power spectrum of the travelling wave]
				{This power spectrum demonstrates the
                  development of nonlinear effects in the
                  $\bhat{z}$-component of the electric field as the
                  amplitude of the wave is increased in a $B_0=B_k$
                  background field. From top to bottom the curves
                  follow the solutions whose amplitude of $E_z$ equals
                  $0.01, 0.08$ and $0.16 B_k$.  The inset focusses on the first
                  and second harmonics.}
	\label{fig:PSvsampE}
\end{figure}

For the strongest waves even the non-orthogonal magnetic field is
strong, so it can generate non-linearities in the electric field.
Although the strongest effect is around the fundamental, it is
completely swamped by the fundamental of the electric field.  On the
other hand, the magnetic field drives a first and second harmonic in
the electric field as seen in figure~\ref{fig:PSvsampE}.  The strength
of these harmonics is approximately given by the formula in
equation~(\ref{eq:3}) or equivalently figure~\ref{fig:ampvsamp} if one
substitutes the value of $B_z$ for $E_z$ and uses result for $E_z$.
This is essentially a sixth-order correction from the effective
Lagrangian.  Because we have used the complete Lagrangian rather than
a term-by-term expansion, all of the corrections up to sixth order
(and further) are automatically included in the calculation.

\begin{figure}
  	\centering
		\includegraphics[width=8.5cm]{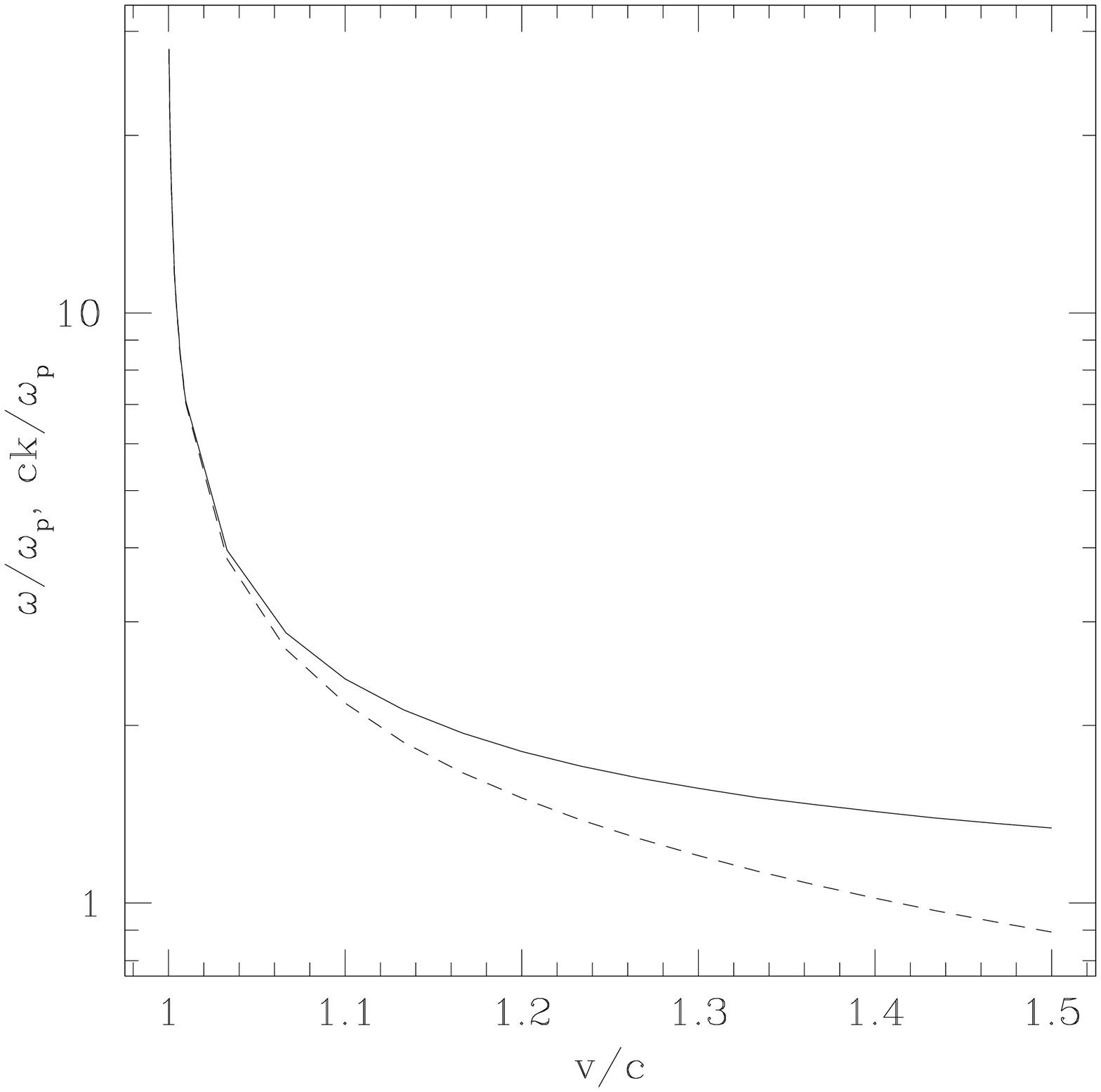}
                \caption[Frequency and wavenumber versus phase velocity]
				{The frequency (solid curve) and wavenumber
                  (dashed curve) of the travelling wave as a
                  function of its phase velocity.  }
	\label{fig:freqvsspeedEz}
\end{figure}

Figure \ref{fig:freqvsspeedEz} demonstrates how the frequency of the
solutions varies with the speed of propagation.  Because there is no
mode information stored directly in our numerical solutions, we take
the frequency to be rate that local minima in the electric field pass a
fixed observer.  In general, we find that the frequency of traveling
waves increases as the phase velocity approaches the speed of light,
very closely following the formula
\be
\left ( \frac{\omega}{\omega_p} \right )^2 =  \frac{v^2}{v^2-1}.
\ee
This formula also results from an analysis of the dielectric tensor,
equation~(\ref{eq:2}).  The vacuum makes a small contribution to the wave
velocity in this regime.
%

\section{Conclusion}

We have discussed techniques for computing electromagnetic waves in a
strongly magnetized plasma which nonperturbatively account for the
field interactions arising from \ac{QED} vacuum effects.  We
applied these methods to the case of travelling waves, which have a
spacetime dependence given by the parameter $S=x-vt$.  Travelling
waves can be described without any decomposition into Fourier modes
and this is ideal for exploring the nonperturbative aspects of waves.

The main result from this analysis is the observation that
electromagnetic waves in a strongly magnetized plasma can
self-stabilize by exciting additional nonorthogonal wave components.
In the cases studied above, a large amplitude excitation of the
electromagnetic field, for example, from the coupling between Alfv\'en
waves to starquakes, can induce nonlinear waves which are stabilized
against the formation of shocks.  The result is a periodic wave train
with distinctly nonlinear characteristics.  Such structures may play a
role in forming pulsar microstructures.

This result demonstrates that shock formation is not a necessary outcome 
for waves in a critically magnetized plasma.  It is possible that 
nonlinear features of a wave can stabilize it against shock formation. 
The shock-wave solutions generally decrease in magnitude, dissipating 
energy along the way unless they travel into a low-field region, so the 
self-stabilizing nonlinear waves are the 
only ones that keep their shape and energy content intact as they propagate.
It is not yet clear what set of conditions will determine 
if a particular wave will self-stabilize or collapse to form a shock or how 
plasma inhomogenaities will affect the propagation of a travelling 
wave train.   These are issues which can be clarified in future work.

\part{Magnetic Flux Tubes in Neutron Stars}
\label{pt:fluxtubes}
\chapter{Magnetic Flux Tubes}
\label{ch:fluxtubes}
\acresetall

\begin{summary}
This chapter serves as an introduction to the work presented in chapters 
\ref{ch:greensfunc} through \ref{ch:periodic}.  Neutron stars have a crust made 
of dense nuclear material which is generally expected to be in a type-II superconducting state. 
We therefore expect the intense magnetic fields in the crust of a neutron star 
to form a lattice of individual flux tubes. This provides the primary motivation for the 
following chapters which study magnetic flux tubes in \ac{QED}. In this chapter we will 
review some of the basic properties of the dense nuclear material of neutron stars that 
may result in the formation of a flux tube lattice.
\end{summary}

\section{Introduction}
In this part, we will explore the \ac{QED} effective action 
in cylindrically symmetric, extended tubes of magnetic flux.
These configurations may be called 
flux tubes, strings, or vortices, depending on the context.  
Flux tubes are of interest in astrophysics 
because they describe magnetic structures near stars and planets, 
cosmic strings~\cite{vilenkin2000cosmic}, and vortices in the superconducting core of neutron stars
\cite{2006pfsb.book..135S, schmitt2010dense}.
Outside of astrophysics, magnetic vortex systems are at the forefront of research in condensed 
matter physics for the role they play in superconducting systems and in \ac{QCD} research 
for their relation to center vortices, a gluonic configuration analogous to magnetic vortices 
which is believed to be important to quark confinement~\cite{tHooft19781, 2003PrPNP..51....1G}.  
In this part of the dissertation, we will primarily discuss the 
roles played by magnetic flux tubes in neutron star cores. 

Our motivation for discussing flux tubes comes from the fact that 
superconductivity is predicted in the nuclear matter of neutron stars
and that some superconducting materials produce a lattice of flux tubes 
when placed in an external magnetic field. Superconductivity is a macroscopic 
quantum state of a fluid of fermions that, most notably,
allows for the resistanceless conduction of charge. In 1933, 
Meissner and Ochsenfeld observed that magnetic fields are repelled 
from superconducting materials~\cite{meisner33}. 
In 1935, F. and H. London described the Meissner effect in terms of 
a minimization of the free energy of the superconducting current~\cite{london35}.
Then, in 1957, by studying the superconducting 
electromagnetic equations of motion in cylindrical coordinates, 
Abrikosov predicted the possible existence of line defects in superconductors 
which can carry quantized magnetic flux through the superconducting material
\cite{abrikosov57}. 

A more complete microscopic description of superconducting materials is given by 
BCS (Bardeen, Cooper, and Schrieffer) theory~\cite{PhysRev.108.1175}. 
However, detailed discussions 
are outside the scope of this chapter. Instead, I will simply outline the main 
features of superconductivity that motivate the study of flux tubes in neutron 
stars. Interested readers may pursue more thorough reviews of superconductivity 
and superfluidity in neutron stars~\cite{2006pfsb.book..135S, schmitt2010dense}.

\section{Superconductivity}

To see how superconductivity is possible, 
consider fermions at zero temperature with chemical potential, $\mu$. The
free energy of this system is 
\be
	\Omega = E - \mu N.
\ee
For non-interacting fermions, we could add fermions at the Fermi surface 
without changing the free energy, since the first and second terms would 
change by the same amount. If, instead, the fermions are interacting, 
the binding energy between the fermions means that the free energy can 
be reduced by adding fermions. So, if there is an attractive interaction 
between the fermions, then there is a new ground state of the fermion fluid
where pairs of fermions are created at the Fermi surface. These pairs of 
fermions are called Cooper pairs and (since a pair of fermions can be viewed 
as a boson) they form a Bose condensate.

The particles in a Cooper pair do not form a bound state, but they are interacting 
through attractive forces, so there is an energy $\Delta$ associated with 
separating them. The consequence of this is that there is an energy gap in 
the dispersion relation for the Cooper pair:

\be
	\epsilon_k = \sqrt{(\sqrt{k^2+m^2} - \mu)^2 + \Delta^2}.
\ee
This energy gap means that a finite amount of energy is required to 
excite a single electron state, even near the Fermi surface. 
An amazing consequence of this is that particles flowing through 
the Bose condensate cannot scatter inelastically 
from the phonons because 
fermions cannot be excited at low energies. For neutral fermions, 
such as neutrons, this can give rise to superfluidity, characterized by 
frictionless flows. If the fermions forming 
the cooper pairs are charge carriers such as protons or electrons, 
this effect gives rise to superconductivity, characterized by the 
resistanceless conduction of charge. 

\section{Energy Interpretation of the \texorpdfstring{\acs{QED}}{QED} Effective Action}

In section \ref{sec:qedea}, I argued that $E[J^\mu, \bar{\eta},\eta]$ was the 
vacuum energy associated with the external source. Thus, equation (\ref{eqn:LegendreTrans})
implies that the effective action is closely related to the vacuum energy of a 
classical field configuration. Importantly, the effective action 
can be thought of as the additive inverse of the energy. 
Since this idea is important for understanding 
the free energy associated with a magnetic flux tube, I will make the relationships 
between the Hamiltonian expectation value, $E[J^\mu, \bar{\eta},\eta]$, 
and $\Gamma[A^0_\mu, \bar{\psi}^0, \psi^0]$ clear and explicit
\footnote{This section follows section 16.3 of Weinberg~\cite{weinberg1966quantum}.}.

To study the vacuum energy of a static electromagnetic field, we would like to 
minimize the expectation value of the Hamiltonian, 
\be
	\mean{H}_\Omega = \bra{\Omega}H\ket{\Omega},
\ee
under the constraint that the vacuum expectation value of the quantum gauge 
field is the classical field,
\be
	\bra{\Omega}A_\mu(\vec{x})\ket{\Omega} = A_\mu^0(\vec{x}).
\ee
We assume that the classical fields, $\bar{\psi}^0$ and $\psi^0$, vanish.
Using the method of Lagrange multipliers, the function which we would like to 
minimize is:
\be
	\mean{H}_\Omega - \alpha\braket{\Omega}{\Omega}
	-\int d^3x \beta^\mu(\vec{x})\bra{\Omega}A_\mu(\vec{x})\ket{\Omega}.
\ee
So, we find that we must satisfy
\be
	\label{eqn:Hmin}
	H\ket{\Omega} = \alpha\ket{\Omega} + \int d^3x 
	\beta^\mu(\vec{x}) A_\mu(\vec{x}) \ket{\Omega}.
\ee
This can be done if the Lagrange multipliers 
$\alpha$ and $\beta^\mu(\vec{x})$ are functionals of $A_\mu^0(\vec{x})$.

Let $\ket{\Psi}_{J^\mu}$ be normalized eigenvectors of the Hamiltonian in the presence of 
external source $J^\mu$. The energy eigenvalue equation is 

\be
	\left[H-\int d^3x\left(J^\mu A_\mu + \bar{\eta}\psi + \bar{\psi}\eta\right)\right]
	\ket{\Psi}_{J^\mu} = \frac{E[J^\mu,\bar{\eta},\eta]}{\mathcal{T}}\ket{\Psi}_{J^\mu},
\ee
where $\mathcal{T}$ is the time extent of the functional integration.
If we turn on the external source $J^{0,\mu}$ adiabatically to put the vacuum into the 
energy eigenstate $\ket{\Psi}_{J^{0,\mu}}$, then equation (\ref{eqn:Hmin}) is satisfied by 
taking 
\be
	\ket{\Omega} = \ket{\Psi}_{J^{0,\mu}},
\ee
\be
	\alpha = E[J^{0, \mu}, 0, 0],
\ee
and
\be
	\beta_\mu(\vec{x}) = J_\mu^0(\vec{x}).
\ee
Under these substitutions, equation (\ref{eqn:Hmin}) provides an expression for the 
vacuum expectation value for the Hamiltonian in the external field in terms of the 
external fields and currents.
\be
	\mean{H}_{A_\mu^0} = \frac{E[J_\mu^0,0,0]}{\mathcal{T}} + \frac{\int d^4x J_\mu^0A^{0,\mu}}{\mathcal{T}}
\ee
The above equation is related simply to the definition of the effective action 
as a Legendre transformation of the energy
\be
	\tag{\ref{eqn:LegendreTrans}}
	\Gamma[A_\mu^0, \bar{\psi}^0, \psi^0] = -E[J^\mu, \bar{\eta},\eta]
	-\int d^4 y (J^\mu(y)A_\mu^0(y) + \bar{\eta}(y)\psi^0(y) +\bar{\psi}^0(y)\eta(y)).
\ee
So, we arrive at the relationship between the expectation value of the 
Hamiltonian in the external field and the effective action.
\be
	\mean{H}_{A_\mu^0} = -\frac{1}{\mathcal{T}} \Gamma[A_\mu^0]
\ee
Because the energy is extensive, the most interesting quantity for magnetic 
flux tube configurations is the total energy per unit length.
\be
	\frac{\mean{H}_{A_\mu^0}}{L_z} = -\frac{\Gamma[A_\mu^0]}{L_z \mathcal{T}}
\ee
Thus, we have explicitly confirmed the expectation that the negative effective 
action is related to the total vacuum energy of a specific field configuration.

\section{Flux Tube Free Energy}

Many of the basic properties of flux tubes in superconductors can 
be understood by examining the free energy in Ginzberg-Landau theory.
The Ginzberg-Landau free energy is given by

\ba
	\label{eqn:GLFE}
	E &=& \int d^2 x \biggr\{\frac{1}{2m}\biggr|\left(\vec{\nabla }
		- iq\vec{A}(\vec{x})\right)\psi(\vec{x})\biggr|^2 \nonumber \\
	 & & - \mu | \psi(\vec{x}) |^2+\frac{a}{2}| \psi(\vec{x}) |^4 
	 + \frac{1}{8\pi}(\vec{\nabla} \times \vec{A} (\vec{x}))^2\biggr\},	
\ea
where $\psi(\vec{x})$ is a complex order field such that the 
density of superconducting fermions is proportional to $|\psi|^2$, $\mu$ 
is the chemical potential and $a$ is related to the scattering length. 
The first term represents the dynamical energy of fluctuations of the 
order field. The next two terms represent the potential energy 
of the order field. The final term is the familiar energy of the 
classical magnetic field.

Incorporating the 1-loop effects from \ac{QED}, the classical term 
must be replaced with the energy of the field, 
$E_{\rm 1-loop} = -\frac{\Gamma[A_\mu^0]}{L_z\mathcal{T}}$.
\ba
	E &=& \int d^2 x \biggr\{\frac{1}{2m}\left|\left(\vec{\nabla }
		- iq\vec{A}(\vec{x})\right)\psi(\vec{x})\right|^2 \nonumber \\
	 & & - \mu|\psi(\vec{x})| ^2+\frac{a}{2}|\psi(\vec{x})|^4 
	 \biggr\} -\frac{\Gamma[A_\mu^0]}{\mathcal{T}}.	
\ea
This addition can be viewed as a correction to the classical 
magnetic field energy in the Ginzburg-Landau free energy.
Many important behaviours of superconductors are determined 
by minimizing the free energy. The technique discussed in section 
\ref{sec:stationaryEA} may lead to a way to perform
this minimization while including 1-loop \ac{QED} effects.

The Ginzburg-Landau free energy plays an important role in 
determining the difference between type-I and type-II superconductors.
For example, the free energy of a pair of flux tubes will contain 
terms characterizing the energy of each flux tube, but also 
cross terms arising from the interaction between these flux tubes.
The sign of this interaction energy determines if the flux tubes 
will repel each other and form a lattice, or attract each other and collapse.
In the classical case, when the fields are separated spatially, there 
is no interaction energy arising from the last term in equation (\ref{eqn:GLFE}).
However, when the 1-loop \ac{QED} effects are taken into account, the 
non-local effects arising due to a nearby flux tube make a contribution 
to the free energy~\cite{Langfeld:2002vy}. 

\subsection{Meissner Effect}

The free energy in a superconductor is minimized if the magnetic 
field obeys its equation of motion, the London equation,

\be
	\vec{\nabla} \times \vec{\nabla} \times \vec{B}(\vec{x})
	= - \lambda_L^2\vec{B}(\vec{x}),
\ee
where $\lambda_L$ is called the London penetration depth,
which is defined as
\be
	\label{eqn:LondonPD}
	\lambda_L = \sqrt{\frac{2m_f c^2}{4\pi (2 q_f)^2 n_0}}
\ee
where $m_f$ and $q_f$ are the fermion mass and charge, 
respectively, and $n_0$ is the fermion density.

The solution to this equation implies that magnetic 
fields are repelled exponentially from the surface
of the superconductor.

\be
	B(x) = e^{-\frac{x}{\lambda_L}}
\ee

Superconducting regions of a material must repel magnetic 
fields according to this Meissner effect. However, there 
can be a large free energy cost to expelling field lines 
around a superconducting region. For some materials, it 
is therefore energetically favourable to allow magnetic field 
lines to penetrate the material through a non-superconducting 
core. When this occurs, tubes of magnetic flux are 
permitted in the material.

\subsection{Interaction Energy Between Vortices}

The interaction energy between two vortices, and therefore 
the force between them, can be found by examining the 
free energy of a two vortex system and subtracting off 
the energy expected from each individual flux tube~\cite{PhysRevB.3.3821}.
The interaction energy is then given by the remaining cross terms

\be
	E_{\rm int} = \int d^2 x \left(J_1^\mu A_\mu^2 - \rho_1 \psi_2\right)
\ee
where $\psi_i$ and $A_\mu^i$ are the fields induced by sources 
$\rho_i$ and $J_i^\mu$~\cite{Speight_staticintervortex}.

In a superconductor, these two terms tend to oppose one 
another. Two neighbouring vortices 
have charged fermion currents in opposite directions 
and are repelled from each other. However, there 
is also an attractive force due to a free energy associated 
with defects in the superconductor which can be reduced 
if two vortices are combined~\cite{PhysRevB.3.3821}. 
So, the vortices can be 
either attractive or repulsive depending on the relative 
sizes of these terms.

This leads to two different behaviours for superconductors 
in the presence of magnetic fields. The behaviour is 
naively determined by the Ginzburg-Landau parameter,

\be
	\label{eqn:GLparam}
	\kappa = \frac{\lambda_L}{\xi}.
\ee

The coherence length, $\xi$, is the length scale associated with density variations of superconducting fermions. This
length scale is given in terms of $a$ in the Ginzburg-Landau theory (see equation (\ref{eqn:GLFE})), 
or in terms of the energy gap, $\Delta$, and the velocity of Cooper pairs, $v_f$ in BCS theory. 
\ba
	\xi &=& \sqrt \frac{\hbar^2}{4 m_f |a|} \\
	 &=& \frac{2 \hbar v_f}{\pi \Delta}
\ea
When $\kappa > \frac{1}{\sqrt{2}}$, the vortices will 
repel, and when $\kappa < \frac{1}{\sqrt{2}}$, the vortices 
will attract. Of course, this analysis has assumed that 
the free energy has been completely accounted for by 
the standard analysis. If there are additional 
considerations in the interaction energy, they will 
influence the properties of the superconducting material.
Previously, the effects of an asymmetry in the scattering 
length between neutron and proton cooper 
pairs~\cite{2004PhRvL..92o1102B},  and the 
effects of currents transported along vortices~\cite{2007PhRvC..76a5801C} have been 
suggested as mechanisms for making a type-II superconductor 
behave like a type-I superconductor. Friction between superconducting and 
normal fluid domains in a type-I proton superfluid 
causes a dissipation which is consistent with precession~\cite{2005PhRvD..71h3003S}.
Thus, a major motivation 
for studying the \ac{QED} contribution to the free energy 
of flux tubes, is to learn how this contribution may affect their 
interaction energies and the rotational dynamics of neutron stars.

If $\kappa$ is large, meaning the vortices repel, 
the superconductor can be in a so-called mixed state where the 
vortices will form a triangular Abrikosov lattice
with each vortex carrying a single quanta of flux~\cite{abrikosov57}. 
The superconductors where this happens allow partial 
penetration of the magnetic field and are called 
type-II superconductors. However, this mixed state only occurs for 
magnetic fields strong enough to form vortices, but not so strong that they 
destroy superconductivity. If $\kappa$ is small, 
the vortices will attract each other and annihilate. 
When this happens, the superconductor does not allow 
magnetic flux to penetrate the material and is called 
a type-I superconductor.

\begin{figure}[h]
	\centering
		\includegraphics[width=8.5cm]{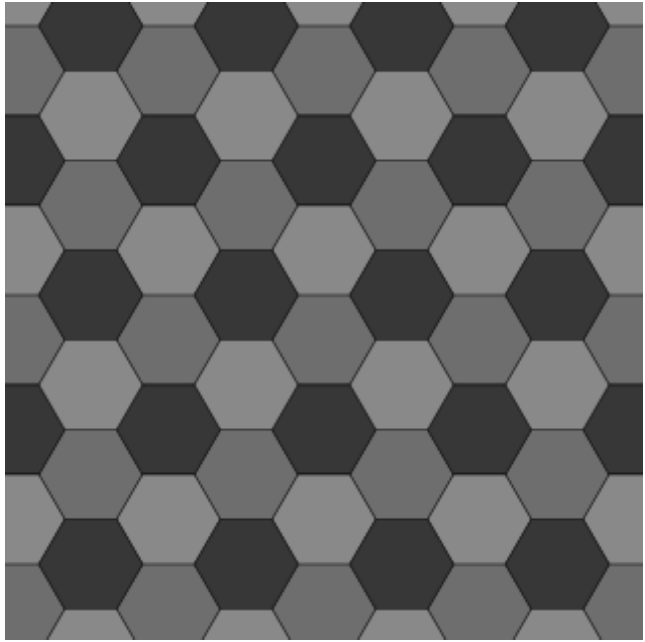}
                \caption[Hexagonal lattice]
				{In type-II superconductors, flux tubes 
				form an Abrikosov lattice. The geometry is 
				shown in the diagram above where each 
				hexagonal space can be imagined as containing 
				a flux tube at its centre}
	\label{fig:hexlatticegs}
\end{figure}

In this section, I have argued that without the electromagnetic 
forces, the vortices in a superconductor will experience an 
attractive force. Naively, this suggests that superfluids 
should be type-I since they lack a coupling to the gauge fields.
This contrasts with the experimental observation that 
superfluids are type-II. The subtlety here is that the 
nature of the self-interaction of the order field is different 
in the absence of gauge fields, and is always repulsive for 
superfluids. We can understand a superconductor as the $q_e \to 0$
limit of a superfluid. In this limit, the London penetration depth, equation 
(\ref{eqn:LondonPD}), becomes infinite and so does the Ginzburg-Landau
parameter, equation (\ref{eqn:GLparam}).
\be
	\lim_{q_e\to0}\kappa = \lim_{q_e\to 0}\frac{\lambda_L}{\xi} 
	\sim \lim_{q_e\to0}\frac{1}{q_e} = \infty
\ee
So, we predict type-II behaviour from superfluids, consistent with 
experimental results. The self-interaction of the order field 
is repulsive in the absence of a gauge field.

\section{Nuclear Superconductivity in Neutron Stars}

In the dense nuclear matter of a neutron star, it may be possible 
to have neutron superfluidity, proton superconductivity, and 
even quark colour superconductivity~\cite{2006pfsb.book..135S, schmitt2010dense}. 
The first prediction of 
neutron-star superfluidity dates back to Migdal in 1959~\cite{1959NucPh..13..655M}.
The arguments which make superfluidity seem likely are based on 
the temperatures of neutron stars. A short time after their 
creation, neutron stars are very cold. The temperature in the 
interior may be a few hundred keV. Studies of nuclear matter 
show that the transition temperature is 
$T_c \gtrsim 500$ keV~\cite{1989tns..conf..457S}.
So, it is expected that the nuclear matter in a neutron star 
forms condensates of cooper pairs.

Because of this, some fraction of neutrons in the inner crust 
of a neutron star are expected to be superfluid. These neutrons make up 
about a percent of the moment of inertia of the star and are 
weakly coupled to the nuclear crystal lattice which makes up the 
remainder of the inner crust. If the neutron vortices in the 
superfluid component move at nearly the same speed as the nuclear 
lattice, the vortices can become pinned to the lattice so that the 
superfluid shares an angular velocity with the crust. 
This pinning between the fluid and solid crust has observable impacts 
on the rotational dynamics of the neutron star, for the same reasons that 
hard-boiling an egg (pinning the yolk to the shell) produces an 
observable difference in the way it spins on its side.

This picture of a neutron superfluid co-rotating with a solid crust 
has been used to interpret several types of pulsar timing anomalies.
Pulsars are nearly perfect clocks, although they gradually spin-down 
as they radiate energy. Occasionally, though, pulsars demonstrate 
deviations from their expected regularity. A glitch is an abrupt 
increase in the rotation and spin-down rate of a pulsar, followed 
by a slow relaxation to pre-glitch values over weeks or years. This 
behaviour is consistent with the neutron superfluid suddenly 
becoming unpinned from the crust and then dynamically relaxing 
due to its weak coupling to the crust until it is pinned once 
again~\cite{1975Natur.256...25A}.

The neutron star in Cassiopeia A has been observed to be rapidly cooling
\cite{2010ApJ...719L.167H}. The surface temperature decreases by about 
4\% every 10 years. This observation is also strong evidence 
of superfluidity and superconductivity in neutron stars
\cite{2011PhRvL.106h1101P, 2011MNRAS.412L.108S}. The observed cooling 
is too fast to be explained by the observed x-ray emissions 
and standard neutrino cooling. However, 
the cooling is readily explained by the emission of neutrinos during 
the formation of neutron Cooper pairs. Based on such a model, 
the superfluid transition temperature of neutron star matter is
$\sim 10^{9}$ K or $\sim 90$ keV. 

Further hints regarding superfluidity in neutron stars 
come from long-term periodic variability
in pulsar timing data. For example, variabilities in PSR B1828-11
were initially interpreted as free 
precession (or wobble) of the star~\cite{2000Natur.406..484S}. 
If neutron stars can
precess, observations could strongly constrain the ratio of the moments
of inertia of the crust and the superfluid neutrons. Moreover, the existence
of flux tubes (\ie type-II superconductivity)
in the crust is generally incompatible with the slow, large 
amplitude precession suggested by PSR B1828-11~\cite{2000Natur.406..484S}. The neutron vortices 
would have to pass through the flux tubes, which should cause a huge dissipation 
of energy and a dampening of the precession which is not observed~\cite{PhysRevLett.91.101101}. 
However,
recent arguments suggest that the timing variability data is not well explained 
by free precession and that it more likely suggests that the star is switching between 
two magnetospheric states~\cite{2010Sci...329..408L}. 
Nevertheless, other authors suggest not being premature in throwing 
out the precession hypothesis without further observations
\cite{2012MNRAS.420.2325J}.

\section{Magnetic Flux Tubes in Neutron Stars}

If a magnetic field is able to penetrate the proton superfluid on a microscopic level,
it must do so by forming a triangular Abrikosov 
lattice with a single quanta of flux in each flux tube. So, the 
density of flux tubes is given simply by the average field strength.
If the distance between flux tubes is $l_f$, the flux in a circular 
region within $l_f/2$ of a flux tube is given by 

\be
	F = \frac{2\pi \mathcal{F}}{e} = 2\pi \int_0^{l_f/2} B \rho d\rho
\ee
where we have introduced a dimensionless measure of flux 
$\mathcal{F} = \frac{e}{2\pi}F$.
So, the distance between flux tubes is
\be
	l_f = \sqrt{\frac{8 \mathcal{F}}{eB}}.
\ee
If the magnetic field is the critical field strength, $B_k$, then 
the flux tubes are separated by a few Compton electron wavelengths.
This is particularly interesting since this is the distance scale 
associated with non-locality in \ac{QED}.

The size of a flux tube profile in laboratory superconductors is 
on the nanometer or micron scale~\cite{poole2007superconductivity}. Because the flux 
is fixed, the size of the tube profile determines the strength of the magnetic field 
within the tube. For laboratory superconductors
the field strengths are small compared to the quantum critical field, and the field is slowly 
varying on the scale of the Compton wavelength.  In this case, the quantum corrections to the 
free energy are known to be much smaller than the classical contribution 
(see section \ref{sec:EAisoflux}).
The size scale for the flux tubes in a superconductor is determined by the London penetration depth. 
In a neutron star, this quantity is estimated to be a small fraction of a 
Compton wavelength, much smaller than in laboratory superconductors
\cite{lrr-2008-10, PhysRevLett.91.101101}. In this case, 
the magnetic field strength at the centre of the 
tube exceeds the quantum critical field strength and the field varies rapidly, 
rendering the derivative expansion description of the effective action unreliable.

The Ginzburg-Landau parameter, equation (\ref{eqn:GLparam}), is the 
ratio of the proton coherence length, $\xi_p \sim 30$ fm, and the 
London penetration depth of a proton super conductor, $\lambda_p \sim 80$~fm
\cite{PhysRevLett.91.101101}. 

\be
	\kappa = \frac{\lambda_p}{\xi_p} \sim 2
\ee
We therefore expect that the proton cooper pairs most likely 
form a type-II superconductor~\cite{Sedrakian:2006xm}. 
However, it is possible that 
physics beyond what is taken into account in the standard 
picture affects the free energy of a magnetic flux tube. In 
that case, the interaction between two flux tubes may indeed 
be attractive in which case the neutron star would be a type-I 
superconductor.

\section{\texorpdfstring{\acs{QED}}{QED} Effective Actions of Flux Tubes}
\label{sec:EAisoflux}

Vortices of magnetic flux have very important impacts on the quantum 
mechanics of electrons. In particular, the phase of the electron's wavefunction 
is not unique in such a magnetic field. This is demonstrated by the Aharonov-Bohm
effect~\cite{0370-1301-62-1-303,PhysRev.115.485}. The first calculations of the 
fermion effective energies of these configurations were for infinitely thin Aharonov-Bohm
flux strings~\cite{Gornicki1990271, 1998MPLA...13..379S}. Calculations for thin strings 
were also performed for cosmic string configurations~\cite{0264-9381-12-5-013}. For these 
infinitely-thin string magnetic fields, the energy density is singular for small radii. So, 
it is not possible to define a total energy per unit length. Another approach was to 
compute the effective action for a finite-radius flux tube where the magnetic flux was 
confined entirely to the radius of the tube~\cite{PhysRevD.51.810}. This approach 
results in infinite classical energy densities.

Physical flux tube configurations would have a finite radius.
The earliest paper to deal with finite radius flux tubes in QED considered the effective 
action of a step-function profiled flux tube using the Jost function of the related 
scattering problem~\cite{1999PhRvD..60j5019B}. One of the conclusions from this research 
was that the quantum correction to the classical energy was relatively 
small for any value of the flux tube size, for the entire range of applicability of \ac{QED}. 
The techniques from this study were 
soon generalized to other field profiles including, a delta-function cylindrical shell magnetic 
field~\cite{PhysRevD.62.085024}, and more realistic flux tube configurations 
such as the Gaussian~\cite{2001PhRvD..64j5011P} and the Nielsen-Olesen 
vortex~\cite{2003PhRvD..68f5026B}. 
Flux tube vacuum energies were also analyzed extensively using a spectral method
\cite{2005NuPhB.707..233G, 2006JPhA...39.6799W, Weigel:2010pf}.

The effective actions of flux tubes have been previously analyzed using worldline 
numerics~\cite{Langfeld:2002vy}.  This research investigated isolated 
flux tubes, but also made use of the loop cloud method's applicability to situations 
of low symmetry to investigate pairs of interacting 
vortices. One conclusion from that investigation was that the fermionic effects resulted 
in an attractive force between vortices with parallel orientations, and a repulsive 
force between vortices with anti-parallel orientations.
Due to the similarity in scope and technique, the latter mentioned 
research is the closest to the research presented in chapter \ref{ch:periodic}.

\section{Conclusion}

In this chapter I have discussed my motivation for studying tubes of magnetic 
flux, and the 1-loop \ac{QED} correction to the energy of these configurations. 
In particular, I am interested in the role that flux tubes play in neutron stars 
where the magnetic fields vary rapidly on the Compton wavelength scale. 
In chapter \ref{ch:greensfunc}, I outline progress toward a new Green's 
function technique for computing effective actions for flux tubes, and which 
yields an expression for the quantum-corrected equations of motion for the 
magnetic field. In chapters \ref{ch:WLNumerics} and \ref{ch:WLError}, I outline 
a numerical method implemented on \acp{GPU} to compute the 1-loop effective action 
for magnetic flux tubes. The physics results and conclusions from this method 
are presented in chapter \ref{ch:periodic}.

\chapter{Green's Function Method for Cylindrically Symmetric Flux Tubes}
\label{ch:greensfunc}
\acresetall

\begin{summary}
	The effective action for arbitrary, non-homogeneous magnetic backgrounds can be expressed in terms 
	of Green's functions. In this chapter, I derive the differential equation 
	for these Green's functions for magnetic flux tubes of arbitrary profiles. This provides a 
	new tool for thinking about the flux tube effective action and suggests a new numerical 
	method that might be applied to the problem. 
	The Green's functions expressions are also useful for examining 
	field configurations with stationary effective actions. The specific case of step-function 
	flux tubes is discussed in detail in the Green's function picture.
\end{summary}

\section{Introduction}
In chapter \ref{ch:fluxtubes}, we discussed the astrophysical motivations for examining 
the effective actions of magnetic flux tubes in \ac{QED}. However, computing the fermion 
determinant for non-homogeneous magnetic fields can be quite involved. For computing magnetic 
flux tube effective actions, there are only a few methods available in the literature, some 
of which do not generalize easily to different flux tube profile shapes. These methods are 
discussed in section \ref{sec:EAisoflux}. Despite the calculational difficulties, 
the quantum correction to the classical energies of flux tube configurations are potentially 
significant, and could impact our conclusions about their stability or other properties. 
It is therefore interesting to explore other methods 
that might provide new tools for performing the calculation, and new insights into the problem.

In this chapter, I discuss some progress toward such a new method for the magnetic flux tube 
problem based on expressing the effective action in terms of integrals and sums over 
Green's functions. Since the Green's functions can be determined numerically, this 
method suggests a possible new numerical technique for computing effective actions. 
A Green's function technique has been used to exactly derive the effective action for a family of 
non-homogeneous magnetic fields in 2+1 dimensions~\cite{PhysRevD.52.R3163} and 3+1 dimensions
\cite{1998PhLB..419..322D}. The family of fields is given by 

\be
	\vec{B}(\vec{x}) = B_0 {\rm sech}{^2\left(\frac{x}{\lambda}\right)}\hat{z}.
\ee
For fields in this form, the effective action can be expressed as a finite integral. This 
result is particularly valuable because there are so few field configurations which are 
exactly solvable for the effective action. So, it provides a good tool for testing our 
understanding of the derivative expansion and other approximate expressions of the effective 
action. Presently, we will use the approach used to discover these exact solutions 
to explore the effective actions of 
flux tube configurations. 

\section{Effective Action in Arbitrary Fields}
\label{sec:arbitraryfields}

%
%
%
 We begin with the effective action, equation (\ref{eqn:generalEA}), 
 in a static, but non-homogeneous magnetic field:
 
 \be
 \Gamma[A^0_\mu] = \Gamma_0 - \frac{i \hbar}{2} {\rm Tr ~ln} \left[\frac{(p_\mu+eA^0_\mu)^2
	+\frac{1}{2}e\sigma^{\mu\nu}F^0_{\mu\nu} + m^2}{p^2 +m^2}\right].
\ee
Since $A_\mu^0$ is time-independent and $A_0^0$ vanishes for static magnetic 
fields, we may express the canonical four-momentum in terms of its angular 
frequency eigenbasis and the canonical three-momentum,
\ba
	(p_\mu+eA^0_\mu)^2 &=& p^2 +e(p^\mu A^0_\mu + A^0_\mu p^\mu) + e (A^0)^2 \\
	& = & -\omega^2 -\vec{\nabla}^2 - ie[(\vec{\nabla}\cdot\vec{A}^0) + 
	(\vec{A}^0\cdot \vec{\nabla})] -e^2 (A^0)^2 \\
	& \equiv & -\omega^2 -\vec{\Pi}^2.
\ea
Then,

\be 
\label{eqn:EffAct}
\Gamma[A^0_\mu] = \Gamma_0 - \frac{i \hbar}{2} {\rm Tr ~ln} \left[\frac{-\omega^2 - \vec{\Pi}^2
	+\frac{1}{2}e\sigma^{\mu\nu}F^0_{\mu\nu} + m^2}{p^2 +m^2}\right] 
\ee
where
 \be 
	\sigma^{\mu \nu} = \frac{i}{2}[\gamma^\mu,\gamma^\nu]
 \ee 

and
\be 
	\Gamma_0 = \int d^4x\left(-\frac{1}{4}F^0_{\mu \nu}F^{0~\mu \nu}\right)
\ee 
is the classical action.

In order to express the effective action in terms of Green's functions, 
we make use of the fact that the trace involves an integration over 
frequency, $\omega$, and integration-by-parts in this integral 
produces the inverse of the operator:
\ba
	\lefteqn{{\rm Tr~ln}\left[-\omega^2-\vec{\Pi}^2+
	\frac{1}{2}e\sigma^{\mu \nu}F^0_{\mu \nu}+m^2\right]} \nonumber \\
	& = & \int_{-\infty}^\infty d\omega {\rm Tr}_3 ~{\rm ln}
	\left[-\omega^2-\vec{\Pi}^2+\frac{1}{2}e\sigma^{\mu \nu}F^0_{\mu \nu}+m^2\right] \\
	&=& 2\int_{-\infty}^\infty \omega^2 d\omega {\rm Tr}_3 
	\left[-\omega^2-\vec{\Pi}^2+\frac{1}{2}e\sigma^{\mu \nu}F^0_{\mu \nu}+m^2\right]^{-1} \\
	& = & \int_{-\infty}^\infty d\omega \int d^3\vec{x} d^3\vec{x}' 
	G_3(\vec{x},\vec{x}')\delta^{(3)}(\vec{x}-\vec{x}').
\ea

So, we may express (\ref{eqn:EffAct}) in terms of time-independent 
Green's functions to evaluate the trace:

\be
	\Gamma[A^0_\mu] =\Gamma_0 - i\hbar \int^{\infty}_{-\infty}\omega^2d\omega\int d^3\vec{x} 
		~\left(G_3(\vec{x},\vec{x})-G_3^0(\vec{x},\vec{x})\right),
\ee
where $G_3(\vec{x},\vec{x}')$ is defined by the Schr\"{o}dinger-like equation

\be 
\label{eqn:Greens}
\left[ -\vec{\Pi}^2 + \frac{1}{2}e\sigma^{\mu \nu}F^0_{\mu \nu} + m^2 
	-\omega^2\right]G_3(\vec{x},\vec{x}') 
	= \delta^{(3)}(\vec{x}-\vec{x}').
\ee 
$G_3^0(\vec{x},\vec{x}')$ is the Green's function in the absence of external fields (i.e. $A_\mu=0$),

\be 
\label{eqn:Green0}
\left[ -\vec{p}^2 + m^2 
	-\omega^2\right]G^0_3(\vec{x},\vec{x}') 
	= \delta^{(3)}(\vec{x}-\vec{x}').
\ee 

\section{Cylindrically Symmetric B-fields}

We now specialize our analysis to the case of a cylindrically symmetric magnetic field, 
pointing in the \bhattext{z}-direction in cylindrical coordinates, the symmetry relevant 
to flux tube configurations.
Making a gauge choice that $A_0 = A_\rho = A_z = 0$, 
the magnetic field is $\vec{B} = B(\rho)$\bhattext{z} with

\be 
\label{eqn:Bgf}
	B(\rho) = \frac{A_\phi(\rho)}{\rho} + \frac{dA_\phi(\rho)}{d\rho}.
\ee

Using the cylindrical symmetry, the Green's function is a function of the radial 
coordinate and can be expressed as a sum over the magnetic quantum number, $m_l$, 
the momentum along the axis of symmetry, $k_z$, the frequency, $\omega$, and 
the spin projection eigenvalues, $\sigma^3$:

\be
	G_3(\vec{x},\vec{x}') = \sum_{m_l,k_z,\omega,\sigma^3} G_{m_l,k_z,\omega,\sigma^3}(\rho,\rho').
\ee
From equation (\ref{eqn:Greens}), we can write the equation governing each term of the 
Green's function,

\be 
\label{eqn:cylGreen}
\left[ -\frac{d^2}{d\rho^2}  - \frac{1}{\rho}\frac{d}{d\rho}+ \frac{1}{\rho^2}+V_{m_l}(\rho) +k_z^2 +m^2-\omega^2
	 +\frac{1}{\rho^2}\right]\cylGreen 
	= \delta(\rho-\rho')
\ee 
with the magnetic potential given by

\be 
V_{m_l}(\rho) = \frac{1}{2}e\sigma^{\mu \nu}F^0_{\mu \nu} +\frac{(m_l^2-1)}{\rho^2} 
	-\frac{2m_l e}{\rho}\Aphi +e^2 (\Aphi)^2.
\ee 

The spin-field coupling is proportional to the spin eigenvalue and the magnetic field,
$\sigma^{\mu \nu}F_{\mu\nu} = 2 \sigma^{1 2}B_z = 
	2 \sigma ^3\left(\frac{\Aphi}{\rho} +\frac{d\Aphi}{d\rho}\right)$.  So, the 
potential term is given by

%
%
%
%
%
%

\be 
\label{eqn:effPot}
V_{m_l}(\rho) = e\sigma^{3} \left(\frac{\Aphi}{\rho}  + 
	\frac{d\Aphi}{d\rho}\right) + \frac{(m_l^2-1)}{\rho^2}+
	e^2 (\Aphi)^2 - \frac{2e m_l}{\rho} \Aphi.
\ee 
$\sigma^3$ is the spin projection eigenvalue $+1/-1$ for spin up/down.

\section{Solution Strategy}

We will attempt to find the effective action by computing the Green's function, 
(\ref{eqn:cylGreen}).  This can be done by
matching the independent solutions of the homogeneous equation,

\be
\left[ -\frac{d^2}{d\rho^2}  - \frac{1}{\rho}\frac{d}{d\rho}+ \frac{1}{\rho^2}+V_{m_l}(\rho) +k_z^2 +m^2-\omega^2
	 +\frac{1}{\rho^2}\right]u_{m_l,k_z,\omega,\sigma^3}(\rho) 
	= 0.
\ee
For a given model of the magnetic field (a choice of $\Aphi$), 
the homogeneous solutions $u_0(\rho)$ and $u_\infty(\rho)$ are regular at $\rho =0$
and $\rho=\infty$ respectively.  The Wronskian of these solutions is defined as

\be 
W(\rho) = u'_0(\rho)u_\infty(\rho) - u_0(\rho)u_\infty'(\rho).
\ee 
Then, the Green's function which solves (\ref{eqn:cylGreen}) is simply given by

\be 
G_{\omega,k_z, m_l} (\rho, \rho') = \theta(\rho' - \rho) \frac{\rho u_0(\rho)u_\infty(\rho')}{W_0} + 
	\theta(\rho-\rho')\frac{\rho u_0(\rho')u_\infty(\rho)}{W_0}
\ee 
where
\be 
W_0 = \rho W(\rho)
\ee 
is constant by Abel's theorem.

\begin{sloppypar}
We can exactly solve for the Green's function in the 
absence of fields, $G_{\subscripts}^0(\rho,\rho)$,
 by noticing that the 
differential equation is Bessel's equation in the dimensionless 
coordinate $x=\sqrt{\omega^2-k_z^2 -m^2}\rho$.
Then, the homogeneous solution that is regular at $\rho=0$, $u_0(\rho)$ is identified 
with $J_{m_l}(\sqrt{\omega^2-k_z^2-m^2}\rho)$ and the homogeneous solution 
that is regular at $\rho=\infty$ is identified with $Y_{m_l}(\sqrt{\omega^2-k_z^2-m^2}\rho)$, 
where $J_n(x)$ and $Y_n(x)$ are Bessel functions of the first and second kind,
respectively.
So, we have

\be 
G_{\subscripts}^0(\rho,\rho) = \frac{\rho}{W^0_0}J_{m_l}(\sqrt{\omega^2-k_z^2-m^2}\rho)
	Y_{m_l}(\sqrt{\omega^2-k_z^2 -m^2}\rho).
\ee 
We can also compute
$W_0^0$ exactly for the field-free case using the Bessel function identity

\be 
\label{eqn:BessID}
J'_n(x)Y_n(x) - J_n(x)Y'_n(x)=-\frac{2}{\pi x}.
\ee 
Then,
\be 
\label{eqn:BkgGreens}
G_{\subscripts}^0(\rho,\rho) = -\frac{\pi \rho}{2}J_{m_l}(\sqrt{\omega^2-k_z^2 -m^2}\rho)
	Y_{m_l}(\sqrt{\omega^2-k_z^2 -m^2}\rho)
\ee 
\end{sloppypar}
We can now write the (unrenormalized) effective action as

\ba 
	\Gamma &=& \Gamma_0 -i\hbar \sum_{\rm \sigma^3= \{\pm1\}}\sum_{m_l=-\infty}^\infty  
	 \int_{-\infty}^\infty \omega^2 d\omega dk_z
	\int_0^{\infty} d\rho \rho (G_{\subscripts}(\rho,\rho) \nonumber \\ 
	& &~~~~~~- G_{\subscripts}^0(\rho,\rho)) \\
\label{eqn:soln}
	&=& \Gamma_0 -i\hbar \sum_{\rm \sigma^3= \{\pm1\}}\sum_{m_l=-\infty}^\infty  
	\int_{-\infty}^\infty \omega^2 d\omega dk_z
	\int_0^{\infty} d\rho \rho^2 \biggr(\biggr[\frac{ 
	u_0(\rho)u_\infty(\rho)}{W_0}\biggr]_{\subscripts} \nonumber \\ 
	& &~~~~~~~+\frac{\pi}{2}J_{m_l}(\sqrt{\omega^2-k_z^2 -m^2}\rho)
	Y_{m_l}(\sqrt{\omega^2-k_z^2 -m^2}\rho)\biggr) \nonumber \\
	&=& \Gamma_0 +\hbar\pi \sum_{\rm \sigma^3= \{\pm1\}}\sum_{m_l=-\infty}^\infty  
	\int_{0}^\infty \chi^3 d\chi
	\int_0^{\infty} d\rho \rho^2 \biggr(\biggr[\frac{ 
	u_0(\rho)u_\infty(\rho)}{W_0}\biggr]_{\csubscripts} \nonumber \\ 
	& &~~~~~~~+\frac{\pi}{2}J_{m_l}(\sqrt{\chi^2 -m^2}\rho)
	Y_{m_l}(\sqrt{\chi^2 -m^2}\rho)\biggr).
\ea 
The final equality is arrived at by analytically 
continuing $\omega\to i\omega$ and observing that 
the value of the integral along a semi-circle which 
closes the contour is unchanged by the rotation 
in the complex plane.  
By Abel's identity, $W_0$ has no dependence on $\rho$.  
It can come outside of the integral.  In the above equations, we have introduced a sum 
over the eigenvalues of the operator $\sigma^{1 2} = \frac{i}{2}(\gamma^1 \gamma^2 - \gamma^2 \gamma^1)$.
So, to account for two pairs of degenerate eigenvalues, 
we introduce an overall factor of $2$ and sum over $\sigma^3=\{+1,-1\}$.

The $u(\rho)_{\csubscripts}$'s are solutions to the differential equation

\be 
\label{eqn:chieqn}
\left(-\frac{d^2}{d\rho^2} - \frac{1}{\rho}\frac{d}{d\rho} + 
	V_{m_l}(\rho) - \chi^2 + m^2 +\frac{1}{\rho^2}\right)u_{[0,\infty]}(\rho)|_{\chi,m_l,\sigma^3} = 0
\ee 
where $[u_0(\rho)]_{\csubscripts}$ and $[u_\infty(\rho)]_{\csubscripts}$ are independent solutions which 
are regular at $\rho = 0$ and $\rho = \infty$, respectively.

%
%

%
%

Equations (\ref{eqn:soln}) and (\ref{eqn:chieqn}) are the most general form for the 
effective action in this method. To use these equations for computations, we must know the 
functions $[u_0(\rho)]_{\csubscripts}$ and $[u_\infty(\rho)]_{\csubscripts}$. Therefore, 
we can proceed either by choosing a form of the potential, $V_{m_l}(\rho)$, and 
solving the differential equation analytically, or by computing the required functions 
numerically for an arbitrary potential. In section \ref{sec:stepfunc}, we will 
make some analytic progress using a step-function flux tube profile. For either 
technique, though, we must establish the boundary conditions on these solutions, 
so we discuss those next.

\section{Boundary Conditions}
We would like to choose suitable boundary conditions to numerically 
compute $[u_0(\rho)]_{\csubscripts}$ and $[u_\infty(\rho)]_{\csubscripts}$.  We begin by 
considering $\Aphi$ in the form

\be 
\Aphi = \frac{F}{2\pi \rho}f_\lambda(\rho)
\ee 
so that
\be 
B_z(\rho)=\frac{F}{2\pi\rho}\frac{df_\lambda(\rho)}{d\rho}.
\ee 
The subscript $\lambda$ refers to a length scale that characterizes the size of the flux 
tube.  In practice, it is inconvenient to discuss the boundary conditions at infinity. So, 
we will introduce a scale, $L_\rho \gg \lambda$, which is finite but 
sufficiently far away from the flux tube. 
The total magnetic flux is
\be 
\Phi=F(f_\lambda(L_\rho)-f_\lambda(0)).
\ee 
It is convenient to express the flux in units of $\frac{2 \pi}{e}$ and define 
a dimensionless quantity

\be 
\mathcal{F}=\frac{e}{2 \pi} F
\ee 
Under these definitions, our operator is

\be 
-\frac{d^2}{d\rho^2} - \frac{1}{\rho}\frac{d}{d\rho} + 
	\frac{\sigma^3\mathcal{F}}{\rho}f'_\lambda(\rho) +
	 \frac{1}{\rho^2}\left(m_l -\mathcal{F}f_\lambda(\rho)\right)^2
	-\left(\chi^2-m^2\right).
\ee 

If we require that $B_z(\rho)$ remains finite at $\rho \to 0$ and that
 $\frac{dB_z(\rho)}{d\rho}\biggr|_{\rho=0}=0$, then we can choose without 
loss of generality that, for small $\rho$, $f_\lambda(\rho) = C_1\rho^2 + C_2\rho^4 +...$ 
and $f_\lambda(L_\rho) = 1$. Then, the total flux, $\Phi$, is simply $F$, and the 
field profile near the center is a quadratic with coefficient given by

\be 
C_1=\frac{eB_z(0)}{2 \mathcal{F}}\equiv\frac{e B_0}{2\mathcal{F}}.
\ee 

With the above observations, we notice that equation (\ref{eqn:chieqn}) takes the form of 
Bessel's equation in the asymptotic limits of small and large $\rho$.  
We use this fact to 
inform the following boundary conditions:

\ba 
\label{eqn:BCa}
 u_0(0)|_{\csubscripts} &=& J_{m_l}(0) \\
 u'_0(0)|_{\csubscripts} &=& \sqrt{\chi^2-m^2-eB_0(\sigma^2-m_l)} J'_{m_l}(0)\nonumber \\
	&=&\frac{\sqrt{\chi^2-m^2-eB_0(\sigma^3-m_l)}}{2} \nonumber \\
 	& & \times(J_{m_l-1}(0)-J_{m_l+1}(0))\\
 u_\infty(L_{\rho})|_{\csubscripts} &=& Y_n(\sqrt{\chi^2-m^2}L_\rho) \\
\label{eqn:BCb}
 u'_\infty(L_{\rho})|_{\csubscripts} &=& \sqrt{\chi^2-m^2-eB_0(\sigma^2-m_l)} Y'_{m_l}(0)\nonumber \\
	&=&\frac{\sqrt{\chi^2-m^2}}{2}
  	(Y_{n-1}(\sqrt{\chi^2-m^2}L_\rho) \nonumber \\
 	& & -Y_{n+1}(\sqrt{\chi^2-m^2}L_\rho)) \\
 n&\equiv&m_l-\mathcal{F}.
\ea 

If we take $F=0$, these boundary conditions reproduce our result for the 
Green's function in the absence of any fields, equation (\ref{eqn:BkgGreens}).

From equation (\ref{eqn:soln}), we can compute the effective action given the homogeneous solutions 
to the Green's function, $u_0(\rho)$ and $u_\infty(\rho)$.
Equations (\ref{eqn:cylGreen}) and (\ref{eqn:effPot}) define an \ac{ODE} that can be numerically 
integrated to find the homogeneous solutions given a cylindrically symmetric vector potential $A_\phi(\rho)$
and suitable boundary conditions, equations (\ref{eqn:BCa}) to (\ref{eqn:BCb}).

\section{Configurations with Stationary Action}
\label{sec:stationaryEA}

The classical motion of a system is determined by the principle of least action: 
classical systems follow paths for which the variation of the action is zero. 
\be 
\delta \Gamma = 0
\ee 
The quantum 
mechanical explanation for this is well-known. Consider the partition functional 
of \ac{QED}, equation (\ref{eqn:qedpartition}),

\ba
\lefteqn{Z[J^\mu, \bar{\eta}, \eta]}\nonumber \\
  & =&\int \mathcal{D}A_\mu \mathcal{D}\psi \mathcal{D} \bar{\psi}
  \exp\left( \frac{i}{\hbar}\int d^4x(\mathcal{L} + J^\mu A_\mu + \bar{\eta}\psi
  +\bar{\psi}\eta)\right).
  \label{eqn:qedpartition2}
\ea
In a classical system, the action in units of $\hbar$ is very large. This means 
that the functional integral is highly oscillatory and for most field configurations 
gives a small contribution. Therefore, the main contributions come from critical 
points where the action is stationary relative to the fields. These critical points 
are simply the solutions to the classical field equations. This observation is the 
basis of the stationary phase method in quantum mechanics, which replaces the functional 
integrals with a sum of integrals over neighbourhoods of stationary action. 

Since the 
effective action can be thought of as a quantum-correction of the classical action, 
the principle of least effective action would give us the quantum-corrected 
equations of motion. The Green's function method allows us to express the 
effective action in terms of functionals of an arbitrary magnetic field profile.
So, we take this opportunity to explore the stationary points of the 
effective action with respect to the flux tube profile shape.

In our geometry, the classical action is

\be 
\Gamma_0=-\int d^4x \frac{B^2}{2} = -\pi \int_{-\infty}^\infty dt dz \int_0^{L_\rho} \rho d\rho
	B_z(\rho)^2.
\ee We would like to minimize the effective action with respect to fluctuations 
of the function $\flr$ and its first derivative $f'_\lambda(\rho)$.  
Taking the entire action, 

\be 
\delta \Gamma = \pderiv{\Gamma}{\flr} \delta \flr + \pderiv{\Gamma}{(f'_\lambda(\rho))}\delta f'_\lambda(\rho).
\ee 

For now, we will neglect the vacuum renormalization terms and reintroduce them at the end of the analysis:

\ba 
\label{eqn:varActa}
\delta \Gamma &=& \biggr[\biggr\{-2\pi \int_{-\infty}^\infty dt dz B_z(\rho) + 
	\hbar \pi \sum_{m_l,\sigma^3} \int_{0}^\infty \chi^3 d\chi W_0^{-1}\nonumber \\
	& &\left(
	\pderiv{u_0(\rho)}{f'_\lambda(\rho)}u_\infty(\rho) +u_0(\rho)\pderiv{u_\infty(\rho)}{f'_\lambda(\rho)}
	\right)\biggr\}\delta \flr \biggr]_{\rho=0}^{L_\rho} \nonumber \\
	& &+\int_0^{L\rho}d\rho\biggr\{2\pi\int_{-\infty}^\infty dtdz\left(\frac{dB_z(\rho)}{d\rho}\right) 
	-\hbar \pi\sum_{m_l,\sigma^3} \int_{0}^\infty \chi^3d\chi 
	 \rho^2 W_0^{-1}\nonumber \\
	& & \times \biggr[u_\infty(\rho)\left(\pderiv{u_0(\rho)}{\flr} - \pderiv{u'_0(\rho)}
	{f'_\lambda(\rho)}\right) + u_0(\rho)\left(\pderiv{u_\infty(\rho)}{\flr} - \pderiv{u'_\infty(\rho)}
	{f'_\lambda(\rho)}\right) \nonumber \\
	& & -\pderiv{u_0(\rho)}{f'_\lambda(\rho)}u'_\infty(\rho) -
	 \pderiv{u_\infty(\rho)}{f'_\lambda(\rho)}u'_0(\rho)
	\biggr]_{\chi, m_l, \sigma^3}\biggr\}\delta\flr = 0.
\ea 

To compute the functional derivatives of the $u(\rho)$'s, we make use of the chain rule.  Here it is 
appropriate since functionals can be defined in terms of integrals over composed functions~\cite{greiner1996field}:

\be 
\frac{\delta u(\rho, [\flr])}{\delta \flr} = \pderiv{u(\rho)}{\rho}\frac{d\rho}{d \flr} = \frac{u'(\rho)}{f'_\lambda(\rho)}.
\ee 

The surface terms for the classical part vanish straightforwardly.  
We can impose conservation of flux by fixing the 
endpoints of the field fluctuations so that $\delta f_\lambda(0)=\delta f_\lambda(L_\rho) =0$.  
In this case, we can ignore the surface terms in (\ref{eqn:varActa}) to arrive at

\ba 
\label{eqn:varAct}
\int_0^{L\rho}d\rho\biggr\{2\pi\int_{-\infty}^\infty dtdz\left(\frac{d B_z(\rho)}{d\rho}\right) 
	-\frac{\hbar \pi}{2}\sum_{m_l,\sigma^3} \int_{0}^\infty \chi^3d\chi 
	 \rho^2  W_0^{-1}& &\nonumber \\
	 \times \biggr[u_\infty(\rho)\left(\frac{u'_0(\rho)}{f'_\lambda(\rho)} - \frac{u''_0(\rho)}
	{f''_\lambda(\rho)}\right) + u_0(\rho)\left(\frac{u'_\infty(\rho)}{f'_\lambda(\rho)} 
	- \frac{u''_\infty(\rho)}
	{f''_\lambda(\rho)}\right) & &\nonumber \\
	 -2\frac{u'_0(\rho)u'_\infty(\rho)}{f''_\lambda(\rho)}
	\biggr]_{\chi,m_l,\sigma^3}\delta \flr\biggr\} = 0. & &
\ea 
Inside the integral, there are no restrictions on the fluctuations $\delta \flr$
meaning that the rest of the integrand must vanish on its own:

\ba 
\label{eqn:stationary}
\frac{d B_z(\rho)}{d \rho}
	-\frac{\hbar}{2\Delta} \int_{0}^\infty \chi^3d\chi 
	 \rho^2 
	 \sum_{m_l,\sigma^3}W_0^{-1}\biggr[u_\infty(\rho)\biggr(\frac{u'_0(\rho)}{f'_\lambda(\rho)} 
	- \frac{u''_0(\rho)}{f''_\lambda(\rho)}\biggr) 
	& & \nonumber \\
	+ u_0(\rho)\left(\frac{u'_\infty(\rho)}{f'_\lambda(\rho)} 
	- \frac{u''_\infty(\rho)}
	{f''_\lambda(\rho)}\right) 
	 -2\frac{u'_0(\rho)u'_\infty(\rho)}{f''_\lambda(\rho)}
	\biggr]_{\chi,m_l,\sigma^3} =0,
\ea 
where $\Delta=\int_{-\infty}^{\infty}d\omega \int_{-\infty}^{\infty}dk_z$ is an infinite constant
arising because the field extends infinitely in time and along the $\hat{z}$-direction.

Finally, we subtract the zero flux 
equivalents of each of the one-loop terms to ensure that the action 
remains finite and vanishes with the magnetic field. Thus, we arrive at the 
 quantum-corrected equations of motion for a magnetic flux tube:

 \ba 
 \label{eqn:stationaryEffAct}
 &&\frac{d B_z(\rho)}{d \rho}
 	-\frac{\hbar}{2\Delta} \int_{0}^\infty \chi^3d\chi 
 	 \rho^2 
 	 \sum_{m_l,\sigma^3}\biggr[W_0^{-1}\biggr\{u_\infty(\rho)\biggr(\frac{u'_0(\rho)}{f'_\lambda(\rho)} 
 	- \frac{u''_0(\rho)}{f''_\lambda(\rho)}\biggr) 
 	 \nonumber \\
 	& &+ u_0(\rho)\left(\frac{u'_\infty(\rho)}{f'_\lambda(\rho)} 
 	- \frac{u''_\infty(\rho)}
 	{f''_\lambda(\rho)}\right) 
 	 -2\frac{u'_0(\rho)u'_\infty(\rho)}{f''_\lambda(\rho)}
 	\biggr\} \nonumber \\
 	& &+\frac{\pi}{2}\biggr\{\BESSO{Y}\biggr(\frac{\sqrt{\chi^2-m^2}\BESSO{J'}}{f'_\lambda(\rho)} \nonumber \\
 	& &- \frac{(\chi^2-m^2)\BESSO{J''}}{f''_\lambda(\rho)}\biggr) 
 	\nonumber \\
 	& & + \BESSO{J}\biggr(\frac{\sqrt{\chi^2-m^2}\BESSO{Y'}}{f'_\lambda(\rho)} \nonumber \\
 	& &- \frac{(\chi^2-m^2)\BESSO{Y''}}
 	{f''_\lambda(\rho)}\biggr)  \nonumber \\
 	& & -2\frac{(\chi^2-m^2)\BESSO{J'}\BESSO{Y'}}{f''_\lambda(\rho)}
 	\biggr\}
 \biggr]_{\chi,m_l,\sigma^3} =0.
 \ea 
 This expression is not guaranteed to be finite because it does not yet account for the field-strength 
 renormalization. It is not clear how this renormalization can be performed for the general case.


Here we see the familiar classical result:  the field which minimizes the classical action is homogeneous 
($\frac{d B_z(\rho)}{d \rho} =0$ everywhere).  However, the second term arising from the one-loop 
QED effects suggests that some non-homogeneous field ($\frac{d B_z(\rho)}{d \rho} \ne 0$)
may be a stationary point of the action for a fixed amount of flux. This equation 
potentially provides a numerical method for computing the quantum-corrected equations 
of motion of a cylindrically symmetric magnetic field. However, exploring 
this equation in more depth is outside the scope of this dissertation.

\section{Step-Function Flux Tube}
\label{sec:stepfunc}
As an example of how the Green's function method is applied to specific magnetic field 
profiles, $f_\lambda(\rho)$, we consider a step-function flux tube. The \ac{QED} effective 
action in this profile has already been computed using other 
methods than the ones shown here~\cite{1999PhRvD..60j5019B}. In this 
profile, there is a uniform field in the cylindrical region $\rho < \lambda$, and 
no field outside of the region:
\be 
B_z(\rho) = \frac{F}{\pi \lambda^2} \theta(\lambda-\rho).
\ee 
The profile function which gives this field is

\be 
f_\lambda(\rho)=\frac{\rho^2}{\lambda^2}\theta(\lambda-\rho) + \theta(\rho-\lambda)
\ee 
and the potential function is

\ba 
V_{m_l}(\rho)&=&\frac{m_l^2-1}{\rho^2} + \theta(\lambda-\rho)\frac{\mathcal{F}}{\lambda^2}
	\left(2(\sigma^3-m_l)+\frac{\mathcal{F}}{\lambda^2}\rho^2\right) \nonumber \\
	& &+ \theta(\rho-\lambda)\frac{\mathcal{F}}{\rho^2}
	\left(\mathcal{F}-2m_l\right) -\frac{2\mathcal{F}}{\lambda^2}\delta(\lambda-\rho).
\ea 

The solutions to equation (\ref{eqn:kchi}) for the inner region ($\rho < \lambda$) of 
this potential are expressed in terms of 
Whittaker $M$ and $W$ functions~\cite{slater1960confluent},

\be 
\label{eqn:Whit1}
[u_0(\rho)]_{\csubscripts}=\frac{W_{\frac{\lambda^2k^2}{4\mathcal{F}},\frac{m_l}{2}}
	\left(\frac{\mathcal{F}}{\lambda^2}\rho^2\right)}{\rho}
\ee 

\be 
\label{eqn:Whit2}
[u_\infty(\rho)]_{\csubscripts}=\frac{M_{\frac{\lambda^2k^2}{4\mathcal{F}},\frac{m_l}{2}}
	\left(\frac{\mathcal{F}}{\lambda^2}\rho^2\right)}{\rho},
\ee 
where 

\be
	\label{eqn:kchi}
	k^2=\chi^2-m^2-\frac{2\mathcal{F}}{\lambda^2}(\sigma^3-m_l).
\ee
The constant $W_0=\rho W(\rho)$ associated with these solutions is~\cite{slater1960confluent}

\be 
W_0^{-1}=-\frac{\lambda^2}{2\mathcal{F}}\frac{\Gamma\left[\frac{1}{2}\left(m_l+1-\frac{k^2\lambda^2}{2\mathcal{F}}\right)\right]}{\Gamma(1+m_l)}.
\ee 

The solutions for the exterior region ($\rho > \lambda$) are the Bessel functions,
\be 
\label{eqn:u0ext}
[u_0(\rho)]_{\csubscripts}=J_n(\sqrt{\chi^2-m^2}\rho)
\ee 
and
\be 
\label{eqn:uinfext}
[u_\infty(\rho)]_{\csubscripts}=Y_n(\sqrt{\chi^2-m^2}\rho),
\ee 
where $n=m_l-\mathcal{F}$.  We note that for large $m_l$ (i.e. $m_l\gg \mathcal{F}$), $n\approx m_l$ and these solutions coincide with the field-free case.

\subsection{Exterior Integral}
\label{sec:extint}

The homogeneous solutions for the exterior ($\rho > \lambda$) region are 
given by equations (\ref{eqn:u0ext}) and (\ref{eqn:uinfext}):

\ba 
	\Gamma^{(1)}_{\rm ext}& =& -\frac{\hbar\pi^2}{2} \sum_{\rm \sigma^3= \{\pm1\}}
	\sum_{m_l=-\infty}^{\infty}  
	\int_{0}^\infty \chi^3 d\chi
	\int_0^{\lambda} d\rho \rho^2 \biggr(\nonumber \\
	& &~~~~~~~J_{m_l-\mathcal{F}}(\sqrt{\chi^2 -m^2}\rho)
	Y_{m_l-\mathcal{F}}(\sqrt{\chi^2 -m^2}\rho)\nonumber \\ 
	& &~~~~~~~-J_{m_l}(\sqrt{\chi^2 -m^2}\rho)
	Y_{m_l}(\sqrt{\chi^2 -m^2}\rho) \biggr).
\ea
This integral is somewhat problematic because it is apparently divergent for large 
values of $\sqrt{\chi^2-m^2}\rho$. The asymptotic expansions of the Bessel functions 
for large argument are
\be
	J_n(x) \approx \sqrt{\frac{2}{\pi x}} \cos{\left(x-\frac{n\pi}{2}-\frac{\pi}{4}\right)}
\ee
and
\be
	Y_n(x) \approx \sqrt{\frac{2}{\pi x}} \sin{\left(x-\frac{n\pi}{2}-\frac{\pi}{4}\right)}.
\ee
So, in this limit, the integrand for an individual term of the $m_l$ sum results in a 
divergent integral:
\ba
	\lefteqn{J_{m_l-\mathcal{F}}(\sqrt{\chi^2 -m^2}\rho)
	Y_{m_l-\mathcal{F}}(\sqrt{\chi^2 -m^2}\rho)}\nonumber \\
	& &-
	J_{m_l}(\sqrt{\chi^2 -m^2}\rho)
	Y_{m_l}(\sqrt{\chi^2 -m^2}\rho) \nonumber \\
	& \approx&\frac{2}{\pi \sqrt{\chi^2-m^2}\rho}\sin{\left(\frac{\pi}{2}\mathcal{F}\right)}
	\cos{\left(\frac{\pi}{2}(\mathcal{F}-2m_l+1)+2\sqrt{\chi^2-m^2}\rho\right)}.
\ea
This seems to be an ultraviolet divergence, but it cannot be easily renormalized 
by a field-strength renormalization because the dependence on the field 
strength, $\mathcal{F}$, is not quadratic like the classical term. Nevertheless, 
since \ac{QED} is a renormalizable theory, the effective action in the exterior 
region must either be finite, or proportional to the classical action. To make 
sense of this divergence, we need to consider the entire sum over $m_l$, and 
not simply the individual terms in isolation.

We note that the exterior integral must be zero when 
$\mathcal{F}$ is an integer. In this case, the Bessel function order numbers 
of the first term are shifted 
by an integer value relative to the second term. But, since the sum is over all integer 
values of $m_l$, there will still be cancellation in pairs. In other words, for any
finite integer, $j$,
\be
	\sum_{i=-\infty}^{\infty}J_i(x)Y_i(x) = \sum_{i=-\infty}^{\infty}J_{i-j}(x)Y_{i-j}(x).
\ee
Physically, integer values of $\mathcal{F}$ correspond to the disappearance of the 
Aharonov-Bohm effect when the Berry's phase of a closed path is a multiple of $2\pi$. 
Since our primary motivation is in flux tubes in superconducting materials where 
$\mathcal{F}$ is naturally quantized to integer values, 
we will focus on integer values of $\mathcal{F}$ for now, and set this 
integral to zero. 

The bulk of the effective action density is expected to be 
in the interior region where the field is, and 
where all of the classical action density is. However, nonlocal quantum effects can
cause diffusion of the effective action density to regions where 
no classical field exists~\cite{Gies:2001zp}. So, there may be interesting 
physical content within this integral.
Nevertheless, evaluating this integral for non-integer values of $\mathcal{F}$ is 
left for future research.

\subsection{Interior Integral}

Here, we consider the integral for the interior ($\rho < \lambda$) region.  
The Whittaker functions in equations (\ref{eqn:Whit1}) and (\ref{eqn:Whit2}) obey the differential
equation

\be 
\frac{d^2 W(\rho)}{d\rho^2} - \frac{1}{\rho}\frac{dW(\rho)}{d\rho} +\left[
	k^2+\frac{(1-m_l^2)}{\rho^2}-\rho^2\left(\frac{\mathcal{F}}{\lambda^2}\right)^2\right]W(\rho)=0.
\ee 
In exploring the asymptotics of the solutions, it is interesting to write this equation in 
canonical form

\be 
\label{eqn:canonical}
y''(\rho) +\left[k^2+\frac{1/4-m_l^2}{\rho^2} -\rho^2\left(\frac{\mathcal{F}}{\lambda^2}\right)^2 \right]y(\rho)=0,
\ee 
with $W(\rho)=\sqrt{\rho}y(\rho)$.

We may neglect the contribution from negative values of $m_l$ since for negative integer values of $n$~\cite{slater1960confluent},

\be 
	\frac{W_{k,\frac{n}{2} }(x) M_{k,\frac{n}{2}}(x)}{\Gamma(1+n)} = 0.
\ee 
For large values of $m_l$ ($m_l \gg \mathcal{F}$), the Whittaker functions correspond to Bessel 
functions and there is cancellation between them and the background terms.  So, we may compute the 
interior integral to arbitrary precision by summing a finite number of positive-$m_l$ terms.

\subsubsection{Field-Strength Renormalization}

There is an ultraviolet divergence for large values of $\chi$ which may be subtracted using a 
field-strength renormalization.  To see this, we consider the differential equation, (\ref{eqn:canonical}), 
in the limit of asymptotically large $k$ using the WKB approximation.  Our asymptotic $u(\rho)$ functions 
in this approximation are

\be 
\label{eqn:WKBu0}
[u_0(\rho)]_{k,m_l,\sigma^3} \approx \frac{1}{\sqrt{\rho}}\cos\left\{k\rho-\frac{1}{2k\rho}\left[(1/4-m_l^2) +
	\frac{\rho^4}{3}\left(\frac{\mathcal{F}}{\lambda^2}\right)^2\right]\right\}
\ee 
and
\be 
\label{eqn:WKBui}
[u_\infty(\rho)]_{k,m_l,\sigma^3} \approx \frac{1}{\sqrt{\rho}}\sin\left\{k\rho-\frac{1}{2k\rho}\left[(1/4-m_l^2) +
	\frac{\rho^4}{3}\left(\frac{\mathcal{F}}{\lambda^2}\right)^2\right]\right\}.
\ee 

The relevant constant for these solutions is $W_0=\rho W[u_0(\rho),u_\infty(\rho)]$, where

\be 
\label{eqn:WKBW0}
W_0\approx k+\frac{(1/4-m_l^2)}{2k\rho^2}-\frac{\rho^2}{2k}
	\left(\frac{\mathcal{F}}{\lambda^2}\right)^2.
\ee 
Taking equations (\ref{eqn:WKBu0}), (\ref{eqn:WKBui}), and 
(\ref{eqn:WKBW0}), and expanding in powers of $\frac{1}{k\rho}$ gives
\ba 
\lefteqn{\left[\frac{u_0(\rho)u_\infty(\rho)}{W_0}\right]_{k,m_l,\sigma^3} \approx
	\frac{\sin{(k\rho)}\cos{(k\rho)}}{k\rho}}\nonumber \\
	& &-\frac{1}{2(k\rho)^2}
	\left[(1/4-m_l^2)+\frac{1}{3}\left(\frac{\rho^2\mathcal{F}}{\lambda^2}\right)^2\right]\cos{(2k\rho)} \nonumber \\
	& & -\frac{(1/4-m_l^2)}{2(k\rho)^3}\sin{(k\rho)}\cos{(k\rho)} \nonumber \\
	& &+\frac{1}{2(k\rho)^3}\left(\frac{\rho^2\mathcal{F}}{\lambda^2}\right)^2\sin{(k\rho)}\cos{(k\rho)} \nonumber \\
	& & +\frac{1}{4(k\rho)^4}\left[(1/4-m_l^2)+\frac{1}{3}\left(\frac{\rho^2\mathcal{F}}{\lambda^2}\right)^2\right] \nonumber \\
	& &\times \left[(1/4-m_l^2)-\left(\frac{\rho^2\mathcal{F}} {\lambda^2}\right)^2\right]\cos{(2k\rho)} \\
	\label{eqn:infcn}
	\lefteqn{\approx \left[\frac{u_0(\rho)u_\infty(\rho)}{W_0}\right]_{k,m_l,\sigma^3} 
	\biggr|_{\mathcal{F}=0} }\nonumber\\
	 & &-\frac{1}{6(k\rho)^2}\left(\frac{\rho^2\mathcal{F}}
	 {\lambda^2}\right)^2\cos{(2k\rho)}\nonumber \\
	 & & +\frac{1}{2(k\rho)^3}
	\left(\frac{\rho^2\mathcal{F}}
	 {\lambda^2}\right)^2\sin{(k\rho)}\cos{(k\rho)} \nonumber \\
	 & & -\frac{1}{6(k\rho)^4}\left(\frac{\rho^2\mathcal{F}}
	 {\lambda^2}\right)^2\biggr[(1/4-m_l^2)  -	\left(\frac{\rho^2\mathcal{F}}
	 {\lambda^2}\right)^2\biggr]\cos{(2k\rho)}.
\ea

The subtraction of the background field corresponds to cancellation of the first term in 
equation (\ref{eqn:infcn}).  The remaining terms represent ultraviolet divergences proportional 
to $\left(\frac{\mathcal{F}}{\lambda^2}\right)^2$.  These terms may be subtracted out from the integral
 by a field-strength renormalization.

The renormalized contribution to the effective action from the interior region is, then


\ba 
	\label{eqn:gfee}
	\Gamma^{(1)}_{\rm int}& =& -\hbar\pi \sum_{\rm \sigma^3= \{\pm1\}}\sum_{m_l=0}^{m_l\gg\mathcal{F}}  
	\int_{0}^\infty \chi^3 d\chi
	\int_0^{\lambda} d\rho \rho^2 \biggr(
	\frac{\lambda^2}{2\mathcal{F}}\nonumber \\
	& &\frac{\Gamma\left(\frac{1}{2}\left(m_l+1-\frac{k^2\lambda^2}{2\mathcal{F}}\right)\right)}
		{m_l!} 
	W_{\frac{\lambda^2k^2}{4\mathcal{F}},\frac{m_l}{2}}
	\left(\frac{\mathcal{F}}{\lambda^2}\rho^2\right) 
	M_{\frac{\lambda^2k^2}{4\mathcal{F}},\frac{m_l}{2}}
	\left(\frac{\mathcal{F}}{\lambda^2}\rho^2\right)\nonumber \\ 
	& &~~~~~~~-\frac{\pi}{2}J_{m_l}(\sqrt{\chi^2 -m^2}\rho)
	Y_{m_l}(\sqrt{\chi^2 -m^2}\rho)\nonumber \\
	& & +\left(\frac{\mathcal{F}}{\lambda^2}\right)^2
	\biggr[ \frac{\rho^3}{2k^2} \sin(\Theta_{m_l,k}(\rho))
	+\frac{\rho^2}{6k^3}\cos(\Theta_{m_l,k}(\rho))\biggr]\biggr),
\ea 
where $\Theta_{m_l,k}(\rho)\equiv 2k\rho -\frac{(1/4-m_l^2)}{k\rho}$ and $k$ is given 
in terms of the relevant integration and summation parameters by 
equation (\ref{eqn:kchi}).

\section{Discussion}
Equation (\ref{eqn:gfee}) is a renormalized expression of the 1-loop correction term to the effective 
action for a step-function flux tube in terms of finite integrals and sums over finitely many terms. 
So, it is straightforward to implement a numerical algorithm for computing the effective action using 
this expression. However,  the integrand is polluted by numerous poles
which make the computations time consuming. For this reason, it is more practical to approach the problem 
using other numerical methods such as those presented in chapter \ref{ch:WLNumerics}. 
Moreover, the effective action for the step function flux tube profile 
has been computed before, arguably in a simpler way~\cite{1999PhRvD..60j5019B}. 
Establishing if the method discussed here produces the same results as 
previous methods is an important next step along this research path.
Nevertheless, the Green's function 
method described in this chapter has proved useful for 
establishing an expression for non-homogeneous 
configurations with stationary effective action, equation 
(\ref{eqn:stationaryEffAct}). Additionally, it may 
prove useful in future research since it is a technique which 
can be applied to arbitrary flux tube 
profiles, and doesn't have the precision limitations of Monte Carlo 
techniques. One could also imagine that 
a clever choice of potential might lead to considerable analytic 
progress. For example, some 
potential function 
may result in functions $u_0(\rho)$ and $u_\infty(\rho)$ for 
which the integrals can be evaluated.

One obstacle with the approach outlined in this chapter is in performing 
the field-strength renormalization in general. In section \ref{sec:extint}, we 
observed that the renormalization proceedure could not be performed on individual 
terms alone, but rather needed to account for the entire angular momentum summation.
This poses certain technical problems that were not addressed here. However, previous 
research applying a WKB phase shift method to fermion determinants in non-homogeneous 
background fields has produced a renormalization scheme which may be generalized to 
the background fields discussed in this chapter~\cite{2004PhLB..600..302D}. 
There are therefore promising prospects in addressing the field-strength renormalization 
in this method by proceeding analogously with the WKB phase shift method.

Perhaps the most interesting result from this section is that 
the Green's function method may provide 
a way to numerically compute the
quantum corrected equations of motion for external 
electromagnetic fields. Making further progress on studying 
equation (\ref{eqn:stationaryEffAct}) may therefore be a
particularly interesting direction for future research.

\chapter{Parallel Worldline Numerics in \texorpdfstring{\acs{CUDA}}{CUDA}} 

\label{ch:WLNumerics}

\acresetall

\begin{summary}
	In this chapter, I give an overview of the \ac{WLN} 
	technique, and discuss the parallel 
	\ac{CUDA} implementation of the algorithm created for this 
	thesis. In the \ac{WLN} technique, we wish to 
	generate an ensemble of representative closed-loop particle 
	trajectories, and use these to compute an approximate 
	average value for Wilson loops. We show how this can be done with 
	a specific emphasis on cylindrically symmetric magnetic fields.
	The fine-grained, massive parallelism provided by 
	the \ac{GPU} architecture results in considerable speedup in 
	computing Wilson loop averages.
\end{summary}

In this chapter, I will give an overview of a numerical technique 
which can be used to compute the effective actions of external 
field configurations.
The technique, called either \acl{WLN} or the Loop Cloud Method,
was first used by Gies and Langfeld~\cite{Gies:2001zp} and has since been
applied to computation of effective actions 
\cite{Gies:2001tj,Langfeld:2002vy,Gies:2005sb,Gies:2005ym,Dunne:2009zz}
and Casimir energies
\cite{Moyaerts:2003ts,Gies:2003cv,PhysRevLett.96.220401}. More recently, 
the technique has also been applied to pair production~\cite{2005PhRvD..72f5001G}
and the vacuum polarization tensor~\cite{PhysRevD.84.065035}.
\ac{WLN} is able to compute quantum effective actions in
the one-loop approximation to all orders in both the coupling and in the
external field, so it is well suited to studying non-perturbative aspects
of \acp{QFT} in strong background fields. Moreover, the technique maintains
gauge invariance and Lorentz invariance.  The key idea of the technique is
that a path integral is approximated as the average of a finite set of $N_l$
representative closed paths (loops) through spacetime. We use a standard
Monte-Carlo procedure to generate loops which have large contributions to
the loop average.

\section{\texorpdfstring{\acs{QED}}{QED} Effective Action on the Worldline}

Worldline numerics is built on the worldline formalism which was initially 
invented by Feynman~\cite{PhysRev.80.440, PhysRev.84.108}. 
Much of the recent interest in this formalism is based on the 
work of Bern and Kosower, who derived it from the infinite string-tension limit of 
string theory and demonstrated that it provided an efficient means for computing 
amplitudes in \ac{QCD}~\cite{PhysRevLett.66.1669}. 
For this reason, the worldline formalism is often referred to as 
`string inspired'. However, the formalism can also be obtained straight-forwardly 
from first-quantized field theory~\cite{1992NuPhB.385..145S}, which is the approach 
we will adopt here.  In this formalism
the degrees of freedom of the field are represented in terms of one-dimensional 
path integrals over an ensemble of closed trajectories.

We begin with the \ac{QED} effective action expressed in the proper-time 
formalism, (\ref{eqn:proptimelag}),

\ba
	\label{eqn:trln}
	{\rm Tr~ln}\left[\frac{\slashed{p}+e\slashed{A}_\mu^0
	-m}{\slashed{p}-m}\right] &=& -\frac{1}{2}\int d^4x \int_0^\infty
	\frac{dT}{T}e^{-iTm^2} \nonumber \\
	& &\times {\rm tr}\biggr( \bra{x}e^{iT(\slashed{p}
	+e\slashed{A}^0_\mu)^2}\ket{x} 
	- \bra{x}e^{iTp^2}\ket{x}\biggr).
\ea

To evaluate $\bra{x}e^{iT(\slashed{p}_\mu + e\slashed{A}_\mu)^2}\ket{x}$, we recognize that it is simply 
the propagation amplitude $\braket{x,T}{x,0}$ from ordinary quantum mechanics 
with $(\slashed{p}_\mu + e\slashed{A}_\mu)^2$ playing the role of the Hamiltonian.  
We therefore express this factor in its path integral form:

\ba
	\bra{x}e^{iT(\slashed{p}_\mu + e\slashed{A}_mu)^2}\ket{x} &=&  \mathcal{N}
	\int \mathcal{D}x_\rho(\tau) e^{-\int_0^T d\tau \left[\frac{\dot{x}^2(\tau)}{4} 
	+ i A_\rho x^\rho(\tau)\right]} \nonumber \\
	& & \times \Paulispin.
\ea

$\mathcal{N}$ is a normalization constant which we can fix by using our renormalization 
condition that the fermion determinant should vanish at zero external field:

\ba
	\bra{x}e^{iTp^2}\ket{x} & = & \mathcal{N}\int \mathcal{D} 
		x_p(\tau)e^{-\int_0^T d\tau\frac{\dot{x}^2(\tau)}{4}} \\
	&=& \int \frac{d^4p}{(2\pi T)^4}\bra{x}e^{iTp^2}\ket{p}\braket{p}{x} \\
	& = & \frac{1}{(4\pi T)^2},	
\ea

\be
	\mathcal{N}\int \mathcal{D}x_\rho(\tau) e^{-\int_0^T d\tau \frac{\dot{x}^2(\tau)}{4}} 
	= \frac{1}{(4\pi T)^2}.
\ee

We may now write
\ba
	\mathcal{N}\int \mathcal{D}x_\rho(\tau) e^{-\int_0^T d\tau[\frac{\dot{x}^2(\tau)}{4}+iA_\rho x^\rho(\tau)]}
	\Paulispin \nonumber \\
	= \frac{1}{(4\pi T)^2} \left\langle e^{-i\int_0^T d\tau A_\rho x^\rho(\tau)}\Paulispin\right\rangle_x ,
\ea
where
\be
	\label{eqn:meandef}
	\mean{\hat{\mathcal{O}}}_x = \frac{\int \mathcal{D}x_\rho(\tau) \hat{\mathcal{O}} 
	e^{-\int_0^T d\tau\frac{\dot{x}^2(\tau)}{4}}}{\int \mathcal{D}x_\rho(\tau)  
	e^{-\int_0^T d\tau\frac{\dot{x}^2(\tau)}{4}}}
\ee
is the weighted average of the operator $\hat{\mathcal{O}}$ over an ensemble of closed particle 
loops with a Gaussian velocity distribution.

Finally, combining all of the equations from this section results in the 
one-loop effective action for \ac{QED} on the worldline:

\ba
	\label{eqn:QEDWL}
	\Gamma^{(1)}[A_\mu] &=& \frac{2}{(4\pi)^2}\int_0^\infty
	\frac{dT}{T^3}e^{-m^2T}\int d^4x_{\rm CM} \nonumber \\
	& & \times \left[\left\langle e^{i\int_0^Td\tau A_\rho(x_{\rm CM}+x(\tau))\dot{x}^\rho(\tau)}
	\Paulispin\right\rangle _x -1\right]. \nonumber \\
	& &
\ea

\section{Worldline Numerics}

The averages, $\mean{\hat{\mathcal{O}}}$, defined by equation (\ref{eqn:meandef})
involve 
functional integration over every possible closed path through spacetime
which has a Gaussian velocity distribution.  
The prescription of the \ac{WLN} technique is to compute 
these averages approximately using a finite set of $N_l$ representative loops 
on a computer.  The worldline average is then approximated as the mean of 
an operator evaluated along each of the worldlines in the ensemble:

\be
	\mean{\hat{\mathcal{O}}[x(\tau)]} \approx 
	\frac{1}{N_l} \sum_{i=1}^{N_l} \hat{\mathcal{O}}[x_i(\tau)].
\ee

\subsection{Loop Generation} 
\label{sec:loopgen}
The velocity distribution for the loops depends on
the proper time, $T$.  However, generating a separate ensemble of loops for 
each value of $T$ would be very computationally expensive.  This problem is alleviated by generating
a single ensemble of loops, $\vec{y}(\tau)$, representing unit proper time,
and scaling those loops accordingly for different values of $T$:

\be 
\vec{x}(\tau) = \sqrt{T}\vec{y}(\tau/T) ,
\ee

\be \int_0^T d\tau \vec{\dot{x}}^2(\tau) \rightarrow \int_0^1 dt
\vec{\dot{y}}^2(t).
\ee

There is no way to treat the integrals as continuous as we generate our loop
ensembles.  Instead, we treat the integrals as sums over discrete points
along the proper-time interval $[0,T]$.  This is fundamentally different
from space-time discretization, however.  Any point on the worldline loop
may exist at any position in space, and $T$ may take on any value.  It is
important to note this distinction because the Worldline method retains
Lorentz invariance while lattice techniques, in general, do not.

The challenge of loop cloud generation is in generating a discrete set
of points on a unit loop which obeys the prescribed velocity distribution.
There are a number of different algorithms for achieving this goal which have
been discussed in the literature.  Four possible algorithms are compared
and contrasted in \cite{Gies:2003cv}.  In this work, we choose a more
recently developed algorithm, dubbed ``d-loops", which was first described
in \cite{Gies:2005sb}. To generate a ``d-loop", the number of points is iteratively 
doubled, placing the new points in a Gaussian distribution between the existing neighbour points.
We quote the algorithm directly:

\begin{quote} 
  \begin{singlespace}
    \begin{itemize} 
	  \item[(1)] Begin with one arbitrary point
		$N_0=1$, $y_{N}$.

	  \item[(2)] Create an $N_1=2$ loop, i.e., add a point $y_{N/2}$ that is
	    distributed in the heat bath of $y_N$ with
	   \begin{equation} 
		 e^{-\frac{N_1}{4} 2 (y_{N/2} -y_{N})^2}. \label{yn2}
	   \end{equation}

      \item[(3)] Iterate this procedure, creating an $N_k=2^k$ppl loop by
        adding $2^{k-1}$ points $y_{{qN}/{N_k}}$, $q=1,3,\dots, N_k-1$ with
        distribution
      \begin{equation} 
	    e^{-\frac{N_k}{4} 2 [y_{qN/N_k} -\frac{1}{2}(y_{(q+1)N/N_k}+
		y_{(q-1)N/N_k})]^2}. \label{ynk}
	  \end{equation}

      \item[(4)] Terminate the procedure if $N_k$ has reached $N_k=N$ for
        unit loops with $N$ ppl.

      \item[(5)] For an ensemble with common center of mass, shift each
        whole loop accordingly.

    \end{itemize}
  \end{singlespace}	
\end{quote}

The above d-loop algorithm was selected since it is simple and about 10\%
faster than previous algorithms, according to its developers, because it
requires fewer algebraic operations.

The generation of the loops is largely independent from the main program.
Because of this, it was simpler to generate the loops using a Matlab 
script. The Matlab function used for generating d-loops using the 
above algorithm can be found in appendix \ref{sec:srcWLgen}.

This function was used to produce text files containing the worldline data for 
ensembles of loops. Then, these text files were read into memory at the 
launch of each calculation. The results of this generation routine can 
be seen in figure \ref{fig:worldlineplot}.

When the \ac{CUDA} kernel is called\footnote{Please see appendix \ref{ch:cudafication} for an overview of \ac{CUDA}.}, 
every thread in every block executes 
the kernel function with its own unique identifier. So, it is best to 
generate worldlines in integer multiples of the number of threads per 
block. The Tesla C1060 device allows up to 512 threads per block.

\begin{figure}
	\centering
		\includegraphics[width=14cm]{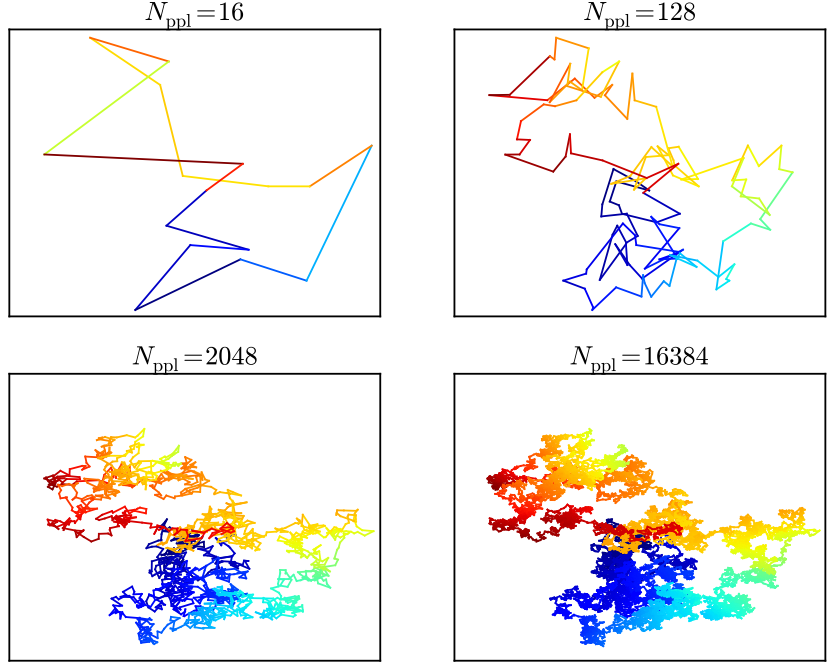}
	\caption[Discretization of the worldline loop]
	{A single discrete worldline loop shown at several levels 
	of discretization. The loops form fractal patterns and have a strong 
	parallel with Brownian motion. The colour
	represents the phase of a particle travelling along the loop, and 
	begins at dark blue, progresses in a random walk through yellow, 
	and ends at dark 
	red. The total flux through this particular worldline at $T=1$ and 
	$B=B_k$ is about $0.08 \pi/e$.}
	\label{fig:worldlineplot}
\end{figure}

\section{Cylindrical Worldline Numerics} 
We now consider cylindrically symmetric external magnetic fields.
In this case, we may simplify (\ref{eqn:QEDWL}),

\be \frac{\Gamma^{(1)}_{\rm ferm}}{T
L_z} = \frac{1}{4\pi} \int_0^\infty \rho_{\rm cm}
  \left[ \int_0^\infty \frac{dT}{T^3}e^{-m^2T}\left\{\langle
  W\rangle_{\vec{r}_{\rm cm}} - 1 -\frac{1}{3}(eB_{\rm
  cm}T)^2\right\}\right]d\rho.
  \label{eqn:cylEA}
\ee 

\subsection{Cylindrical Magnetic Fields} 

We have $\vec{B} =
B(\rho)$\bhattext{z} with

\be \label{eqn:BWLN} B(\rho) = \frac{A_\phi(\rho)}{\rho} +
\frac{dA_\phi(\rho)}{d\rho} \ee if we make the gauge choice that $A_0 =
A_\rho = A_z = 0$.

We begin by considering $\Aphi$ in the form

\be \Aphi = \frac{F}{2\pi \rho}f_\lambda(\rho) \ee so that \be
B_z(\rho)=\frac{F}{2\pi\rho}\frac{df_\lambda(\rho)}{d\rho} \ee and the total
flux is \be \Phi=F(f_\lambda(L_\rho)-f_\lambda(0)).  \ee It is convenient
to express the flux in units of $\frac{2 \pi}{e}$ and define a dimensionless
quantity

\be 
	\mathcal{F}=\frac{e}{2 \pi} F.
\ee







\subsection{Wilson Loop}

The quantity inside the angled brackets in equation (\ref{eqn:QEDWL}) is a 
gauge invariant observable called a Wilson loop. We note that the proper time
integral provides a natural path ordering for this operator.
The Wilson loop expectation value is

\be
\label{eqn:wilsonloop}
\langle W\rangle_{\vec{r}_{\rm cm}}= \left \langle e^{ie\int_0^T d\tau
\vec{A}(\vec{r}_{{\rm cm}} + \vec{r}(\tau)) \cdot \dot{\vec{r}}}
 \frac{1}{4} {\rm tr}  e^{\frac{e}{2}\int_0^T d\tau
  \sigma_{\mu \nu}F_{\mu \nu}(\vec{r}_{{\rm cm}} + \vec{r}(\tau))}\right
  \rangle_{\vec{r}_{\rm cm}},
\ee
which we look at as a product between a scalar part ($e^{ie\int_0^T d\tau
\vec{A}(\vec{r}_{{\rm cm}} + \vec{r}(\tau)) \cdot \dot{\vec{r}}}$)
and a fermionic part ($\frac{1}{4} {\rm tr}  e^{\frac{e}{2}\int_0^T d\tau
  \sigma_{\mu \nu}F_{\mu \nu}(\vec{r}_{{\rm cm}} + \vec{r}(\tau))}$).

\subsubsection{Scalar Part}

In a magnetic field, the scalar part is related to the flux through 
the loop, $\Phi_B$, by Stokes theorem:

\ba
 e^{ie\int_0^T d\tau
 \vec{A} \cdot \dot{\vec{r}}} &=&
 e^{ie\oint \vec{A}\cdot d\vec{r}} = e^{ie\int_{\vec{\Sigma}} \vec{\nabla}\times\vec{A} \cdot d\vec{\Sigma}}\\
 & = & e^{ie\int_{\vec{\Sigma}} \vec{B} d\vec{\Sigma}} = e^{ie\Phi_B}.
\ea
So, this factor accounts for the Aharonov-Bohm phase aquired by 
particles in the loop.

The loop discretization results in the following approximation of the
scalar integral:

\be
  \oint \vec{A}(\vec{r})\cdot d\vec{r} =  \sum_{i=1}^{N_{ppl}}
  \int_{\vec{r}^i}^{\vec{r}^{i+1}}\vec{A}(\vec{r})\cdot d\vec{r}.
\ee 
Using a linear parameterization of the positions, the line integrals are

\be 
  \int_{\vec{r}^i}^{\vec{r}^{i+1}}\vec{A}(\vec{r})\cdot d\vec{r} =
  \int_0^1dt \vec{A}(\vec{r}(t))\cdot(\vec{r}^{i+1} - \vec{r}^i).
\ee
Using the same gauge choice outlined above ($\vec{A}=A_\phi \hat{\phi}$),
we may write

\be 
	\vec{A}(\vec{r}(t)) = \frac{\mathcal{F}}{e\rho^2} 
	f_\lambda(\rho^2)(-y,x,0),
\ee 
where we have chosen $f_\lambda(\rho^2)$ to depend on $\rho^2$ instead
of $\rho$ to simplify some expressions and to 
avoid taking many costly square roots in the worldline numerics.
We then have

\be \int_{\vec{r}^i}^{\vec{r}^{i+1}}\vec{A}(\vec{r})\cdot
d\vec{r} = \mathcal{F} (x^iy^{i+1}-y^i x^{i+1})\int_0^1
dt\frac{f_\lambda(\rho_i^2(t))}{\rho_i^2(t)}.  \ee The linear interpolation
in Cartesian coordinates gives

\be 
	\label{eqn:rhoi} \rho_i^2(t) = A_i + 2B_it + C_i t^2,
\ee
where
\ba 
	A_i &=& (x^i)^2 + (y^i)^2 \\ 
	B_i&=&x^i(x^{i+1} - x^i) + y^i(y^{i+1}-y^i)\\ 
	C_i &= &(x^{i+1}-x^i)^2 + (y^{i+1}-y^i)^2. 
	\label{eqn:Ci} 
\ea

In performing the integrals along the straight lines connecting
each discretized loop point, we are in danger of violating gauge invariance.
If these integrals can be performed analytically, than gauge invariance
is preserved exactly.  However, in general, we wish to compute these integrals
numerically.  In this case, gauge invariance is no longer guaranteed, but
can be preserved to any numerical precision that's desired.

\subsubsection{Fermion Part} 

For fermions, the Wilson loop
is modified by a factor,

\ba 
W^{\rm ferm.} &=& \frac{1}{4}{\rm tr}\left(e^{\frac{1}{2}
	e\int_0^T d\tau \sigma_{\mu \nu}F^{\mu \nu}}\right)\\ 
	&=& \frac{1}{4}{\rm tr}\left(e^{\sigma_{x y} 
	e\int_0^T d\tau B\left(x(\tau)\right)}\right) \\	
	&=& \cosh{\left(e \int_0^T d\tau B\left(x(\tau)\right)\right)}\\
	&=& \cosh{\left(2\mathcal{F}\int_0^T
	d\tau f'_\lambda(\rho^2(\tau))\right)},
	\label{eqn:Wfermfpl}
\ea 
where I have used the
relation 
\be 
	eB = 2\mathcal{F}\frac{d f_\lambda(\rho^2)}{d \rho^2} =
	2\mathcal{F}f'_\lambda(\rho^2).  
\ee 

This factor represents an additional contribution to the 
action because of the spin interaction with the magnetic field. 
Classically, for a particle with a magnetic moment $\vec{\mu}$ 
travelling through a magnetic field in a time $T$, the 
action is modified by a term given by
\be
	\Gamma^0_{\rm spin} = \int_0^T \vec{\mu} \cdot \vec{B}(\vec{x}(\tau)) d\tau.
\ee
The magnetic moment is related to the electron spin 
$\vec{\mu} = g\left(\frac{e}{2m}\right)\vec{\sigma}$, 
so we see that the integral in the above quantum fermion factor is 
very closely related to the classical action 
associated with transporting a magnetic moment through a magnetic field:
\be
	\Gamma^0_{\rm spin} = g\left(\frac{e}{2m}\right) \sigma_{x y} \int_0^T B_z(x(\tau))d\tau.
\ee
Qualitatively, we could write
\be 
	W^{\rm ferm} \sim \cosh{\left(\Gamma^0_{\rm spin}\right)}.
\ee

As a possibly useful aside, we may want to express 
the integral in terms of $f_\lambda(\rho^2)$ instead of its derivative.
We can do this by integrating by parts:
\ba 
	\int_0^T d\tau f'_\lambda(\rho^2(\tau)) &=& 
	\frac{T}{N_{\rm ppl}}\sum_{i=1}^{N_{\rm ppl}}\int_0^1 dt f'_\lambda(\rho^2_i(\tau))
	\\ & = & 
	\frac{T}{N_{\rm ppl}}\sum_{i=1}^{N_{\rm ppl}}\biggr[
	\frac{f_\lambda(\rho^2_i(t))}{2(B_i+C_it)}\biggr|^{t=1}_{t=0} \nonumber \\
  & & +\frac{C_i}{2}\int_0^1 \frac{f_\lambda(\rho^2_i(t))}{(B_i+C_i
  t)^2} dt\biggr] \\ & = &  \frac{T}{N_{\rm ppl}}\sum_{i=1}^{N_{\rm
  ppl}}\frac{C_i}{2}\int_0^1 \frac{f_\lambda(\rho^2_i(t))}{(B_i+C_i t)^2} dt,
\ea 
with $\rho_i^2(t)$ given by equations (\ref{eqn:rhoi}) to (\ref{eqn:Ci}).
The second equality is obtained from integration-by-parts.  In the third
equality, we use the loop sum to cancel the boundary terms in pairs:

\be 
\label{eqn:Wfermfl}
W^{{\rm ferm.}} = \cosh{\left(\frac{\mathcal{F}T}{N_{\rm
ppl}}\sum_{i=1}^{N_{{\rm ppl}}}
  C_i \int_0^1 dt \frac{f_\lambda(\rho^2_i(t))}{(B_i+C_i t)^2}\right)}.
\ee
In most cases, one would use equation (\ref{eqn:Wfermfpl}) to compute the fermion factor 
of the Wilson loop. However, 
equation (\ref{eqn:Wfermfl}) may be useful in cases where $f'_\lambda(\rho^2(\tau))$
is not known or is difficult to compute. 



\subsection{Renormalization}

The field strength renormalization counter-terms result from the small $T$
behaviour of the worldline integrand.  In the limit where $T$ is very small,
the worldline loops are very localized around their center of mass.  So,
we may approximate their contribution as being that of a constant field
with value $\vec{A}(\vec{r}_{{\rm cm}})$.  Specifically, we require that the field
change slowly on the length scale defined by $\sqrt{T}$. This condition on 
$T$ can be written

\be 
T \ll \left|\frac{m^2}{e B'(\rho^2)}\right| 
= \left| \frac{m^2}{2\mathcal{F}f''_\lambda(\rho^2_{\rm cm})}\right|.
\ee 

When this limit is satisfied, we may use the exact expressions for the
constant field Wilson loops to determine the small $T$ behaviour of the
integrands and the corresponding counter terms.

The Wilson loop averages for constant magnetic fields in scalar and 
fermionic \ac{QED} are

\be
	\mean{W}_{\rm ferm} = eBT\coth{(eBT)}
\ee
and
\be
	\mean{W}_{\rm scal} = \frac{eBT}{\sinh{(eBT)}}.
\ee
So, the integrand for fermionic \ac{QED} in the limit of small $T$ is

\ba \label{eqn:fermI} I_{\rm ferm}(T) &=&
\frac{e^{-m^2T}}{T^3}\left[eB(\vec{r}_{cm})T\coth{(eB(\vec{r}_{cm})T)}
- 1 -\frac{e^2}{3} B^2(\vec{r}_{cm})T^2\right] \nonumber \\ &\approx&-\frac{(eB)^4
T}{45}+\frac{1}{45} (eB)^4 m^2 T^2+\left( \frac{2 (eB)^6}{945}-\frac{(eB)^4
m^4}{90}\right)T^3 \nonumber \\
  & & +\frac{(7 (eB)^4 m^6-4 (eB)^6 m^2)T^4 }{1890}+O(T^5).
\ea
For \ac{ScQED}, we have
\ba 
\label{eqn:scalI} I_{\rm scal}(T) &=&
\frac{e^{-m^2T}}{T^3}\left[eB(\vec{r}_{cm})T/\sinh{(eB(\vec{r}_{cm})T)}
- 1 +\frac{1}{6} (eB)^2(\vec{r}_{cm})T^2\right] \nonumber \\ &\approx&\frac{7
(eB)^4 T}{360}-\frac{7  (eB)^4 m^2 T^2}{360}+\frac{(147 (eB)^4 m^4-31
(eB)^6)T^3 }{15120} \nonumber \\ & &+\frac{ (31 (eB)^6 m^2-49 (eB)^4
m^6)T^4}{15120}+O(T^5). 
\ea

Beyond providing the renormalization conditions, these expansions can be used
in the small $T$ regime to avoid a problem with the Wilson loop uncertainties
in this region.  Consider the uncertainty in the integrand arising from the
uncertainty in the Wilson loop:

\ba 
\delta I(T) &=& \frac{\partial I}{\partial W} \delta W \\
  &=& \frac{e^{-m^2 T}}{T^3} \delta W.
\ea In this case, even though we can compute the Wilson loops for small $T$
precisely, even a small uncertainty is magnified by a divergent factor when
computing the integrand for small values of $T$.  So, in order to perform
the integral, we must replace the small $T$ behaviour of the integrand with
the above expansions (\ref{eqn:fermI}) and (\ref{eqn:scalI}).  Our worldline
integral then proceeds by analytically computing the integral for the small
$T$ expansion up to some small value, $a$, and adding this to the remaining
part of the integral~\cite{MoyaertsLaurent:2004}:

\be 
	\int_0^{\infty} I(T) dT = \underbrace{\int_0^a I(T) dT}_{\rm small ~T} 
	+ \underbrace{\int_a^\infty I(T)dT}_{\rm worldline ~numerics}.
\ee
Because this normalization proceedure uses the constant field expressions for small values of 
$T$, this scheme introduces a small systematic uncertainty. To improve on the 
method outlined here, the derivatives of the background field can 
be accounted for by using the analytic forms of the heat kernel expansion to perform the 
renormalization~\cite{Gies:2001tj}. 

\begin{figure}[ht] 
  \begin{center}
    \includegraphics[width=5in]{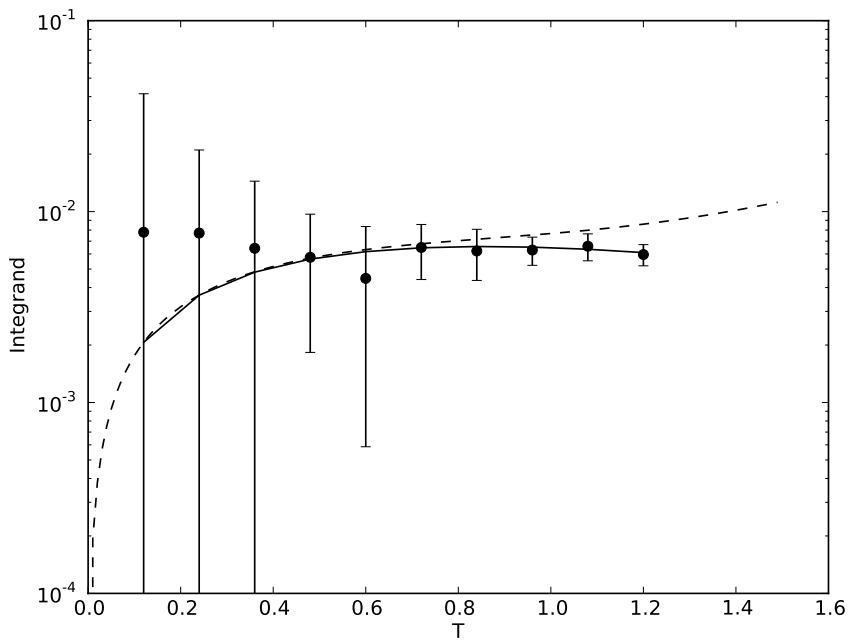} 
	    \caption[small $T$ behaviour of worldline numerics]
		{The small $T$
		behaviour of \ac{WLN}.  The data points represent the numerical
		results, where the error bars are determined from the jackknife analysis described 
		in chapter \ref{ch:WLError}.
		The solid line represents the exact solution while the dotted line represents
		the small $T$ expansion of the exact solution.  Note the amplification of
		the uncertainties.} 
  \end{center} 
\end{figure}

\section{Computing an Effective Action}

\begin{figure}
	\centering
		\includegraphics[width=12cm]{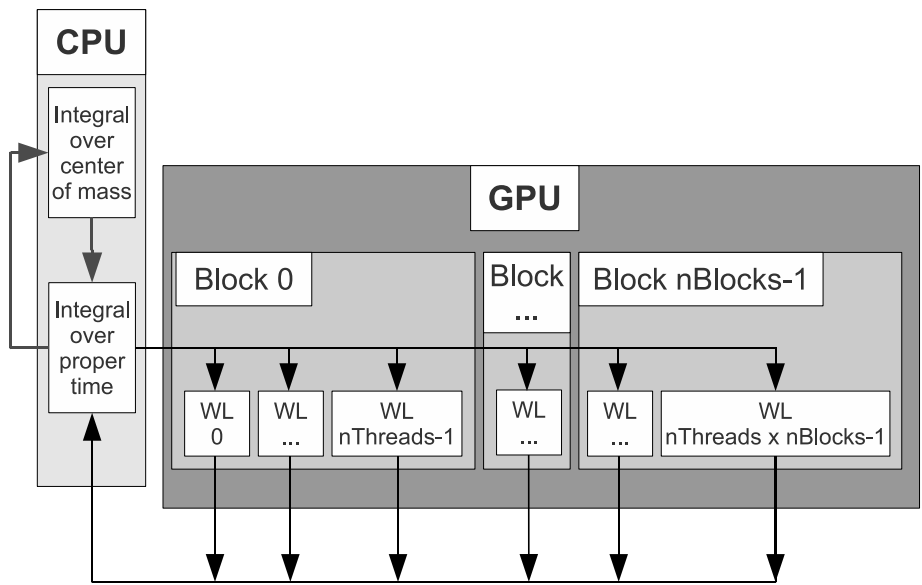}
                \caption[Heterogeneous processing scheme for worldline numerics]
				{The \ac{CPU} manages the loops which compute the 
					integrals over center of mass and proper time. 
					For each proper time integral, we require the results 
					from a large number of individual worldlines. The \ac{GPU}
					is used to compute the integral 
					over each worldline in parallel, and the 
					results are returned to the \ac{CPU} for use in the effective action
					calculation.}.
	\label{fig:coprocessing}
\end{figure}

The ensemble average in the effective action is simply the sum over the
contributions from each worldline loop, divided by the number of loops in
the ensemble.  Since the computation of each loop is independent of the
other loops, the ensemble average may be straightforwardly parallelized by
generating separate processes to compute the contribution from each loop.
For this parallelization, four Nvidia Tesla C1060 \acp{GPU} were used through
the \ac{CUDA} parallel programing framework.  Because \acp{GPU} can spawn
thousands of parallel processing threads\footnote{Each Tesla C1060 device has 30 multiprocessors
which can each support 1,024 active threads.  So, the number of threads available at a time 
is 30,720 on each of the Tesla devices.  Billions of threads may be scheduled on each 
device~\cite{cudazone}.} 
with much less computational overhead
than an \ac{MPI} cluster, they excel at handling a very large number of parallel 
threads, although the clock speed is slower and fewer memory resources are typically available.
In contrast, parallel computing on a cluster using \acs{CPU} 
tends to have a much higher speed per thread, but there are typically fewer 
threads available.  The worldline technique is exceedingly parallelizable, 
and it is a straightforward matter to divide the task into thousands or tens of thousands of 
parallel threads.  In this case,  one should expect excellent performance 
from a \ac{GPU} over a parallel \ac{CPU} implementation, unless thousands 
of \acp{CPU} are available for the program. The \ac{GPU} architecture has 
recently been used by another group for computing Casimir forces using 
worldline numerics~\cite{2011arXiv1110.5936A}.
Figure \ref{fig:coprocessing} 
illustrates the coprocessing and parallelization scheme used here for the 
worldline numerics.

In an
informal test, a Wilson loop average was computed using an ensemble of 5000
loops in 4.7553 seconds using a serial implementation 
while the \ac{GPU} performed the same calculation
in 0.001315 seconds using a \ac{CUDA} implementation.  
A parallel \ac{CPU} code would require a cluster with thousands of 
cores to achieve a similar speed, even if we assume linear (ideal) 
speed-up.  So, for \ac{WLN} computations,
a relatively inexpensive \ac{GPU} can be expected to outperform a small
or mid-sized cluster.  This increase in computation speed has enabled the
detailed parameter searches discussed in this dissertation.

Of course, there are also tradeoffs from using the 
\ac{GPU} architecture with the \ac{WLN} technique. The most significant 
of these is the limited availability of memory on the device. A \ac{GPU} device 
provides only a few GB of global memory (4GB on the Tesla C1060). This limit 
forces compromises between the number of points-per-loop and the number of 
loops to keep the total size of the loop cloud data small. The limited availability of memory 
resources also limits the number of threads which can be executed concurrently on the device.
Because of the overhead associated with transfering data to and from the device, 
the advantages of a \ac{GPU} over a \ac{CPU} cluster are most pronounced 
on problems which can be divided into several hundred or thousands of processes. 
Therefore, the \ac{GPU} may not offer great performance advantages 
for a small number of loops. Finally, there is some additional complexity involved 
in programming for the \ac{GPU} in terms of learning specialized libraries and 
memory management on the device. This means that the code may take longer to develop
and there may be a learning curve for researchers. However, this problem 
is not much more pronounced with \ac{GPU} programming than with other 
parallelization strategies.

Once the ensemble average of the Wilson loop has been computed, computing
the effective action is a straightforward matter of performing numerical
integrals.  The effective action density is computed by performing the
integration over proper time, $T$.  Then, the effective action is computed
by performing a spacetime integral over the loop ensemble center of mass.
In all cases where a numerical integral was performed, Simpson's method was
used~\cite{burden2001numerical}. Integrals from 0 to $\infty$ were mapped to the 
interval $[0,1]$ using substitutions of the form $x = \frac{1}{1+T/T_{\rm max}}$, 
where $T_{\rm max}$ sets the scale for the peak of the integrand. In the constant field 
case, for the integral over proper time,
 we expect $T_{\rm max} \sim 3/(eB)$ for large fields and $T_{\rm max} \sim 1$ for 
fields of a few times critical or smaller. In chapter \ref{ch:WLError}, we present 
a detailed discussion of how the statistical and discretization uncertainties 
can be computed in this technique.

In this implementation, the numerical integrals are done using a serial CPU
computation. 
This serial portion of the algorithm tends to limit the speedup that can be achieved with 
the large number of parallel threads on the \ac{GPU} device\footnote{By Amdahl's law, 
for a program with a ratio, $P$, of parallel to serial computations on $N$ processors, 
the speedup is given by $S < \frac{1}{(1-P)+\frac{P}{N}}$. This law predicts 
rapidly deminishing returns from increasing the number of processors when $P>0$ 
for a fixed problem size.}. 
However, an important 
benefit of the large number of threads available on the \ac{GPU} is that the 
number of worldlines in the ensemble can be increased without limit, as long 
as more threads are available, without significantly increasing the computation time. 
If perfect occupancy could be achieved on a Tesla C1060 device, an ensemble 
of up to 30,720 worldlines could be computed concurrently.
Thus, the \ac{GPU} provides an excellent architecture for improving the 
statistical uncertainties. Additionally, there is room for further optimization of 
the algorithm by parallelizing the serial portions of the 
algorithm to achieve a greater speedup.

More details about implementing this algorithm on the \ac{CUDA} architecture 
can be found in appendix \ref{ch:cudafication}. A listing of the 
\ac{CUDA} \ac{WLN} code appears in appendix \ref{ch:sourcecode}.

\section{Verification and Validation}

The \ac{WLN} software can be validated and verified by making sure that it 
produces the correct results where the derivative expansion is a good approximation, 
and that the results are consistent with previous numerical calculations of flux tube 
effective actions. For this reason, the validation was done primarily with flux tubes with 
a profile defined by the function
\be
	f_\lambda(\rho^2) = \frac{\rho^2}{(\lambda^2 + \rho^2)}.
\ee
For large values of $\lambda$, this function varies slowly on the Compton wavelength 
scale, and so the derivative expansion is a good approximation. Also, flux tubes 
with this profile were studied previously using \ac{WLN}
\cite{Moyaerts:2003ts, MoyaertsLaurent:2004}.

Among the results presented in~\cite{MoyaertsLaurent:2004} is a comparison of 
the derivative expansion and \ac{WLN} for this magnetic field 
configuration. The result is that the next-to-leading-order term 
in the derivative expansion is only a small correction to the the 
leading-order term for $\lambda \gg \lambda_e$, where the derivative 
expansion is a good approximation. The derivative expansion breaks
down before it reaches its formal validity limits 
at $\lambda \sim \lambda_e$. For this reason,
we will simply focus on the leading order derivative expansion (which we call the
\ac{LCF} approximation).
The effective action of \ac{QED} in the \ac{LCF} approximation is given 
in cylindrical symmetry by
\ba
	\label{eqn:LCFferm}
	\Gamma^{(1)}_{\rm ferm} &=& \frac{1}{4\pi}\int_0^\infty dT 
	\int_0^\infty \rho_{\rm cm} d\rho_{\rm cm}\frac{e^{-m^2T}}{T^3} \nonumber \\
	& &\left\{eB(\rho_{\rm cm})T\coth{(eB(\rho_{\rm cm})T)} 
	- 1 -\frac{1}{3}(eB(\rho_{\rm cm})T)^2\right\}.
\ea

\begin{figure}
	\centering
		\includegraphics[width=12cm]{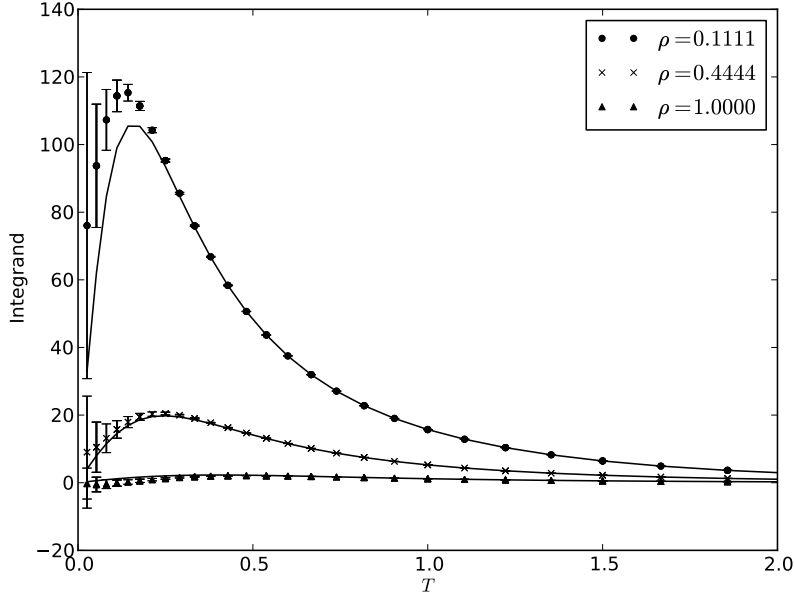}
	\caption[Comparison with derivative expansion for $T$ integrand]
	{The integrand of the proper time, $T$, integral for three different values 
	of the radial coordinate, $\rho$ for a $\lambda = 1$ flux tube. The solid lines 
	represent the zeroth-order derivative expansion, which, as expected, is a good approximation 
	until $\rho$ becomes too small.}
	\label{fig:igrandvsT}
\end{figure}

Figure \ref{fig:igrandvsT} shows a comparison between the proper time integrand,
\be
	\frac{e^{-m^2T}}{T^3}\left[\langle W\rangle_{\vec{r}_{\rm cm}} 
		- 1 -\frac{1}{3}(eB_{\rm cm}T)^2\right],
\ee
and the \ac{LCF} approximation result for a flux tube with $\lambda = \lambda_e$ and
$\mathcal{F} = 10$. In this case, the \ac{LCF} approximation is only appropriate far from the 
center of the flux tube, where the field is not changing very rapidly. In the figure, we can 
begin to see the deviation from this approximation, which gets more pronounced closer to the 
center of the flux tube (smaller values of $\rho$).

\begin{figure}
	\centering
		\includegraphics[width=12cm]{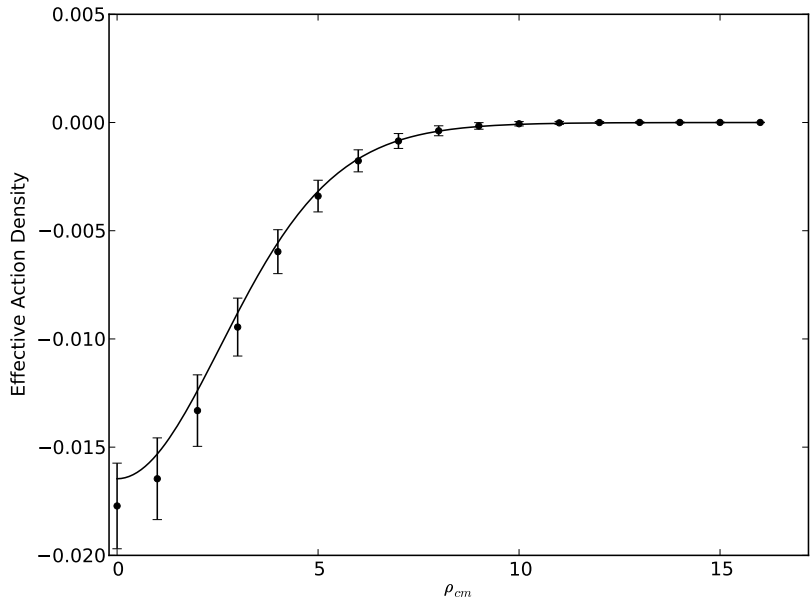}
	\caption[Comparison with \acs{LCF} approximation for action density]
	{The fermion term of the effective action density as a function 
	of the radial coordinate for a flux tube with width $\lambda = 10 \lambda_e$. }
	\label{fig:vsrhocm}
\end{figure}

The effective action density for a slowly varying flux tube is plotted in figure
\ref{fig:vsrhocm} along with the \ac{LCF} approximation. In this case, 
the \ac{LCF} approximation agrees within the statistics of the \ac{WLN}.



\section{Conclusions}

In this chapter, I have reviewed the \ac{WLN} numerical technique with a focus 
on computing the effective action of \ac{QED} in nonhomogeneous, cylindrically symmetric 
magnetic fields. The method uses a Monte Carlo generated ensemble of worldline loops 
to approximate a path integral in the worldline formalism. These worldline loops are 
generated using a simple algorithm and encode the information about the magnetic 
field by computing the flux through the loop and the action aquired from transporting a 
magnetic moment around the loop.
This technique preserves Lorentz symmetry exactly and can preserve 
gauge symmetry up to any required precision. 

I have discussed implementing this technique on \ac{GPU} architecture using 
\ac{CUDA}. The main advantage of this architecture is that it allows for a 
very large number of concurrent threads which can be utilized with 
very little overhead. In practice, this means that a large ensemble 
of worldlines can be computed concurrently, thus allowing for a considerable 
speedup over serial implementations, and allowing for the precision of the 
numerics to increase according to the number of threads available. 

\chapter{Uncertainty analysis in worldline numerics} 
\label{ch:WLError}

\acresetall

\begin{summary}
	This chapter gives a brief overview of uncertainty analysis in the \ac{WLN} method.
	There are uncertainties from discretizing each loop, and from using a statistical ensemble of 
	representative loops. The former can be minimized so that the latter dominates. However, 
	determining the statistical uncertainties is complicated by two subtleties. Firstly, 
	the distributions generated by the worldline ensembles are highly non-Gaussian, and 
	so the standard error in the mean is not a good measure of the statistical uncertainty.
	Secondly, because the same ensemble of worldlines is used to compute the Wilson loops 
	at different values of $T$ and $x_{\rm cm}$, the uncertainties associated with each computed 
	value of the integrand are strongly correlated. I recommend a form of jackknife analysis 
	which deals with both of these problems.
\end{summary}

So far in the \ac{WLN} literature, 
the discussions of uncertainty analysis has been unfortunately brief. 
It has been suggested that the standard deviation of the worldlines provides 
a good measure of the statistical error in the worldline method~\cite{Gies:2001zp, Gies:2001tj}. 
However, the distributions produced by the worldline ensemble are 
highly non-Gaussian, and therefore the standard error in the mean is not a good measure of the 
uncertainties involved. Furthermore, the use of the same worldline ensemble to 
compute the Wilson loop multiple times in an integral results in strongly correlated 
uncertainties. Thus, propagating uncertainties through integrals can be 
computationally expensive due to the complexity of computing correlation 
coefficients.

The error bars on worldline 
calculations impact the conclusions that can be drawn from 
calculations, and also have important implications for the fermion problem, which limits 
the domain of applicability of the technique (see section \ref{sec:fermionproblem}).  
It is therefore important that the error 
analysis is done thoughtfully and transparently.
The purpose of this chapter is to contribute a more 
thorough discussion of uncertainty analysis in the \ac{WLN} technique to the 
literature in hopes of avoiding any confusion associated with the above-mentioned 
subtleties.

There are two sources of uncertainty in the worldline technique: the discretization error in 
treating each continuous worldline as a set of discrete points, and the statistical error of 
sampling a finite number of possible worldlines from a distribution.  In this chapter, I will 
discuss each of these sources of uncertainty.

\section{Estimating the Discretization Uncertainties}
\label{sec:discunc}

The discretization error arising from the integral over $\tau$ in the exponent of each Wilson loop 
(see equation (\ref{eqn:wilsonloop})) is 
difficult to estimate since any number of loops could be represented by each choice of discrete points.
The general strategy is to make this estimation by computing the Wilson loop using several different 
numbers of points per worldline and observing the convergence properties.

The specific procedure adopted for this work involves dividing each discrete worldline into several 
worldlines with varying levels of discretization.
Since we are using the d-loop 
method for generating the worldlines (section \ref{sec:loopgen}), 
a $\frac{N_{ppl}}{2}$ sub-loop consisting of every other 
point will be guaranteed to contain the prescribed distribution of velocities.

\begin{figure}
	\centering
		\includegraphics[width=5cm]{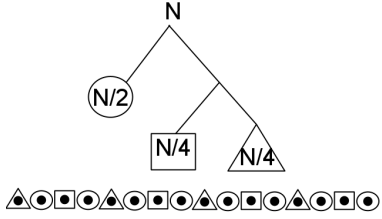}
	\caption[Illustration of convergence testing scheme]
	{Diagram illustrating the division of a worldline into three smaller interleaved worldlines}
	\label{fig:Division}
\end{figure}

To look at the convergence for the loop discretization, 
each worldline is divided into three groups.  One group of $\frac{N_{ppl}}{2}$ points, and two groups of 
$\frac{N_{ppl}}{4}$.  
This permits us to compute the average holonomy factors at three levels of 
discretization:

\be 
\mean{W}_{N_{ppl}/4} = \mean{e^{\frac{i}{2}\triangle} e^{\frac{i}{2}\Box}},
\ee 

\be 
\mean{W}_{N_{ppl}/2} = \mean{e^{i\circ}},
\ee 
and
\be 
\mean{W}_{N_{ppl}} = \mean{e^{\frac{i}{2}\circ} e^{\frac{i}{4}\Box} e^{\frac{i}{4}\triangle}},
\ee 
where the symbols $\circ$, $\Box$, and $\triangle$ denote the worldline integral, 
$\int_0^T d\tau A(x_{CM}+x(\tau))\cdot \dot x$, computed using the sub-worldlines 
depicted in figure \ref{fig:Division}.

\begin{figure}
	\centering
		\includegraphics[width=12cm]{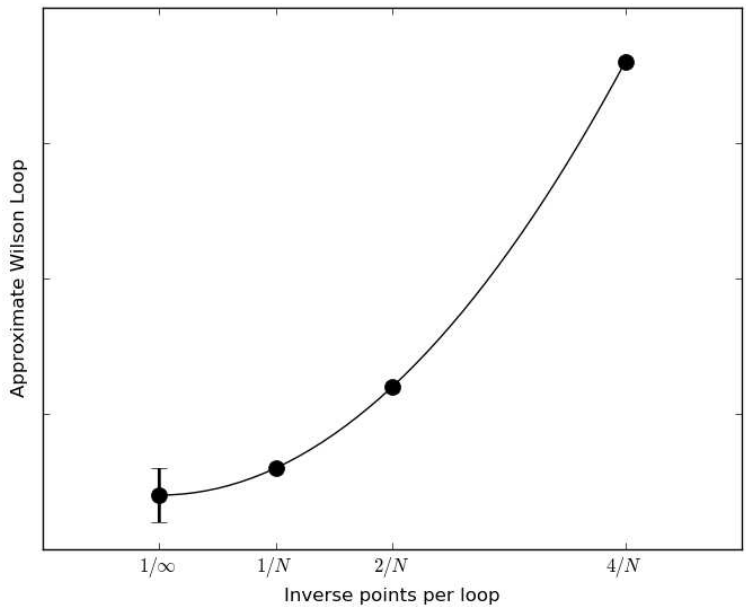}
	\caption[Illustration of extrapolation to infinite points per loop]
	{This plot illustrates the method used to extrapolate the Wilson loop to 
	infinite points per loop and the uncertainty estimate in the approximation.}
	\label{fig:DiscErr}
\end{figure}

We may put these factors into the equation of a parabola to extrapolate the result to an infinite 
number of points per line (see figure \ref{fig:DiscErr}):
\be 
\mean{W}_{\infty} \approx \frac{8}{3}\mean{W}_{N_{ppl}} - 2 \mean{W}_{N_{ppl}/2} +\frac{1}{3}\mean{W}_{N_{ppl}/4}.
\ee 
So, we estimate the discretization uncertainty to be
\be 
\delta \mean{W}_{\infty} \approx |\mean{W}_{N_{ppl}} -  \mean{W}_{\infty}|.
\ee 
Generally, the statistical uncertainties are the limitation in the precision of the 
\ac{WLN} technique. Therefore, $N_{ppl}$ should be chosen to be 
large enough that the discretization uncertainties are small relative to the 
statistical uncertainties.

\section{Estimating the Statistical Uncertainties}

We can gain a great deal of insight into the nature of the statistical uncertainties 
by examining the specific case of the uniform magnetic field since we know the 
exact solution in this case. 

\subsection{The Worldline Ensemble Distribution is Non-Gaussian}

\begin{figure}
	\centering
		\includegraphics[width=12cm]{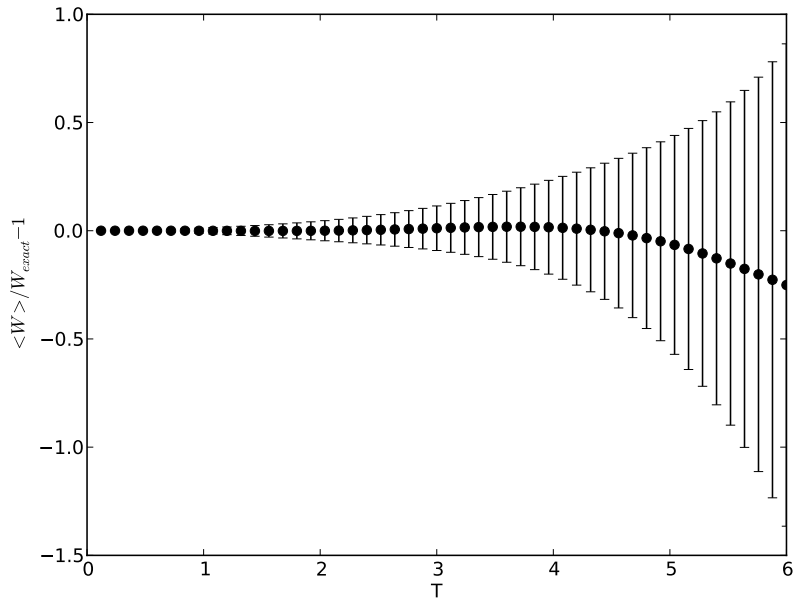}
	\caption[Standard errors in the mean for worldline numerics]
	{Residuals of worldline calculations and the corresponding standard 
	errors in the mean.  For reasons discussed in this section, these error bars 
	overestimate the uncertainties involved.}
	\label{fig:resids}
\end{figure}

A reasonable first instinct for estimating the error bars is to use the standard 
error in the mean of the collection of individual worldlines:
\be 
{\rm SEM}( W ) = \sqrt{\sum_{i=1}^{N_l}\frac{(W_i - \mean{W})^2}{N_l(N_l-1)}}.
\ee 
This approach has been promoted in early papers on \ac{WLN}~\cite{Gies:2001zp, Gies:2001tj}.
In figure \ref{fig:resids}, I have plotted the residuals and the corresponding 
error bars for several values of the proper time parameter, $T$.  From this plot, 
it appears that the error bars are quite large in the sense that we appear to 
produce residuals which are considerably smaller than would be implied by the 
sizes of the error bars.  This suggests that we have overestimated the size of 
the uncertainty.

\begin{figure}
	\centering
		\includegraphics[width=11cm]{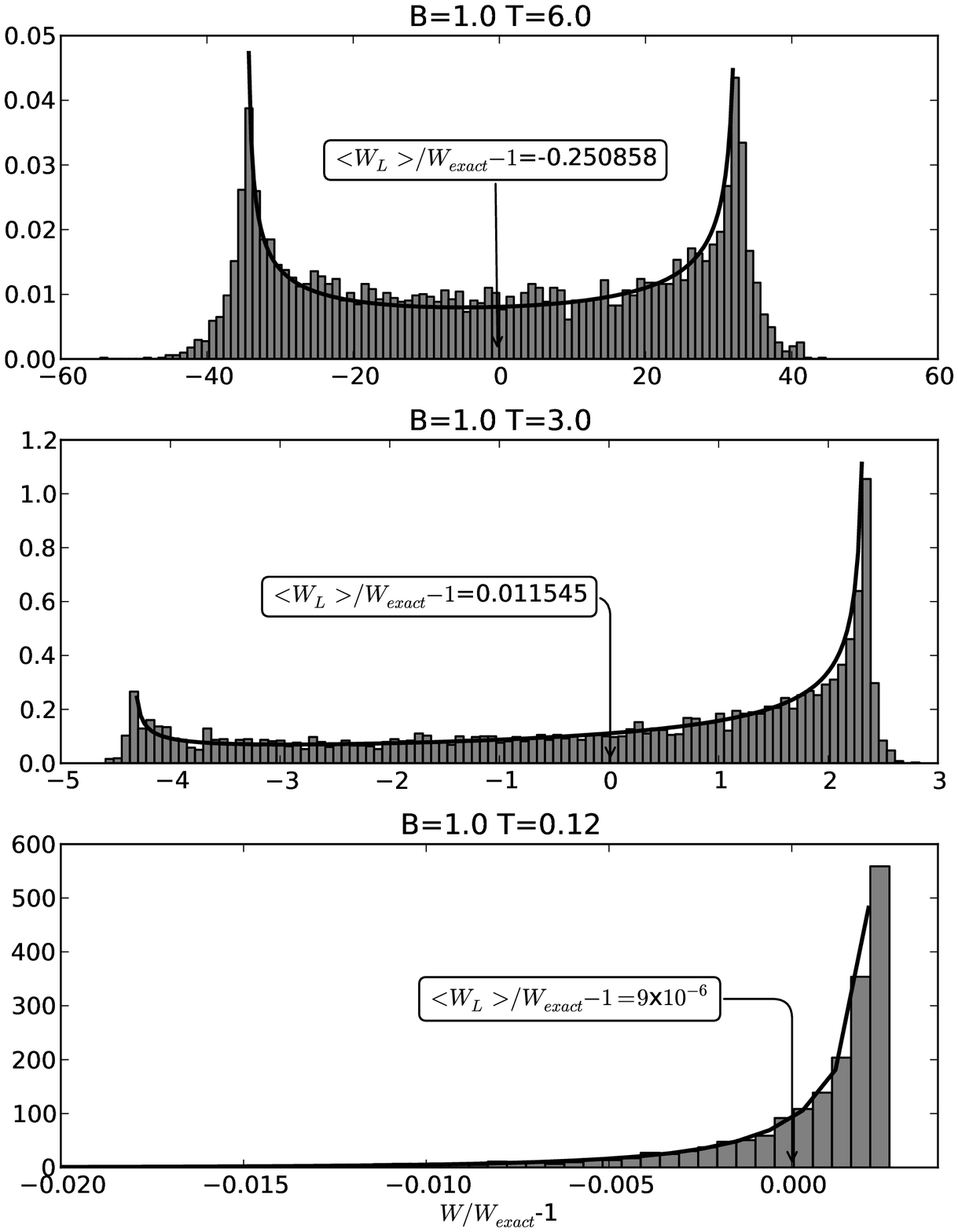}
	\caption[Histograms showing the worldline distributions]
	{Histograms showing the worldline distributions of the residuals 
	for three values of $T$ in the constant magnetic field case. 
	Here, we are neglecting the fermion factor. The dark 
	line represents the exact distribution computed using equation 
	\ref{eqn:exactdist}.  The worldline means are indicated with an arrow, 
	while the exact mean in each case is 0.  There are 5120 worldlines in 
	each histogram.  The vertical axes are normalized to a total area of unity.}
	\label{fig:hists}
\end{figure}

We can see why this is the case by looking more closely at the distributions 
produced by the worldline technique.  An exact expression for these distributions 
can be derived in the case of the constant magnetic field~\cite{MoyaertsLaurent:2004}:
\be 
\label{eqn:exactdist}
w(y)=\frac{W_{\rm exact}}{\sqrt{1-y^2}}\sum_{-\infty}^{\infty}[f(\arccos(y)+2n\pi)+f(-\arccos(y)+2n\pi)]
\ee 
with
\be 
f(\phi)=\frac{\pi}{4BT\cosh^2(\frac{\pi \phi}{2BT})}.
\ee 
Figure \ref{fig:hists} shows histograms of the worldline results along with the 
expected distributions.  These distributions highlight a significant hurdle in 
assigning error bars to the results of \ac{WLN}.

Due to their highly non-Gaussian nature, the standard error in the mean is not a good 
characterization of the distributions that are produced.  We should not interpret each individual 
worldline as a measurement of the mean value of these distributions; for large values of $BT$, 
almost all of our worldlines will produce answers which are far away from the mean of the 
distribution. This means that the variance of the distribution will be very large, even 
though our ability to determine the mean of the distribution is relatively precise
because of the increasing symmetry about the mean as $T$ becomes large.

\subsection{Correlations between Wilson Loops}

\begin{figure}
	\centering
		\includegraphics[width=12cm]{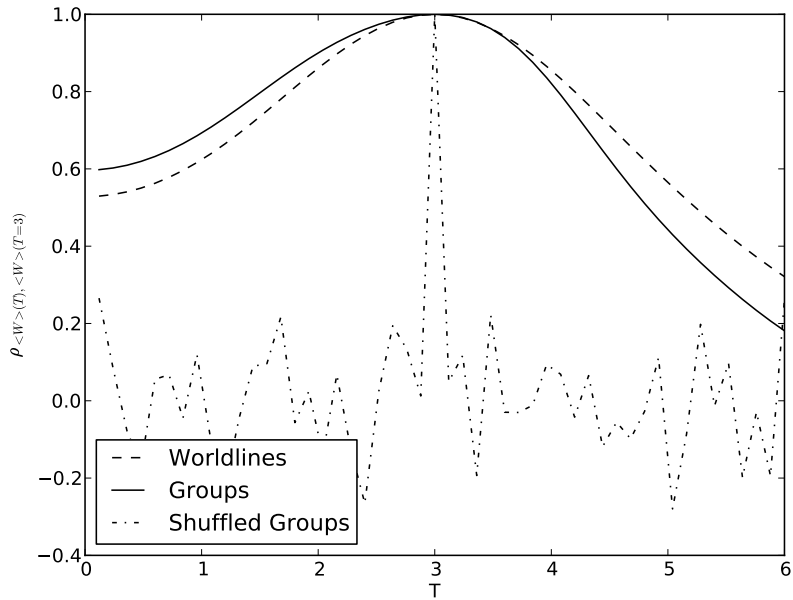}
	\caption[Correlation coefficients between different evaluations of the integrand]
	{Correlation coefficients, equation (\ref{eqn:corrcoef}), between $\mean{W(T)}$ 
		and $\mean{W(T=3)}$ computed using individual worldlines, groups of worldlines, and 
		shuffled groups of worldlines.}
	\label{fig:corr}
\end{figure}

Typically, numerical integration is performed by replacing the integral with a sum over a finite set 
of points from the integrand.  So, I will begin the present discussion by considering the uncertainty 
in adding together two points (labelled $i$ and $j$) in our integral over $T$. Two terms of the 
sum representing the numerical integral will involve a function of $T$ times the two 
Wilson loop factors,
\be 
I = g(T_i)\mean{W(T_i)} + g(T_j)\mean{W(T_j)}
\ee 
with an uncertainty given by
\ba 
	\delta I &=& \left | \pderiv{I}{\mean{W(T_i)}} \right|^2 (\delta \mean{W(T_i)})^2 + 
	\left | \pderiv{I}{\mean{W(T_j)}} \right|^2 (\delta \mean{W(T_j)})^2 \\ \nonumber
	& & + 2 \left | \pderiv{I}{\mean{W(T_i)}} \pderiv{I}{\mean{W(T_j)}} \right | \rho_{ij} 
	(\delta \mean{W(T_i)}) (\delta \mean{W(T_j)}) \\
	&=& g(T_i)^2 (\delta \mean{W(T_i)})^2 + g(T_j)^2 (\delta \mean{W(T_j)})^2 \nonumber \\
	& &+ 2 \left | g(T_i)g(T_j) \right | \rho_{ij} (\delta \mean{W(T_i)}) (\delta \mean{W(T_j)})
\ea 
and the correlation coefficient $\rho_{ij}$ given by
\be 
	\label{eqn:corrcoef}
	\rho_{ij} = \frac{\mean{(W(T_i) - \mean{W(T_i)})(W(T_j)-\mean{W(T_j)})}}{\sqrt{(W(T_i)
	-\mean{W(T_i)})^2}\sqrt{(W(T_j)-\mean{W(T_j)})^2}}.
\ee 
The final term in the error propagation equation takes into account correlations between the 
random variables $W(T_i)$ and $W(T_j)$. Often in a Monte Carlo computation, one can 
treat each evaluation of the integrand as independent, and neglect the uncertainty
term involving the correlation coefficient. However, in \ac{WLN}, 
the evaluations are related because the same worldline ensemble is reused 
for each evaluation of the integrand.
So, the correlations are significant (see figure \ref{fig:corr}), and this term 
can't be neglected. Computing each correlation coefficient takes 
a time proportional to the square of the number of worldlines. So, it may 
be computationally expensive to formally propagate uncertainties through an 
integral.

The point-to-point correlations were originally pointed out by Gies and Langfeld 
who addressed the problem by updating (but not replacing or regenerating)
the loop ensemble 
in between each evaluation of the Wilson loop average~\cite{Gies:2001zp}. 
This may be a good way of addressing the problem. However, in the following 
section, I promote a method which can bypass the difficulties presented by 
the correlations by treating the worldlines as a collection of worldline groups.

\subsection{Grouping Worldlines}

Both of the problems explained in the previous two subsections can be overcome 
by creating groups of worldline loops within the ensemble. Each group of worldlines 
then makes a statistically independent measurement of the Wilson loop average 
for that group. The statistics between the groups of measurements are Gaussian 
distributed, and so the uncertainty is the standard error in the mean of the 
ensemble of groups (in contrast to the ensemble of worldlines).

For example, if we divide the $N_l$ worldlines into $N_G$ groups of $N_l/N_G$ 
worldlines each, we can compute a mean for each group:
\be
	\mean{W}_{G_j} = \frac{N_G}{N_l}\sum_{i=1}^{N_l/N_G}W_i.
\ee
Provided each group contains the same number of worldlines, 
the average of the Wilson loop is unaffected by this grouping:
\ba
	\mean{W} & = & \frac{1}{N_G} \sum_{j=1}^{N_G} \mean{W}_{G_j} \\
		& = & \frac{1}{N_l} \sum_{i=1}^{N_l} W_i.
\ea
However, the uncertainty is the standard error in the mean of 
the groups,
\be
	\delta \mean{W} = \sqrt{\sum_{i=1}^{N_G} 
		\frac{(\mean{W}_{G_i} - \mean{W})^2}{N_G(N_G-1)}}.
\ee

Because the worldlines are unrelated to one another, the choice of how to 
group them to compute a particular Wilson loop average is arbitrary. For example, 
the simplest choice is to group the loops by the order they were generated, so that 
a particular group number, $i$, contains worldlines $iN_l/N_G$ through $(i+1)N_l/N_G -1$. 
Of course, if the same worldline groupings are used to compute different Wilson 
loop averages, they will still be correlated. We will discuss this problem in a moment.

\begin{figure}
	\centering
		\includegraphics[width=12cm]{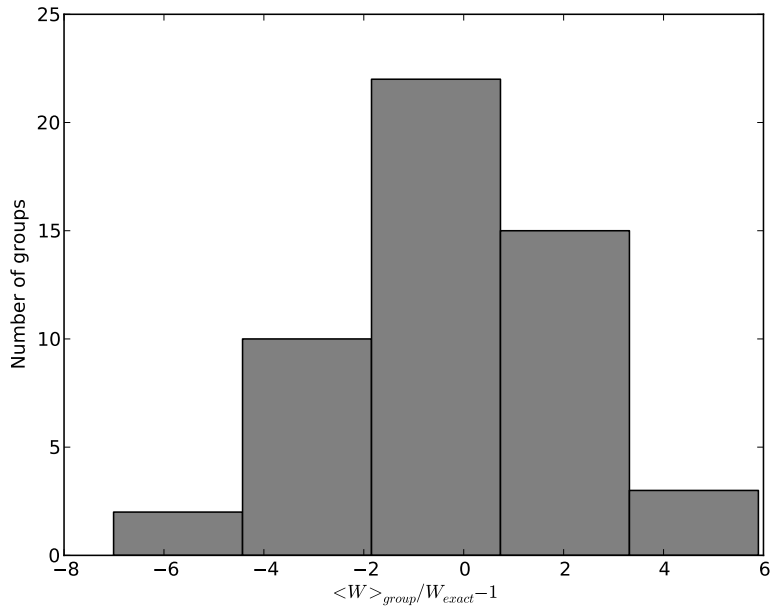}
	\caption[Histogram for reproducing measurements with groups of worldlines]
	{The histogram demonstrating the precision with which we can reproduce 
		measurements of the mean using different groups of 100 worldlines at $BT=6.0$.  
		In this case, the distribution is Gaussian-like and meaningful error bars can 
		be placed on our measurement of the mean.}
	\label{fig:uncinmean}
\end{figure}

The basic claim of the worldline technique is that the mean of the worldline 
distribution approximates the holonomy factor. However, from the distributions 
in figure \ref{fig:hists}, we can see that the individual worldlines themselves 
do not approximate the holonomy factor. So, we should not think of 
an individual worldline as an estimator of the mean of the distribution. 
Thus, a resampling technique is required to determine the precision of our 
statistics. We can think of each group of worldlines 
as making an independent measurement of the mean of a distribution. As expected, 
the groups of worldlines produce a more Gaussian-like distribution (see figure 
\ref{fig:uncinmean}), and 
so the standard error of the groups is a sensible measure of the uncertainty 
in the Wilson loop value. 

\begin{figure}
	\centering
		\includegraphics[width=12cm]{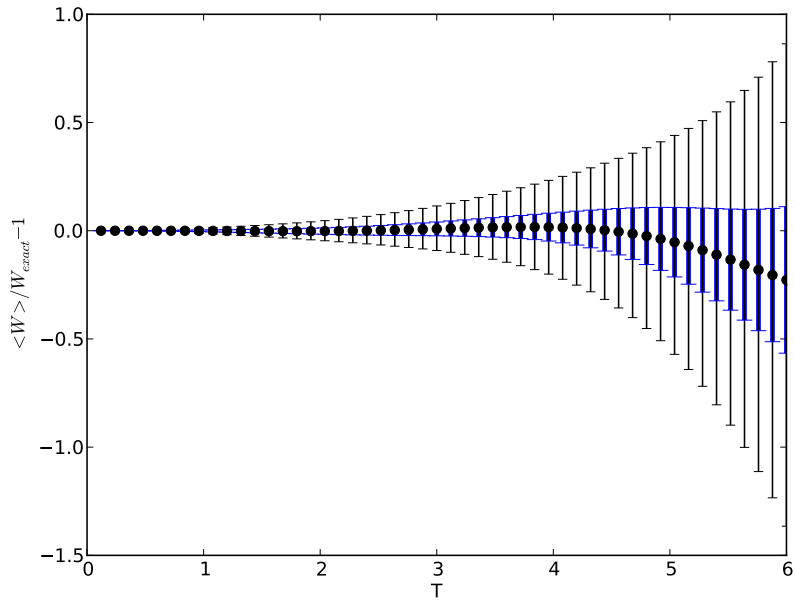}
	\caption[Comparison of error bars between standard error in the mean and 
	jackknife analysis]
	{The residuals of the Wilson loops for a constant magnetic field showing 
		the standard error in the mean (thin error bars) and the uncertainty in determining 
		the mean (thick blue error bars).  The standard error in the mean overestimates the 
		uncertainty by more than a factor of 3 at each value of $T$.}
	\label{fig:resids2}
\end{figure}

We find that the error bars are about one-third 
as large as those determined from the standard error in the mean of the 
individual worldlines, and the smaller error bars better characterize 
the size of the residuals in the constant field case (see figure \ref{fig:resids2}).
The strategy of using subsets of the available data to 
determine error bars is called jackknifing. 
Several previous papers on \ac{WLN} have mentioned using jackknife 
analysis to determine the uncertainties, but without an explanation of the 
motivations or the procedure employed
\cite{2005PhRvD..72f5001G, PhysRevLett.96.220401, Dunne:2009zz, PhysRevD.84.065035}.

The grouping of worldlines alone does not address the problem of correlations 
between different evaluations of the integrands. 
Figure \ref{fig:corr} shows that the uncertainties for groups of worldlines are also 
correlated between different points of the integrand. 
However, the worldline grouping
does provide a tool for bypassing the problem. One possible strategy is to 
randomize how worldlines are assigned to groups between each evaluation of the 
integrand. This produces a considerable reduction in the correlations, as is 
shown in figure \ref{fig:corr}. Then, errors can be propagated through the 
integrals by neglecting the correlation terms. Another strategy is to separately 
compute the integrals for each group of worldlines, and then consider the statistics of 
the final product to determine the error bars. This second strategy is the 
one adopted for the work presented in this thesis. Grouping in this way reduces the 
amount of data which must be propagated through the integrals by a factor 
of the group size compared to a delete-1 jackknife scheme, for example. 
In general, the error 
bars quoted in this thesis are obtained by computing the standard error in the 
mean of groups of worldlines.

\section{Uncertainties and the Fermion Problem}
\label{sec:fermionproblem}

The fermion problem of \ac{WLN} is a name given to an enhancement of the uncertainties 
at large $T$~\cite{Gies:2001zp, MoyaertsLaurent:2004}. It should not be confused with the 
fermion doubling problem associated with lattice methods. In a constant magnetic field, the scalar 
portion of the calculation produces a factor of $\frac{BT}{\sinh{(BT)}}$, while the fermion 
portion of the calculation produces an additional factor $\cosh{(BT)}$. Physically, this contribution 
arises as a result of the energy required to transport the electron's magnetic moment around the 
worldline loop.
At large values of 
$T$, we require subtle cancellation between huge values produced by the fermion portion 
with tiny values produced by the scalar portion.  However, for large $T$, the scalar portion acquires 
large relative uncertainties which make the computation of large $T$ contributions to the integral very imprecise.

This can be easily understood by examining the worldline distributions shown in figure
\ref{fig:hists}. Recall that the scalar Wilson loop average for these histograms is given 
by the flux in the loop, $\Phi_B$:
\be
	\mean{W} = \left<\exp{(ie\int_0^Td\tau \vec{A}(\vec{x}_{\rm cm} 
	+ \vec{x}(\tau))\cdot d\vec{x}(\tau))}\right> = \left< e^{ie\Phi_B}\right>.
\ee
For constant fields, the flux through the worldline loops obeys the distribution function
\cite{MoyaertsLaurent:2004}
\be
	f(\Phi_B) = \frac{\pi}{4BT\cosh^2\left(\frac{\pi \Phi_B}{2BT}\right)}.
\ee
For small values of $T$, the worldline loops are small and the 
amount of flux through the loop is correspondingly small. Therefore, the 
flux for small loops is narrowly distributed about $\Phi_B = 0$. Since zero 
maximizes the Wilson loop ($e^{i0}=1$), 
this explains the enhancement to the right of the distribution for small values of $T$. 
As $T$ is increased, the flux through any given worldline becomes very large and the 
distribution of the flux becomes very broad. 
For very large $T$,  the width of the distribution is many 
factors of $2\pi/e$. Then, the phase ($e \Phi_B\mod{2\pi}$) is nearly 
uniformly distributed, and the 
Wilson loop distribution reproduces the Chebyshev distribution (\ie 
the distribution obtained from projecting uniformly distributed points on 
the unit circle onto the horizontal axis),

\be
	\lim_{T\to\infty}w(y) = \frac{1}{\pi\sqrt{1-y^2}}.
\ee

The mean of the Chebyshev distribution is zero due to its symmetry. 
However, this symmetry is not 
realized precisely unless we use a very large number of loops. Since the 
width of the distribution is already 100$\times$ the value of the mean at 
$T=6$, any numerical asymmetries in the distribution result in very large 
relative uncertainties of the scalar portion. Because of these uncertainties, 
the large contribution from the fermion factor are not cancelled precisely.

This problem makes it very difficult to compute 
the fermionic effective action unless the fields are well localized
\cite{MoyaertsLaurent:2004}. For example, the fermionic factor for 
non-homogeneous magnetic fields oriented along the z-direction is
\be
	\cosh{\left(e\int_0^T d\tau B(x(\tau))\right)}.
\ee
For a homogeneous field, this function grows exponentially with $T$ and 
is cancelled by the exponentially vanishing scalar Wilson loop.
For a localized field, 
the worldline loops are very large for large values of $T$, and they primarily 
explore regions far from the field. Thus, the fermionic factor grows more slowly 
in localized fields, and is more easily cancelled by the rapidly vanishing scalar part.

\section{Conclusions}

In this chapter, I have identified two important considerations in 
determining the uncertainties associated with \ac{WLN} computations. 
Firstly, the computed points within the integrals over proper time, $T$, 
or center of mass, $\vec{x}_{\rm cm}$, are highly correlated because 
one typically reuses the same ensemble of worldlines to compute each point. 
Secondly, the statistics of the worldlines are not Gaussian distributed and 
each individual worldline in the ensemble may produce a result which 
is very far from the mean value. So, in determining the uncertainties in the 
\ac{WLN} technique, one should consider how precisely the mean of the ensemble 
can be measured from the ensemble 
and this is not necessarily given by the standard error in the mean. 

These issues can be addressed simultaneously
using a scheme where the worldlines from 
the ensemble are placed into groups and the effective action or the 
effective action density is evaluated separately for each group. The 
uncertainties can then be determined by the statistics of the groups. 
This scheme is less computationally intensive than a delete-1 jackknife 
approach because less data (by a factor of the group size) needs to be
propagated through the integrals. It is also less computationally 
intensive than propagating the uncertainties through the numerical 
integrals because it avoids the computation of numerous 
correlation coefficients.

\chapter{Magnetic Flux Tubes in a Dense Lattice}
\label{ch:periodic}
\acresetall

\begin{summary}
If the superconducting nuclear material of a neutron star contains magnetic flux tubes, 
the magnetic field is likely to vary rapidly on the scales where \acs{QED} effects 
are important.
In this chapter, I construct a cylindrically symmetric toy model of a flux 
tube lattice in which the influence of neighbouring flux tubes is taken into account.
We compute the effective action densities using the \acs{WLN} technique.
The numerics predict a greater effective energy density in the 
region of the flux tube, but a smaller energy density in the regions 
between the flux tubes compared to a locally-constant-field approximation. 
We also compute the interaction energy between a flux tube and its neighbours 
as the lattice spacing is reduced from infinity. Because our flux tubes 
exhibit compact support, this energy is entirely non-local and
predicted to be zero in local approximations
such as the derivative expansion. This quantity can take 
positive or negative values depending on the magnetic field profile and the 
specific definition of the interaction energy. 
\end{summary}

\section{Introduction}

Computing the quantum effective action for magnetic flux tube 
configurations is a problem that has generated considerable interest
and has been explored through a variety of approaches~\cite{Gornicki1990271, 
1998MPLA...13..379S, 0264-9381-12-5-013, PhysRevD.51.810, 1999PhRvD..60j5019B,
PhysRevD.62.085024, 2001PhRvD..64j5011P, 2003PhRvD..68f5026B, 
2005NuPhB.707..233G, 2006JPhA...39.6799W, Weigel:2010pf, Langfeld:2002vy}
(see section \ref{sec:EAisoflux}). 
Partly, this 
is because it is a relatively simple problem for analyzing non-homogeneous 
generalizations of the Heisenberg-Euler action and for exploring 
limitations of techniques such as the derivative expansion. But, this 
is also a physically important problem because tubes of magnetic flux 
are very important for the quantum mechanics of electrons due 
to the Aharonov-Bohm effect, and they appear 
in a variety of interesting physical scenarios such as stellar astrophysics, 
cosmic strings, in superconductor vortices, and quark confinement
\cite{2003PrPNP..51....1G}.

In the present context, we are concerned with the role that magnetic 
flux tubes play in the superconducting nuclear material of compact stars.
In this scenario, the \ac{QED} effects are particularly interesting because 
the magnetic flux tubes, if they exist, are confined to tubes which may be 
only a few percent of the Compton wavelength, $\lambda_{\rm C}$, in radius. Specifically, 
the flux must be confined to within the London penetration depth of the superconducting 
material, which for neutron stars has been estimated to be 80 fm $=$
0.032 $\lambda_e$~\cite{PhysRevLett.91.101101}. Moreover, the 
flux tube density is expected to be proportional to the 
average magnetic field. For a background field near the quantum critical 
strength, $B_k$, such as in a neutron star, the distance between flux tubes 
is comparable to a Compton wavelength. This Compton wavelength scale is 
also the scale at which the non-locality of \ac{QED} becomes important and 
at which powerful local techniques like the derivative expansion 
are no longer appropriate for computing the effective action.

The free energy associated with these flux tubes is a factor in determining 
whether the nuclear material of a neutron star is a type-I or type-II 
superconductor. The free energy of a flux tube is determined by 
looking at the energies associated with the magnetic field, with the 
creation of a non-superconducting region in the superconductor, and 
with interactions between the flux tubes. Flux tubes can only form 
if it is energetically favourable to do so compared to expelling the 
field due to the Meissner effect. In this chapter, I would 
like to explore the contribution from \ac{QED} to this free 
energy. For an isolated flux tube, this is an additional source 
of energy for creating the magnetic field. For a lattice of 
flux tubes, there is also an energy contribution from the 
presence of neighbouring flux tubes because of the non-local 
nature of quantum field theory.

The energy of two flux tubes has been previously
computed using \ac{WLN} methods and for flux tubes with aligned fields, the 
energy is larger than twice the energy of a single flux tube when the flux tubes 
are closely spaced~\cite{Langfeld:2002vy}. This result implies that there is a 
repulsive interaction between the flux tubes due to \ac{QED} effects,
strengthening the likelihood of the type-II scenario in neutron stars. This interaction 
energy increases as the flux tubes are placed closer together, and can have a similar 
magnitude as the \ac{QED} correction to the energy when the flux tubes are closely spaced.

In this chapter, I will further explore the nature of this 
phenomenon in \ac{QED} using
 a highly parallel implementation of the \ac{WLN} 
technique implemented on a \ac{GPU} architecture. Specifically, I explore 
cylindrically symmetric magnetic field profiles for isolated flux 
tubes and periodic profiles designed to model properties of a 
triangular lattice. This 
algorithm cannot be applied to spinor \ac{QED} calculations 
in our model lattice because 
of the well-known fermion problem of \ac{WLN}. However, the problem 
does not affect \ac{ScQED} calculations. Therefore, 
we explore the quantum-corrected energies of 
isolated flux tubes for both scalar and spinor 
electrons and use this comparison to speculate 
about the relationship of our cylindrical lattice model 
and the spinor \ac{QED} energies of an Abrikosov lattice 
of flux tubes that may be found in neutron stars.

\subsection{Cylindrical Magnetic Fields} 

We consider our flux tubes to have cylindrical symmetry so that 
the field points along the \bhattext{z}-direction with a 
profile that depends on the radial coordinate, $\rho$. 
In this case, we have $\vec{B} = B(\rho)$\bhattext{z}. 
We can find the magnetic field in terms of the vector potential 
from $\vec{B} = \curl{\vec{A}}$ and the gauge choice that $A_0 =
A_\rho = A_z = 0$. Then,

\be 
	\label{eqn:Bperiodic} B(\rho) = \frac{A_\phi(\rho)}{\rho} +
	\frac{dA_\phi(\rho)}{d\rho}. 
\ee 

The vector potential, $\Aphi$, can be characterized by a profile function, 
$f_\lambda$, and a dimensionless flux parameter, $\mathcal{F}=\frac{e}{2\pi}F$:

\be 
	\Aphi = \frac{\mathcal{F}}{e \rho}f_\lambda(\rho).
\ee 
If we choose $f_\lambda(\rho=0) = 0$, then the total flux within
in a radius of $L_\rho$ is 
\be 
	\Phi=\frac{2\pi}{e}\mathcal{F} f_\lambda(L_\rho).  
\ee 

In terms of the profile function, the magnetic field is written
\be
	\label{eqn:Bzfromfl}
	B_z(\rho)=\frac{\mathcal{F}}{e\rho}\frac{df_\lambda(\rho)}{d\rho} 
		=\frac{2\mathcal{F}}{e}\frac{df_\lambda(\rho^2)}{d(\rho^2)}. 
\ee

\section{Isolated Flux Tubes}

To explore isolated magnetic flux tubes, we consider the following 
profile function:

\be
	f_\lambda(\rho^2) = \frac{\rho^2}{\rho^2 + \lambda^2}.
\ee
From equation (\ref{eqn:Bzfromfl}), this give a magnetic field 
with a profile
\be
	\label{eqn:isoB}
	B_z(\rho^2) = \frac{2\mathcal{F}}{e}\frac{\lambda^2}{(\rho^2+\lambda^2)^2}.
\ee
This profile is a smooth flux tube representation that can be evaluated quickly.
Moreover, flux tubes 
with this profile were studied previously using \ac{WLN}
\cite{Moyaerts:2003ts, MoyaertsLaurent:2004}.

\section{Cylindrical Model of a Flux Tube Lattice}

\begin{figure}
	\centering
		\includegraphics[width=10cm]{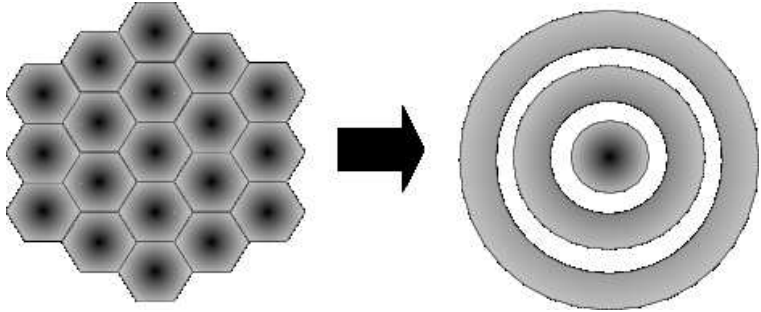}
	\caption[Cylindrical model of a hexagonal lattice]
	{In a type-II superconductor, there are neighbouring flux 
	tubes arranged in a hexagonal Abrikosov lattice which have a nonlocal 
	impact on the effective action of the central flux tube (left). In our model, 
	we account for the contributions from these neighbouring flux tubes in 
	cylindrical symmetry by including surrounding rings of flux (right).}
	\label{fig:latticesymmetry}
\end{figure}

In a neutron star, we do not have isolated flux tubes. The tubes are likely arranged in a dense lattice with the 
spacing between tubes on the order of the Compton wavelength, with the size of a flux tube a few percent of 
the Compton wavelength. Specifically, the maximum size of a flux tube is on the order of the 
coherence length of the superconductor, which for neutron stars has been estimated to be 
$\xi \approx 30$ fm~\cite{PhysRevLett.91.101101}. 
This situation can be directly computed in the \ac{WLN} technique. However, 
this requires us to integrate over two spatial dimensions instead of one. Moreover, it requires the use of more 
loops to more precisely probe the spatial configurations of the magnetic field. Despite these problems, 
it is very interesting to consider a dense flux tube lattice. Unlike the isolated flux tube, the wide-tube limit 
of the configuration doesn't have zero field, but an average, uniform background field. If this background field is 
the size of the critical field, there are interesting quantum effects even in the wide-tube limit. 

In this section, I build a cylindrically symmetric toy model of a hexagonal flux tube lattice. We focus 
on one central flux tube and treat the surrounding six flux tubes as a continuous ring with six units 
of flux at a distance $a$ from the central tube. The next ring will contain twelve units of flux at a 
distance of $2a$, etc (see figure \ref{fig:latticesymmetry}). 
Because of this condition, the average strength of the field is fixed, and the field becomes uniform in the 
wide tube limit instead of going to zero. For small values of $\lambda$, we will have nonlocal 
contributions from the surrounding rings in addition to the local contributions from the central 
flux tube.

It is difficult to construct a model of this scenario if the flux tubes bleed into one another 
as they are placed close together. For example, with Gaussian flux tubes or flux tubes 
with the profile used in the previous section, it is difficult to increase the width of the flux tubes 
while accounting for the magnetic flux that bleeds out of their regions. Moreover, it is 
difficult to integrate these schemes to find the profile function $f_\lambda(\rho)$ which is needed 
to compute the scalar part of the Wilson loop.  In order to keep each tube as a distinct entity which 
stays within its assigned region, we assign 
a smooth function with compact support to represent each tube. 
This is most easily done with the bump function, $\Psi(x)$, defined as
\be
	\Psi(x) = 
	\begin{cases}
	e^{-1/(1-x^2)} & \mbox{ for } |x| < 1\\
	0 & \mbox{ otherwise} 
	\end{cases}.
\ee
This function can be viewed as a rescaled Gaussian.

We start by defining the magnetic field outside of the central flux tube. Here, the magnetic field 
is a constant background field, with the flux ring contributing a bump of width $\lambda$. The height 
of the bump must go to zero as the width of the flux tube approaches the distance between flux tubes, 
and should become infinite as the flux tube width goes to zero:

\be
	B_z(\rho>\frac{a}{2}) = B_{\rm bg} + A\left(\frac{a-\lambda}{\lambda}\right)\left[\Psi(2(\rho-n a)/\lambda)-B\right].
\ee
with $n\equiv \lfloor\frac{\rho+a/2}{a}\rfloor$.

If we require 6 units of flux in the first outer ring, 12 in the second, 
and so on (see figure \ref{fig:latticesymmetry}), the size of the background field 
is fixed to $B_{\rm bg} = \frac{6 \mathcal{F}}{ea^2}$. The total flux contribution due to the $\lambda$-dependent 
terms must be zero:

\be
	\int_{(n-1/2)a}^{(n+1/2)a} \rho A\left(\frac{a-\lambda}{\lambda}\right)
		\left[\Psi(2(\rho-n a)/\lambda)-B\right] d\rho = 0
\ee

\be
	\frac{\lambda}{2}\int_{-1}^1 \left(\frac{\lambda}{2}x+na\right)\Psi(x)dx-Ba^2n = 0.
\ee
This fixes the value of the constant $B$ to
\be
	B=\frac{q_1}{2}\frac{\lambda}{a}.
\ee
The numerical constant $q_1$ is defined by
\be
	q_1 = \int_{-1}^{1}\Psi(x)dx \approx 0.443991.
\ee

For a given bump amplitude, $A$, the magnetic field will become negative if $\lambda$ becomes small enough.
Therefore, we replace $A$ with its maximum value for which the field is positive if $\lambda > \lambda_{\rm min}$
for some choice of minimum flux tube size:

\be
	A=\frac{12 \mathcal{F}}{e a q_1 (a-\lambda_{\rm min})}.
\ee
The choice of $\lambda_{\rm min}$ sets the tube width at which the field between the flux tubes vanishes. If 
$\lambda < \lambda_{\rm min}$, the magnetic field between the flux tubes will 
point in the $-\hat{\vec{z}}$-direction.

Because we are trying to fit a hexagonal peg into a round hole, we must treat the central flux tube differently. 
For example, the average field inside the central region for a unit of flux, is different than the average 
field in the exterior region. Therefore, even when $\lambda \rightarrow a$, the field cannot be quite uniform. 
We consider the field in the central region to be a constant field with a bump centered at $\rho = 0$:

\be
	B_z(\rho < \frac{a}{2}) = A_0 \Psi(2\rho/\lambda) + B_0.
\ee
The constant term, $B_0$ is fixed by requiring continuity with the exterior field at $\rho = a/2$:

\be
	B_0 = \frac{6\mathcal{F}}{e a^2}\left(1-\frac{a-\lambda}{a-\lambda_{\rm min}}\right).
\ee
The bump amplitude, $A_0$, is fixed by fixing the flux in the central region,
\be
	\int_0^{a/2}\rho\left[A_0\Psi(2\rho/\lambda) + B_0\right]d\rho = \frac{\mathcal{F}}{e}:
\ee

\be
	A_0 \left(\frac{\lambda}{2}\right)^2\int_0^1 x \Psi(x)dx + 
		\frac{B_0}{2}\left(\frac{a}{2}\right)^2 = \frac{\mathcal{F}}{e}
\ee

\be
	A_0 = \frac{4 \mathcal{F}}{\lambda^2 e q_2}\left(1-\frac{3}{4}
		\left(1-\frac{a-\lambda}{a-\lambda_{\rm min}}\right)\right),
\ee
where the numerical constant, $q_2$, is defined by

\be
	\label{eqn:q2}
	q_2 \equiv \int_0^1 x \Psi(x) dx \approx 0.0742478.
\ee

\begin{figure}
	\centering
		\includegraphics[width=10cm]{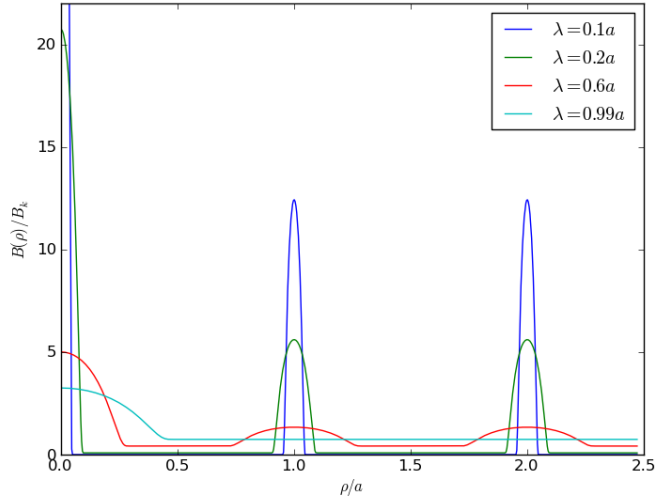}
	\caption[Magnetic field in a cylindrical lattice model]
	{The cylindrical lattice flux tube model for several 
	values of the width parameter $\lambda$. Here we have taken $a =\sqrt{8}\lambda_e$
	and $\lambda_{\rm min} = 0.1a$. Note that in the limit $\lambda\rightarrow a$
	the field is nearly uniform with a mound in the central region. This is a 
	consequence of the flux conditions in cylindrical symmetry requiring different
	fields in the internal and external regions. The height of the $\lambda=0.1a$
	flux tube extends beyond the height of the graph to about $61.9B_k$. 
	A 3D surface plot of the $\lambda=0.6a$ field profile is shown in figure 
	\ref{fig:B3d}.}
	\label{fig:periodicB}
\end{figure}

\begin{figure}
	\centering
		\includegraphics[width=12cm]{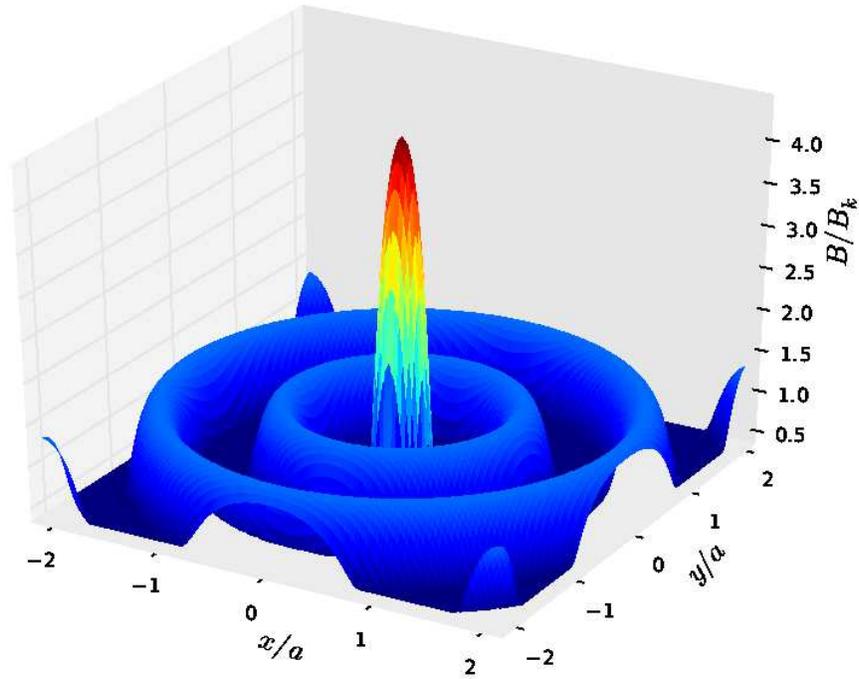}
	\caption[Surface plot of magnetic field model]
	{A 3D surface plot of the $\lambda = 0.6a$ (red) magnetic field profile from 
	figure \ref{fig:periodicB}.}
	\label{fig:B3d}
\end{figure}

\begin{figure}
	\centering
		\includegraphics[width=12cm]{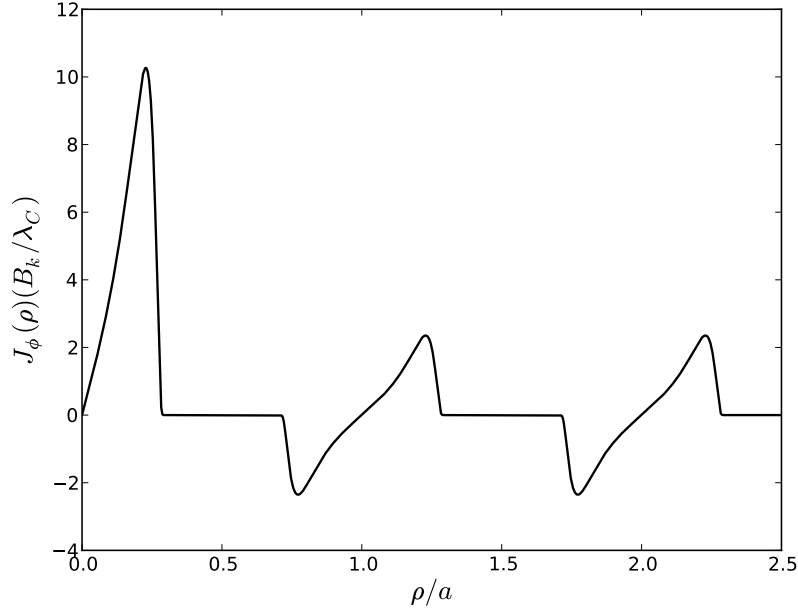}
	\caption[Classical current required to create magnetic field model]
	{The current densities required to create the $\lambda=0.6a$ (red) magnetic field profile. 
	The current is given by the curl of the magnetic field, $J_\psi(\rho) = -\frac{d B_z(\rho)}{d\rho}$.
	The conversion to SI units is $1B_k/\lambda_e \approx 6\times 10^{38}{\rm A/m}^3$.}
	\label{fig:current}
\end{figure}

Finally, collecting together the important expressions, 
the cylindrically symmetric flux tube lattice model is

\ba
\label{eqn:ffbless}
B_z(\rho\le\frac{a}{2}) &=& \frac{4 \mathcal{F}}{\lambda^2 e q_2}
	\left(1-\frac{3}{4}\left(\frac{\lambda-\lambda_{\rm min}}{a-\lambda_{\rm min}}\right)\right)\Psi(2\rho/\lambda) \nonumber \\
	& & + \frac{6\mathcal{F}}{ea^2}\left(\frac{\lambda-\lambda_{\rm min}}{a-\lambda_{\rm min}}\right)
\ea

\be
\label{eqn:Bzext}
B_z(\rho>\frac{a}{2}) = \frac{6 \mathcal{F}}{e a^2}\left(\frac{\lambda-\lambda_{\rm min}}{a-\lambda_{\rm min}}\right)
	+ \frac{12 \mathcal{F}}{q_1ea\lambda}\left(\frac{a-\lambda}{a-\lambda_{\min}}\right)
	\Psi(2(\rho-na)/\lambda)  
\ee
with
\be
	\Psi(x) = 
	\begin{cases}
	e^{-1/(1-x^2)} & \mbox{ for } |x| < 1\\
	0 & \mbox{ otherwise} 
	\end{cases},
\ee

\be
	n=\left \lfloor\frac{\rho+a/2}{a}\right\rfloor,
\ee

\be
	\label{eqn:q1}
	q_1 \equiv \int_{-1}^{1}\Psi(x)dx \approx 0.443991,
\ee
and
\be
	q_2 \equiv \int_0^1 x \Psi(x) dx \approx 0.0742478.
\ee
The magnetic field profile defined by these equations is shown in figures
\ref{fig:periodicB} and \ref{fig:B3d}. The current density required to 
created fields with this profile is shown in figure \ref{fig:current}.

\subsection{The Classical Action}

The classical action, $\Gamma^0$, is infinite for this configuration. To obtain 
a finite result, we must look at the action per unit length in the z-direction, 
per unit time, and per flux tube region (\ie for $\rho < a/2$). 
The action for such a region in 
cylindrical coordinates is given by 

\be
	\frac{\Gamma^0}{T L_z} = -\pi \int_0^{a/2}\rho B_z(\rho)^2d\rho.
\ee
We substitute in equation (\ref{eqn:ffbless}), the magnetic field in the interior 
region:
\ba
	\frac{\Gamma^0}{T L_z} &=& -\pi \int_0^{a/2}\rho \biggr[
	\frac{4\mathcal{F}}{\lambda^2e^2q_2}\left(1-\frac{3}{4}
	\left(\frac{\lambda-\lambda_{\rm min}}{a-\lambda_{\rm min}}\right)\right)
	\Psi(2\rho/\lambda)\nonumber \\
	& & + \frac{6\mathcal{F}}{ea^2}\left(\frac{\lambda-\lambda_{\rm min}}{a-\lambda_{\rm min}}\right)
	\biggr]^2 d\rho.
\ea

After some algebra, we are left with an expression for the classical action,

\ba
	\frac{\Gamma_0}{TL_z} &=& \pi \int_0^{a/2}\rho B_z(\rho)^2d\rho \\
	& = & -\frac{\pi \mathcal{F}^2}{e^2a^2}\biggr\{4\frac{a^2q_3}{\lambda^2q_2^2}
	+\left(\frac{\lambda-\lambda_{\rm min}}{a - \lambda_{\rm lmin}}\right)
	\biggr[ \nonumber \\
	& &\left(\frac{9}{4}\frac{a^2q_3}{\lambda^2q_2^2}-\frac{9}{2}\right)
	\left(\frac{\lambda-\lambda_{\rm min}}{a - \lambda_{\rm lmin}}\right) 
	-6\frac{a^2q_3}{\lambda^2q_2^2}+12 \biggr]\biggr\},
\ea
with $q_3$ being another numerical constant related to integrating the bump function:

\be
	\label{eqn:q3}
	q_3 \equiv \int_0^1 x\left(\Psi(x)\right)^2 dx 
	= \int_0^1 x e^{-\frac{2}{1-x^2}}dx \approx 0.0187671.
\ee

\subsection{Integrating to Find the Potential Function}

To compute the Wilson loops, it is generally required to use the 
vector potential which describes the magnetic field. For us, this means that 
we must find $f_\lambda(\rho)$ for our magnetic field model. 
This could always be done numerically, but can be computationally costly since 
it is evaluated by every discrete point of every worldline in the ensemble.
For computations on the \ac{CUDA} device, an increase in the complexity of the 
kernel often means that less memory 
resources are available per processing thread, limiting 
the number of threads that can be computed concurrently.
It is therefore preferable to find an analytic expression for this function.
From equation (\ref{eqn:Bzfromfl}), this function is related to the integral 
of the magnetic field with respect to $\rho^2$. For the inner region, we have

\ba
	f_\lambda(\rho < a/2) &=& \frac{e}{2\mathcal{F}}\biggr[\frac{4\mathcal{F}}{\lambda^2eq_2}
		\left(1-\frac{3}{4}\left(\frac{\lambda-\lambda_{\rm min}}{a-\lambda_{\rm min}}\right)\right)
		\int_0^{\rho^2}\Psi\left(\frac{2\rho'}{\lambda}\right)d \rho'^2 \nonumber \\
		& &+\frac{6 \mathcal{F}}{ea^2}\left(\frac{\lambda-\lambda_{\rm min}}{a-\lambda_{\rm min}}\right)\rho^2\biggr].
\ea
The integral over the bump function can be computed in terms of the exponential integral 
$E_i(x) = \int_{-\infty}^x \frac{e^t}{t} dt$:
\ba
	\int_0^{\rho^2}\Psi\left(\frac{2\rho'}{\lambda}\right)d\rho'^2 = 
	\begin{cases}
	\left(\frac{\lambda}{2}\right)^2\biggr[2q_2+\left(\frac{4\rho^2}{\lambda^2}-1\right) 
		e^{-\frac{1}{1-\frac{4\rho^2}{\lambda^2}}} & {}\\
		~~~~~~~~~~- E_i\left(-\frac{1}{1-\frac{4\rho^2}{\lambda^2}}\right)\biggr]& \mbox{ for } \rho < \lambda/2\\
	\frac{q_2\lambda}{2} & \mbox{ for } \rho \ge \lambda/2 
	\end{cases}.
\ea
So, our expression for the profile function in the inner region is
\be
	f_\lambda(\rho\le a/2) = \left(1-\frac{3}{4}\left(\frac{\lambda-\lambda_{\rm min}}{a-\lambda_{\rm min}}\right)\right)
		\Phi(2\rho/\lambda) + \frac{3\rho^2}{a^2}\left(\frac{\lambda-\lambda_{\rm min}}{a-\lambda_{\rm min}}\right),
\ee
with
\be
	\Phi(x) \equiv
	\begin{cases}
	1 + \frac{1}{2q_2}\left(x^2-1\right)e^{-\frac{1}{1-x^2}}
		-\frac{1}{2q_2}E_i\left(-\frac{1}{1-x^2}\right)& \mbox{ for } x < 1\\
	1 & \mbox{ for } x \ge 1 
	\end{cases}.
\ee

The exterior integral is a bit more challenging, but we can make significant progress 
and obtain an approximate expression. The first term is a constant given by the value of the profile 
function at $\rho = a/2$. This value is given by the flux in the central flux tube, which 
we have already chosen to be 1,
\be
	f_\lambda(\rho>a/2) = 1 + \frac{e}{\mathcal{F}}\int_{a/2}^{\rho}\rho'B(\rho'>a/2)d\rho'.
\ee
We may plut the magnetic field, equation (\ref{eqn:Bzext}), into this expression to get
\ba
	f_\lambda(\rho>a/2) &=& 1 + \frac{3}{4}\left(\frac{4\rho^2}{a^2}-1\right)
		\left(\frac{\lambda-\lambda_{\rm min}}{a-\lambda_{\rm min}}\right) \nonumber \\
		& &+\frac{12}{q_1a\lambda}\left(\frac{a-\lambda}{a-\lambda_{\rm min}}\right)
		\int_{a/2}^{\rho}\rho'\Psi\left(\frac{2(\rho-na)}{\lambda}\right)d\rho'.
\ea
The remaining integral is over every bump between $\rho'=a/2$ and $\rho'=\rho$. We express the result 
as a term which accounts for each completely integrated bump, and an integral over the partial bump if $\rho$ is 
within a bump:
\ba
	f_\lambda(\rho>a/2) &=& 1 + \frac{3}{4}\left(\frac{4\rho^2}{a^2}-1\right)
		\left(\frac{\lambda-\lambda_{\rm min}}{a-\lambda_{\rm min}}\right)
		+3n(n-1)\left(\frac{a-\lambda}{a-\lambda_{\rm min}}\right) \nonumber \\
		& &+\frac{3\lambda}{q_1a}\left(\frac{a-\lambda}{a-\lambda_{\rm min}}\right)\chi(2(\rho-na)/\lambda),
\ea
where
\be
	\chi(x_0) = 
	\begin{cases}
	0 & \mbox{ for } x_0 \le -1  \\
	\int_{-1}^{x_0}xe^{-\frac{1}{1-x^2}}dx 
			+ \frac{2na}{\lambda}\int_{-1}^{x_0}e^{-\frac{1}{1-x^2}}dx& \mbox{ for } |x_0| < 1\\
	 \frac{2naq_1}{\lambda} & \mbox{ for } x_0 \ge 1 	  
	\end{cases}.
\ee
One of the integrals in $\chi(x_0)$ can be expressed in terms of the exponential integral:
\be
	\int_{-1}^{x_0}xe^{-\frac{1}{1-x^2}}dx =\frac{1}{2}\left[(x_0^2-1)e^{-\frac{1}{1-x_0^2}}
		-E_i\left(-\frac{1}{1-x_0^2}\right)\right].
\ee

The remaining integral can't be simplified analytically. To use this integral in our numerical model, 
it must be computed for each discrete point on each loop for each $\rho_{\rm cm}$ and $T$ value. 
Therefore, it is worthwhile to consider an approximate expression which models the integral, and 
can be computed faster than performing a numerical integral each time. To find this approximation, 
I computed the numerical result at 300 values between $x_0=-1.2$ and $x_0=1.2$. The data was then 
input into Eureqa Formulize, a symbolic regression program which uses genetic algorithms to find 
analytic representations of arbitrary data~\cite{formulize}. 
A similar technique has been used to produce 
approximate analytic solutions of ODEs~\cite{geneticODEs}.
The result is a model of the numerical data points with a maximum error of 0.0001 on the range $|x_0| < 1.0$:

\be
	\int_{-1}^{x_0} e^{-\frac{1}{1-x^2}}dx \approx 
	\frac{0.444}{1+\exp{\left[ -3.31x_0 - 
	\frac{5.25x_0^3 - 3.31x_0^2\sin{(x_0)}\cos{(-0.907x_0^2-1.29x_0^8)}}{\cos{(x_0)}}\right]}}.
\ee

This function evaluates a factor of ten faster than the numerical integral evaluated at the 
same level of precision with the \ac{GSL} Gaussian quadrature library functions and 
with fewer memory registers. 
Using this approximation
introduces a systematic uncertainty  which is small compared to that associated with the 
discretization of the loop integrals, and considerably smaller than the statistical error bars.

Using these expressions for the integrals, we can express $f_\lambda(\rho)$ in any region in terms of 
exponential integrals, exponential, and trigonometric functions with suitable precision:

\small
\ba
	\chi(x_0) \approx 
	\begin{cases}
	0 & \mbox{ for } x_0 \le -1 \\
	\frac{1}{2}\left[(x_0^2-1)e^{-\frac{1}{1-x_0^2}}
		-E_i\left(-\frac{1}{1-x_0^2}\right)\right] +
		\frac{2na}{\lambda}\biggr[\\
		~~~~~\frac{0.444}{1+\exp{\left( -3.31x_0 - 
	\frac{5.25x_0^3 - 3.31x_0^2\sin{(x_0)}\cos{(-0.907x_0^2-1.29x_0^8)}}
		{\cos{(x_0)}}\right)}}\biggr] & \mbox{ for } |x_0| < 1\\
	 \frac{2naq_1}{\lambda} & \mbox{ for } x_0 \ge 1 	  
	\end{cases}.
\ea
\normalsize
For computation, we may express the exponential integral 
as a continued fraction:

\be
E_i(-x) = -E_1(x) = -e^x\left[ \cfrac{1}{x+1-\cfrac{1}{x+3-\cfrac{4}{x+5-\cfrac{9}{x+7-...}}}}\right].
\ee
This expression converges very rapidly in the range of 
interest (i.e. $x>1$) using Lentz's 
algorithm~\cite{nr3rd}.

\begin{figure}
	\centering
		\includegraphics[width=10cm]{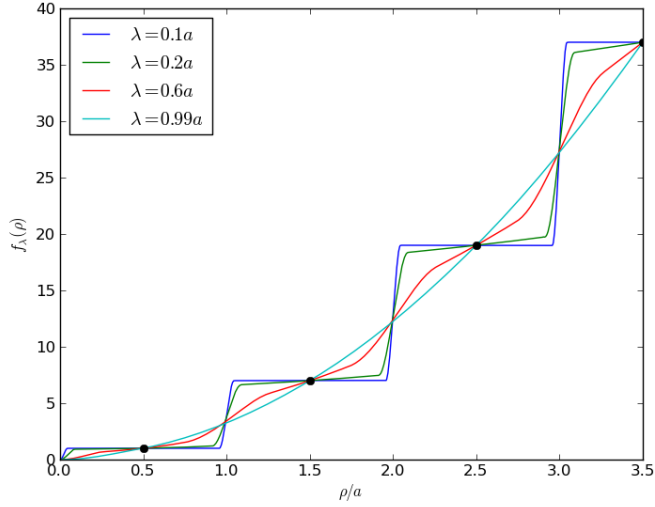}
	\caption[The profile function in a cylindrical lattice model]
	{The function $f_\lambda(\rho)$ for the above described magnetic 
	field model. The flux conditions require the function to pass through the 
	black dots. A quadratic function corresponds to a uniform field while 
	a staircase function corresponds to delta-function flux tubes. The parameter 
	$\lambda$ smoothly transitions between these two extremes.
	Each of these functions corresponds to a magnetic field profile in 
	figure \ref{fig:periodicB}.}
	\label{fig:periodicfl}
\end{figure}

The profile function, $f_\lambda(\rho)$, is plotted in figure \ref{fig:periodicfl}.

\section{Results}
\label{sec:periodic_results}

\subsection{Comparing Scalar and Fermionic Effective Actions}

Because of the fermion problem of \ac{WLN} 
(see section \ref{sec:fermionproblem}), the 1-loop effective 
action for the cylindrical flux tube lattice model could not be 
computed for the case of spinor \ac{QED}. 
The fermion problem does not 
affect the scalar case. So, we are confined to analyzing this 
model for \ac{ScQED}. 

\begin{figure}
	\centering
		\includegraphics[width=10cm]{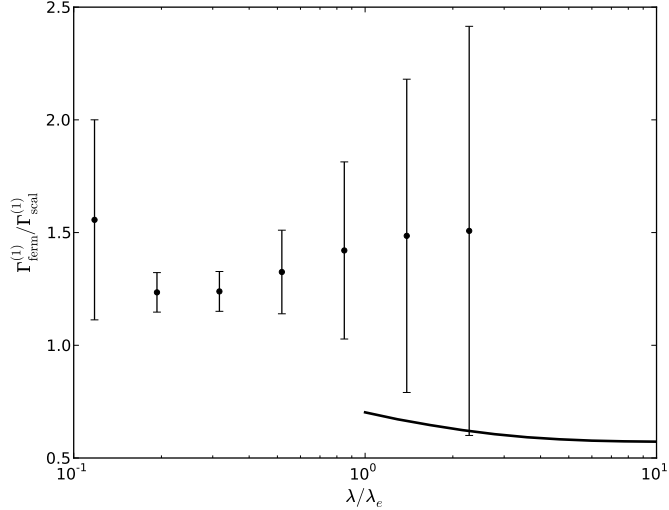}
	\caption[Comparison of 1-loop term in \acs{QED} and \acs{ScQED}]
	{The ratio of the 1-loop term in \ac{QED} to the 1-loop term 
	in \ac{ScQED}. The solid line is the \ac{LCF} approximation, 
	while the points are the result of \ac{WLN} calculations. Note that the 
	\ac{LCF} approximation breaks down near $\lambda = \lambda_e$ and 
	that the statistics from point to point are strongly correlated. This plot shows that 
	the 1-loop correction in \ac{ScQED} differs from the \ac{QED} correction by a factor 
	close to unity for a wide range of flux tube widths.}
	\label{fig:scalfermiratio}
\end{figure}

For isolated flux tubes, the decay of the magnetic field for large distances
protects the calculations from the fermion problem (see section 
\ref{sec:fermionproblem}). Therefore, the
effective action can be computed for both scalar and spinor 
\ac{QED}. In figure \ref{fig:scalfermiratio}, we plot the 
ratio of the spinor to scalar 1-loop correction term for 
identical magnetic fields, along with the prediction of the 
\ac{LCF} approximation for large values of $\lambda$. The 
\ac{LCF} approximation in \ac{ScQED} is given by
\ba
	\label{eqn:LCFscal}
	\Gamma^{(1)}_{\rm scal} &=& -\frac{1}{2\pi}\int_0^\infty dT 
	\int_0^\infty \rho_{\rm cm} d\rho_{\rm cm}\frac{e^{-m^2T}}{T^3} \nonumber \\
	& &\left\{\frac{eB(\rho_{\rm cm})T}{\sinh{(eB(\rho_{\rm cm})T)} }
	- 1 +\frac{1}{6}(eB(\rho_{\rm cm})T)^2\right\}.
\ea
This can be compared to the spinor \ac{QED} approximation, 
equation (\ref{eqn:LCFferm}).

There are two important notes to make about figure \ref{fig:scalfermiratio}. 
Firstly, the \ac{LCF} approximation is only a good approximation 
for $\lambda \gg \lambda_e$, and isn't accurate when pushed near its formal 
validity limits~\cite{MoyaertsLaurent:2004}. The second note is that the 
statistics of the \ac{WLN} points are strongly correlated. So, we conclude 
that the \ac{ScQED} 1-loop correction is larger than the \ac{QED} correction 
for large $\lambda$, and this appears to be reversed for small $\lambda$. 
However, the large \ac{WLN} error bars and the invalidity of the 
\ac{LCF} approximation near $\lambda = \lambda_e$ prevent us from 
seeing how this transition happens. Nevertheless, the main 
conclusion from this figure is that the scalar 1-loop correction 
reflects the behaviour of the full \ac{QED} 1-loop correction 
to within a factor of about 2 over a wide range of flux tube widths 
for isolated flux tubes.

Besides using a finite field profile, the fermion problem can also be circumvented by increasing
the electron mass. 
From equation (\ref{eqn:cylEA}), we can see that the square of the electron mass 
sets the scale for the exponential suppression of the large proper time Wilson loops that 
contribute to the fermion problem. However, if we increase the fermion mass, we are 
reducing the Compton wavelength of our theory so that the flux tube lattice is no longer 
dense in terms of the (new) Compton wavelength. It is the Compton wavelength of the theory that 
determines what is meant by `dense'. We therefore cannot avoid the fermion 
problem for dense lattice models by changing the electron mass.

Based on the results presented in figure \ref{fig:scalfermiratio}, we conclude that the coupling between the electron's spin 
and the magnetic field do not have a dramatic effect on the vacuum 
energy for isolated flux tubes. Therefore, we expect that \ac{ScQED} provides a good model of the 
underlying vacuum physics near these flux tubes, at least at the level our toy model 
flux tube lattice.

\subsection{Flux Tube Lattice}

The \ac{WLN} technique computes an effective action density which is then integrated to 
obtain the effective action. This quantity differs from the Lagrangian in that it is 
not determined by local operators, but encodes information about the field everywhere
through the worldline loops.
Like the classical action, the 1-loop term of the effective action per unit length 
is infinite for a flux tube lattice because the field extends infinitely far. 
For this reason, we define the effective action to be the action density integrated 
over the region of a central flux tube $(0<\rho<a/2)$:

\small
\ba
	\frac{\Gamma}{\mathcal{T}L_z} &=& -\pi \int_0^{a/2}\rho B_z(\rho)^2d\rho \nonumber \\
	& &-\frac{1}{2\pi}\int_0^{a/2}\rho_{\rm cm} d\rho_{\rm cm}\int_0^\infty \frac{dT}{T^3}e^{-m^2T}
	\nonumber \\
	& &\times \left\{ \mean{W}_{\rho_{\rm cm}} - 1 +\frac{1}{6}(eB(\rho_{\rm cm})T)^2\right\}.
\ea
\normalsize

\begin{figure}
	\centering
		\includegraphics[width=10cm]{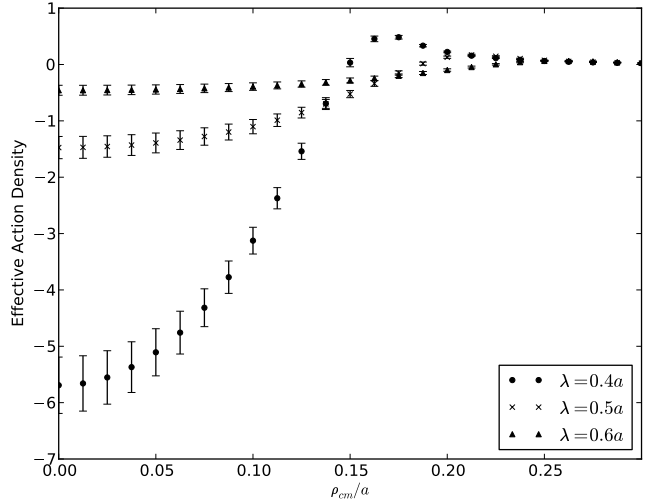}
	\caption[The 1-loop action density for a cylindrical lattice in \acs{ScQED}.]
	{The \ac{ScQED} effective action density for the central 
	flux tube in our cylindrical lattice model for several tube widths, 
	$\lambda$. The average field strength is the critical field, $B_K$. 
	The effective action is positive in between flux tubes due to nonlocal 
	effects.}
	\label{fig:EAscffpeek}
\end{figure}

The 1-loop 
term of the effective action density is plotted in figure \ref{fig:EAscffpeek} for the 
cylindrical flux tube lattice model. The most pronounced feature of this density is 
that there is a negative contribution from the regions where the field is strong. 
This contribution has the same sign as the classical term. Therefore, quantum correction 
tends to reinforce the classical action. A less pronounced feature is that there is a 
positive contribution arising from the $\rho > \lambda/2$ region, in between 
the lumps of magnetic field which represent the flux tubes. 
In this region, the local magnetic field is positive, but small.

\begin{figure}
	\centering
		\includegraphics[width=10cm]{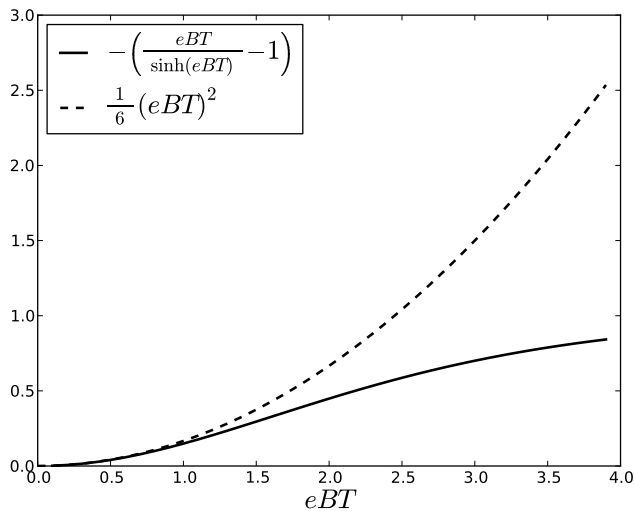}
	\caption[Wilson loop and counter term contributions for a constant field]
	{The Wilson loop and counter term contributions to the integrand 
	of the effective action for a constant field in \ac{ScQED}. For constant fields, the 
	effective action is always negative due to the domination of the counter 
	term over the Wilson loop. For non-homogeneous fields, 
	A positive effective action density signifies 
	that non-local (\ie $T > 0$) effects dominate the counter term.}
	\label{fig:CounterTermDomination}
\end{figure}

To interpret this feature, we consider the relative contributions between the 
Wilson loops and the counter term. These terms are shown in figure 
\ref{fig:CounterTermDomination} for the constant field case. 
For all values of proper time, $T$, the counter terms dominate, giving 
an overall negative sign. In order for the action density to be positive, there 
must be a greater contribution from the Wilson loop average than from the 
counter term, since this term tends to give a positive contribution to the 
action. In our flux tube model, this seems to occur in the regions between the 
flux tubes. In these regions, the local contribution from the counter term is 
relatively small because the field is small. However, the contribution from the 
Wilson loop average is large because the loop cloud is exploring the nearby 
regions where the field is much larger. The effect is largest where the field 
is small, but becomes large in a nearby region. We therefore interpret the positive 
contributions to the 1-loop correction from these regions as a non-local 
effect. A similar example of such an effect from fields which vary on 
scales of the Compton wavelength has been observed previously using the 
\ac{WLN} technique~\cite{PhysRevD.84.065035}.

\begin{figure}
	\centering
		\includegraphics[width=10cm]{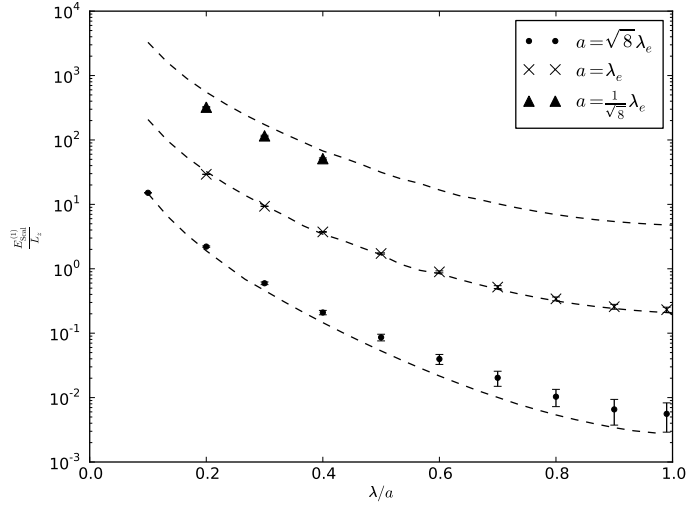}
	\caption[The 1-loop \acs{ScQED} term versus flux tube width]
	{The 1-loop \ac{ScQED} term of the effective action as a function 
	of flux tube width, $\lambda/a$, for several values of the flux tube spacing, $a$.
	The dotted lines are computed from the \ac{LCF} approximation.}
	\label{fig:EAscffvsl}
\end{figure}

\begin{figure}
	\centering
		\includegraphics[width=10cm]{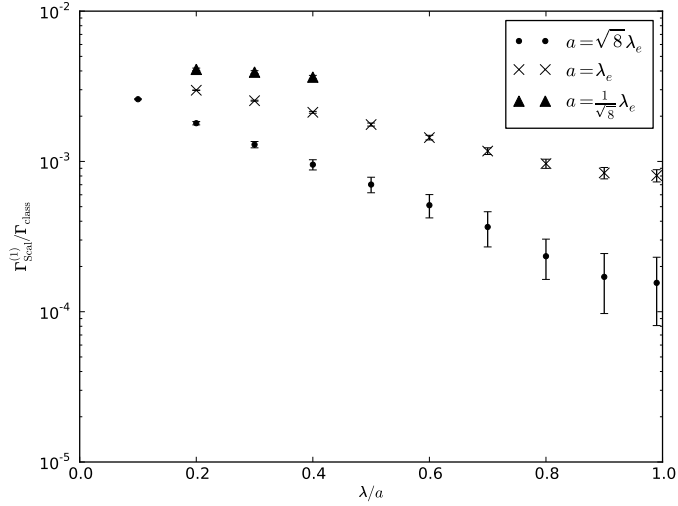}
	\caption[The 1-loop \acs{ScQED} quantum to classical ratio versus flux tube width]
	{The 1-loop \ac{ScQED} term divided by the classical term of the effective action as a function 
	of flux tube width, $\lambda/a$ for several values of the flux tube spacing, $a$.}
	\label{fig:ActRatscff}
\end{figure}

In figure \ref{fig:EAscffvsl}, we plot the magnitude of the 1-loop 
\ac{ScQED} term of the effective 
action as a function of the flux tube width. As the flux tubes become smaller, there 
is an amplification of the 1-loop term, just as there is for the classical action. 
Similarly, for more closely spaced flux tubes, $a$ is smaller, and the 1-loop term 
increases in magnitude. The ratio of the 1-loop term to the classical term is plotted 
in figure \ref{fig:ActRatscff}. The quantum contribution is greatest for closely spaced, 
narrow flux tubes, but does not appear to become a significant fraction of the total action.

\begin{figure}
	\centering
		\includegraphics[width=10cm]{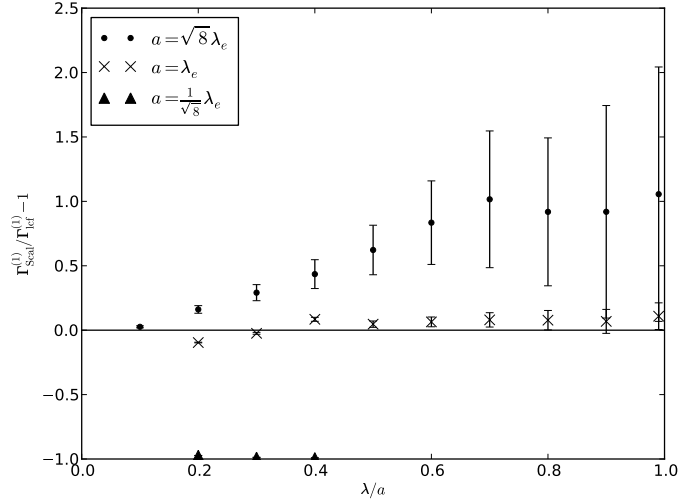}
	\caption[The deviation of the cylindrical flux tube lattice from the \ac{LCF}
	approximation]
	{The residuals between the \ac{WLN} results and the \ac{LCF} approximation 
	for the cylindrical flux tube lattice. The level of agreement is not expected because 
	the field varies rapidly on the Compton wavelength scale. This agreement is believed to 
	be due to an averaging effect of integrating over the electron degrees of freedom making a
	mean-field approximation appropriate.}
	\label{fig:EAscffresid}
\end{figure}

We observe that the \ac{LCF} approximation is surprisingly good despite the 
fact that the magnetic field is varying rapidly on the Compton wavelength scale of the electron.
We plot the residuals showing the deviations between the \ac{WLN} results and the \ac{LCF}
approximation in figure \ref{fig:EAscffresid}. To understand this, recall the discussion 
surrounding figure \ref{fig:CounterTermDomination}. The Wilson loop term is sensitive to the 
average magnetic field through the loop ensemble, $\mean{B}_{\rm e}$. In contrast, the counter 
term is sensitive to the magnetic field at the center of mass of the loop, $B_{\rm cm}$. 
Since these terms carry opposite signs, we can understand the difference from the 
constant field approximation in terms of a competition between these terms. When 
$B_{\rm cm} < \mean{B}_{\rm e}$, such as when the center of mass is in a local minimum of the 
field, there is a reduction of the energy relative to the locally constant field case, with 
a possibility of the quantum term of the energy density becoming negative. 
However, when $B_{\rm cm} > \mean{B}_{\rm e}$, such as in a local maximum of the field, 
there is an amplification of the energy relative to the constant field case. We can put a 
bound on the difference between the mean field through a loop and the field at the 
center of mass for small loops (\ie small $T$),
\be
	|\mean{B} - B_{\rm cm}| \lesssim |B''(\rho_0)| T
\ee
where $|B''(\rho_0)| \ge B''(\rho)$ for all $\rho$ in the loop. This expression is proved the 
same way as determining the error in numerical integration using the midpoint rectangle 
rule.

If the field varies rapidly about some mean value 
on the Compton wavelength scale, the various contributions 
from local minima and local maxima are averaged out and the mean field approximation 
provided by the \ac{LCF} method becomes appropriate. A similar argument applies in the 
fermion case, where the important quantity is the mean magnetic field along the circumference 
of the loop. This quantity is also well served by a mean-field approximation when integrating 
over rapidly varying fields.

Another interesting feature of figure \ref{fig:EAscffresid} is that the \ac{LCF} 
approximation appears to describe narrower flux tubes better than wider ones, even when the 
spacing between the flux tubes is held constant. This effect is likely a result of the compact 
support given to the flux tube profiles. For narrow tubes, we are guaranteed to have many more 
center of mass points outside the flux tube than inside, giving a smaller energy contribution than 
for isolated flux tubes without compact support where the distinction between inside and outside is 
not as abrupt. This also explains why narrow, closely spaced tubes produce a lower energy than 
is predicted by the \ac{LCF} approximation.

\begin{figure}
	\centering
		\includegraphics[width=10cm]{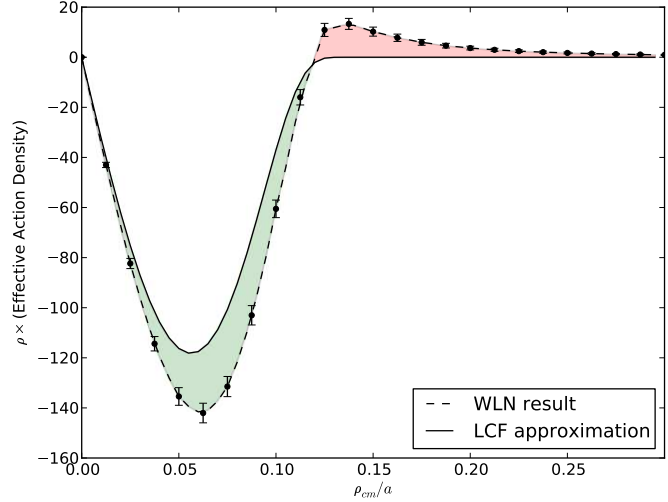}
	\caption[A comparison between the action density in the \ac{LCF} approximation and \ac{WLN}]
	{The action density in \ac{WLN} and the \ac{LCF} approximation, scaled by $\rho$ so 
	an area on the figure is proportional to a volume. The approximation 
	is poor everywhere, however, when there are regions of local minima and local maxima of the field about 
	a mean value, the effective action approximately agrees between these methods. This is due to 
	a partial cancellation between regions where the estimate provided by the approximation is too 
	large (green) and other regions where the estimate is too small (red).}
	\label{fig:EAscffvsconst}
\end{figure}

This argument does not apply to the smooth isolated flux tubes given 
by equation (\ref{eqn:isoB}). 
For these flux tubes, the only region where there is a large discrepancy between $\mean{B}_{\rm e}$ and
$B_{\rm cm}$ is near the center of the flux tube. This 
is a global maximum of the field, and the only maximum of $|B''(\rho)|$. 
There are no regions where the average field in the loop ensemble is 
much stronger than the center of mass magnetic field. So, we expect an amplification of the 
energy near the flux tube relative to the constant field case. 
In the flux tubes with compact support, however, 
there is such a region just outside the flux tube. We can understand the surprisingly close agreement 
of these results to the \ac{LCF} approximation in our model 
in terms of competition between these 
regions of local minima and maxima of the field (see figure \ref{fig:EAscffvsconst}).

Finally, I find that the quantum term remains small compared to the 
classical action for the range of parameters investigated. This is shown 
in figure \ref{fig:ActRatscff} where we plot the ratio of the \acs{ScQED}
term of the action to the classical action. The relative smallness of this correction 
is consistent with the predictions from homogeneous fields and the derivative 
expansion, as well as with previous studies on flux tube 
configurations~\cite{1999PhRvD..60j5019B}.

\subsection{Interaction Energies}

Using this model, we may also investigate the energy associated with interactions 
between the flux tubes. Since the flux tubes in our model exhibit compact support, 
the interaction energy is entirely due to a nonlocal interaction between nearby 
flux tubes. Thus, it contrasts with previous research which has investigated the 
interaction energies between flux tubes which have 
overlapping fields~\cite{Langfeld:2002vy}. In this 
case, there is a classical interaction energy ($\propto B_1 B_2$) as well as 
a quantum correction ($\propto (B_1 + B_2)^4 - B_1^4 - B_2^4$ in the weak-field limit). 
Even when these field overlap interactions are not present, 
there are also nonlocal energies in the 
vicinity of a flux tube due to the presence of other flux tubes. For example, 
the energy from nearby flux tubes can interact with a flux tube through the 
quantum diffusion of the magnetic field. Because of this 
phenomenon, we expect an interaction energy in the region of the central flux tubes 
due to the proximity of neighbouring flux tubes, even though no changes are made 
to the field profile or its derivatives in the region of interest. 
Since this interaction represents a force due to 
quantum fluctuations under the influence 
of external conditions, it is an example of a Casimir force. The Casimir force between 
two infinitely thin flux tubes in \ac{ScQED} has previously been found to be attractive~\cite{duru1993}. Our model 
can shed light specifically on this interaction, which is not predicted by 
local approximations such as the derivative expansion.

Consider a central flux tube with a width $\lambda = 0.5\lambda_e$. 
When $\lambda_{\rm min} = \lambda$, 
the magnetic field outside of the flux tube, $B_{\rm bg}$, is zero. Then, if the distance 
between flux tubes, $a$, is set very large, the energy density will be 
localized to the central flux tube and there will be no nonlocal interaction 
energy due to neighbouring flux tubes.
We define the interaction energy, $E_{\rm int}$, as the 
difference in energy within a distance $a/2$ of the central flux tube
between a configuration with a given value of $a$ 
and a configuration with $a = \infty$. In practice, we use $a = 10,000\lambda_e$ as 
a suitable stand-in for $a = \infty$:

\be
	\frac{E_{\rm int}}{L_z} = -\frac{\Gamma_{\rm scal}(a)}{L_z \mathcal{T}} 
	+\frac{\Gamma_{\rm scal}(a=1\times 10^4 \lambda_e)}{L_z \mathcal{T}}.
\ee
With this definition, the interaction energy is the energy associated with 
lowering the distance between flux rings from infinity. This the the analogue 
in our model of reducing the lattice spacing of the flux tubes.

One complication of this definition is that there is no clear distinction between 
energy density which `belongs' to the central flux tube and energy density which 
`belongs' to the neighbouring flux tubes. We continue to use our convention that the total energy 
for the central flux tube is determined by the integral over the non-local action density
in a region within a radius 
of $a/2$ of the flux tube. As $a$ is taken smaller and smaller, some energy from nearby 
flux tubes is included within this region, but also, some energy associated with the central 
flux tube is diffused out of the region. This ambiguity is unavoidable within this 
model. We can't numerically compute the energy over all of space and subtract off 
different contributions, because these energies are infinite.

\begin{figure}
	\centering
		\includegraphics[width=10cm]{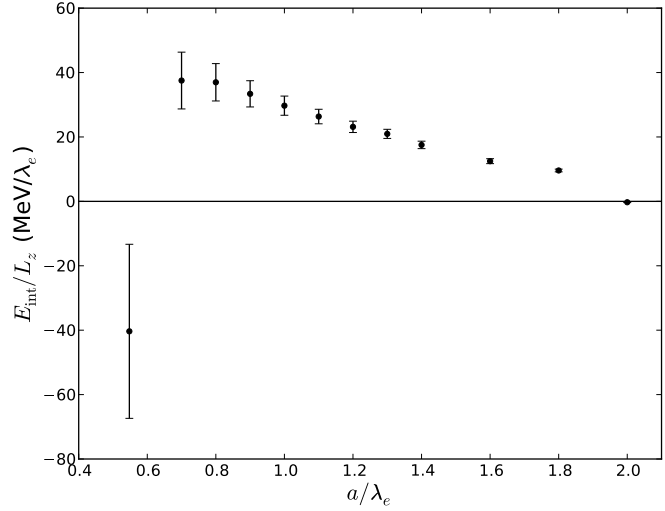}
	\caption[The interaction energy versus the flux tube spacing]
	{The interaction energy per unit length of flux tube as a function 
	of the flux tube spacing, $a$. The energy density of a critical 
	strength magnetic field is 17 GeV/$\lambda_e^3$, so this 
	energy density is small in comparison to the classical magnetic field energies, 
	or the 1-loop corrections. However, the local interactions are constant in $a$.
	At $a<\lambda_e/2$, 
	the bump functions from neighbouring tubes overlap. This approximately corresponds
	to the critical magnetic field which destroys superconductivity.}
	\label{fig:interaction}
\end{figure}

The interaction energy is plotted in figure \ref{fig:interaction}. In this plot, the 
error bars are 1-sigma error bars that account for the correlations in the means 
computed for each group of worldlines,

\be
	\sigma_{E_{\rm int}} = \sqrt{\sigma_{E_a}^2 + \sigma_{E_{a=10000}}^2 
		- 2\operatorname{Cov}(E_a, E_{a=10000})},
\ee
where $\operatorname{Cov}(a,b)$ is the covariance between random variables $a$ and $b$.
Recall from figure \ref{fig:EAscffpeek} that there is a positive contribution 
to the effective action, and therefore a negative contribution to the energy 
from the region between flux tubes. As we reduce $a$, bringing the flux tubes 
closer together, two considerations become important. Firstly, we are increasing 
the average field strength meaning there tends to be more flux through the 
worldline loops which tends to give a negative contribution to the interaction 
energy. Secondly, we are reducing the spatial volume over which we integrate the 
energy since we only integrate $\rho$ from $0$ to $a/2$. This effect makes a positive 
contribution to the interaction energy since we include less and less of the 
region of negative energy density in our integral. 

In figure \ref{fig:interaction}, there appears to be a landscape with both 
positive and negative interaction energies at different values of $a$. 
These appear to be consistent with the interplay between positive and 
negative contributions described in the previous paragraph. This is consistent with 
the usual expectation of attractive Casimir forces~\cite{duru1993}. The dominant contribution 
for the positive energy values is caused by less of the negative energy region contributing 
as the domain assigned to the flux tube is reduced. However, the point at $a/\lambda_e = 2$ is 
negative ($-0.3\pm0.1$MeV/$\lambda_e$) indicating that an attractive interaction from 
nearby flux tubes is dominant. 
We are continuing to compute points at larger values of $a$. 

At $a/\lambda_e = 0.5$, 
the flux tubes are positioned right next to one another, and the negative 
contribution from the increase in the mean field appears to be 
slightly larger than the positive contribution from the loss of the 
region of negative energy density from the integral. Beyond
this, the flux tubes would overlap each other, which approximately corresponds 
to the critical background field which destroys superconductivity.

Based on the above explanation, it appears that the non-local 
interaction energy between magnetic fields has a strong dependence on 
the specific profile of the classical magnetic field that was used. 
This makes it difficult to predict if it will result in attractive 
or repulsive forces
in a more realistic model of a flux tube lattice. 

The energy density of a 
critical strength magnetic field is about $17{\rm GeV}/\lambda_e^3$, 
so the energy associated with this interaction is relatively small. 
However, there are no other interactions which affect the 
energy of moving the flux tubes closer together when they are 
separated by many coherence lengths. Here, the characteristic 
distance associated with the interaction, $\lambda_e$, is considerably 
larger than coherence length or London penetration depth, so the 
interactions between flux tubes through the order field are heavily 
suppressed.

\section{Discussion and Conclusions}

In this chapter, I have developed a cylindrically symmetric magnetic field 
model which reproduces some of the 
features of a flux tube lattice: for a given central flux tube, there are 
nearby regions of large magnetic field which interact nonlocally, and 
the large flux tube size limit goes to a large uniform magnetic field
instead of to zero field. I have investigated the 1-loop effects from 
\ac{ScQED} in this model using the \ac{WLN} technique for various 
combinations of flux tube size, $\lambda$, and flux tube spacing, $a$.

In contrast to isolated flux tubes, I find that there are some regions where the 
\ac{WLN} results are greater than the \ac{LCF} approximation and other 
regions where they are less than the \ac{LCF} approximation. This can be understood 
by thinking of the difference from the \ac{LCF} approximation as a competition 
between the local counter term and the Wilson loop averages. 
For magnetic fields which vary on the Compton wavelength scale 
about some mean field strength, the \ac{LCF} approximation 
provides a poor approximation of the energy density, but may provide a 
good approximation to the total energy density of the field due to it 
being a good mean field theory approximation to the energy density. 
The appropriateness of the \ac{LCF} approximation in this case
can be understood as an approximate balance between regions where the 
field is a local maximum and the magnitude of the quantum correction to the 
action density is larger than in the constant field case, and regions where 
the field is a local minimum and the quantum corrections predict a smaller 
action density than the constant field case. This 
washing out of the field structure due to non-local effects has also 
been observed in \ac{WLN} studies of the vacuum polarization 
tensor~\cite{PhysRevD.84.065035}. 

There is a force between nearby magnetic flux tubes due 
to the quantum diffusion of the energy density. This interaction is 
non-local and is not predicted by the local derivative expansion.
It is an example of a Casimir force (\ie a force resulting from 
quantum vacuum fluctuations) and it is computed in a very similar 
way as the Casimir force between conducting bodies in the \ac{WLN} 
technique~\cite{Gies:2003cv}.
The size of the energy densities involved in this force are small even 
compared to the 1-loop corrections to the energy densities, which are
in turn small compared to the classical magnetic energy density. 

Although this interaction energy is small, 
the interactions between flux tubes in a neutron star due to the 
order field of the superconductor are suppressed because the distance between the 
tubes is considerably larger than the coherence length and 
London penetration depth. Therefore, this force is possibly important 
for the behaviour of flux tubes in neutron star crusts and interiors.
For example, in our lattice model, this force could contribute 
to a bunching of the worldlines, producing regions where flux tubes 
are separated by $\sim2\lambda_e$ and other regions which have 
no flux tubes. Consequently, this force may have important implications 
for neutron star physics. However, investigating these implications is 
outside the scope of this thesis.

The nature of this 
interaction energy is expected to depend on the model of the 
magnetic field profile for the reasons discussed in section 
\ref{sec:periodic_results}. It is therefore reasonable that 
forces of either sign, attractive or repulsive, may be possible 
depending on the particular landscape formed by the 
magnetic field and the particular definition used of the interaction 
energy. In a superconductor, the profiles of the magnetic 
flux tubes are determined by the minimization of the free 
energy for the interacting system formed by the magnetic 
and order parameter fields. Therefore, investigating this 
phenomenon using more realistic models is an interesting 
direction for future research. In particular, it would be interesting 
to determine if certain conditions allowed for a non-local interaction 
between magnetic flux tubes to be experimentally observable.

The conclusions from this chapter are applicable to \ac{ScQED}. 
However, we have also investigated the relationship between spinor 
and scalar \ac{QED} for isolated flux tubes where the \ac{WLN} technique 
can be applied to both cases. We find that both theories have the same 
qualitative behaviour, and agree within a factor of order unity quantitatively. The arguments 
and explanations given for the \ac{ScQED} results have strong parallels in the spinor 
\ac{QED} case. The spinor case can also be understood in terms of a competition between 
the Wilson loop averages and the local counter term. We therefore speculate that 
the results from this chapter will hold in the spinor case, at least qualitatively. However, 
addressing the fermion problem so that the spinor case can be studied explicitly for 
flux tube lattices would be valuable progress in this area of research.

\part{Conclusions}
\chapter{Conclusions}
\label{ch:conclusions}
\acresetall



The goal of this dissertation has been to explore the role played by 
the quantum vacuum in extreme astrophysical environments. Specifically, 
we have explored three specific physical scenarios: (Part \ref{pt:cosmology}) 
the generation of particles 
(and ultimately large-scale structure) during the universe's inflationary epoch, 
(Part \ref{pt:EMwaves}) the propagation of strong electromagnetic waves through 
the magnetosphere of 
a magnetar, and (Part \ref{pt:fluxtubes}) densely packed flux tubes inside the 
nuclear matter in a pulsar.
In each case, I developed models of quantum field theories coupled
non-perturbatively to their environment and used these models to explore the 
possible role that the quantum vacuum might play in these astrophysical systems. 
This research has produced new insights into the physics of both the quantum 
vacuum and of the astrophysical scenarios studied. Additionally, this 
dissertation also makes contributions to the methods available for studying 
quantum field theories under external conditions. In this conclusion, 
we will review the main scientific contributions of this dissertation, and 
suggest some possible directions for future research which could draw on this work.

\section{Inflation}
In chapter \ref{ch:inflaton}, we generalized a particle picture of quantum 
fluctuations during inflation to include the effects of nonlinear interactions.
Thus, we were able to simulate the quantum mechanical evolution of the field 
treating both the gravitational interaction and the self-interaction nonperturbatively.
This model was useful for tracking exactly the entanglement entropy between modes.
This new insight allowed us to make a comparison between the entanglement entropy and 
other measures of decoherence used in the literature.
Decoherence arising from non-linear interactions 
is not efficient enough to explain classicality in general, but it leaves 
open the exciting possibility that scalar fields not participating in 
reheating may keep some quantum characteristics that may some day be measurable.
It is possible that the techniques of the particle picture developed for this 
dissertation may be applicable to studying entanglement entropy in other systems.  

One possible example is to use the particle picture to address the apparent
loss of information from evaporating black holes (\ie the information 
loss problem). The final state 
of an evaporated black hole is expected to be described by a mixed, thermal 
density matrix, even in the case where the initial state was a pure 
state~\cite{susskind2005introduction}.
This evolution from a pure state to a mixed state is impossible in a closed 
system through unitary evolution, according to the postulates of quantum mechanics.
It has been shown that the entanglement entropy of scalar fields in
their ground state near the black hole horizon can reproduce the form
of the Beckenstein-Hawking entropy~\cite{PhysRevD.34.373,PhysRevLett.71.666}. 
This provides a
hint that the black hole entropy may be due to quantum mechanical
degrees of freedom.  This idea has led to recent interest in
investigating the von Neumann entropy contained in scalar fields near
the black hole's horizon, tracing over the region which is inaccessible 
due to the horizon~\cite{2008CaJPh..86..653D,2006JHEP...01..098B}. 
The degrees of freedom from such a field
give an entanglement entropy to the system, and have been
shown to lead to corrections to the Beckenstein-Hawking entropy in
certain circumstances~\cite{2008PhRvD..77f4013D}. 
Therefore, one expects that by studying 
the black hole spacetime in a conformally flat Euclidean path-integral 
formalism, a Schr\"{o}dinger-like equation will emerge describing 
the unitary evolution of quantum states, but with an entanglement entropy 
consistent with the Beckenstein-Hawking entropy due to entanglement 
between modes inside and outside of the horizon. So, the particle picture 
may provide a new way of looking at black hole information loss, and may 
bring new insights to the problem.

Another potential application of the particle picture is simulating 
the quantum coherence of relic neutrinos.  Neutrinos have the interesting 
property that their mass eigenstates are not coincident with their 
flavour eigenstates~\cite{mohapatra2004massive}.  
One of the consequences of this is that 
the flavour eigenstates of the cosmological relic neutrino background 
become spread out to very large cosmological scales because they are 
in a superposition of different mass states which travel at different speeds 
and become nonrelativistic at different cosmological 
epochs.  It is possible then that a flavour 
eigenstate can be in a large-scale 
coherent superposition of relativistic and nonrelativistic 
mass states at the epoch when neutrinos begin to collapse into dark matter 
gravitational wells.  So, this leads to the question of whether gravitational 
tidal stresses can lead to decoherence between the mass eigenstates of 
relic neutrinos~\cite{PhysRevLett.102.201303}.  It may be possible to use 
the particle picture methods to address decoherence in this scenario.

\section{Electromagnetic Waves Near a Magnetar}
In chapter \ref{ch:travellingwaves}, I developed
a computational method for searching for travelling wave solutions 
to the 1-loop \ac{QED} corrected Maxwell's equations in a plasma.  This 
tool has uncovered a new class of waves which form long wave 
chains with highly non-sinusoidal profiles.  These waves may have 
an effect on energy transport near neutron stars which may have 
observable implications for pulsar microstructure. So, incorporating
this phenomenon into a holistic model of the pulsar 
magnetosphere and determining if it results in observable 
predictions is a future step in this line of research.

One weakness of the findings from this section is that the 
travelling wave conditions were assumed {\it a priori}. So, 
while the results demonstrate that the stable travelling waves 
are a possible outcome for electromagnetic waves coupled to a plasma, 
it is not possible 
to tell from the analysis if they are a likely outcome compared to 
waves which form shocks. So, determining what conditions 
will produce waves which are stable compared to ones that collapse 
is an interesting problem which follows from this research.

This study used a relatively simple plasma model consisting of a 
relativistic pair plasma at zero temperature. This model is 
appropriate considering the intense magnetic fields involved. However, it is 
straightforward to extend the model to more detailed or realistic 
plasma models. Typically, these can be incorporated by modifying the 
vacuum dielectric and permeability tensors appropriately to account for 
a multi-species plasma, non-linearities, and/or thermal effects.

One of the 
original motivations for developing the methods for this study
was to explore the possibility that solitons may be formed in 
the \ac{QED} vacuum + plasma system. These solutions would 
represent stable bundles of electromagnetic energy that would 
travel through the magnetosphere, and a novel structure of the 
\ac{QED} quantum vacuum coupled to a plasma. While no solitonic waves have 
yet been discovered in such a system,
this remains a possible application for the 
techniques that can be further explored in the future.

\section{Flux tubes}
In general, evaluating the fermion determinants 
in non-homogeneous fields is a challenging problem without many 
techniques and approaches available to researchers.
Through my investigations of \ac{QED}  
magnetic flux tubes, I have derived a new expression for 
the effective action of cylindrically symmetric magnetic fields 
based on a Green's function technique 
which is suitable for numerical calculations. This technique is the 
subject of chapter \ref{ch:greensfunc}. While the expressions 
derived in this section are difficult to work with and 
were not ultimately employed, they may 
prove useful for future research into the effective actions of flux 
tube configurations. 
An interesting result from this line of research 
is that an expression can be derived which extremizes the effective action 
with respect to the function that describes the magnetic field profile. 
This expression includes contributions from \ac{QED} which suggests that some 
nonhomogeneous magnetic field extremizes the effective action in contrast to the 
purely classical case. This result begs the question of what effect the quantum corrections 
have to the stability of flux tubes and flux tube lattices.

In order to compute the effective actions of flux tubes, I ultimately 
turned to the numerical 
worldline (or loop cloud) method. Two contributions to this technique 
are made in this dissertation and are discussed in chapters \ref{ch:WLNumerics} and 
\ref{ch:WLError}.  Firstly, I made the observation that the non-Gaussian nature of the 
worldline distributions and the correlations between the computed points in the 
integrals should both be taken into account in the uncertainty analysis. 
Chapter \ref{ch:WLError} is the first thorough discussion of error analysis 
for the \ac{WLN} technique.
Such a discussion is important because the 
subtleties can be easily missed or ignored by new adopters of the technique.
Secondly, by implementing the technique on \acp{GPU}, I have 
been able to achieve a 3600-fold speed increase over a serial 
implementation in computing Wilson loop averages.  This likely makes 
my implementation of the Wilson loop algorithm 
an order of magnitude faster than a 
parallel implementation on \ac{CPU} clusters with hundreds of processors
and demonstrates that the \acp{GPU} architecture is very well suited to the 
technique.
Dramatic speed increases in computational physics such as this 
may open up entirely new avenues of research. Since the worldline 
numerical technique has been useful in a variety of applications, 
developing the technique for \acs{GPU} may help research in 
computing effective actions, Casimir energies~\cite{Gies:2003cv, Moyaerts:2003ts}, 
or problems in \ac{QCD}. 

Further development of the \ac{WLN} code could produce increased performance, 
robustness, and greater flexibility in terms of the types of scenarios which could be 
computed. For example, more highly developed integration software such as that 
provided by \ac{GSL}, would improve the robustness and reliability of the code. 
These libraries cannot be used directly for \ac{GPU} 
device code, but would be simple to implement for the integrals over proper time and 
centre of mass which are currently computed by the \ac{CPU}.
The most significant performance bottleneck in the algorithm is the large portion 
of the computation which is performed serially. Parallelizing parts of this 
serial portion with additional 
\ac{CUDA} kernels would help to accelerate the algorithm. There an
opportunity for this kind of optimization, but this can only be done 
for portions of code which do not depend on the Wilson loop kernel. 
One could also straightforwardly
parallelize the algorithm onto several \ac{GPU} devices, with each one responsible for a 
different centre of mass point, for example. Finally, there is also some room for optimizing the 
memory throughput to accelerate the Wilson loop calculations themselves for kernels which can make 
frequent use of the fast shared memory, or for \acp{GPU} which can store the worldline 
data in constant memory. 

In chapter \ref{ch:periodic}, I have developed a cylindrically symmetric toy model 
intended to reproduce some aspects of a triangular lattice of flux tubes. I have 
used this model, and the worldline numerics technique to compute the \ac{ScQED} effective 
action of a flux tube in a situation where nonlocal effects are expected to be significant. 
The quantum corrections to the energies in these configurations are small and are 
close to the \ac{LCF} approximation because the energy density is 
diffused by the worldline loops so that the total energy is close to the mean field 
approximation. Nevertheless, I find that there is a nonlocal interaction between flux tubes 
which is not predicted by local techniques such as a derivative expansion. While the 
Casimir force between two infinitely thin solenoids has been previously computed~\cite{duru1993}, 
the \ac{WLN} technique opens up the possibility of exploring this force for other less 
idealized field configurations.

Although this interaction is generally small, it occurs between flux tubes separated by 
a Compton wavelength, while the distance scales associated with flux tube interactions 
in a neutron star are typically much smaller. Therefore, if the flux tube density is such 
that the distance between tubes is comparable to a Compton wavelength, this interaction 
may have a considerable impact on the  of the flux tubes. If this force is attractive 
at distances much greater than the coherence length and London penetration depths, it may 
contribute to a bunching of flux tubes.
Investigating the consequences of this force for neutron star physics, and possibly for 
laboratory experiments is a very interesting goal for future research.

The \ac{WLN} technique does not require any special symmetries, so it is easily generalizable 
to a more realistic flux tube lattice geometry. However, this would be significantly more computationally 
expensive because there is an additional spatial 
dimension to integrate, and more worldlines consisting of more 
points per line may be required. We were not able to compute effective actions for fermionic \ac{QED}
because of the fermion problem: when the fields are not well localized, one must compute the ratio 
of two very large numbers which results in amplified uncertainties. Future research 
should focus on addressing this problem so that the worldline technique can be applied to lattice 
configurations, which have large magnetic fields distributed throughout space.

A very compelling prediction of \acp{QFT} is that the laws of physics are played out on a 
lively background called the quantum vacuum. In this thesis, I have developed new physical 
models designed to illuminate specific aspects of how this quantum vacuum influences three 
different astrophysical scenarios: the emergence of classicality in the large scale structure of the 
universe, the behaviour of travelling electromagnetic waves near a magnetar, and the 
energies of narrow, densely spaced flux tubes in a superconducting neutron star. For each of these 
scenarios, I have performed novel mathematical and numerical analyses which have lead to contributions 
to the tools available to physicists, and to new physical insights about the systems studied. 
This thesis has contributed to our understanding of the
interesting relationships which can exist between the vacuum 
and extreme astrophysical objects. Understanding these relationships is critical for 
understanding these objects in detail, and is also crucial to learning how observing these 
objects can contribute to our knowledge of the microscopic physics of the quantum vacuum.

\sloppy
\begin{singlespace}
\raggedright
\bibliographystyle{abbrvnat}
\bibliography{references}
\end{singlespace}

\appendix
\lstset{language=C, otherkeywords={__device__, __global__, float4,
	__launch_bounds__, __CUDA_ARCH__, cudaError_t, define, endif, ifndef, \#}}
\phantomsection	
\addcontentsline{toc}{part}{Appendices}

\chapter{\texorpdfstring{\acs{CUDA}}{CUDA}fication of Worldline Numerics}
\label{ch:cudafication}

My implementation of the \ac{WLN} technique used a ``co-processing"
(or heterogeneous)
approach where the \ac{CPU} and \ac{GPU} are both used in unison to 
compute different aspects of the problem.  The \ac{CPU} was used 
for computing the spatial and proper-time integrals while the 
\ac{GPU} was employed to compute the contributions to the Wilson 
Loop from each worldline in parallel for each value of $\rho_{\rm cm}$ and $T$.
This meant that each time the integrand was to be computed, it could 
be computed thousands of times faster than on a serial implementation.
A diagram illustrating the co-processing approach is shown in figure 
\ref{fig:coprocessing}

Code which runs on the \ac{GPU} device must be implemented in the 
\ac{CUDA}-C programming language~\cite{cudazone, CUDAGuide3.2}.
The \ac{CUDA} language is an extension of the C programming language
and compiles under a proprietary compiler, nvcc, which is based off 
of the \acs{GNU} C compiler, gcc. 

\section{Overview of \texorpdfstring{\acs{CUDA}}{CUDA}}

\ac{CUDA} programs make use of special functions, called kernel functions, 
which are executed many times in parallel. Each parallel thread executes 
a copy of the kernel function, and is provided with a unique identification 
number, \lstinline{threadIdx}. \lstinline{threadIdx} may be a one, two, or three-dimensional vector
index, allowing the programmer to organize the threads logically
according to the task.

A program may require many thousands of threads, which are organized into a 
series of organizational units called blocks. For example, the Tesla C1060 
\ac{GPU} allows for up to 1024 threads per block. These blocks are further 
organized into a one or two-dimensional structure called the grid. 
\ac{CUDA} allows for communication between threads within a block, 
but each block is required to execute independently of other blocks.

\ac{CUDA} uses a programming model in which the \ac{GPU} and \ac{CPU} 
share processing responsibilities. The \ac{CPU} runs a host process 
which may call different kernel functions many times over 
during its lifetime. When the host process encounters a kernel 
function call, the \ac{CUDA} device takes control of the processing 
by spawning the designated number of threads and blocks to evaluate 
the kernel function many times in parallel. After the kernel function 
has executed, control is returned to the host process which may then 
copy the data from the device to use in further computations.

The \ac{GPU} device has separate memory from that used by the \ac{CPU}. 
In general, data must be copied onto the device before the kernel is executed 
and from the device after the kernel is executed. 
The \ac{GPU} has a memory hierarchy containing
several types of memory which can be utilized by threads. Each thread 
has access to a private local memory. A block of threads may all access 
a shared memory. Finally, there is memory that can be accessed by any 
thread. This includes global memory, constant memory, and texture memory. 
These last three are persistent across kernel launches, meaning that 
data can be copied to the global memory at the beginning of the program 
and it will remain there throughout the execution of the program.

\section{Implementing \texorpdfstring{\acs{WLN}}{WLN} 
in \texorpdfstring{\acs{CUDA}}{CUDA}-C}

In the jargon of parallel computing, an embarrassingly parallel problem 
is one that can be easily broken up into separate tasks that do not 
need to communicate with each other. The worldline technique is one 
example of such a problem: the individual contributions from each worldline
can be computed separately, and do not depend on any information from 
other worldlines. 

Because we may use the same ensemble of worldlines throughout 
the entire calculation, the worldlines can be copied into the 
device's global memory at the beginning of the program. The global memory 
is persistent across all future calls to the kernel function.
This helps to reduce overhead compared to 
parallelizing on a cluster where the worldline data would need 
to be copied many times for use by each \ac{CPU}.
The memory copy can be done from 
\ac{CPU} code using the built-in function \lstinline{cudaMemcpy()} and 
the built-in flag \lstinline{cudaMemcpyHostToDevice}.

\begin{lstlisting}
//Copy worldlines to device
errorcode = cudaMemcpy(worldlines_d, worldlines_h,
	nThreads*nBlocks*Nppl*sizeof(*worldlines_h),
	cudaMemcpyHostToDevice);
if(errorcode > 0) printf("cudaMemcpy WLs: 
	\%s\n", cudaGetErrorString(errorcode));
\end{lstlisting}
The full source listing appears in appendix \ref{sec:srcEffAct}.
In the above, \lstinline{worldlines_d} and \lstinline{worldlines_h} are 
pointers of type \lstinline{float4} (discussed below) which point to the worldline 
data on the device and host, respectively. \lstinline{nThreads},
\lstinline{nBlocks}, and \lstinline{Nppl} are integers representing the 
number of threads per block, the number of blocks, and the number 
of points per worldline. 
So, \lstinline{nThreads*nBlocks*Nppl*sizeof(*worldlines_h)}
is the total size of memory to be copied. The variable \lstinline{errorcode}
is of a built-in \ac{CUDA} type, \lstinline{cudaError_t} which returns 
an error message through the function \lstinline{cudaGetErrorString(errorcode)}.

The global memory of the device has a very slow bandwidth compared to 
other memory types available on the \ac{CUDA} device. 
If the kernel must access the worldline 
data many times, copying the worldline data needed by the block of threads 
to the shared memory of that block will provide a performance increase. 
If the worldline data is not too large for the device's constant memory, 
this can provide a performance boost as well since the constant memory 
is cached and optimized by the compiler. 
However, these memory optimizations are not used here 
because the worldline data is too large for constant memory and 
is not accessed many times by the kernel. This problem can 
also be overcome by generating the loops on-the-fly directly on the 
\ac{GPU} device itself without 
storing the entire loop in memory~\cite{2011arXiv1110.5936A}.

In order to compute the worldline Wilson loops, we must create 
a kernel function which can be called from \ac{CPU} code, but 
which can be run from the \ac{GPU} device.  Both must 
have access to the function, and this is communicated to the compiler
with the \ac{CUDA} function prefix \lstinline{__global__}.  

\begin{lstlisting}
#define THREADS_PER_BLOCK 256
#if __CUDA_ARCH__ >= 200
    #define MY_KERNEL_MAX_THREADS (2 * THREADS_PER_BLOCK)
    #define MY_KERNEL_MIN_BLOCKS 3
#else
    #define MY_KERNEL_MAX_THREADS (2 * THREADS_PER_BLOCK)
    #define MY_KERNEL_MIN_BLOCKS 2
#endif

__global__ void 
__launch_bounds__(MY_KERNEL_MAX_THREADS, MY_KERNEL_MIN_BLOCKS)
__global__ void ExpectValue(float4 *Wsscal, float4 *Wsferm, 
	float4 *worldlines, float4 xcm, float F, 
	float l2, float rtT, int Nl, int Nppl, int fermion)
//Each thread computes the Wilson loop value for a single 
//worldline identified by inx.
{
        int inx = blockIdx.x * blockDim.x + threadIdx.x;        
        WilsonLoop(worldlines, Wsscal, Wsferm, 
			xcm, inx, F, l2, rtT, Nppl, fermion);
            
}
\end{lstlisting}
The above listing is an excerpt from the file listed in appendix 
\ref{sec:srcintEV}. The preprocessor commands (\ie the lines beginning with \#) 
and the function \lstinline{__launch_bounds__()} provide 
the compiler with information that helps it minimize the registers needed and prevents 
spilling of registers into much slower local memory. More information can be found in 
section B.19 of the \ac{CUDA} programming guide~\cite{CUDAGuide3.2}.
The built-in variables \lstinline{blockIdx}, \lstinline{blockDim}, and \lstinline{threadIdx} 
can be used as above to assign a unique index, \lstinline{inx}, 
to each thread.
\ac{CUDA} contains a native vector data type called \lstinline{float4} 
which contains a four-component vector of data which can be copied 
between host and device memory very efficiently 
\footnote{
float4 is defined in the \ac{CUDA} header file \lstinline{builtin_types.h}}.  
This is clearly 
useful when storing coordinates for the worldline points or the 
center of mass.  These coordinates are accessed using 
C's usual structure notation: \lstinline{xcm.x}, \lstinline{xcm.y}, \lstinline{xcm.z}, \lstinline{xcm.w}.
I also make use of this data type to organize the output 
Wilson loop data, \lstinline{Wsscal} and \lstinline{Wsferm}, into the groups discussed in section 
\ref{sec:discunc}.

The function \lstinline{WilsonLoop()} contains code which only the 
device needs access to, and this is denoted to the compiler by the 
\lstinline{__device__} function prefix. For example, from appendix \ref{sec:srcintEV},
we have,

\begin{lstlisting}
extern "C"
__device__ void WilsonLoop(float4 *worldlines, float4 *Wsscal, 
	float4 *Wsferm, float4 xcm, int inx, float F, float l2, 
	float rtT, int Nppl, int fermion)
//Compute the WilsonLoops for the thread inx and store the 
//results in Wsscal[inx] (scalar part) 
//and Wsferm[inx] (fermion part)
{
	...
}
\end{lstlisting}
Note that we pass $\sqrt{T}$ to the function (\lstinline{float rtT}) instead of 
$T$ so that we only compute the square root once instead of once 
per thread.  Avoiding the square root is also why I chose to 
express the field profile in terms of $\rho^2$.

As mentioned above, the \ac{CUDA} 
device is logically divided into groups of 
threads called blocks.  The number of blocks, nBlocks, 
and the number of threads per block, \lstinline{nThreads}, which are to 
be used must be specified when calling \ac{CUDA} kernel functions 
using the triple angled bracket notation.

\begin{lstlisting}
MyKernel<<<nBlocks,nThreads>>>(void* params)
\end{lstlisting}

In the following snippet of code, we call the \ac{CUDA} device 
kernel, \lstinline{ExpectValue()} from a normal C function. 
We then use the \lstinline{cudaMemcpy()} function to copy 
the Wilson loop results stored as an array 
in device memory as \lstinline{params.Wsscal_d}
to the host memory with pointer \lstinline{params.Wsscal_h}.  The 
contents of this variable may then be used by the \ac{CPU} 
using normal C code, as is done in appendix \ref{sec:srcintEV}.

\begin{lstlisting}
//Call to CUDA device
ExpectValue<<<params.nBlocks, params.nThreads>>>(
	params.Wsscal_d, params.Wsferm_d, 
	params.worldlines, params.xcm, params.F, 
	params.l2, rtT,params.Nl, params.Nppl
	);
//Check for errors during kernel execution
errorcode = cudaGetLastError();
if (errorcode > 0) printf(
	"cuda getLastError EV(): %s\n",
	cudaGetErrorString(errorcode)
	);
//Copy device memory back to host
errorcode = cudaMemcpy(
	params.Wsscal_h, params.Wsscal_d,
	params.Nl*sizeof(params.Wsscal_h[0]), 
	cudaMemcpyDeviceToHost
	);
//Check for memory copy errors
if(errorcode > 0) printf(
	"CUDA memcpy scal Error EV(): %s\n",
	cudaGetErrorString(errorcode)
	);	
if(params.fermion == 1) //if fermionic calculation
 {
	//Copy fermion data from device to host
  	errorcode = cudaMemcpy(
		params.Wsferm_h, params.Wsferm_d,
		params.Nl*sizeof(params.Wsferm_h[0]), 
		cudaMemcpyDeviceToHost
		);
	//Check for memory copy errors
	if(errorcode > 0) printf(
		"CUDA memcpy ferm Error EV(): %s\n",
		cudaGetErrorString(errorcode)
		);
 };
\end{lstlisting}

\section{Compiling \texorpdfstring{\acs{WLN}}{WLN} 
\texorpdfstring{\acs{CUDA}}{CUDA} Code}

Compilation of \ac{CUDA} kernels is done through the 
Nvidia \ac{CUDA} compiler driver \lstinline{nvcc}. \lstinline{nvcc} can be provided 
with a mixture of host and device code. It will compile the device 
code and send the host code to another compiler for processing. On Linux 
systems, this compiler is the \acs{GNU} C compiler, \lstinline{gcc}. In general,
\lstinline{nvcc} is designed to mimic the behaviour of \lstinline{gcc}. So the interface and 
options will be familiar to those who have worked with gcc. 

There are two dynamic libraries needed for compiling \ac{CUDA} code. They are called 
\lstinline{libcuda.so} and \lstinline{libcudart.so} and are
located in the \ac{CUDA} toolkit install path. Linking with these libraries is handled by the \lstinline{nvcc} options 
\lstinline{-lcuda} and \lstinline{-lcudart}. The directory containing the \ac{CUDA} libraries 
must be referenced in the \lstinline{\$LD_LIBRARY_PATH} 
environment variable, or the compiler will produce library not found errors.

Originally, \ac{CUDA} devices did not support double precision floating point numbers. 
These were demoted to \lstinline{float}. More recent devices do support \lstinline{double}, however. This 
is indicated by the compute capability of the device being equal or greater than 
1.3. This capability is turned off by default, and must be activated by supplying 
the compiler with the option \lstinline{-arch sm_13}. The kernel presented in section \ref{sec:srcintEV}
primarily uses \lstinline{float} variables because this reduces the demands on register memory 
and allows for greater occupancy.

Another useful compile option is \lstinline{--ptxas-options="-v"}. This option provides verbose information 
about shared memory, constants, and registers used by the device kernel. The kernel code for my project 
became sufficiently complex that I ran out of registers and suffered some baffling behaviour from the 
program that I wasn't initially error checking for properly. This can be avoided by paying attention 
to the number of registers in use, and using the \ac{CUDA} Occupancy Calculator 
spreadsheet provided by Nvidia to determine the maximum number of threads per block that can 
be supported~\cite{cudaOC}. 
However, the \ac{CUDA} error checking used in this chapter is sufficient to discover the 
problem if it arises.

A Makefile showing the compilation and linking of a program defined by 
multiple \lstinline{.c}, \lstinline{.cu} and \lstinline{.h} files 
is given in appendix \ref{sec:srcmake}.
\chapter{\texorpdfstring{\acs{WLN}}{WLN} Source Code}
\label{ch:sourcecode}

In this appendix, I collect a listing of the source code 
used for the calculations in chapters \ref{ch:WLNumerics}, 
\ref{ch:WLError} and \ref{ch:periodic}. To obtain compileable code, 
these listings can be copy and pasted, taking care to fix strings 
which have been split onto two lines, and removing page numbers. 
\begin{itemize}
\item Section \ref{sec:srcintT} 
contains a header file used by many of the .c and .cu files. 
\item The main driver for the program and the 
routines for performing the integral over center of mass 
is listed in section \ref{sec:srcEffAct}. 
\item The routines for 
integrating over proper time are listed in section \ref{sec:Tint}.
\item The functions for computing the integrand of the proper time and 
center of mass integrals are provided in section \ref{sec:srcIntegrand}.
\item Section \ref{sec:srcintEV} contains the \ac{CUDA} kernel 
and the \ac{CUDA} device functions which compute the scalar and fermionic 
contributions to the Wilson loops.
\item The routine for reading worldline data from an ASCII file is listed 
in section \ref{sec:srcgetwl}.
\item Section \ref{sec:srcmake} contains the Makefile which 
compiles software using the Nvidia proprietary compiler \lstinline{nvcc}.
\item Two Matlab functions which are used to generate worldline data 
and save them to ASCII files are listed in section \ref{sec:srcWLgen}.
\end{itemize}

\lstset{language=C, otherkeywords={__device__, __global__, float4,
	__launch_bounds__, __CUDA_ARCH__, cudaError_t, define, endif, ifndef, \#}}

\section{Common Header File - intT.h}
\label{sec:srcintT}
A header file shared by all of the .c and .cu files.
\begin{lstlisting}[language=C, title=intT.h]
struct Wparams 
{
  float4 *worldlines;   //worldline points
  float4 *Wsscal_d;     //scalar Wilson exponents on device
  float4 *Wsscal_h;     // '' on host
  float4 *Wsferm_d;     //fermion Wilson exponents on device
  float4 *Wsferm_h;     // '' on host
  float4 xcm;           //center of mass of loop cloud
  double F;              //flux parameter
  double l2;             //radius squared of flux tube
  int Nppl;             //Number of points per line
  int Nl;               //Number of loops
  int ng;               //Number of groups for jackknife
  int fermion;          //Flag for scalar/spinor QED calculation
  int nBlocks;          //Number of blocks on cuda device
  int nThreads;         //Number of threads per CUDA block
  double SE;             //Standard Error
  float *rho2sp;        //rho^2 spline points for periodic field
  float4 *flcoefs;      //spline coefficients for periodic field
};

//Function prototypes
int getwl(float4 *worldlines, char *filename, int Nl, 
	int Nppl, int Dim);

__global__ void ExpectValue(float4 *Wsscal, float4 *Wsferm, 
	float4 *worldlines, float4 xcm, float F, float l2, 
	float rtT, int Nl, int Nppl, float4 *flcoefs, 
	float *rho2sp, int fermion);

double* integrateT(int n, double tmax, void * p);

double* Tfunc(double T, void* p, int* order);

//Constant definitions
#ifndef intT_h
  #define intT_h
  #define m2 1.0  //electron mass squared
  #define e 0.302822121  //electron charge (alpha = e^2/(4 pi))
  #define pi 3.14159265
  #define verbosity 2 //determines amount of output
#endif

//Profile definitions and constants
#ifndef profile
  enum proftype {
	step,
	smooth,
	quadratic,	//untested
	gaussian,
	periodic,	//untested
	spline,		//untested
	fixflux
  };
  //Select the profile here
  enum proftype profile = smooth;
  #define Nspline 10000 //for spline
  #define tubedist 8.0 //for fixflux
  #define q1 0.443991 //bump function const. for fixflux
  #define q2 0.0742478 //bump function const.for fixflux
  #define lmin (0.1f*tubedist) //for fixflux
#endif




\end{lstlisting}

\section{Main Program Driver - EffAct.c}
\label{sec:srcEffAct}
Contains the \lstinline{main()} function and routines for 
performing the integral over $\rho_{\rm cm}$.
\begin{lstlisting}[language=C, title=EffAct.c]
#include <stdio.h>
#include <stdlib.h>
#include <math.h>
#include <builtin_types.h> 
#include "intT.h" 

double MagField( double rho2, void* p );

float4* getspline( float* rho2, double lambda2, 
	double a, int Npoints );
double testspline( float rho2, double lambda2, double a, 
	int Npoints, float *rho2sp, float4 *flcoefs);

double* rhogrand(double rho, void* p)
//The Lagrangian Density Function
{
  int i, Npoints = 100;
  double *intout;
  double tmax, Bcm;

  struct Wparams *params = (struct Wparams *) p;
  //by symmetry, we may integrate rho along the x-axis
  params->xcm.x = rho;
  params->xcm.y = 0.0;
  params->xcm.z = 0.0;
  //Determine the T value where the constant 
  //field integrand has a maximum for large fields
  Bcm = MagField(rho*rho, p);
  if (e*Bcm > 100.0)
	tmax = 3.0/(e*Bcm); //from large field limit
  else
	tmax = 1.0;
  //perform the proper time integration
  if(verbosity >= 2)
  	printf("Beginning T integration\n");
  //cudaSetDevice();
  intout = integrateT(Npoints, tmax, p);
  if(verbosity >= 2)
  	printf("Returning from T integration\n");
  //for each group of worldlines
  for(i = 0; i < params->ng; i++)
  {
	//multiply the integrand by rho
	intout[i] = rho*intout[i];
  }
  return intout;
}

void testvalrho(double *val, int n, double t)
//Outputs diagnostic information about the rho integrand
{
  int i;
  double avg=0.0; //average
  double SEM=0.0; //standard error in the mean
  //compute the average
  for(i = 0; i < n; i++)
  {
	if( isnan(val[i]) ) 
		printf("for rho=%f val[%d]=nan\n", 
			(1.0-t)/t, i);
	avg += val[i];
  }
  avg /= n;
  //compute the standard error
  for(i = 0; i < n; i++)
  {
	SEM += (val[i]-avg)*(val[i]-avg);
  }
  SEM = sqrt(SEM/((double)n*((double)n-1.0)));
  //rho Leff SEM-Leff
  printf("testvalrho: %f %f %f\n", (1.0-t)/t, avg, SEM);
}

double* transformRho(double rhot, void* p)
//Maps the interval [0,inf) to [0,1] using rho=(1-rhot)/rhot
{
	struct Wparams *params = (struct Wparams *) p;
	double* rhofunc;
	double rho = (1.0-rhot)/rhot;
	int i;
	
	rhofunc = rhogrand(rho, p);
	for(i = 0; i < params->ng; i++)
	{
		rhofunc[i] = (rhofunc[i]/rhot)/rhot;

	}
	return rhofunc;

}

double* integrateRhotoa(int n, void * p)
//integrates rho from 0 to a/2 for periodic flux tubes
//This algorithm is the composite Simpson's rule as
//described in Burden and Faires "Numerical Analysis", 7th ed.
{
        double* val;
	double* XI0;
	double* XI1;
	double* XI2;
	double* rg;
	double* rg2;
	struct Wparams *params = (struct Wparams *) p;
	double rho;
	int inx, ig, i;
	val = (double *)calloc(params->ng, sizeof(*val));
	XI0 = (double *)calloc(params->ng, sizeof(*XI0));
	XI1 = (double *)calloc(params->ng, sizeof(*XI1));
	XI2 = (double *)calloc(params->ng, sizeof(*XI2));
	double h = tubedist/((double) 2*n);
	double rhot;
	if(!val) printf("integrate(): malloc failed\n");
	rg = rhogrand(0.0, p);
	rg2 = rhogrand(tubedist/2.0, p);
	for(ig = 0; ig < params->ng; ig++)
	{
		val[ig] = 0.0;
		XI1[ig] = 0.0;
		XI2[ig] = 0.0;
		XI0[ig] = rg[ig] + rg2[ig];
	}
	for(inx = 0; inx < n; inx++)
	{
		printf("rho integrate inx=%d\n", inx);
		rho = h*((double)inx + 1.0);
		rg = rhogrand(rho, p);
		for(ig = 0; ig < params->ng; ig++)
		{
			if(inx%2 == 0) XI2[ig] += rg[ig];
			else XI1[ig] += rg[ig];
		}
		free(rg);
		
	}
	for(ig = 0; ig < params->ng; ig++)
	{
		val[ig] = h/3.0*(XI0[ig] + 2.0*XI2[ig] + 4.0*XI1[ig]);
	}	
	free(XI0); free(XI1); free(XI2); free(rg2);
        return val;
            
} 

double* integrateRho(int n, void * p)
//integrates rho from 0 to infinity for isolated flux tubes
//This algorithm is the composite Simpson's rule as
//described in Burden and Faires "Numerical Analysis", 7th ed.
{
        double* val;
	double* rhort;
	double* rhortphalf;
	double* rhortp1;
	struct Wparams *params = (struct Wparams *) p;
	int inx, ig, i;
	val=(double *)calloc(params->ng, sizeof(*val));
	double h = 1.0/((double) n);
	double rhot;
	if(!val) printf("integrate(): malloc failed\n");
	for(ig = 0; ig<params->ng; ig++)
	{
		val[ig] = 0.0;
	}
	for(inx = 0; inx < n; inx++)
	{
		printf("rho integrate inx=%d\n", inx);
		rhot = h*((double)inx+1.0);
		rhort = transformRho(rhot, p);
		rhortphalf = transformRho(h*((double)inx + 0.5), p);
		rhortp1 = transformRho(h*((double) inx+1.0), p);
		
		
		for(ig = 0; ig < params->ng; ig++)
		{
			//check for infinities and nans
			if(isinf(rhort[ig])) 
			{
			  printf("Warning: rhort[%d]=%f.  
			    Setting to zero.\n", ig, rhort[ig]);
			  rhort[ig]=0.0;
			}
			if(isinf(rhortphalf[ig])) 
			{
			  printf("Warning: rhortphalf[%d]=%f.  
			    Setting to zero.\n",ig,rhortphalf[ig]);
			  rhortphalf[ig]=0.0;
			}
			if(isinf(rhortp1[ig])) 
			{
			  printf("Warning: rhortp1[%d]=%f.  
			    Setting to zero.\n",ig,rhortp1[ig]);
			  rhortp1[ig]=0.0;
			}
//This equation extrapolates the three different 
//Nppl spacings to Nppl=infinity
        		val[ig] += h/6.0*(
			  rhort[ig] 
			  + 4.0*rhortphalf[ig]	
			  +rhortp1[ig]); 
			if(isinf(val[ig])){
			  printf("rhort[%d]=%f\n", ig, rhort[ig]);
			  printf("rhortphalf[%d]=%f\n", ig, 
			    rhortphalf[ig]);
			  printf("rhortp1[%d]=%f\n", ig, rhortp1[ig]);
			  printf("val[%d]=%f\n", ig, val[ig]);	
			}
		}
		
	}
	free(rhort);
	free(rhortphalf);
	free(rhortp1);
	
        return val;         
} 

double PeriodicCA(double l2)
//Computes the spline Classical action
{
  double a = tubedist;
  double l = sqrt(l2);
  double expal = exp(a/l);
  double expalp1 = 1.0 + expal;
  double PCA;
  PCA = 2.0*a + 4.0*a/(expalp1*expalp1*expalp1)
	 -l + 4.0*l/expalp1
	-2.0*(3.0*a+2.0*l)/(expalp1*expalp1) 
	+ l*log(16.0)-4.0*l*log(expalp1);
  return PCA;
}

double FixFluxCA(double l2)
//Computes the fixflux (periodic) classical action
{
  double q3 = 0.0187671;
  double a = tubedist;
  double l = sqrt(l2);
  double A = q3*(a*a/(l2*q2*q2));
  double lmlmin = (l-lmin)/(a-lmin);
  double FFCA;
  FFCA = 4.0*A + lmlmin*((9.0/4.0*A-9.0/2.0)*lmlmin
	 - 6.0*A + 12.0);
  return FFCA;
}

double ClassAction(double F, double l2)
//Returns the Classical action for the given flux tube profile
{
  double CA;
  switch(profile){
	case step:
	  CA=-2.0*pi*F*F/(e*e*l2);
	  break;
	case smooth:
	  CA=-2.0*pi*F*F/(3.0*l2*e*e);
	  break;
	case quadratic:
	  CA=-8.0*pi*F*F/(3.0*l2*e*e);
	  break;
	case gaussian:
	  CA=-pi*F*F/(l2*e*e);
	  break;
	case periodic:
	  CA=-pi*sqrt(l2)/(24.0*l2*l2*log(2.0)*log(2.0))*PeriodicCA(l2);
	  break;
	case spline:
	  CA=1.0;
	  break;
	case fixflux:
	  CA = -pi*F*F/(e*e*tubedist*tubedist)*FixFluxCA(l2);
  }	
   return CA;
}

double *EAction(int Npoints, void *p)
//Calls the integral over rho and returns 
//the effective action
//Npoints is the number of points used in rho integral
{
  int i;
  double *intout;
  //The average and standard error for the effective action
  double result, resultSE;
  struct Wparams *params = (struct Wparams *) p;

  //Determine which class of rho integral is appropriate
  if(profile == periodic || profile == spline 
    || profile == fixflux)
	  intout=integrateRhotoa(2*Npoints, p);
  else
  	  intout=integrateRho(Npoints, p);
  result = 0.0;
  for(i = 0; i < params->ng; i++)
  {
	if(params->fermion == 1)
        	intout[i] /= 4.0*pi;
	else
		intout[i] /= -2.0*pi;
	result += intout[i];
	printf("rho: intout[%d]=%e\n", i, intout[i]);	
  }
  result /= params->ng;
  resultSE = 0.0;
  for(i = 0; i < params->ng; i++)
  {
	resultSE += (intout[i]-result)*(intout[i]-result);
  }
  resultSE = sqrt(resultSE/((double)params->ng
	*((double)params->ng-1.0)));
  if (verbosity >= 1)
  {
    printf("EAvsLambda: %e %e %e %e \n",sqrt(params->l2), 
	ClassAction(params->F,params->l2), result, resultSE);
    printf("ActRatio: %e %e %e\n",sqrt(params->l2), 
	result/ClassAction(params->F,params->l2), 
	resultSE/ClassAction(params->F,params->l2));
  }
  return intout;
}

int main (int argc, char *argv[])
{
  //Number of worldlines, points per line, and 
  //dimensions in the worldline file
  const int Nl = 5120, Nppl = 1024, Dim = 2;
  const int fermion = 0; //1 => fermion, 0 => scalar
  //nThreads set to 256 later for periodic fields.
  int i, nBlocks, nThreads = 512; 
  //worldline text file 
  char *filename = "worldlines.5120.1024.2.dat"; 
  //Arrays to hold values of the Wilson 
  //loops on the device and host
  float4 *Wsscal_d, *Wsferm_d;
  float4 *Wsscal_h, *Wsferm_h;
  float4 xcm;
  double exactX;
  //F=0.5*e*B_0*l2
  double F = 1.0, l2 = 1.0;
  double texe, T,X;
  float4 *worldlines_h; //worldlines on host
  float4 *worldlines_d; //worldlines on CUDA device
  cudaError_t errorcode;
  //mean and standard error of effective action
  double result, resultSE; 
  double* intout;
  float4 *flcoefs_d; //spline coefficients on device
  float4 *flcoefs_h; //spline coefficients on host
  float *rho2sp_d;  //splined rho values on device
  float *rho2sp_h;  //splined rho values on host

  if(fermion==0)  printf("Warning: performing 
	scQED calculation\n");
  switch(profile){
	case step: printf("F_lambda(rho2) = rho^2/lambda^2*theta(lambda^2-rho^2) + theta(rho^2-lambda^2)\n");
		break;
	case smooth: printf("F_lambda(rho2) = rho^2/(lambda^2 + rho^2)\n");
		break;
	case quadratic: printf("F_lambda(rho2) = 2*rho^2/lambda^2*(1-0.5*rho^2/lambda^2)*theta(lambda^2-rho^2)\n");
	        printf("+theta(rho^2-lambda^2)\n");
		break;
	case gaussian: printf("f_lambda(rho2) = 1-exp(-rho2/lambda2)\n");
		break;
	case periodic: printf("f_lambda(rho2) = Periodic B field\n");
		printf("a=%f\n",tubedist);
		nThreads = 256; //Periodic kernel requires more registers, so cannot support large nThreads.
		break;
	case spline: printf("f_lambda(rho2) = Spline data for periodic field\n");
	        printf("a=%f\n",tubedist);
		nThreads = 256; //Spline kernel requires more registers, so cannot support large nThreads.
	case fixflux: printf("f_lambda(rho2) = A*bump(2rho/lambda) + B\n ");
		printf("a=%f\n",tubedist);
		nThreads = 128; //Fixed Flux kernel requires more registers, so cannot support large nThreads.
		break;
	default: printf("Warning: Invalid field profile set\n");
		break;	
  }

  //determine the number of CUDA blocks required
  if(Nl%nThreads == 0)
	nBlocks = Nl/nThreads; 
  else 
	nBlocks = Nl/nThreads + 1;

  //parameters for the integration function
  struct Wparams params;

  //allocate memory for worldlines on host and device
  worldlines_h = (float4*)malloc(Nppl*nBlocks
	*nThreads*sizeof(*worldlines_h));
  cudaMalloc((void **)&worldlines_d,Nppl*nBlocks
	*nThreads*sizeof(*worldlines_d));
  printf("allocating memory \n");
  if(worldlines_h == NULL | worldlines_d == NULL)
  {
	fprintf(stderr, "out of memory 1\n");
	return(1);
  }
  //read worldlines in from file
  getwl(worldlines_h, filename, Nl, Nppl, Dim);
  printf("Copying worldlines to GPU device \n");
  //Copy worldlines to device
  errorcode=cudaMemcpy(worldlines_d, worldlines_h,
		nThreads*nBlocks*Nppl*sizeof(*worldlines_h), cudaMemcpyHostToDevice);
  if(errorcode > 0) printf("cudaMemcpy WLs: %s\n",
	cudaGetErrorString(errorcode));
  //allocate memory on the host
  Wsscal_h=(float4 *)malloc(nThreads*nBlocks*sizeof(*Wsscal_h));
  Wsferm_h=(float4 *)malloc(nThreads*nBlocks*sizeof(*Wsferm_h));
  //allocate memory on the device
  errorcode=cudaMalloc((float4**)&Wsscal_d, nThreads
	*nBlocks*sizeof(*Wsscal_d));
  if(errorcode > 0) printf("cudaMalloc Ws: %s\n",
	cudaGetErrorString(errorcode));
  errorcode=cudaMalloc((float4**)&Wsferm_d, 
	nThreads*nBlocks*sizeof(*Wsferm_d));
  if(errorcode > 0) printf("cudaMalloc Ws: %s\n",
	cudaGetErrorString(errorcode));
  if(Wsscal_h==NULL | Wsscal_d==NULL 
	|Wsferm_h==NULL | Wsferm_d==NULL)
  {
	fprintf(stderr,"out of memory 1\n");
	return(1);
  }


  //define some integration parameters
  params.Nl = Nl;
  params.Nppl = Nppl;
  params.F = F; 
  params.l2 = l2;
  params.xcm = xcm;
  params.Wsscal_h = Wsscal_h;
  params.Wsscal_d = Wsscal_d;
  params.Wsferm_h = Wsferm_h;
  params.Wsferm_d = Wsferm_d;
  params.worldlines = worldlines_d; 
  params.SE = 0.0;
  params.nBlocks = nBlocks;
  params.nThreads = nThreads;
  params.ng = 40;
  params.fermion = fermion;

  printf("parameters:\n");
  printf("F=%f B_0=%f Bk\n l2=%f \n fermion=%d \n a=%d \n", F, 
	e*MagField(0.0,&params), l2, fermion, tubedist); 
  //set up spline data
  if(profile == spline)
  {
	rho2sp_h = (float *)malloc(Nspline*sizeof(*rho2sp_h));	
	flcoefs_h = getspline(rho2sp_h, l2, tubedist, Nspline);
	for(i = 0; i < Nspline; i++)
	{
		if(isnan(rho2sp_h[i]) || isnan(flcoefs_h[i].x) || 
		  isnan(flcoefs_h[i].y) ||isnan(flcoefs_h[i].y) ||
		  isnan(flcoefs_h[i].y) )
			  printf("is nan detected\n");
	        else if(isinf(rho2sp_h[i]) || isinf(flcoefs_h[i].x)
		  || isinf(flcoefs_h[i].y) ||isinf(flcoefs_h[i].y) 
		  ||isinf(flcoefs_h[i].y) )
			  printf("isinf detected\n");
	}
	errorcode = cudaMalloc((float**)&rho2sp_d, 
		Nspline*sizeof(*rho2sp_d));
	if(errorcode > 0) printf("cudaMalloc rho2sp: %s\n",
		cudaGetErrorString(errorcode));
	errorcode = cudaMemcpy(rho2sp_d, rho2sp_h,
		Nspline*sizeof(*rho2sp_d),cudaMemcpyHostToDevice);
  	if(errorcode > 0) printf("cudaMemcpy rho2sp: %s\n",
		cudaGetErrorString(errorcode));

	errorcode = cudaMalloc((float4**)&flcoefs_d, 
		Nspline*sizeof(*flcoefs_d));
	if(errorcode > 0) printf("cudaMalloc flcoefs: %s\n",
		cudaGetErrorString(errorcode));
	errorcode = cudaMemcpy(flcoefs_d, flcoefs_h,
		Nspline*sizeof(*flcoefs_d), cudaMemcpyHostToDevice);
  	if(errorcode > 0) printf("cudaMemcpy flcoefs: %s\n",
		cudaGetErrorString(errorcode));
	printf("TESTSPLINE: %f\n", testspline(1.0f, l2, 
		tubedist, Nspline, rho2sp_h, flcoefs_h));

  }
  else
  //Spline data will not be used
  {
	flcoefs_d = NULL;
	rho2sp_d = NULL;
  }
  params.flcoefs = flcoefs_d;
  params.rho2sp = rho2sp_d;

  //compute the effective action in groups of worldlines
  //first argument is the number of points for the rho integral
  intout = EAction(100, &params);
  
  //compute the average
  result = 0.0;
  for(i = 0; i < params.ng; i++)
  {
	result += intout[i];
	printf("rho: intout[%d]=%e\n", i, intout[i]);	
  }
  result /= params.ng;
  //Compute the standard error
  resultSE = 0.0;
  for(i = 0; i < params.ng; i++)
  {
	resultSE += (intout[i]-result)*(intout[i]-result);
  }
  resultSE = sqrt( resultSE/((double)params.ng
	*((double)params.ng-1.0)) );
  printf("EAction: %e +/- %e\n", result, resultSE);

  //Free the device memory
  cudaFree(Wsscal_d);
  cudaFree(Wsferm_d);
  cudaFree(worldlines_d);
  //Free the host memory
  free(Wsscal_h);
  free(Wsferm_h);
  free(worldlines_h);
  free(intout);

  if(profile == spline)
  {
   	free(flcoefs_h);
	free(rho2sp_h);
	cudaFree(flcoefs_d);
	cudaFree(rho2sp_d);
  } 
  return 0;
}
\end{lstlisting}
 
\section{Integral over T - Tint.c}
\label{sec:Tint}
Contains routines for performing the integral over proper time.
\begin{lstlisting}[language=C, title=Tint.c]
#include <stdio.h>
#include <stdlib.h>
#include <math.h>
#include <gsl/gsl_randist.h>
#include <gsl/gsl_rng.h>
#include <builtin_types.h> 
#include "intT.h" 

double* transformT(double t, double tmax, void* p, int* order)
//Transforms the integral [0,infinity) to [0,1]
//x=tmax*(1-t)/t
{
	struct Wparams *params = (struct Wparams *) p;
	double* tfunc;
	double x = tmax*(1.0-t)/t;
	int i;
	
	tfunc = Tfunc(x,p, order);
	//loop over groups of worldlines
	for(i = 0; i < params->ng; i++)
	{
		tfunc[i] = tmax*(tfunc[i]/t)/t;

	}
	return tfunc;

}

void testval(double *val, int n, double t)
//used for outputting debugging information
{
  int i;
  double avg = 0.0;
  for(i = 0; i < n; i++)
  {
	if(isnan(val[i])) printf("for t=%f val[%d]=nan\n",t, i);
	avg += val[i];
  }
  avg /= n;


  printf("mean for Tt=%f, %f\n",t,avg);
}

void Tintout(double *val, int n, double rho, int fermion)
//Produces output about the T-integral
{
	int i;
	double avg = 0.0, SEM = 0.0;
  	for(i = 0; i < n; i++)
  	{
		if(isnan(val[i])) printf("for rho=%f val[%d]=nan\n",
			rho, i);
		avg += val[i];
  	}
  	avg /= n;
	for(i = 0; i < n; i++)
 	{
		SEM += (val[i]-avg)*(val[i]-avg);
  	}
  	SEM = sqrt(SEM/((double)n*((double)n-1.0)));
  	//rho Leff SEM-Leff
	
	if(fermion == 1)
  		printf("Leffvsrho: %e %e %e\n", rho, 
			avg/(4.0*pi), SEM/(4.0*pi));
	else
  		printf("Leffvsrho: %e %e %e\n", rho, 
			avg/(-2.0*pi), SEM/(2.0*pi));
}


int* shuffleWL(int Nl)
//Returns a vector representing a shuffled order of worldlines
//uncomment the gsl functions to enable worldline shuffling.
{
  gsl_rng * r;
  int i;
  
  // select random number generator 
  //r = gsl_rng_alloc (gsl_rng_mt19937);

  int* order=(int*)malloc(Nl*sizeof(*order));
          
  for (i = 0; i < Nl; i++)
  {
  	order[i] = i;
  }
          
  //gsl_ran_shuffle (r, order, Nl, sizeof (int));
  //free(r);

  return order;
}



double* integrateT(int n, double tmax, void * p)
//Performs the proper time integral using Simpson's Method
//n = number of points to use for Simpson's method
//p = Wparams parameters for the integrand
{
	double* val;    //value of the integral for each group
	double* trt;    //Integrands evaluated at three points
	double* trtphalf;
	double* trtp1;
	int* order;     //Stores the order of worldlines
	struct Wparams *params = (struct Wparams *) p;
	int inx, ig, i; 
	val = (double *)calloc(params->ng,sizeof(*val));
	double h = 1.0/((double) n);   //
	double t;
	if(!val) printf("integrate(): malloc failed\n");
	//initialize val to zero
	for(ig = 0; ig < params->ng; ig++)
	{
		val[ig] = 0.0;
	}
	//main integration loop
	for(inx = 0; inx < n; inx++)
	{
		printf("T integrate inx=%d\n",inx);
		//shuffle worldlines
		order = shuffleWL(params->Nl);
		t = h*((double)inx+1.0);
		trt = transformT(t, tmax, p,order);
		trtphalf = transformT(h*((double)inx + 0.5), tmax, p, order);
		trtp1 = transformT(h*((double) inx+1.0), tmax, p, order);
		
		//A kludge to ignore possible infinities or nans
		for(ig = 0; ig < params->ng; ig++)
		{
			if(isinf(trt[ig])) 
			{
				printf("Warning: trt[%d]=%f.  
				  Setting to zero.\n", ig, trt[ig]);
				trt[ig] = 0.0;
			}
			if(isinf(trtphalf[ig])) 
			{
				printf("Warning: trtphalf[%d]=%f.  
				  Setting to zero.\n", ig, trtphalf[ig]);
				trtphalf[ig] = 0.0;
			}
			if(isinf(trtp1[ig])) 
			{
				printf("Warning: trtp1[%d]=%f.  
				  Setting to zero.\n", ig, trtp1[ig]);
				trtp1[ig] = 0.0;
			}

        		val[ig]+=h/6.0*(
				trt[ig] 
				+ 4.0*trtphalf[ig]	
				+trtp1[ig]); 
			if(isinf(val[ig])){
				printf("trt[%d]=%e\n",ig, trt[ig]);
				printf("trtphalf[%d]=%e\n",ig,trtphalf[ig]);
				printf("trtp1[%d]=%e\n",ig,trtp1[ig]);
				printf("val[%d]=%e\n",ig,val[ig]);
				
			}
			//printf("val[%d]=%e\n", ig, val[ig]);
		}
		testval(val, params->ng, t);
		free(order);
		
	}
	Tintout(val, params->ng, params->xcm.x, params->fermion);
	free(trt);
	free(trtphalf);
	free(trtp1);
	
	return val;
            
} 
\end{lstlisting}

\section{Calculation of the Integrand - Integrand.c}
\label{sec:srcIntegrand}
Contains functions for computing the integrand.
\begin{lstlisting}[language=C, title=Integrand.c]
#include <stdio.h>
#include <stdlib.h>
#include <math.h>
#include <gsl/gsl_randist.h>
#include <gsl/gsl_rng.h>
#include <builtin_types.h> 
#include "intT.h" 
#include <fenv.h>
#include <signal.h>


double EV(double T, void * p, int* WLlist);

double Exact(double T, double B, int fermion)
//Return the exact integrand for constant fields
{
  double TB = e*T*B;
  double Exact;
  if(fermion == 1)
	Exact = exp(-m2*T)/(T*T*T)*(TB/tanh(TB)-1.0-1.0/3.0*TB*TB);
  else
	if (TB < 50.0)
		Exact = exp(-m2*T)/(T*T*T)*(TB/sinh(TB)
			-1.0+1.0/6.0*TB*TB);
	else
		Exact = exp(-m2*T)/(T*T*T)*(1.0/6.0*TB*TB-1.0);
  return Exact;
}

double bumpd(double x)
//The bump function
{
  if (abs(x) < 1.0)
	return exp(-1.0/(1.0-x*x));
  else
	return 0.0;	  
}

double fixfluxB(double rho2, double lambda2)
//Magnetic field strength for the fixflux profile
{
  double l = sqrt(lambda2);
  double lmlmin = (l - lmin)/(tubedist - lmin);
  if(sqrt(rho2) > tubedist/2.0) 
	printf("Warning: bad B field. rho=%f\n", sqrt(rho2));
  return  2.0/(lambda2*q2)*(1.0-0.75*lmlmin)
	*bumpd(2.0*sqrt(rho2/lambda2))
	+ 3.0/(tubedist*tubedist)*lmlmin;
}

double periodicB(double rho2, double lambda2)
//magnetic field for the periodic flux tube profile
{
  double expr, rho, lambda, flraf, f, ceilraf;
  double expral, exprapl;
  rho = sqrt(rho2);
  lambda = sqrt(lambda2); 
  expr = exp(-rho/lambda);
  flraf = floor(rho/tubedist);
  ceilraf = ceil(rho/tubedist);
  if(rho < tubedist)
  {
	expral = 0.0;
	if((tubedist-rho)/lambda < 10.0)
		expral = exp(-(rho-tubedist)/lambda);
	f = expr/(tubedist*(1.0 + expr)*(1.0 + expr)); 
	if(rho > 0.0)
		f += expral/(rho*(1.0+expral)*(1.0+expral));
  }
  else
  {
        if(abs(fmod(rho,tubedist)) < 1.0e-6)
 	{
		f = flraf/(4.0*rho);
	}
	else
	{
	  expral = exp(-(rho-flraf*tubedist)/lambda);
	  exprapl = exp(-(rho-ceilraf*tubedist)/lambda);
          f = flraf*expral/(rho*(1.0+expral)*(1.0+expral)) +
		ceilraf*exprapl/(rho*(1.0+exprapl)*(1.0+exprapl));
        }
  }
  f = tubedist/(2.0*log(2.0)*lambda2)*f;
  return f;
}

double MagField(double rho2, void* p)
//Compute the magnetic field. 
{
  double B;
  double sinrho;
  struct Wparams *params = (struct Wparams *) p;
  switch(profile){
	case step:
	  if(rho2 < params->l2)
		B = 1.0/params->l2;
	  else
		B = 0.0;
	  break;
	case smooth:
	  B  =params->l2/((params->l2+rho2)*(params->l2+rho2));
	  break;
	case quadratic:
	  if(rho2 < params->l2)
		B=2.0/params->l2*(1.0-rho2/params->l2);
	  else
		B=0.0;
	  break;
	case gaussian:
	  if(rho2 < 100.0*params->l2)
	  	B = 1.0/(params->l2*exp(rho2/params->l2));
	  else
		B = 0.0;
	  break;
	case periodic:
	  B = periodicB(rho2, params->l2);
	  break;
	case spline:
	  sinrho = sin(pi*sqrt(rho2)/tubedist);
	  B = 1.0/(params->l2*exp(sinrho*sinrho/params->l2));
	  break;
	case fixflux:
	  B = fixfluxB(rho2,params->l2);
	  break;
  }
  B = 2.0*params->F/e*B;
  
  return B;
}

double smTscal(double B, double T)
//Effective action for T<<1.0 for scalar QED
{
	double B2=e*e*B*B;
	double B4=B2*B2;
	double B6=B2*B4;
	double T2=T*T;
	double T3=T2*T;
	double T4=T3*T;
	double result;
	result = 7.0/360.0*B4*T*(1.0-m2*T)
		+(147.0*B4*m2*m2-31.0*B6)/15120.0*T3
		+(31.0*B6*m2-49.0*B4*m2*m2*m2)/15120.0*T4;
	return result;
}

double smTferm(double B, double T)
//Effective action for T<<1.0 for fermionic QED
{
	double B2=e*e*B*B;
	double B4=B2*B2;
	double B6=B2*B4;
	double T2=T*T;
	double T3=T2*T;
	double T4=T3*T;
	double result;
	result = 1.0/45.0*B4*T*(m2*T-1.0) 
		+ (2.0*B6/945.0 - B4*m2*m2/90.0)*T3
		+(7.0*B4*m2*m2*m2-4.0*B6*m2)/1890.0*T4;
	return result;
}

double meanigrand(double* igrand, int ngroups, double* igranderr)
//compute the mean and std. err. for igrand
{
  double meanig = 0.0;
  int i;
  *igranderr = 0.0;
  for(i = 0; i < ngroups; i++)
  {
	meanig += igrand[i];
  }
  meanig /= ngroups;
  for(i = 0; i < ngroups; i++)
  {
	*igranderr += (igrand[i]-meanig)*(igrand[i]-meanig);
  }
  *igranderr = sqrt(*igranderr);
  *igranderr /= ngroups;
  return meanig;
}


double* Tfunc(double T, void* p, int* order)
//The integrand of the proper time, T, integral
{
  struct Wparams *params = (struct Wparams *) p;
  double rho2 = params->xcm.x*params->xcm.x 
	+ params->xcm.y*params->xcm.y;
  double smallT;
  double B;
  double TB;
  double* igrand = malloc(params->ng*sizeof(*igrand));
  int groupsize = 128; //loops per group
  int* WLlist = malloc(groupsize*sizeof(*WLlist));
  int i, j;
  double rho = params->xcm.x;
  double l2 = params->l2;
  double sin2rho = sin(2.0*pi*rho/tubedist);
  double denominator;
  double igmean; double igerr;
  if(!igrand || !WLlist) 
	printf("error Tfunc():  malloc failed\n");

  B = MagField(rho2, p);
  TB = e*T*B;
  //determine a suitable definition for small T
  switch(profile)
  {
	case step:
	  smallT = abs(rho*rho-l2);
          break;
        case smooth:
	  smallT = 0.5*(l2+rho*rho);
          break;
        case quadratic:
	  smallT = 0.5*(l2+rho*rho);
          break;
        case gaussian:
	  smallT = l2;
	  break;
        case periodic:
	  smallT = 0.5*(l2+rho*rho);
          break;
	case spline:
	  if(rho/tubedist > 1.0e-8)
	  	smallT = tubedist*tubedist*l2/(pi*pi*sin2rho*sin2rho);
	  else
		smallT = 1.0e12;
	  break;
	case fixflux:
	  //small if T << l2, or if the loop is 
	  //far from any bump functions.
	  if(rho*rho<l2/4.0)
		smallT = 0.25*l2;
	  else
		smallT = 0.25*l2+(rho-sqrt(l2)/2.0)
			*(rho-sqrt(l2)/2.0);
  }
  for(i = 0; i < params->ng; i++)
  {      
	if(T < 0.1 && T < 0.01*smallT)
  	{
		if(verbosity >= 2)
		  printf("using small T expression: %f %f %f\n", 
		  rho2, T, smallT);
		if(params->fermion == 1)
			igrand[i] = smTferm(B, T);
		else
			igrand[i] = smTscal(B, T);
  	}
  	else
  	{
		for(j = 0; j < groupsize; j++)
        	{
			//printf("func j=%d\n",j);
			WLlist[j] = order[j + i*groupsize];
        	}
		if(verbosity >= 3)
		  printf("EV= %e Exact = %e\n",
		    EV(T, p, WLlist), T/sinh(T));
		if(params->fermion == 1)
			igrand[i]= exp(-m2*T)/(T*T*T)
			  *(EV(T, p, WLlist)-1.0-1.0/3.0*TB*TB);
		else
			igrand[i]= exp(-m2*T)/(T*T*T)
			  *(EV(T, p, WLlist)-1.0+1.0/6.0*TB*TB);
  	}
 
  }
  if(T > 0.0f && verbosity >= 2){
	igmean = meanigrand(igrand, params->ng, &igerr);
	printf("WLvconstExpression: ");
  		printf("%e %e %e %e %e %e\n",
  		  rho2, T, Exact(T, B, params->fermion), 
		  smTscal(B, T), igmean, igerr);
  }
  free(WLlist);
  return igrand;
}

void signal_handler(int sig)
//Floating point signal handler
{
  printf( "Error: Floating point signal detected.\n");
  exit(1);
}

void EVMean(double *EV, float4 *Wsscal_h, float4 *Wsferm_h, 
	int n, int *WL, double T, int fermion)
//Compute the expectation value and error of the Wilson loops
//WL is a list representing a group of wordline indices 
//to compute the expectation value of
{
  int i, WLi;
  double EVWlNb2, EVWlN, EVWlNb4;
  double SEWlNb2, SEWlN, SEWlNb4;
  double *EVpart1;
  double *EVpart2;
  double *EVpart4;
  double Nby4part;

  //Turn on floating point exceptions:                                                                                                                      
  feenableexcept(FE_DIVBYZERO | FE_INVALID 
	| FE_OVERFLOW | FE_UNDERFLOW);

  // Set the signal handler:
  signal(SIGFPE, signal_handler);

  *EV = 0.0;
  EVWlNb2 = 0.0; EVWlN = 0.0; EVWlNb4 = 0.0;
  //allocate arrays to store the wilson loop of each worldline
  EVpart1 = (double *) malloc(n*sizeof(*EVpart1));
  EVpart2 = (double *) malloc(n*sizeof(*EVpart2));
  EVpart4 = (double *) malloc(n*sizeof(*EVpart4));
  if(!EVpart1 || !EVpart2 || !EVpart4) 
	printf("EVonly(): malloc failed\n");
  //determine the sum of the contributions
  for(i = 0; i < n; i++){
	WLi=WL[i];
	//Compute the scalar part
	Nby4part = Wsscal_h[WLi].y + Wsscal_h[WLi].z;
	EVpart1[i] = cos(0.5*Wsscal_h[WLi].x+0.25*(Nby4part));
	EVpart2[i] = cos(Wsscal_h[WLi].x);
	EVpart4[i] = cos(0.5*(Nby4part));
	if(fermion == 1)
	{
		if(isinf(Wsferm_h[WLi].x) !=0 
		  || isinf(Wsferm_h[WLi].y) !=0 
		  || isinf(Wsferm_h[WLi].z) !=0 
		  || isnan(Wsferm_h[WLi].x) !=0 
		  || isnan(Wsferm_h[WLi].y) !=0 
		  || isnan(Wsferm_h[WLi].z) !=0 )
		{
			printf("Warning: WSferm is infinite. 
			  Worldline=%d\n", WLi);
		}
		//Compute the fermion part
		Nby4part = Wsferm_h[WLi].y+Wsferm_h[WLi].z;
		if(abs(Nby4part) > 600.0 | abs(Wsferm_h[WLi].x) > 600.0)
		{
			printf("Error: Large cosh argument.  
				T=%f, WLi=%d\n",T,WLi);
			EVpart1[i] *= 1.0e10;
			EVpart2[i] *= 1.0e10;
			EVpart4[i] *= 1.0e10;
		}
                else
		{
			EVpart1[i] *= cosh(0.5*(double)Wsferm_h[WLi].x
				+0.25*((double)Nby4part));
			EVpart2[i] *= cosh((double)Wsferm_h[WLi].x);
			EVpart4[i] *= cosh(0.5*((double)Nby4part));
		}
		if(isnan(EVpart1[i]) != 0 
		  || isnan(EVpart2[i]) != 0 
		  || isnan(EVpart4[i]) != 0 
		  || isinf(EVpart1[i]) != 0 
		  || isinf(EVpart2[i]) != 0 
		  || isinf(EVpart4[i]) != 0) 
		    printf("Warning: Infinity detected: WL# %d\n", WLi);
	}
	EVWlNb4 += EVpart4[i];
	EVWlNb2 += EVpart2[i];
	EVWlN   += EVpart1[i];
  }
  EVWlNb4 /=n;
  EVWlNb2 /=n;
  EVWlN  /=n;
  //Extrapolate to N=infinity
  *EV=8.0/3.0*EVWlN - 2.0*EVWlNb2 + 1.0/3.0*EVWlNb4;
  
  free(EVpart1);
  free(EVpart2);
  free(EVpart4);
  
}


\end{lstlisting}

\section{\texorpdfstring{\acs{CUDA}}{CUDA} Kernel and Kernel Call - intEV.cu}
\label{sec:srcintEV}
Contains the function call to the \ac{CUDA} kernel, as well as 
all functions which compute Wilson loops on the \ac{CUDA} device.
\begin{lstlisting}[language=C, title=intEV.cu]
#include <builtin_types.h>
#include <cuda.h>
#include <stdio.h>
#include "intT.h"
#include <math.h>

extern "C" void EVMean(double *EV, float4 *Wsscal_h, float4 *Wsferm_h, int n, int *WL, double T, int fermion);
#if profile == fixflux
 #define THREADS_PER_BLOCK 128
 #define MY_KERNEL_MAX_THREADS THREADS_PER_BLOCK
 #define MY_KERNEL_MIN_BLOCKS 4
#else
 #define THREADS_PER_BLOCK 256
 #if __CUDA_ARCH__ >= 200
  #define MY_KERNEL_MAX_THREADS (2 * THREADS_PER_BLOCK)
  #define MY_KERNEL_MIN_BLOCKS 3
 #else
  #define MY_KERNEL_MAX_THREADS (2 * THREADS_PER_BLOCK)
  #define MY_KERNEL_MIN_BLOCKS 2
 #endif
#endif

extern "C"
__device__ float expint(const float x)
//Evaluates the exponential integral Ei(x)=-E1(-x)
// assuming x<-1. This algorithm is an abbreviated 
//version of Numerical Recipes expint().
//See Chapter 6 on Special Functions. We assume 
//(x > 1.0) and include only the relevant code. 
//This bit of code is Lentz's algorithm 
//(section 5.2 of NR).
{
  const int MAXIT = 400;
  const float EPS = 1.0e-6;
  const float BIG = 1.0e10;
  int i;
  float a, b, c, d, del, h, ans;
  b = -x + 1.0f;
  c = BIG;
  d = 1.0f/b;
  h = d;
  for (i = 1; i <= MAXIT; i++)
  {
	a = -(float)i*(float)i;
	b += 2.0f;
	d = 1.0f/(a*d+b); //Denominators cannot be zero
	c = b + a/c;
	del = c*d;
	h *= del;
	if (fabsf(del-1.0f) <= EPS)
	{
		ans = -h*__expf(x);
		return ans;
	}
  }
  return 0.0f;

}


extern "C"
__host__ __device__ float interp(float rho2, 
	float *rho2sp, float4 *coefs)
//interpolation function for periodic spline profile
{
  int j;
  int upperi = Nspline-1, loweri=0;
  float rho2diff = 0.0f;
  float flambda;

    //Discover which interval to look in using a binary search
  if(rho2 < rho2sp[Nspline-1] && rho2 > rho2sp[0])
  {
   	while(upperi-loweri > 1)
	{
		if(rho2 >= rho2sp[(upperi+loweri)/2]) 
			loweri=(upperi+loweri)/2;
		else upperi = (upperi+loweri)/2;
  	}
  	//interpolate using the jth interval
  	j = loweri;
	rho2diff = rho2-rho2sp[j];
  	flambda = coefs[j].x+rho2diff
		*(coefs[j].y+rho2diff*(coefs[j].z
		+rho2diff*coefs[j].w));
  }
  else
  {
	flambda = 0.0f;
  }
  return flambda;
}



extern "C"
__host__ double EV (double T, void * p, int* WLlist) {
//Function for calling the Kernel, then computing the 
//Expectation value from the results of each worldline
  struct Wparams params = *(struct Wparams *) p;
  double EV;
  const int groupsize = 128;
  double rtT = sqrt((double)T);
  cudaError_t errorcode;
  // call to integrate the function func
  if( verbosity >= 5)
  	printf("call to CUDA device\n");
  ExpectValue<<<params.nBlocks, params.nThreads>>>
	(params.Wsscal_d, params.Wsferm_d,
	params.worldlines,params.xcm, (float)params.F, 
	(float)params.l2, (float)rtT, params.Nl, 
	params.Nppl, params.flcoefs, params.rho2sp, 
	params.fermion);
  errorcode = cudaGetLastError();
  if ( errorcode>0) printf("cuda getLastError EV(): %s\n", 
	cudaGetErrorString(errorcode));
  if (verbosity >= 6)
  	printf("return from CUDA\n");
  // Make certain that all threads are idle before proceeding
  errorcode = cudaThreadSynchronize();
  if ( errorcode>0) printf("cuda Synchronize EV(): %s\n", 
	cudaGetErrorString(errorcode));
  //Copy device memory back to host
  errorcode = cudaMemcpy(params.Wsscal_h, params.Wsscal_d,
	params.Nl*sizeof(params.Wsscal_h[0]), 
	cudaMemcpyDeviceToHost);
  if(errorcode > 0) printf("cuda memcpy scal Error EV(): %s\n",
	cudaGetErrorString(errorcode));
  if(params.fermion == 1)
  {
  	errorcode = cudaMemcpy(params.Wsferm_h, 
		params.Wsferm_d,
		params.Nl*sizeof(params.Wsferm_h[0]),
		cudaMemcpyDeviceToHost);
	if(errorcode > 0) 
	  printf("cuda memcpy ferm Error EV(): %s\n",
		cudaGetErrorString(errorcode));
  }
  //Compute the expectation value from the Wilson Loop data
  EV = 0.0;
  EVMean(&EV, params.Wsscal_h, params.Wsferm_h, 
	groupsize,WLlist, T, params.fermion);
  return EV;
}

extern "C"
__device__ float bump(const float x)
//Device version of the bump function
{
  //the 0.999 makes no numerical difference compared to 1.0
  //but seems to prevent some unpredictable, "unspecified CUDA errors"
  if(x*x < 0.999f) return __expf( -1.0f/(1.0f-x*x) );
  else return 0.0f;
}

extern "C"
__device__ float phi(const float x)
//Computes the \Phi function which is defined in the thesis
{
  const float onemx2 = 1.0f-x*x;
  //the 0.999 makes no numerical difference compared to 1.0
  //but seems to prevent some unpredictable, 
  //unspecified CUDA errors
  if( x < 0.999f )
  {
	return 1.0f - ( 0.5f/q2 )
	  *( onemx2*__expf(-1.0f/onemx2)
	  + expint(-1.0f/onemx2) );
  }
  else
	return 1.0f;
}

extern "C"
__device__ float chi(const float x, const float n, const float lambda)
//Computes the \Chi function which is defined in the thesis
{
  float ans;
  const float onemx2 = 1.0f-x*x;
  const float x2 = x*x;
  const float x4 = x2*x2;
  if(x <= -0.999f) ans = 0.0f;
  else if(x2 < 0.999f)
  {
	ans = 0.5f*(-onemx2*__expf(-1.0f/onemx2) 
		- expint(-1.0f/onemx2));
	ans += 2.0f*n*tubedist/lambda*
		0.444f/( 1.0f + __expf(-(3.31f*x +
		  5.25f*x*x2*sin(x)
		  *cos( -0.907f*x2 - 1.29f*x4*x4 ))
		/cos(x)) );
  }
  else //for x >= 1.0f
  	ans = 2.0f*n*tubedist*q1/lambda;
  return ans;
}


extern "C"
__device__ float flperi(const float rho2, const float lambda2)
//f_lambda for periodic flux tube profile
{
  float expr, exprp1, exprap1, rho, lambda, flraf, f;
  int N = 2;
  int i;
  rho = sqrtf(rho2);
  lambda = sqrtf(lambda2);

  flraf = floorf(rho/tubedist);
  f = 0.0f;
  if((int)flraf-N > 0)
  {
	f=0.5f*(flraf-(float)N-1.0f)*(flraf-(float)N);
  }
  for(i = (int)flraf-N; i <= (int)flraf+N; i++)
  {
	if(i == 0)
	{
		expr = __expf(-rho/lambda);
		exprp1 = 1.0f+expr;
		f = -rho/tubedist*(expr/exprp1) 
			- lambda/tubedist*log(exprp1);
	}
        else if(i > 0)
	{
		exprap1 = 1.0f
		  +__expf(-(rho-(float)i*tubedist)/lambda);
		f = f+(float)i/exprap1;
	}
  }
  return tubedist/(lambda*log(2.0f))*f + 1.0f;
}



extern "C"
__device__ float ffixflux(const float rho2, const float lambda2)
//f_lambda for the fixflux profile
{
  const float lam = sqrtf(lambda2);
  const float lmlmin = (lam-lmin)/(tubedist-lmin);
  const float aml = (tubedist-lam)/(tubedist-lmin);
  const float n = floorf((sqrt(rho2)+tubedist/2.0f)/tubedist);
  float ans;
  if(rho2 <= tubedist*tubedist/4.0f)
  {
	ans = (1.0f-0.75f*lmlmin)*phi(2.0f*sqrt(rho2/lambda2)) 
		+ 3.0f*rho2/(tubedist*tubedist)*lmlmin;

  }
  else
  {
	ans = 1.0f+0.75f*(4.0f*rho2/(tubedist*tubedist)
	  -1.0f)*lmlmin + 3.0f*n*(n-1.0f)*aml; 
	ans += 3.0f*lam/(q1*tubedist)*aml
	  *chi(2.0f*(sqrt(rho2)-n*tubedist)/lam, n, lam);
  }
  return ans;
}


extern "C"
__device__ float flambda(float rho2, float lambda2, float4 *flcoefs, float *rho2sp)
//flambda(rho^2,lambda^2) defines the magnetic vector potential in cylindrical coordinates
{
  float f;

  switch(profile){
	case step: 
	  if(rho2 < lambda2)
		f = rho2/lambda2;
	  else
		f = 1.0f;
	  break;
	case smooth:
  	  f = rho2/(lambda2+rho2);
	  break;
	case quadratic:
	  if(rho2 < lambda2)
	  	f = rho2/lambda2*(2.0f-rho2/lambda2);
	  else
		f = 1.0f;
	  break;
	case gaussian:
	  f = 1.0f-exp(-rho2/lambda2);
	  break;
	case periodic:
	  f = flperi(rho2, lambda2);
	  break;
	case spline:
	  f = interp(rho2, rho2sp, flcoefs);
	  break;
	case fixflux:
	  f = ffixflux(rho2, lambda2);
	  break;
  }
  return f;
} 


extern "C"
__device__ float fplperi(float rho2, float lambda2)
//f'_lambda for periodic profile
{
  float expr,exprp1,rho,lambda,flraf,f;
  float expral, expralp1;
  int N = 2;
  int i;
  rho = sqrtf(rho2);
  lambda = sqrtf(lambda2); 
  flraf = floorf(rho/tubedist);
  f=0.0f;
  for(i = (int)flraf-N; i <= (int)flraf+N; i++)
  {
	if(i == 0)
	{
		expr = __expf(-rho/lambda);
		exprp1 = 1.0f+expr;
		f = expr/(tubedist*exprp1*exprp1);
	}
	else if(i > 0)
	{
		expral = __expf(-(rho-(float)i*tubedist)/lambda);
		expralp1 = 1.0f+expral;
		f = f +((float)i/rho)*expral/(expralp1*expralp1);
	}
  }
  return tubedist/(2.0f*lambda2*log(2.0f))*f;
}




extern "C"
__device__ float fpfixflux(const float rho2, const float lambda2)
//f'_lambda(rho^2) for the fixflux field profile
{
  const float lam = sqrtf(lambda2);
  const float lmlmin = (lam-lmin)/(tubedist-lmin);
  const float rho = sqrtf(rho2);
  const float n = floorf((rho+tubedist/2.0f)/tubedist);
  const float a2 = tubedist*tubedist;
  const float aml = (tubedist - lam)/(tubedist - lmin);
  float ans;
  if(rho <= tubedist/2.0f)
  {
	ans = 2.0f/(lambda2*q2)*(1.0f-0.75f*lmlmin)
	  *bump(2.0f*rho/lam);
	ans += 3.0f/a2*lmlmin;
  }
  else
  {
	ans = 3.0f/a2*lmlmin 
	  + 6.0f/(q1*lam*tubedist)*aml
	  *bump(2.0f*(rho-n*tubedist)/lam);
  }
  return ans;
}


extern "C"
__device__ float fplambda(const float rho2, 
	const float lambda2, float4 *flcoefs, float *rho2sp)
//fplambda(rho^2,lambda^2) defines the 
//magnetic vector potential derivative 
// wrt rho^2 in cylindrical coordinates
{
  float f, fjunk;
  f=1.0f;
  switch(profile){
	case step: 
	  if(rho2 < lambda2)
		f = 1.0f/lambda2;
	  else
		f = 0.0f;
	  break;
	case smooth:
  	  f = lambda2/((lambda2+rho2)*(lambda2+rho2));
	  break;
	case quadratic:
	  if(rho2<lambda2)
	  	f = 2.0f/lambda2*(1.0f-rho2/lambda2);
	  else
		f = 0.0f;
	  break;
	case gaussian:
	  f = 1.0f/lambda2*exp(-1.0f*rho2/lambda2);
	  break;
	case periodic:
	  f = fplperi(rho2, lambda2);
	  break;
	case spline:
	  fjunk = __sinf(sqrtf(rho2)*pi);
	  if(fjunk>100.0f)
		f = 0.0f;
	  else
	  {
		f = __expf(-1.0f*fjunk*fjunk/lambda2)/lambda2;
	  }
	  break;
	case fixflux:
	  f = fpfixflux(rho2, lambda2);
	  break;
  }
  return f;
} 

extern "C"
__device__ void Idt(float *scalI, float *fermI, float4 Ai, 
	const float l2, float4 *flcoefs, float *rho2sp, int fermion)
//Computes the integral over t from 0 to 1 in the scalar and fermion Wilson loop factors
{
	int i;
	//number of points in point-to-point proper time integral
	const int n = 50; 
	float t,  rhoi2; //proper time and rho squared
	//distance between points in integral
	const float h = 1.0f/((float) n);  
	float4 xiscal, xiferm;     //scalar and fermi integrands
	if (Ai.x<1.0e-8) Ai.x = 1.0e-8;
	if (Ai.y<1.0e-8) Ai.y = 1.0e-8;
	if (Ai.z<1.0e-8) Ai.z = 1.0e-8;
	float Aip1 = Ai.x+2.0f*Ai.y+Ai.z;  //rho^2 for the final point
	if(Aip1<1.0e-8) Aip1 = 1.0e-8;
	//Begin the Simpson's method algorithm
	xiscal.x = flambda(Ai.x,l2,flcoefs,rho2sp)/Ai.x 
		 + flambda(Aip1,l2,flcoefs,rho2sp)/Aip1;
	xiscal.y = 0.0f;
	xiscal.z = 0.0f;
	if (fermion == 1)
	{
	  xiferm.x = fplambda(Ai.x, l2, flcoefs, rho2sp)
		+ fplambda(Aip1, l2, flcoefs, rho2sp);
	  xiferm.y = 0.0f;
	  xiferm.z = 0.0f;
	}
	for(i = 1; i < n; i++)
	{
		t = (float)i*h;
		//rho2 at the point
		rhoi2 = Ai.x + 2.0f*Ai.y*t + Ai.z*t*t;
		//prevent singularities
		if(rhoi2 < 1.0e-10) rhoi2 = 1.0e-10;

		if(i%2==0) 
		{
			xiscal.z += flambda(rhoi2, l2, 
				flcoefs, rho2sp)/rhoi2;
			if(fermion == 1)
			  xiferm.z += fplambda(rhoi2, l2, 
				flcoefs, rho2sp);
		}
		else 
		{
			xiscal.y += flambda(rhoi2, l2, 
				flcoefs, rho2sp)/rhoi2;
			if(fermion == 1)
			  xiferm.y += fplambda(rhoi2, l2, 
				flcoefs, rho2sp);
		}
	
	}
	*scalI = (xiscal.x + 2.0f*xiscal.z + 4.0f*xiscal.y)*h/3.0f;
	if(fermion == 1)
	  *fermI = (xiferm.x + 2.0f*xiferm.z + 4.0f*xiferm.y)*h/3.0f;
}

extern "C"
__device__ void getzp1(float4 *zip1, float4 *worldlines, 
	float rtT, float4 xcm, int i, int inx, int Nppl)
//Function for determining the next point on the 
//worldline loop for each of the sub loops
{
  int inxp1;
  //get the next worldline index for the N/2 group
  if(i%2 == 1){
	if(i == Nppl-1)
	{
		inxp1 = inx*Nppl+1;
	}
	else
	{
		inxp1 = inx*Nppl+i+2;
	}
  }
  //get the next worldline index for the first N/4 group
  else if(i%4 == 0){
	if(i == Nppl-4)
	{
		inxp1 = inx*Nppl;
	}
	else
	{
		inxp1 = inx*Nppl+i+4;
	}
  }
  //get the next worldline index for the second N/4 group
  else if((i-2)%2 == 0){
	if(i == Nppl-2)
	{
		inxp1 = inx*Nppl+2;
	}
	else
	{
		inxp1 = inx*Nppl+i+4;
	}
  }
  //compute the next point
  zip1->x = xcm.x + rtT*worldlines[inxp1].x;
  zip1->y = xcm.y + rtT*worldlines[inxp1].y;
  zip1->z = xcm.z + rtT*worldlines[inxp1].z;

}

extern "C"
__device__ void WilsonLoop(float4 *worldlines, float4 *Wsscal, 
	float4 *Wsferm, float4 xcm, int inx, float F, 
	float l2, float rtT, int Nppl, float4 *flcoefs, 
	float *rho2sp, int fermion)
//Returns the Wilson loop value
{
	int i;
	float4 WLstemp, WLftemp;
	float4 zi, zip1;
	float4 Ai;
	float xyyx;
	float scalI, fermI;
	//Compute the scalar contribution
	WLstemp.x = 0.0f; WLstemp.y = 0.0f; WLstemp.z = 0.0f;
	WLftemp.x = 0.0f; WLftemp.y = 0.0f; WLftemp.z = 0.0f;
	for(i = 0; i < Nppl; i++){
		//Compute the scaled, shifted coordinate
		zi.x = xcm.x + rtT*worldlines[inx*Nppl+i].x;
		zi.y = xcm.y + rtT*worldlines[inx*Nppl+i].y;
		getzp1(&zip1, worldlines, rtT, xcm, i, inx, Nppl);
		//Ai Bi and Ci coefficients for the rho2 polynomial
		Ai.x = zi.x*zi.x + zi.y*zi.y;
		Ai.y = zi.x*(zip1.x-zi.x)+zi.y*(zip1.y-zi.y);
		Ai.z = (zip1.x-zi.x)*(zip1.x-zi.x) 
			+ (zip1.y-zi.y)*(zip1.y-zi.y);
		Idt(&scalI, &fermI, Ai, l2, flcoefs, rho2sp, fermion);
		//Compute the contribution to the N/2 integral
		xyyx = (zi.x*zip1.y-zi.y*zip1.x);
		if(i%2 == 1){
			WLstemp.x += xyyx*scalI;
			WLftemp.x += fermI;
		}
		//Compute the contribution to the first N/4 integral
		else if(i%4 == 0){
			WLstemp.z += xyyx*scalI;
			WLftemp.z += fermI;
		}
		//Compute the contribution to the second N/4 integral
		else if((i-2)%2 == 0){
			WLstemp.y += xyyx*scalI;
			WLftemp.y += fermI;
		}
	}
	Wsscal[inx].x = F*WLstemp.x;
	Wsscal[inx].y = F*WLstemp.y;
	Wsscal[inx].z = F*WLstemp.z;
	if( fermion == 1)
	{
	  Wsferm[inx].x = 2.0f*F*WLftemp.x*rtT*rtT/(Nppl/2.0f);
	  Wsferm[inx].y = 2.0f*F*WLftemp.y*rtT*rtT/(Nppl/4.0f);
	  Wsferm[inx].z = 2.0f*F*WLftemp.z*rtT*rtT/(Nppl/4.0f);
	}
}

__global__ void 
__launch_bounds__(MY_KERNEL_MAX_THREADS, MY_KERNEL_MIN_BLOCKS)
ExpectValue(float4 *Wsscal, float4 *Wsferm, float4 *worldlines, 
	float4 xcm, float F, float l2, float rtT, int Nl, int Nppl, float4 *flcoefs, float *rho2sp, int fermion)
//Each thread computes the Wilson loop value for a single 
//worldline
{
	//index to identify which CUDA thread we are in
        int inx = blockIdx.x * blockDim.x + threadIdx.x;       
        WilsonLoop(worldlines, Wsscal, Wsferm, xcm, 
		inx, F, l2, rtT, Nppl, flcoefs, 
		rho2sp, fermion);     
}




\end{lstlisting}

\section{Reading Worldlines from ASCII - getwl.c}
\label{sec:srcgetwl}
Routines for reading the worldlines from an ASCII file.
Matlab routines for producing ASCII files with the 
required formatting are listed in appendix \ref{sec:srcWLgen}.
\begin{lstlisting}[language=C, title=getwl.c]
#include <builtin_types.h>
#include <stdio.h>
int getwl(float4 *worldlines, char *filename, 
	int Nl, int Nppl, int Dim)
//Reads worldlines from filename and stores the results in worldlines.
//Worldline files may be prepared with wlGen.m in matlab
{
	//loop variables
	int i, j, k;
	FILE *fp;
	printf("getting worldlines from %s\n",filename);
	if((fp = fopen(filename,"rb"))==NULL){
		printf("cannot open %s\n", *filename);
	}
	else{
		//loop over lines
		for(i = 0; i < Nl; i++){
			//printf("getting worldline %d \n", i);
			//loop over points in line, x coordinate
			for(j = 0; j < Nppl; j++){
				fscanf(fp,"%f ", 
				  &(worldlines[i*Nppl + j].x));
			}
			if(Dim>1){
				//loop over points in line, y coordinate
				for(j=0;j<Nppl;j++){
					fscanf(fp,"%f ",
					  &(worldlines[i*Nppl + j].y));
				}
				if(Dim>2){
					//loop over points in line, z coordinate
					for(j=0;j<Nppl;j++){
						fscanf(fp,"%f ",
						  &(worldlines[i*Nppl + j].z));
					}
				}
			}	
		}
	}
	if(fp) fclose(fp);  //Close file
	return 0;
}
\end{lstlisting}


\lstset{language=make, morekeywords={all, clean}}
\section{Compiling Instructions - Makefile}
\label{sec:srcmake}
Makefile for compiling the software with \lstinline{nvcc}.
Provided the other source files are in the directory along 
with the Makefile, the program can be compiled with the \acs{GNU} 
make command. The correct path of \lstinline{nvcc} must be 
provided if it is different than the one in the makefile.
\begin{lstlisting}[language=make, title=Makefile]
all: EffAct

Tint.o:  Tint.c intT.h
	/usr/local/cuda/bin/nvcc -c -g Tint.c -L/home/mazur/software/lib -lgsl -lgslcblas -lm

Integrand.o: Integrand.c intT.h Makefile
	/usr/local/cuda/bin/nvcc -c -g Integrand.c -L/home/mazur/software/lib -lgsl -lgslcblas -lm

getwl.o: getwl.c Makefile
	/usr/local/cuda/bin/nvcc -c getwl.c

intEV.o: intEV.cu intT.h Makefile
	/usr/local/cuda/bin/nvcc -c -g -arch sm_13 -prec-div=false --ptxas-options="-v" -prec-sqrt=false intEV.cu -lm

EffAct.o: EffAct.c intT.h Makefile
	/usr/local/cuda/bin/nvcc -c -g EffAct.c


EffAct: EffAct.o intEV.o getwl.o Tint.o Integrand.o Makefile
	/usr/local/cuda/bin/nvcc -arch sm_13 -prec-div=false -prec-sqrt=false --ptxas-options="-v" -g -o EffAct EffAct.o intEV.o getwl.o Tint.o Integrand.o -L/home/mazur/software/lib -lcuda -lcudart -lgsl -lgslcblas -lm

clean:
	rm *.o EffAct
\end{lstlisting}

\section{Worldline Generation}
\label{sec:srcWLgen}
Matlab function for generating a worldline loop using the d-loop algorithm
described in section \ref{sec:loopgen}. 
\lstset{language=Matlab, morekeywords={random, normrnd}}
\begin{lstlisting}
function points = genLoop(N, Dim);
%Construct a d-loop worldline path
    points = zeros(N, Dim);
    SS = 1; %Arbitrary range for first point
    %first point is arbitrary
    points(end, :) = random('uniform', 0, SS, 1, Dim);
    %Construct an N point closed loop.
    %We assume that log2(N) is an integer.
    for k = 1:log2(N)
    %Add 2^(k-1) points to the line
        Nk = 2^k;
        for q = 1:2:Nk
        %for each point, sample from the distribution
        %set y_(q+1)
            if(q == Nk) %periodic boundary 
               yqp1 = points(end, :);
            else
               yqp1 = points((q+1)*N/Nk, :);
            end
			%set y_(q-1)
            if(q == 1)  %periodic boundary 
                yqm1 = points(end, :);
            else
                yqm1 = points((q-1)*N/Nk, :);
            end
            %compute the mean and standard deviation
            %of the normal distribution
            sigma = 1.0/sqrt(Nk)*ones(1, Dim);
            mu = 0.5*(yqp1+yqm1);
            %Sample a point randomly from 
            %a normal distribution
            points(q*N/Nk, :) = normrnd(mu, sigma);
        end
    end
    %shift the loop mean to the origin
    points = points-ones(N, 1)*mean(points);
end
\end{lstlisting}

The following function calls the previous function and outputs 
an ascii file containing the data representing an 
ensemble of worldlines. This data is in 
the format required by the function, getwl(), 
described in appendix \ref{sec:srcgetwl},
which reads the worldlines into the main program.
Since only entire blocks of threads can be called 
by \ac{CUDA}, it is best to generate worldlines in 
multiples of the number of threads per block.

\begin{lstlisting}
  function genCloud(filename,Nl,ppl,dim)
  %Produces an ascii file, filename, containing
  %NL dim-dimensional worldlines with ppl points per line.
    worldlines=zeros(ppl,dim,Nl);
    for i=1:Nl %loop over worldlines
		%generate a loop and append the result
		%to worldlines
        if i==1
            worldlines=genLoop(ppl,dim)';
        else
            worldlines=[worldlines; genLoop(ppl,dim)'];
        end
    end
	%save worldlines to file: filename
    save(filename, 'worldlines', '-ascii');
end
\end{lstlisting}


\backmatter

\end{document}